\documentclass[prb,aps,amssymb,twocolumn,nobibnotes]{revtex4-2}
\usepackage{amsmath,graphicx,hyperref,color,amssymb,float}
\usepackage[justification=raggedright,singlelinecheck=false]{caption}
\usepackage{ulem} 

\usepackage{subcaption}
\usepackage{apptools}
\AtAppendix{}
\usepackage{float}

\newcommand{\be}{\begin{equation}}
\newcommand{\ee}{\end{equation}}
\newcommand{\bg}{\begin{figure}}
\newcommand{\eg}{\end{figure}}

\definecolor{pink}{RGB}{255, 10, 203}

\definecolor{brown}{RGB}{255, 165, 0} 

\definecolor{norman}{RGB}{160, 30, 110}

\definecolor{HYS}{RGB}{0,130,209}

\DeclareMathOperator{\sech}{sech}
\newcommand{\bs}[1]{\boldsymbol{#1}}

\begin{document}

\title{Conformations, Dynamics, and Looping Kinetics of Partially Active Polymers}

\author{Koushik Goswami*}
\email{goswamikoushik10@gmail.com, 
https://orcid.org/0000-0002-0632-9654}
\affiliation{Physics Division, National Center for Theoretical Sciences, 
Taipei 106319, Taiwan
}
\author{Norman Hsia}
\affiliation{Department of Physics, National Taiwan University, Taipei 106319, Taiwan}
\author{Cheng-Hung Chang*}
\email{chchang@nycu.edu.tw, https://orcid.org/0000-0002-0333-8897}
\affiliation{Institute of Physics, National Yang Ming Chiao Tung University, Taiwan}
\affiliation{Physics Division, National Center for Theoretical Sciences, 
Taipei 106319, Taiwan}
\author{Hong-Yan Shih*}
\email{hongyan@gate.sinica.edu.tw,https://orcid.org/0000-0002-3578-7785}
\affiliation{Institute of Physics, Academia Sinica, Taiwan}
\affiliation{Physics Division, National Center for Theoretical Sciences, 
Taipei 106319, Taiwan}

\begin{abstract}
We investigate the conformational and dynamical properties of a partially active Rouse chain, where activity is localized within a specific segment, positioned at various locations along the chain and spanning any given length. Through analytical methods and simulations, we reveal how the location and size of the active segment influence polymer swelling patterns. Likewise, we observe that the mean squared distance between two points along the polymer, as well as the mean squared displacement of a tagged point, are notably affected by the local activity. In addition, the reconfiguration and looping dynamics show anomalous scaling behaviors, particularly in intermediate chain lengths, capturing the interplay between persistence of active motion and polymer relaxation dynamics. Our model, relevant to spatially varying activity observed in active biopolymeric systems,  provides a basis for exploring more realistic models of chromatin behavior, especially those incorporating heterogeneous activity.
\end{abstract}


\maketitle


\section{Introduction} \label{sec-intro}

Active matter, characterized by systems operating far from equilibrium due to nonthermal fluctuations \cite{fodo16,dabelow2019irreversibility,goswami2019heat,o2022time,gosw23}, spans macroscopic to mesoscopic scales and exhibits a plethora of phenomena, including motility-induced phase separation and flocking \cite{rama10mech,bech16,vics95,liebchen2017collective}.  Active polymers, a key subset of active matter,  play a crucial role in understanding biological and synthetic systems at mesoscopic scales. These polymers encompass diverse structures such as chromatin \cite{weber2012nonthermal,saintillan2018extensile}, cytoskeleton networks\cite{koenderink2009active,fletcher2010cell}, flagella, and motile cilia \cite{chelakkot2014flagellar,meng2021conditions,chakrabarti2022multiscale}. For instance, microorganisms such as bacteria self-propel through the rotary motion of flagella, driven by motor proteins powered by ion pumps or hydrolysis of Adenosine triphosphate  \cite{chelakkot2014flagellar}. Egg cells navigate through fluids using the rhythmic beating of motile cilia, resulting from the bending of filaments mediated by dynein motor proteins \cite{chakrabarti2022multiscale}. Within living cells, chromatin regulates nuclear functions, while the cytoskeleton dictates cellular shape and structural organization \cite{fletcher2010cell,weber2012nonthermal,saintillan2018extensile}.

From a theoretical standpoint, an active polymer can be conceptualized in several ways: as a chain of active particles with elastic interactions between neighboring beads (an active Rouse chain), as semi-flexible filaments with finite rigidity, or as a passive chain embedded in an active medium \cite{doi1988theory,elge15rev,winkler2020physics,goswami2022reconfiguration,yadav2024passive}. In biopolymers, activity can mimic thermal motion with enhanced dynamics when modeled as a chain of beads at an elevated effective temperature \cite{loi2011non}. Alternatively, self-propulsion can be directed along the polymer backbone, as in models of cilia beating, or active forces can be randomly oriented at each site \cite{anand2018structure,tejedor2024progressive}. The above descriptions outline a fully active polymer, where each monomer is subjected to active fluctuations.
Over the past decades, extensive studies on fully active polymers have revealed intriguing properties, including chain swelling, enhanced monomer dynamics, slowed internal reconfiguration, and coil-globule transitions \cite{eisenstecken2017internal,bianco2018globulelike}.

Recent advances in high-precision imaging techniques now enable the precise tracking of specific polymer segments, allowing for a detailed investigation of fine structural features \cite{bornfleth1999quantitative,lippincott2003development,shaban2018formation,mccord2020chromosome}. Of particular interest is the modeling of chromatin, which is composed of compact DNA wrapped around histone proteins within the cell nucleus \cite{misteli2020self}. Recent experiments suggest that chromosomal loci at different positions along the chromosome exhibit varying mobilities, indicating nonuniform activity along the chromatin backbone \cite{javer2013short,ganai2014chromosome}. This variation is likely influenced by factors such as local energy consumption, enzymatic activity, and motor protein binding rates \cite{saito2018inferring,put2019active,belitsky2019stationary,jiang2022phase,sahoo2024nonequilibrium}. Chromatin can therefore be viewed as a heteropolymer, particularly with two distinct regions: transcriptionally active regions (euchromatin), which appear more extended, and inactive regions (heterochromatin), which are more compact \cite{solovei2016rule,goychuk2023polymer}. This distinction aligns with data from genomic analysis techniques, where contact frequencies reveal frequent interactions within the same regions and segregation between different regions \cite{jiang2022phase,brahmachari2024temporally}.

Inspired by the heterogeneous nature of biopolymers, we extend the Rouse model to describe a heteropolymer with one or more active segments.  Here, ``active" refers not to thermal-like fluctuations with enhanced diffusivity, but to additional temporally and spatially correlated active forces acting on specific segments. Our model is adjustable, allowing active segments of any length to be placed anywhere along the polymer backbone. Although we focus on a single active segment for simplicity, the model can easily accommodate multiple active regions. This framework describes a ``partially active polymer", in which only specific portions of the polymer are active.

Recent studies have investigated active polymers with inhomogeneous activities both theoretically and through simulations \cite{goychuk2024delayed,osmanovic2017dynamics,osmanovic2018properties,vatin2024conformation,dutta2024effect}. Simulation-based research has explored how randomly positioned active blocks impact the shape and size of the polymer, with active forces aligned parallel to the backbone \cite{vatin2024conformation}. Other studies have examined active forces at specific point segments as local excitations to capture multiscale dynamics \cite{osmanovic2017dynamics,dutta2024effect} or have modeled small, uniformly sized actively driven regions distributed along the chain \cite{osmanovic2018properties}. Integrating these scenarios, our work aims to bridge the gap by providing a comprehensive theoretical understanding of how the size and location of a single active block on the chain contour influence the polymer’s local and global behavior. Specifically, we examine conformation and dynamics, where each segments within the active block are subjected to additional athermal fluctuations that are exponentially (or delta-) correlated in time.

The organization of biopolymers within confined cellular environments, such as chromatin in the nucleus, often relies on looping mechanisms  \cite{marenduzzo2010biopolymer,vilar2005dna,goychuk2023polymer}. In chromatin, active loop extrusion by motor proteins creates topologically associating domains, which are essential for processes like gene expression and DNA repair  \cite{banigan2020loop,chan2023theory,cao2024motorized,dixon2012topological,banigan2020loop,goychuk2023polymer,rippe2001making,stadler2017regulation}. Beyond chromatin, looping is a fundamental mechanism in various biochemical processes, including protein folding and aggregation \cite{lapidus2000measuring,adamcik2018amyloid,chakravarty2023distinguishing}. Although such processes involve intricate looping mechanisms—where looping between two segments is complex—a specific subset focuses on looping between the two ends of a polymer, known as cyclization. Cyclization process is influenced by factors such as molecular crowding, activity, internal friction, and intermolecular interactions \cite{bhattacharyya2012confinement,samanta2014looping,shin15,yan2024attractive}. Depending on these factors, looping times follow different scaling laws as a function of the number of monomers $N$ in the polymer. For a Rouse chain, the looping time scales as time $\propto N^2$ \cite{toan2008kinetics}.
Previous studies on the looping behavior between the two end beads of a polymer have shown that, in the presence of self-propelled particles, the looping of a passive polymer is facilitated \cite{shin15}, whereas it is delayed for an isolated active polymer due to spontaneous exploration dynamics \cite{ghosh2022active}. However, to our knowledge, the impact of partial activity on looping kinetics in a partially active polymer has not been previously considered, and this also forms a focus of our study.

The remainder of the paper is organized as follows: The detailed specifications of the polymer model, including the characterization of active forces, are outlined in the next section (Sec. \ref{sec-mod}). In  Sec. \ref{sec-conform}, we examine the conformational properties of the model using both analytical approaches and simulations. Sections \ref{sec-dyn} and \ref{sec-loop} are dedicated to the dynamical behavior and looping kinetics, respectively. A summary, along with potential extensions, is provided in Sec. \ref{sec-con}. Additional technical details and analytical derivations are included in the appendices.

\begin{figure}[htp]
    \centering
        \centering
        \begin{subfigure}[b]{0.35\textwidth}
            \centering
            \includegraphics[width=\textwidth]{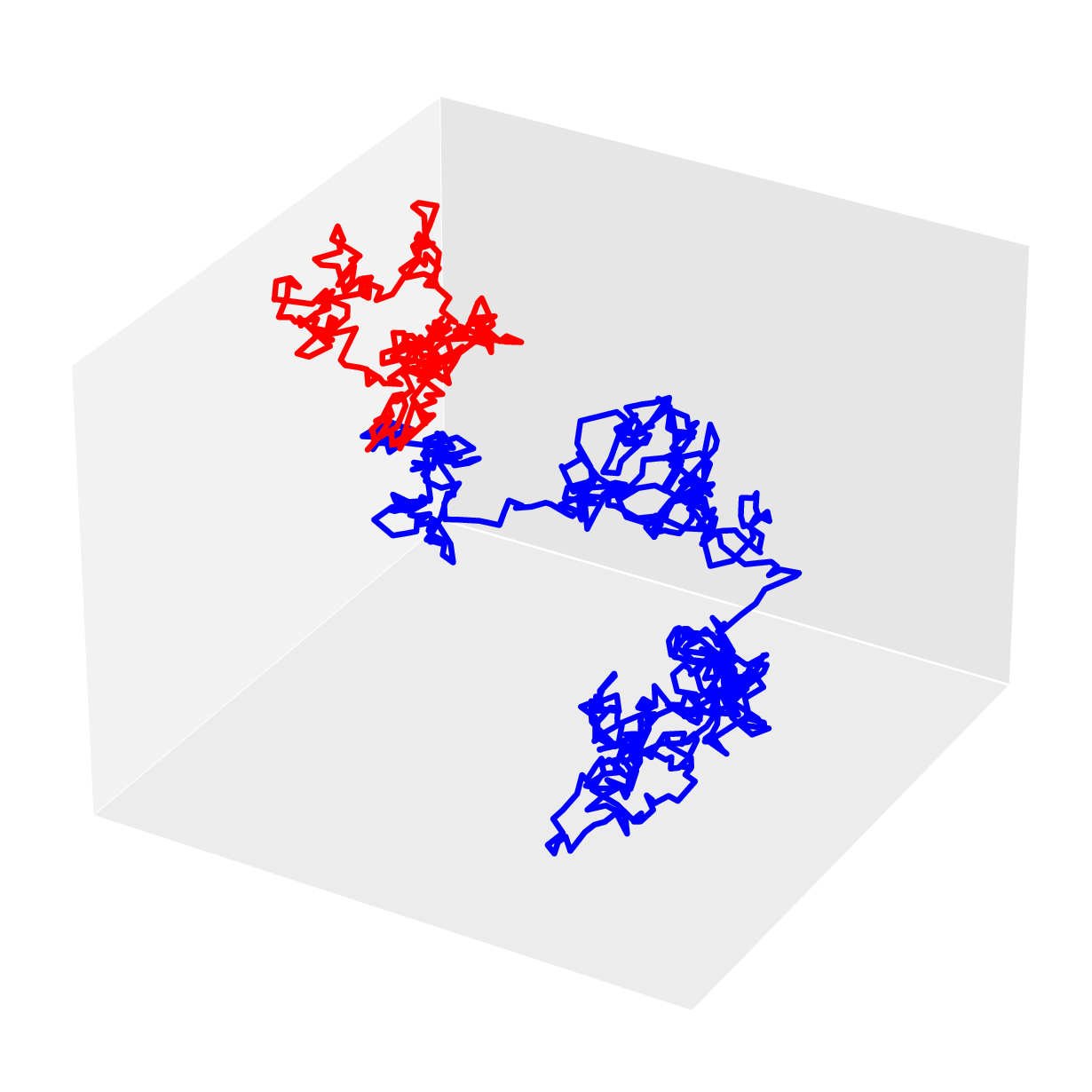}
            \caption{ $(f_1,f_2)=(0,1/3)$.}
        \end{subfigure}
        \begin{subfigure}[b]{0.35\textwidth}
            \centering
            \includegraphics[width=\textwidth]{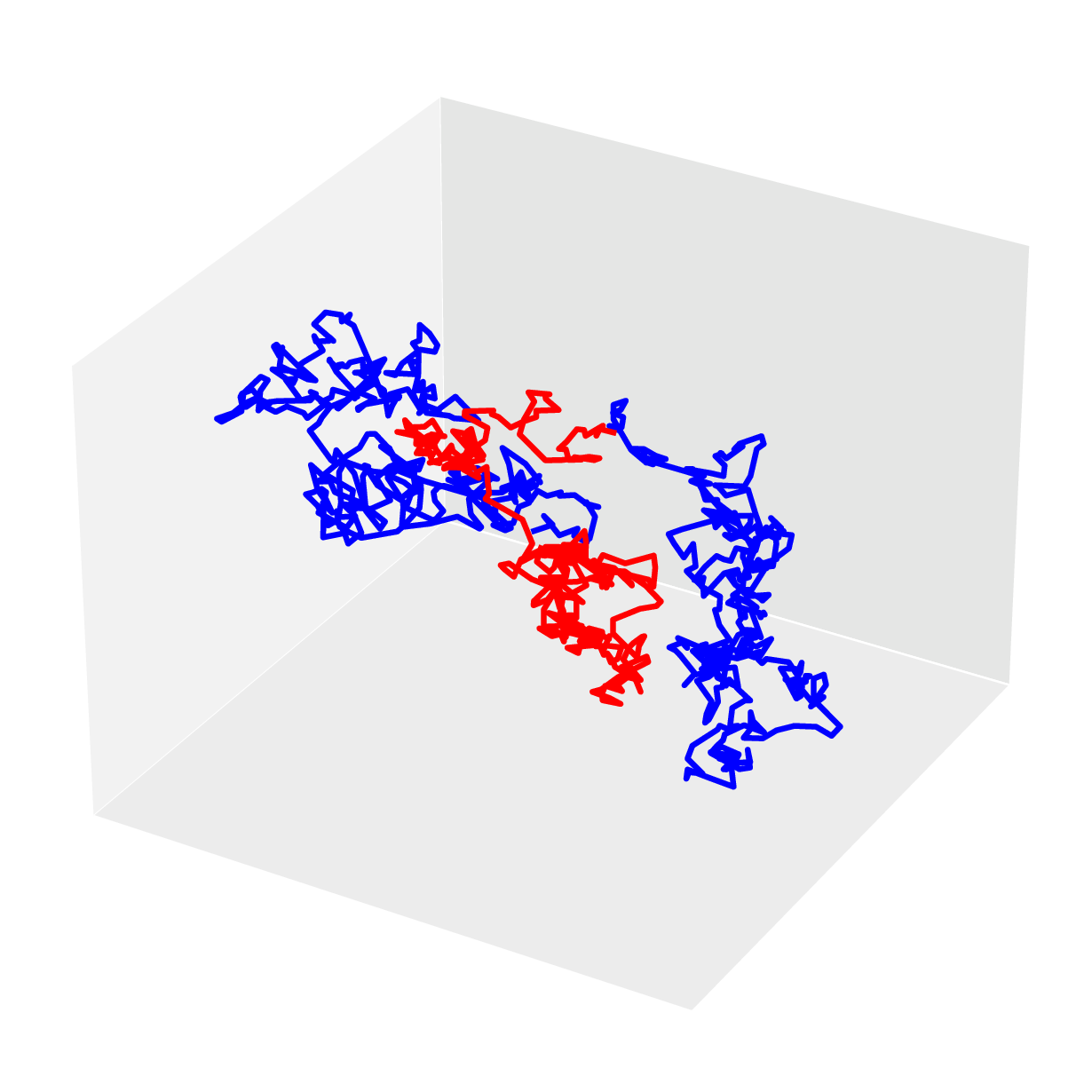}
            \caption{ $(f_1,f_2)=(1/3, 2/3)$.}
        \end{subfigure}
    \vspace{0.2cm}
    \begin{minipage}{0.35\textwidth}
        \centering
        \begin{subfigure}[b]{\textwidth}
            \centering
            \includegraphics[width=\textwidth]{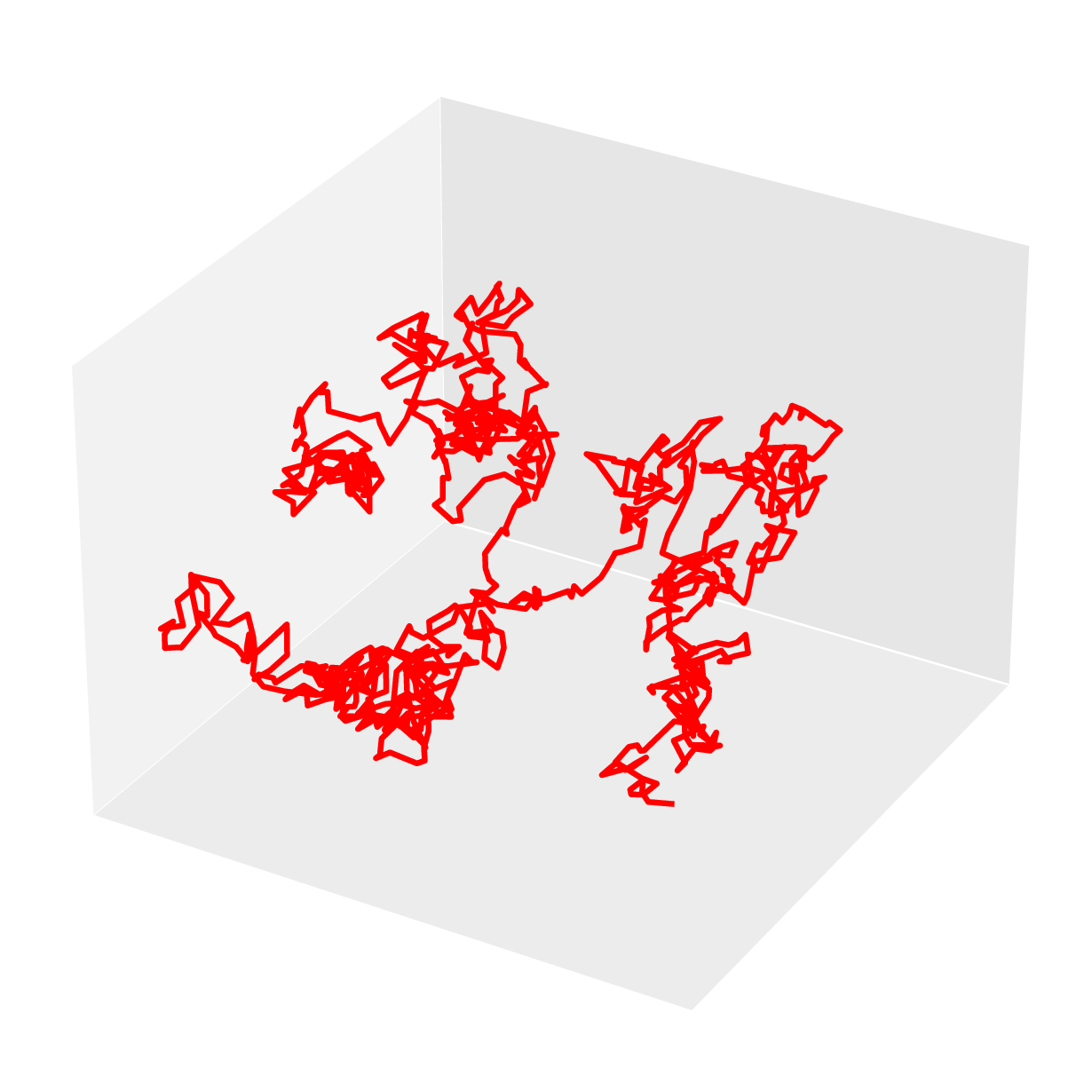}
            \caption{$(f_1,f_2)=(0,1)$.}
        \end{subfigure}
    \end{minipage}

      \captionsetup{width=1.\linewidth}

    \caption{Illustrations of partially active polymers, with active segments colored in red and passive segments in blue: (a) an end segment is active, (b) the middle segment is active, and (c) the entire polymer chain is active.}
    \label{fig:polymer}
\end{figure}

\section{Model  \label{sec-mod}} 

\subsection{Dynamics of  polymer} 
Consider an active Rouse polymer of length $L$, represented as a time-dependent space curve $\bs{r}(s,t)$. Here, the parameter $s$ denotes the arc length along the polymer contour and is treated as a continuous variable within the range $s\in(0,L)$. This continuum representation is an extension of the discrete Rouse model, where a chain of $N$ monomers, indexed by $n \in (1,N)$, are connected by harmonic springs with a fixed spring constant $\kappa$ and Kuhn length $b_0$. Additionally, each monomer experiences both thermal fluctuations, $\bs{\eta}^{\rm T}(t)$, and an active force, $\bs{\sigma}^{\rm A}(t)$.

In the continuum framework, any point along the contour can be indexed by its arc length, taking $b_0 n \rightarrow s$. Thus, under the assumptions of overdamped and free-draining conditions, the dynamics of a point at arc length $s$ (the $s^{\text{th}}$ point) can be described as follows \cite{winkler2020physics,goswami2022reconfiguration}:

\begin{align}
\zeta \frac{\partial}{\partial t}\bs{r}(s,t)=\kappa\frac{\partial^2 \bs{r}(s,t)}{\partial s^2}+\bs{\eta}^{\rm T}(s,t)+\bs{\sigma}^{\rm A}(s,t).\label{polymer_dynamics}
 \end{align}
with $\zeta$ being the drag coefficient on each point. Both ends of the polymer chain are free to move and are not constrained by any external forces. Consequently, force-free boundary conditions are applied, which are expressed as \cite{doi1988theory}
\begin{align}
\frac{\partial \bs{r}(s,t)}{\partial s}|_{s=0}=\frac{\partial \bs{r}(s,t)}{\partial s}|_{s=L}=0.\label{bcs_rouse}
\end{align}

The thermal noise, a characteristic of the heat bath that adheres to the fluctuation-dissipation theorem, is modeled as zero-mean Gaussian noise, viz.
\begin{subequations}
 \begin{align}
  \langle \eta^{\rm T}(s, t)\rangle & =0,\\
 \langle \eta_i^{\rm T}(s_2, t_2) \eta_j^{\rm T}(s_1, t_1)\rangle & =2\zeta k_B T \delta_{ij} \delta(s_2-s_1)\delta(t_2-t_1), 
 \end{align}   
\end{subequations}
where $ (i,\,j) \in\{x,y,z\}$, and  the Kronecker delta $\delta_{ij}$ ensures that the noise is uncorrelated between different spatial directions.

The active force arising from the active processes drives the polymer chain out of equilibrium and can exhibit spatio-temporal correlations \cite{osmanovic2017dynamics}. In this context, we define the spatial correlation $\mathcal{Z}(s_2,s_1)$ between $s_1$ and $s_2$ along the chain, and $\mathcal{K}(t_1, t_2)$ as the correlation between the times $t_1$ and $t_2$.
Therefore, the mean and covariance of the  active force can be expressed as follows:
\begin{subequations}
\begin{align}
\langle \sigma^{\rm A}(s, t)\rangle&=0,\\
\langle \sigma_i^{\rm A}(s_2, t_2) \sigma_j^{\rm A}(s_1, t_1)\rangle&=\delta_{ij}\mathcal{Z}(s_2,s_1) \mathcal{K}(t_2,t_1)\label{active_corr}.
\end{align}
\end{subequations}

As is common practice, the configuration $\bs{r}(s,t)$ can be decomposed as 
\begin{align}
\bs{r}(s,t)=\bs{\chi}_0(t)+2L\sum_{p=1}^{\infty}\mathcal{U}_p(s)\,\bs{\chi}_p(t),\label{r(s,t)}
\end{align} 
 where $\mathcal{U}_p(s)$ denotes the $p^{\text{th}}$ orthonormal eigenmode, which is the eigenfunction of the equation 
\begin{equation}
 -\kappa\frac{\partial^2 \mathcal{U}_p(s)}{\partial s^2}= \epsilon_p \mathcal{U}_p(s) \label{eigen_eqn}
\end{equation}
and $\bs{\chi}_p(t)$ stands for its mode amplitude.
By virtue of Eqs. (\ref{bcs_rouse}) and (\ref{eigen_eqn}), the eigenmodes are found to be  
\begin{align}
    \mathcal{U}_p(s)=\frac{1}{L}\text{cos}\Bigg(\frac{p\pi s}{L}\Bigg)
\end{align}
for $p=1,2,3,\cdots,$ with their corresponding eigenvalues  
\begin{equation}
    \epsilon_p=\kappa \Bigg(\frac{p \pi}{L}\Bigg)^2,
\end{equation}
 where $L$ is the contour length of the polymer.
Thus, the $p^{\text{th}}$ Rouse mode amplitudes, along with the amplitudes of the thermal force and the active force, can be expressed as
\begin{subequations}
 \begin{align}
\bs{\chi}_p(t)&=\int_{0}^{L} ds\, \mathcal{U}_p(s) \bs{r}(s,t),\\
\bs{\eta}^{\rm T}_p(t)&= \int_{0}^{L} ds\, \mathcal{U}_p(s) \bs{\eta}^{\rm T}(s,t),\\
\bs{\sigma}^{\rm A}_p(t)& = \int_{0}^{L} ds\, \mathcal{U}_p(s) \bs{\sigma}^{\rm A}(s,t).
 \end{align}    
\end{subequations}
So from Eqs. (\ref{polymer_dynamics}) and (\ref{r(s,t)}),  the evolution of the $p^{\text{th}}$ mode amplitude in time can be expressed as 
 \begin{align}
 \frac{\partial}{\partial t}\bs{\chi}_p(t)=-\frac{1}{\tau_p}\bs{\chi}_p(t)+\bs{\eta}_p^{\rm T}(t)+\bs{\sigma}^{\rm A}_p(t),\label{rouse_dynamics}
 \end{align}
 for $p=1,2,3,\cdots.$ 
 
 The zeroth Rouse mode follows   \begin{align}
 \frac{\partial}{\partial t}\bs{\chi}_0(t)=\bs{\eta}^{\rm T}_0(t)+\bs{\sigma}^{\rm A}_0(t),\label{rouse_dynamics0}
 \end{align}
which describes the dynamics of the polymer's center of mass. 
In Eq. (\ref{rouse_dynamics}), the parameter $\tau_p$ denotes the relaxation time of the $p^{\text{th}}$ mode, given by 
\begin{align}
\tau_p=\frac{\zeta}{\epsilon_p}=\frac{\tau_1}{p^2},
\label{taup}
\end{align}
where the longest relaxation time or the Rouse time can be defined as 
\begin{align}
\tau_1=\frac{\zeta L^2}{\kappa \pi^2}.   \label{tau1} 
\end{align}

The covariance of the thermal components at different modes can be computed as 
\begin{subequations}
 \begin{align}
\langle \eta_{p,i}^{\rm T}(t_2) \eta_{q,j}^{\rm T}(t_1) \rangle &= \int_{0}^{L} ds_1\, \int_{0}^{L} ds_2 \,\mathcal{U}_p(s_2) \mathcal{U}_q(s_1) \nonumber\\
  & \qquad\qquad \times\langle \eta_i^{\rm T}(s_2, t_2) \eta_j^{\rm T}(s_1, t_1)\rangle \nonumber\\
  &=\Bigg(\frac{k_B T}{L\zeta}\Bigg)\delta_{pq}\delta_{ij}\delta(t_1-t_2),\label{corr_thermal_modes}\\
  \langle \eta_{0,i}^{\rm T}(t_2) \eta_{0,j}^{\rm T}(t_1) \rangle &= \Bigg(\frac{2 k_B T}{L\zeta}\Bigg)\delta_{ij}\delta(t_1-t_2),\label{corr_thermal_modes0}
 \end{align}
 \end{subequations}
where $p, q=1,2,3,\cdots.$
Let us consider that each component of the active force can be expressed as a product of its spatial and temporal parts, as follows:
\begin{align}
    \sigma^{\rm A}_i(s,t) = \sigma_{s,i}^{\rm A}(s) \sigma_{\tau, i}^{\rm A}(t).
\end{align}
Thus, the covariance between different mode amplitudes of the active force in each direction can be expressed as
\begin{align}
 \langle \sigma^{\rm A}_i(s_2,t_2) \sigma_j^{\rm A}(s_1,t_1) \rangle = \langle \sigma_{s,i}^{\rm A}(s_2) \sigma_{s,j}^{\rm A}(s_1) \rangle \langle \sigma_{\tau,i}^{\rm A}(t_2) \sigma_{\tau,j}^{\rm A}(t_1) \rangle. \label{active_corr1}   
\end{align}
Using Eqs. (\ref{active_corr}) and  (\ref{active_corr1}), one can obtain 
\begin{subequations}
 \begin{align}
  &\langle \sigma_{p,i}^{\rm A}(t_2) \sigma_{q,j}^{\rm A}(t_1) \rangle \nonumber \\
  &\;\;=\int_{0}^{L} ds_1\, \int_{0}^{L} ds_2 \,\mathcal{U}_p(s_2) \mathcal{U}_q(s_1)\langle \sigma^{\rm A}_i(s_2, t_2) \sigma^{\rm A}_j(s_1, t_1)\rangle\nonumber\\
  &\;\;= \frac{\mathcal{I}_{pq}}{L^2\zeta^2}\delta_{ij}\mathcal{K}(t_2,t_1),
  \label{active_corr2} \\
  & \langle \sigma_{p,i}^{\rm A}(t_2) \sigma_{0,j}^{\rm A}(t_1) \rangle = \frac{\mathcal{I}_{p0}}{L^2\zeta^2}\delta_{ij}\mathcal{K}(t_2,t_1),\label{active_corr20}\\
  &\langle \sigma_{0,i}^{\rm A}(t_2) \sigma_{0,j}^{\rm A}(t_1) \rangle = \frac{\mathcal{I}_{00}}{L^2\zeta^2}\delta_{ij}\mathcal{K}(t_2,t_1), \label{active_corr21}
 \end{align}
 \end{subequations}
where 
 \begin{subequations}
\begin{align}
&\mathcal{I}_{p0}=\mathcal{I}_{0p}=\int_{0}^{L} ds_1\, \int_{0}^{L} ds_2 \,\text{cos}\Bigg(\frac{p\pi s_2}{L}\Bigg)\mathcal{Z}(s_2,s_1),\\
 &\mathcal{I}_{pq}=\int_{0}^{L} ds_1\, \int_{0}^{L} ds_2 \,\text{cos}\Bigg(\frac{p\pi s_2}{L}\Bigg)\text{cos}\Bigg(\frac{q\pi s_1}{L}\Bigg)\mathcal{Z}(s_2,s_1).   \label{I_pq}
\end{align}
 \end{subequations}
 The Kronecker delta $\delta_{ij}$ in Eqs. (\ref{corr_thermal_modes})-(\ref{corr_thermal_modes0}) and Eqs. (\ref{active_corr2})-(\ref{active_corr21}) ensures that the covariance between two Rouse modes not aligned along the same  axis is zero. Thus, we  simplify notations by dropping subscripts $i$ and $j$ and denoting each component of the thermal noise and the active noise mode as simply $\eta_{p}^{\rm T}$ and $\sigma_{p}^{\rm A}$, respectively. 
 
\subsection{Form of $\mathcal{K}(t_1,t_2)$}
As a common model for active noise, we assume that the active force follows an Ornstein-Uhlenbeck process (OUP),
 described by the equation
 \cite{mar21st,gos23eff}
\begin{align} \frac{d \sigma_{\tau}^A(t)}{dt} = -\frac{\sigma_{\tau}^A(t)}{\tau_A} + \sqrt{\frac{2}{\tau_A}}\eta_{a}(t),\label{ouprocess} \end{align}
where $\eta_{a}(t)$ represents Gaussian white noise with zero mean and unit variance. As a result, the active forces are Gaussian-distributed and exhibit exponential temporal correlations of the form
\begin{align}
\mathcal{K}(t_1,t_2)=  \exp\left(-\frac{|t_1-t_2|}{\tau_A}\right),\label{tempo_corr}    
\end{align}
with $\tau_A$ being the persistence time  of the noise.
\subsection{Form of $\mathcal{Z}(s_2,s_1)$}
Let us consider  that a specific segment $s_A$ of the polymer is active,  and  the segment $s_A$ lies within the range $s \in [f_1 L, f_2 L ]$ and $0\le f_1 \le (s/L) \leq f_2\le 1$,  where $0\leq f_1< f_2\leq 1$.
Different realizations of such a polymer are depicted in Fig. \ref{fig:polymer}. The amplitude of the active force $\sigma_s^A$ at each point of this segment is randomly assigned   by sampling from a white noise distribution along the polymer contour.
Consequently, the monomers in the active regions are uncoupled, implying that 
\begin{subequations}
 \begin{align}
\sigma_s^A(s)&=\mathbf{1}_{s_A}(s) \xi_A(s), \label{sigma_s} \\
\langle \sigma_s^A(s_1) \sigma_s^A(s_2) \rangle & = \mathbf{1}_{s_A}(s_1)\mathbf{1}_{s_A}(s_2) \langle \xi_A(s_1) \xi_A(s_2) \rangle  \label{sigma_s_corr}.
\end{align}   
\end{subequations}
 Here, $\xi_A(s)$ is a delta-correlated random variable with some distribution $P(\xi_A)$, and $\mathbf{1}_{s_A}$ denotes the indicator function for the region $s_A$. As a simple and realistic model,  $\xi_A(s)$ can be either Gaussian white noise or Poisson noise. In the Gaussian case, values of $\xi_A(s)$ are sampled from a Gaussian distribution $P(\xi_A)$ with variance $F_G^2$ at each $s$ in $s_A$. In the Poisson case,  $\xi_A(s)$ can be expressed as $\xi_A(s)= \Sigma_{i=0}^{N_a} F_i \delta(s-s_i).$  Here  $N_a$  represents the number of active sites distributed within the active segment of length $l_A=(f_2-f_1)L,$ according to a Poisson statistics as follows: $P(n=N_a,l_A)=\frac{(\nu_l l_A)^n}{n!}\,e^{-\nu_l l_A},$ where  the Poisson rate $\nu_l$ determines the average number of active sites per unit length. 
 The term $F_i$ is the amplitude of the force at site  $s_i$, which can be either constant $F_P$ or sampled from a distribution such as the Laplace distribution of the form: $P(F_i)=\frac{1}{2F_P}e^{-\frac{|F_i|}{F_P}}$, with  $F_P>0$ being the mean amplitude. In any case, the autocorrelation  of $\xi_A$ is found to be 
 $\langle \xi_A(s_1) \xi_A(s_2) \rangle= F_0^2 \delta(s_1-s_2),$ where for the Gaussian case, $F_0^2 =2 F_G^2,$ while for the Poisson case, $F_0^2=\nu_l F_P^2$.  Thus one can write the spatial correlation function as 
 \begin{align}
\mathcal{Z}(s_2,s_1)&=\langle \sigma_s^A(s_1) \sigma_s^A(s_2) \rangle \nonumber\\
& = F_0^2 \delta(s_2-s_1) \mathbf{1}_{s_A}(s_2)\mathbf{1}_{s_A}(s_1)  \label{sigma_s_corr1}.   
 \end{align}
As a simple yet commonly used model, we adopt Gaussian noise for $\xi_A(s)$ in our simulation of the partially active polymer.
 Inserting  Eq. (\ref{sigma_s_corr1}) into Eq. (\ref{I_pq}), one can find
\begin{align}
&\mathcal{I}_{p'q'} = \int_{0}^{L} ds_1 \int_{0}^{L} ds_2 \cos\left(\frac{p'\pi s_2}{L}\right) \cos\left(\frac{q'\pi s_1}{L}\right) \mathcal{Z}(s_2,s_1) \nonumber \\
&= \begin{cases}
& F_A^2(f_2 - f_1),  \text{if } p' = q' = 0, \\
& F_A^2 \left[\frac{\sin(f_2 q' \pi) - \sin(f_1 q' \pi)}{q' \pi}\right],  \text{if } p' = 0,\, q' \neq 0, \\
& F_A^2\left[\frac{f_2 - f_1}{2} + \frac{\sin(2 f_2 p' \pi)-\sin(2 f_1 p' \pi)}{4 p' \pi}\right],  \text{if } p' = q',\, p' \neq 0, \\
& F_A^2\left[\frac{\sin(f_2 \pi (p' - q'))-\sin(f_1 \pi (p' - q'))}{2 \pi (p' - q')} \right. \nonumber \\
&\quad \left. + \frac{\sin(f_2 \pi (p' + q'))-\sin(f_1 \pi (p' + q'))}{2 \pi (p' + q')} \right], \text{otherwise},
\end{cases}\\ \label{Ipq}
\end{align}
where $F_A^2= F_0^2 L$.

\section{Conformational properties \label{sec-conform}}
We aim to determine the conformational properties of the polymer at steady state. The key quantities considered here are the radius of gyration (RG), the end-to-end distance (ED), and the separation between two segments on a polymer chain.

Using Eq. (\ref{r(s,t)}), the conformational observables can be defined as follows \cite{doi1988theory,eisenstecken2017internal}:

\begin{enumerate} \item  Center of Mass (CM): \begin{align} \bs{r}_0(t) = \frac{1}{L} \int_{0}^{L} ds \, \bs{r}(s,t) = \bs{\chi}_0(t), \label{r_com} \end{align}

\item End-to-End Distance: \begin{align} \bs{R}_e(t) = \bs{r}(L,t) - \bs{r}(0,t) = -4 \sum_{p=1}^{\infty} \bs{\chi}_{2p-1}(t). \end{align}

\item Mean-square radius of gyration (MSRG):
\begin{align} \langle R_g^2 \rangle = \frac{1}{L} \int_{0}^{L} ds \, \langle \left[\bs{r}(s,t) - \bs{r}_0(t)\right]^2 \rangle = 6 \sum_{p=1}^{\infty} \langle \chi_p(t)^2 \rangle. \label{rg} \end{align}

\item Mean square end-to-end distance (MSED): \begin{align} \langle R_e^2 \rangle =\langle \left[\bs{r}(L,t) - \bs{r}(0,t)\right]^2 \rangle = 48 \sum_{p,q=1}^{\infty} \langle \chi_{2p-1}(t) \chi_{2q-1}(t) \rangle. \label{phi_corr0} \end{align} 

\item Contour length separation between the  $s_1^{\text{th}}$ point and the  $s_2^{\text{th}}$ point on the contour.:
\begin{align}
\bs{R}_{s_1s_2}(t)& = \bs{r}(s_2,t) - \bs{r}(s_1,t)\nonumber\\
&= 2\sum_{p=1}^{\infty} \bs{\chi}_{p}(t)\left[\text{cos}\Bigg(\frac{p\pi s_2}{L}\Bigg)-\text{cos}\Bigg(\frac{p\pi s_1}{L}\Bigg)\right].    
\end{align}
\item Mean square separation (MSS) between two positions: 
\begin{align}
&\langle R_{s_1s_2}^2(t) \rangle  =\langle \left[\bs{r}(s_2,t) - \bs{r}(s_1,t)\right]^2 \rangle  = 12 \sum_{p,q=1}^{\infty} \langle \chi_p(t)\chi_q(t) \rangle\nonumber\\
&\times \Bigg[\text{cos}\Bigg(\frac{p\pi s_2}{L}\Bigg)\text{cos}\Bigg(\frac{q\pi s_2}{L}\Bigg) +\text{cos}\Bigg(\frac{p\pi s_1}{L}\Bigg)\text{cos}\Bigg(\frac{q\pi s_1}{L}\Bigg) \nonumber\\
& -\text{cos}\Bigg(\frac{p\pi s_2}{L}\Bigg)\text{cos}\Bigg(\frac{q\pi s_1}{L}\Bigg)-\text{cos}\Bigg(\frac{p\pi s_1}{L}\Bigg)\text{cos}\Bigg(\frac{q\pi s_2}{L}\Bigg) \Bigg].\label{Rmn}    
\end{align}

\end{enumerate}

\begin{figure}[htp]
\centering
\includegraphics[width=1\linewidth]{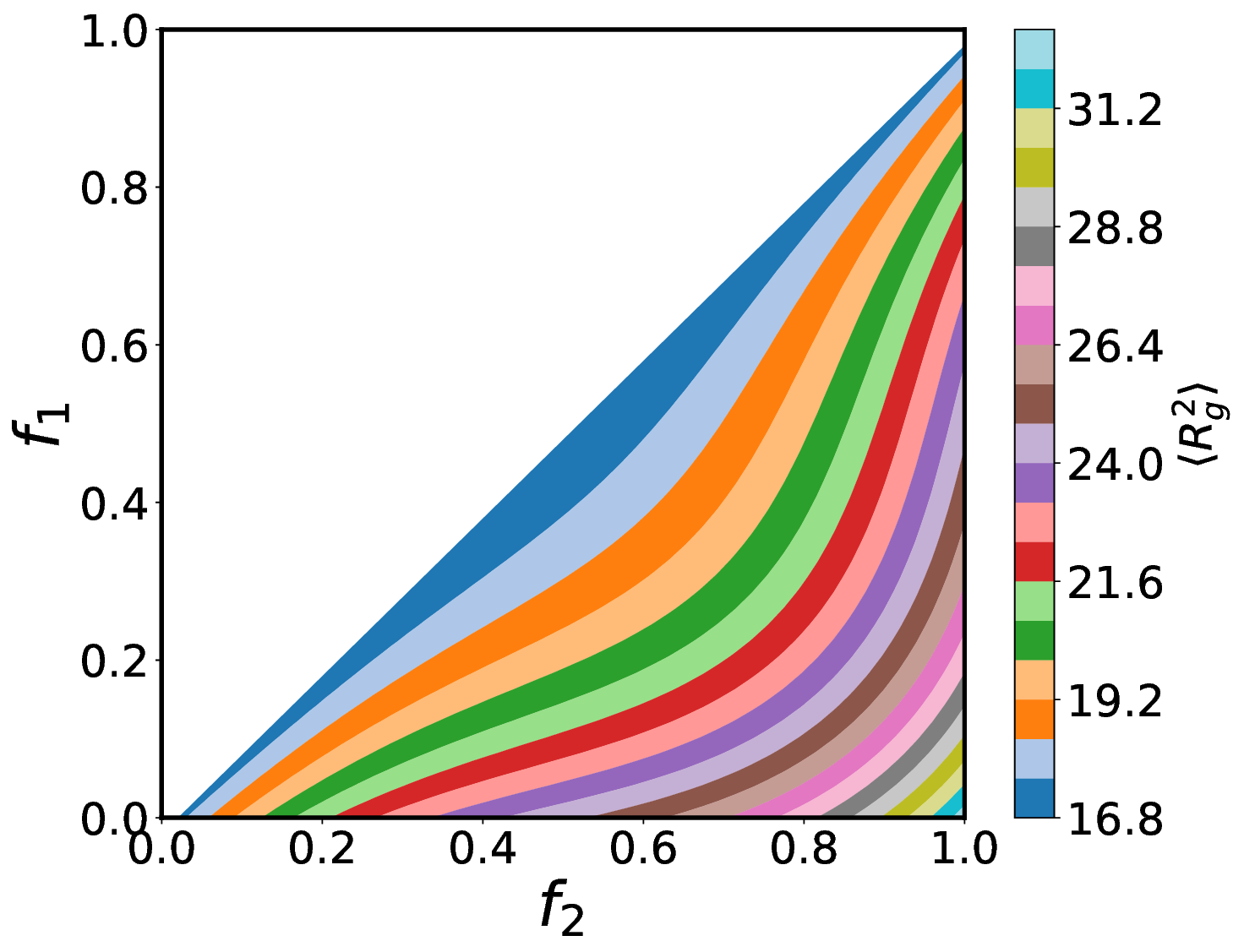}
\caption{Contour plot of MSRG $\langle R_g^2 \rangle$ [Eq. (\ref{rg_1})]  as functions of $f_1$ and  $f_2$  for $\mathcal{K}(t_1,t_2)= \exp(-(|t_1-t_2|)/\tau_A)$. The values of other parameters are $\kappa=3.0$, $\zeta=1.0$, $k_BT=1.0$, $\tau_A=1.0$, $L=100,$ and $F_0=1.0.$
}
\label{fig:rg}
\end{figure}
\begin{figure}[H]
\centering
\includegraphics[width=1\linewidth]{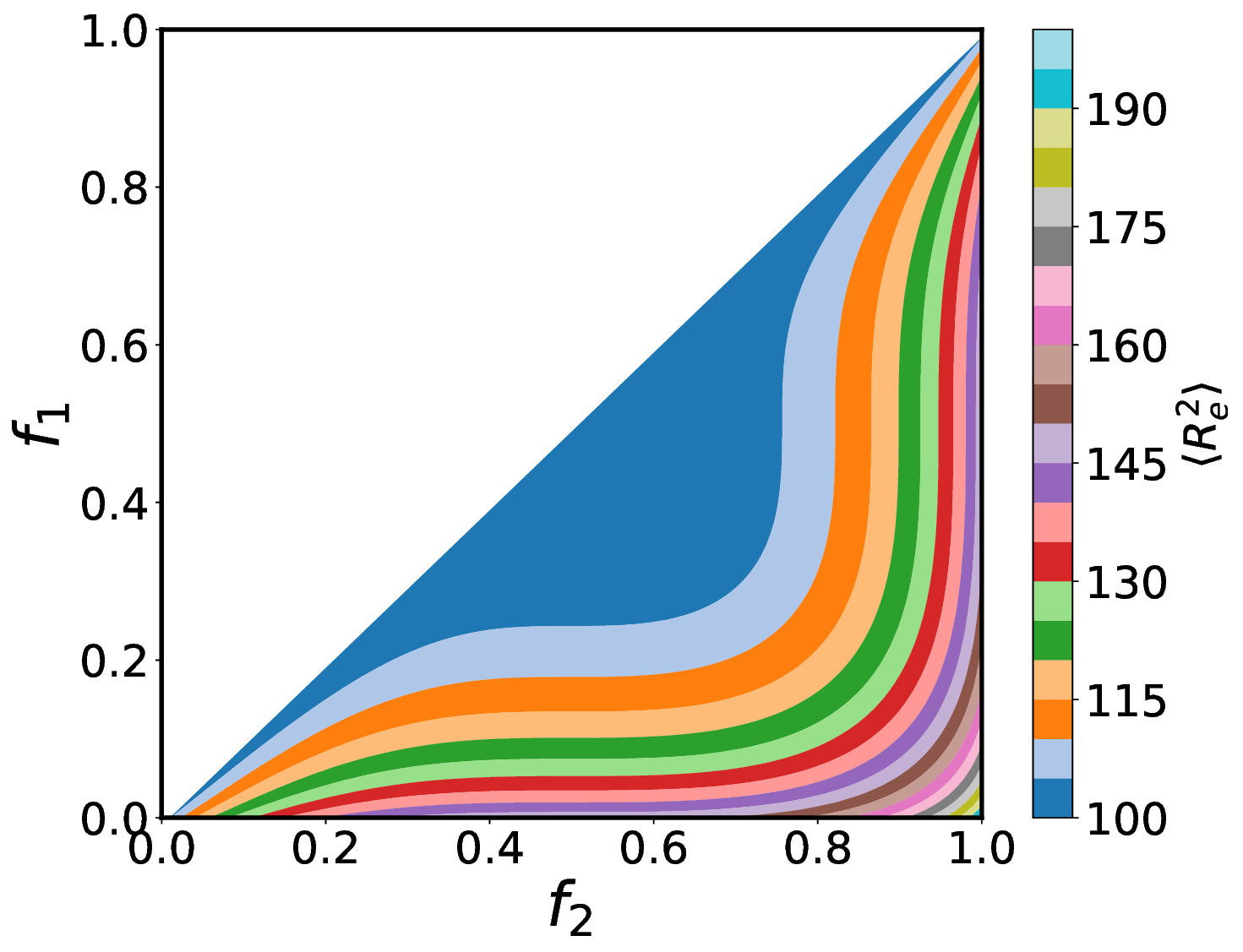}
\caption{Contour plot of MSED $\langle R_e^2 \rangle$ [Eq. (\ref{phi_corr0_1})]  as functions of $f_1$ and  $f_2$  for $\mathcal{K}(t_1,t_2)= \exp(-(|t_1-t_2|)/\tau_A)$. The values of other parameters are the same as in Fig. \ref{fig:rg}.
}
\label{fig:r2e}
\end{figure}

To elucidate the effects of local activity arising from the active segment on the conformations of the entire polymer of length $L$, we consider varying the segment location and size.
The MSRG, $\langle R_g^2 \rangle$, is depicted in Fig. \ref{fig:rg} for a particular polymer length with differential activity. The steady-state values $\langle R_g^2 \rangle$ obtained from simulations, as shown in  Fig. \ref{fig:Rg_sim}, agree well with the analytical predictions, within the numerical error margin. Notably, as activity increases by extending the size of the active region from one end $(f_1=0)$ along the contour, $\langle R_g^2 \rangle$ grows monotonically, indicating that the polymer swells [see Fig. \ref{fig:rg_alls}]. This observation is consistent with findings reported in Ref. \cite{osmanovic2018properties}. A similar trend is observed for the mean square end-to-end distance,  $\langle R_e^2 \rangle$, as shown in Fig. \ref{fig:r2e} [cf. Fig. \ref{fig:Re_sim}]. However, as shown in Fig. \ref{fig:rgre_f10f2}, the extent of swelling varies along the chain, following a three-step growth with respect to $f_2$. In the initial phase (up to  $f_2=0.2$) and the final phase (from $f_2=0.8$), lengthening the active segment results in a significant increase in the MSRG and MSED. In contrast, at intermediate values of $f_2$, its effect on the polymer size is minimal.

Consider an active segment of fixed length that can be positioned anywhere along the polymer chain. To investigate how its location affects polymer conformation, we initially place it at one end and gradually move it along the chain up to the opposite end. This corresponds to following a  straight line which is  parallel to the line  $f_1=f_2$ in the contour plots shown in Figs. \ref{fig:rg}-\ref{fig:r2e}.   Since both ends are equivalent in the absence of activity, the variations are symmetric relative to the midpoint of the polymer. From the variations in MSRG and MSED, we see that the polymer shrinks when the active segment is moved from either end to the center.
These observations are corroborated by Brownian dynamics simulations, as discussed in Appendix   \ref{appen:simulation}.

The generic expression for the MSRG of a polymer with an active segment $s_A = (f_1 L,f_2 L)$ is given by Eq. (\ref{rg_1}), and is provided below for convenience,
\begin{align}
 \langle R_g^2 \rangle_{(f_1,f_2)} =  \frac{k_B T L}{2\kappa}+\frac{6  \tau_1^2}{L^2\zeta^2}\sum_{p=1}^{\infty}\frac{\mathcal{I}_{pp}}{p^2(p^2+\tau_1/\tau_A)}.\label{rg_1_maintext}
\end{align}
Here, the first term on the right-hand side, denoted as $\langle R_g^2 \rangle^{\rm passive}$,
represents the contribution arising from thermal fluctuations. In other words, this term is the same as the MSRG of a polymer without an active segment. The other term, denoted as $\langle R_g^2 \rangle_{(f_1,f_2)}^{\rm active}$, represents the extra contribution arising from the active forces. For specific values of $f_1$ and $f_2$, the sum in Eq. \eqref{rg_1_maintext} takes a simple closed form, providing us with additional physical insights. As a first example, for a fully active polymer (where $f_1=0,\,f_2=1$), the exact expression for $\langle R_g^2 \rangle$ is given by Eq. (\ref{rg_full}), which we provide below,
\begin{align}
\langle R_g^2 \rangle_{(0,1)}  & =  \frac{k_B T L}{2\kappa}+ \frac{\pi^2  F_A^2 \tau_A \tau_1}{2L^2\zeta^2} \nonumber\\
& +\frac{3  F_A^2 \tau_A^2}{2L^2\zeta^2}\left[1-\sqrt{\frac{\pi^2 \tau_1}{\tau_A}}\coth{\sqrt{\frac{\pi^2\tau_1}{\tau_A}}}\right]. \label{rg_full1} 
\end{align}
Equation (\ref{rg_full}) is equivalent to the results previously reported for fully active polymers \cite{kaiser2015does,goswami2022reconfiguration}. 
For a polymer with half of its length active, extending from one end to the midpoint ($f_1=0,\,f_2=0.5$), one could infer from Equations \eqref{Ipq} and \eqref{rg_1_maintext} that 
\begin{align}
\langle R_g^2 \rangle_{(0,0.5)}^{\rm active} = \frac12 \langle R_g^2 \rangle_{(0,1)}^{\rm active},    
\end{align} so the MSRG in this case is the sum of the passive part and half of the active part for the fully active polymer, namely,
\begin{align}
\langle R_g^2 \rangle_{(0,0.5)} = \langle R_g^2 \rangle^{\rm passive}+\frac12 \langle R_g^2 \rangle_{(0,1)}^{\rm active}. \label{rg_half1} 
\end{align}

One can introduce partial activity by assigning activity to the middle region, with $f_1=0.25$ and $f_2=0.75$. For this configuration, the MSRG is given by
\begin{align}
& \langle R_g^2 \rangle_{(0.25,0.75)}  \nonumber\\
& = \frac{k_B T L}{2\kappa}+ \frac{3  F_A^2 \tau_A^2 }{2L^2\zeta^2}+ \frac{5\pi^2  F_A^2 \tau_A \tau_1}{32 L^2\zeta^2} \nonumber\\
& -\frac{3  F_A^2 \tau_A^2}{4L^2\zeta^2}\Bigg[\sqrt{\frac{\pi^2 \tau_1}{\tau_A}}\coth{\sqrt{\frac{\pi^2\tau_1}{\tau_A}}}+\sech{\sqrt{\frac{\pi^2\tau_1}{4\tau_A}}}\Bigg]. \label{rg_half_middle} 
\end{align}  
In Eqs. (\ref{rg_full1})-(\ref{rg_half_middle}), aside from the hyperbolic functions, other terms 
vary linearly or sublinearly with $L$.  In the limit $\tau_1/\tau_A \gg 1,$  the contribution from the hyperbolic parts becomes negligible, resulting in $\langle R_g^2 \rangle_{(f_1,f_2)} \propto L.$ On the other hand, in the limit $\tau_1/\tau_A \ll 1,$ the hyperbolic  terms dominate and can be approximated to yield
\begin{subequations}   
\begin{align}
& \langle R_g^2 \rangle_{(0.0,0.5)}
 \approx   \frac{k_B T L}{2\kappa}+ \frac{\pi^4  F_A^2 \tau_1^2}{60 L^2\zeta^2}\approx \frac{k_B T L}{2\kappa}+ \frac{F_0^2 L^3}{60\kappa^2}, \label{rg0_05}\\\
 & \langle R_g^2 \rangle_{(0.25,0.75)}
 \approx   \frac{k_B T L}{2\kappa}+ \frac{53 F_0^2 L^3}{7680\kappa^2}.\label{rg0_051}
\end{align}
\end{subequations}
From the above analysis, we could reinforce the following observations: 
\begin{itemize}
    \item  $\langle R_g^2 \rangle_{(0.25,0.75)} < \langle R_g^2 \rangle_{(0.0,0.5)}$, which analytically confirms the numerical result in Fig. \ref{fig:rg}, showing that polymers with their active segment at the end swell to a greater extent.
\item  For long polymers, $\langle R_g^2 \rangle_{(f_1,f_2)} \propto L$ for any values of $f_1$ and $f_2$, following the typical Rouse scaling. However, for small values of $\tau_1/\tau_A$ (or equivalently for small $L$ values), an anomalous scaling of MSRG emerges, where $\langle R_g^2 \rangle_{(f_1,f_2)} \propto L^{\alpha}$ with $\alpha >1$ [also, see Fig. \ref{fig:rg_alls}]. So, from Eqs. (\ref{rg0_05})-(\ref{rg0_051}) we can see that   $\alpha \approx 3$ for large values of $\tau_A$ and very short chains, as also  shown in Fig. \ref{fig:rg_alls}, while for long chains where $\tau_1 \gg \tau_A,$ linear scaling $(\alpha \approx 1)$ is recovered.  A similar trend is seen for the MSED, as shown in  Fig. \ref{fig:re_alls}.
\end{itemize}

All results obtained from Brownian dynamics simulations, discussed in Appendix  \ref{appen:simulation}, are consistent with the above analytical results.  From Fig. \ref{fig:E_sim},  we observe that polymers with longer active segments exhibit higher energy, and polymers with active segments of equal length have the same energy. Specifically, the total energy of the polymer is defined in terms of bond fluctuations between consecutive segments and depends solely on the length of the active segment, irrespective of its location along the polymer.  As the size of $s_A$ increases, 
 both the polymer's energy and its size  gradually increase, leading to swelling. The change in conformation when we fix the length of $s_A$ and vary its position can be explained as follows:  When $s_A$ is positioned at the middle of the polymer, it loses some freedom compared to when $s_A$ is located at one end, resulting in a reduced  size in the former scenario.

The conformational properties observed in the OUP model are compared using a simplified model in which the active force is delta-correlated in time, as illustrated in Figs. \ref{fig:rga}-\ref{fig:rea}. For this model, we obtain
\begin{subequations}
 \begin{align}
&  \langle R_g^2 \rangle_{(0,0.5)}^{\rm active} =\frac{ F_0^2 L}{8\zeta \kappa},  \\
&   \langle R_g^2 \rangle_{(0.25,0.75)}^{\rm active}=\frac{5}{8}\frac{F_0^2 L}{8\zeta \kappa}.
\end{align}   
\end{subequations}
For this model, $\langle R_g^2 \rangle_{(f_1,f_2)}^{\rm active} \propto L$, following the Rouse scaling for any $L$, and $\langle R_g^2 \rangle_{(0,0.5)}^{\rm active} > \langle R_g^2 \rangle_{(0.25,0.75)}^{\rm active}$, indicating that the behavior of $\langle R_g^2 \rangle_{(f_1,f_2)}^{\rm active}$ in the  $f_2-f_1$ plane, as shown in Fig.  \ref{fig:rga}, resembles the trends observed in the OUP model.
This indicates that the anomalous scaling emerges specifically due to the presence of persistence, whereas the swelling pattern is primarily determined by local activity and does not inherently depend on persistent motion. 

\begin{figure*}[htp]
    \centering
    \begin{subfigure}[b]{0.495\textwidth}
    \caption{ }
        \centering
        \includegraphics[width=\textwidth]{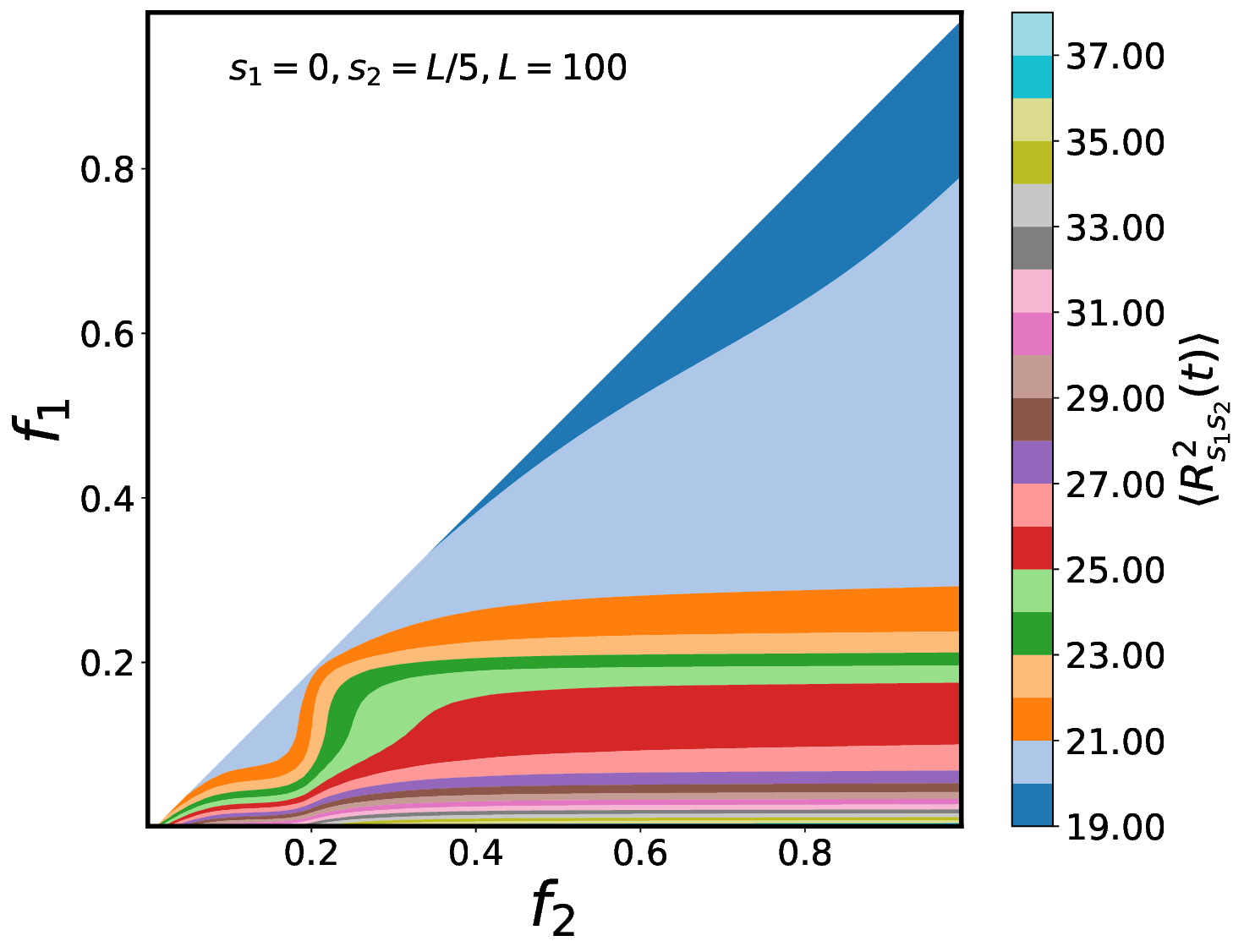}    
    \end{subfigure}
    \hfill
    \begin{subfigure}[b]{0.495\textwidth}
        \caption{}
        \centering
        \includegraphics[width=\textwidth]{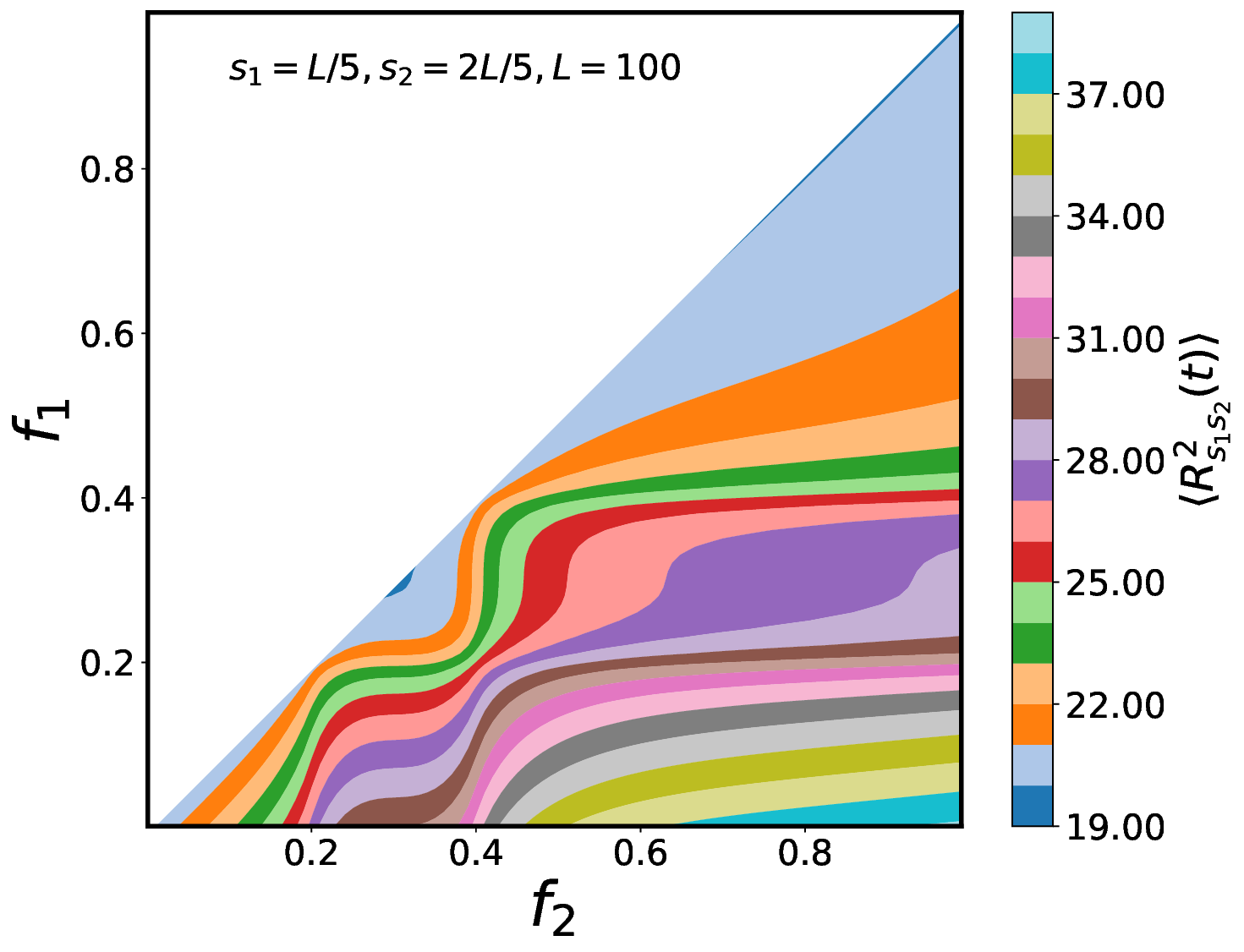}
    \end{subfigure}
  \vspace{0.2cm}  
  
    \centering
    \begin{subfigure}[b]{0.495\textwidth}
    \caption{ }
        \centering
        \includegraphics[width=\textwidth]{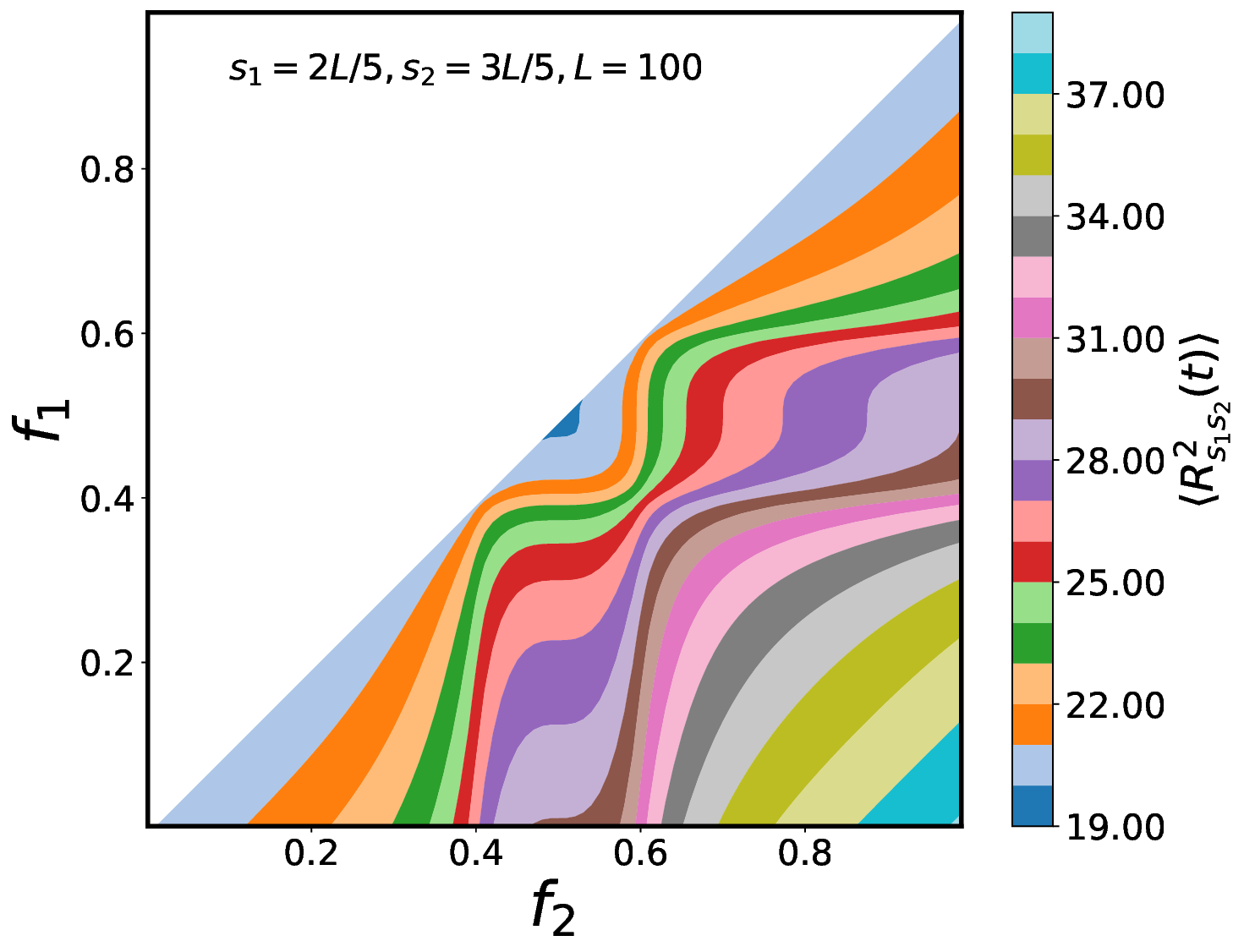}
    \end{subfigure}
    \hfill
    \begin{subfigure}[b]{0.495\textwidth}
             \caption{}
        \centering
        \includegraphics[width=\textwidth]{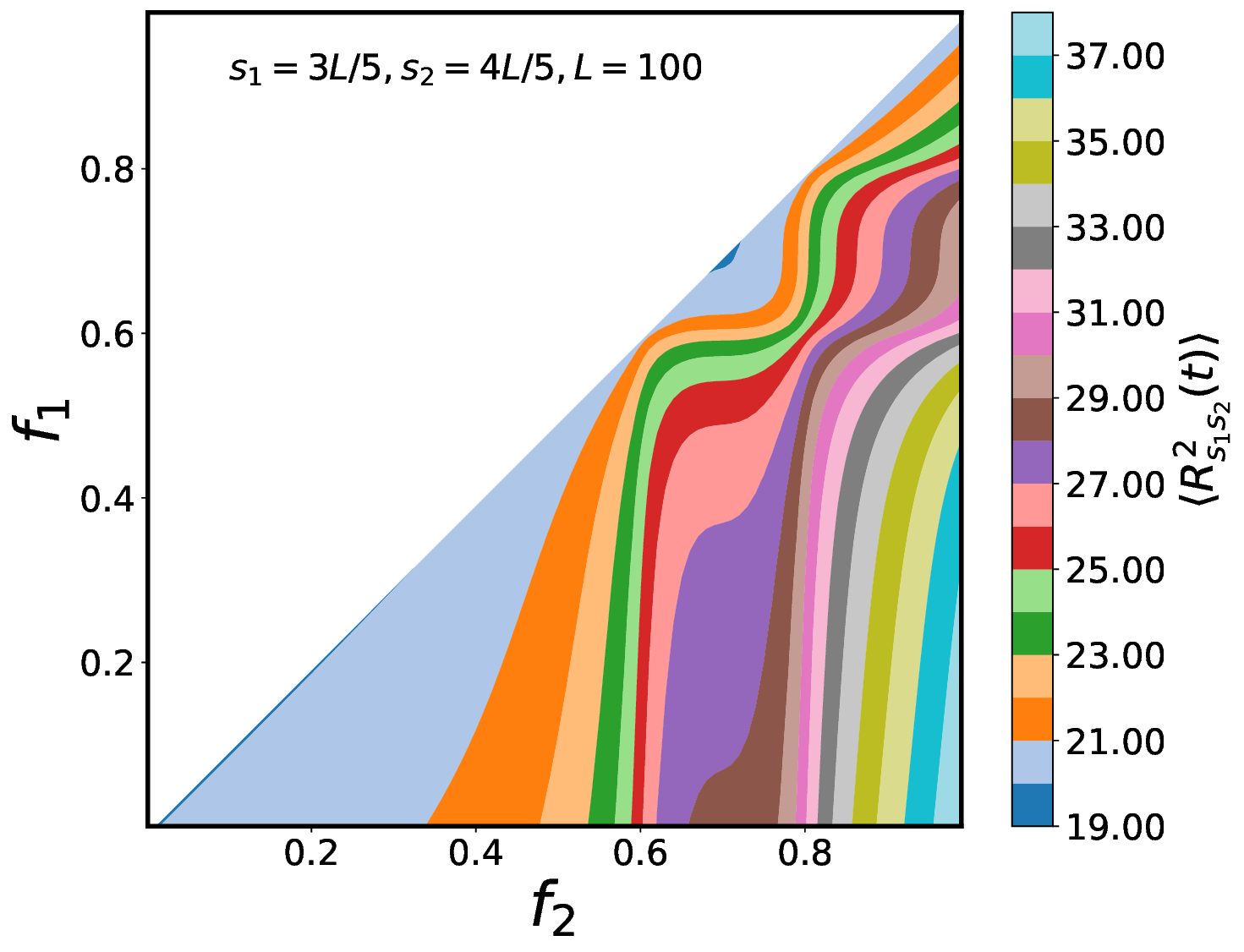}
    \end{subfigure}
    \caption{Contour plot of MSS
    $\langle R_{s_1s_2}^2(t) \rangle$ [Eq. (\ref{Rmn_1})] for different segments of the same contour separation  and $\mathcal{K}(t_1,t_2)= \exp(-(|t_1-t_2|)/\tau_A)$. The values of other parameters are the same as in Fig. \ref{fig:rg}.}
    \label{fig:R2mn}
\end{figure*}

To illustrate the peculiar local contribution of active segments to polymer swelling described above, the MSS $\langle R_{s_1s_2}^2 \rangle $ for a segment of arc length $0.2 L$  is computed  in Eq. (\ref{Rmn_1}) and shown for different fractions of active segments in Fig. \ref{fig:R2mn}. These analytical results are supported by simulations, as presented in  Fig. \ref{fig:Rmn_sim}. In Fig. \ref{fig:R2mn}, the plots reveal several intriguing patterns, allowing us to identify at least three distinct regimes for different $f_1$ and $f_2$ values. Note that the MSS between two points is an experimentally relevant quantity, as live imaging techniques enable the tracking of two fluorescently labeled points on chromatin \cite{child2021live}.

In Fig. \ref{fig:R2mn}(a), when tracing the MSS $\langle R_{s_1s_2}^2 \rangle$ for an observation window $R_o$ between $s_1$ and $s_2$ that is far from the polymer's midpoint, strong asymmetric patterns emerge only when the active segment $s_A$, between $f_1L$ and $f_2L$, is moved close to $R_o$.
Otherwise, the activity does not strongly affect fluctuations in this part, $R_o$, of the chain. 
In Figs. \ref{fig:R2mn}(b) and (c), when we shift $R_o$ closer to the middle, the influence of the location of $s_A$ on the MSS becomes more symmetric, resulting in a more balanced patterns across the $f_1-f_2$ space.
As expected, Fig. \ref{fig:R2mn}(d) is symmetric to Fig. \ref{fig:R2mn}(b).

\begin{figure*}[htp]
\centering
\includegraphics[width=\linewidth]{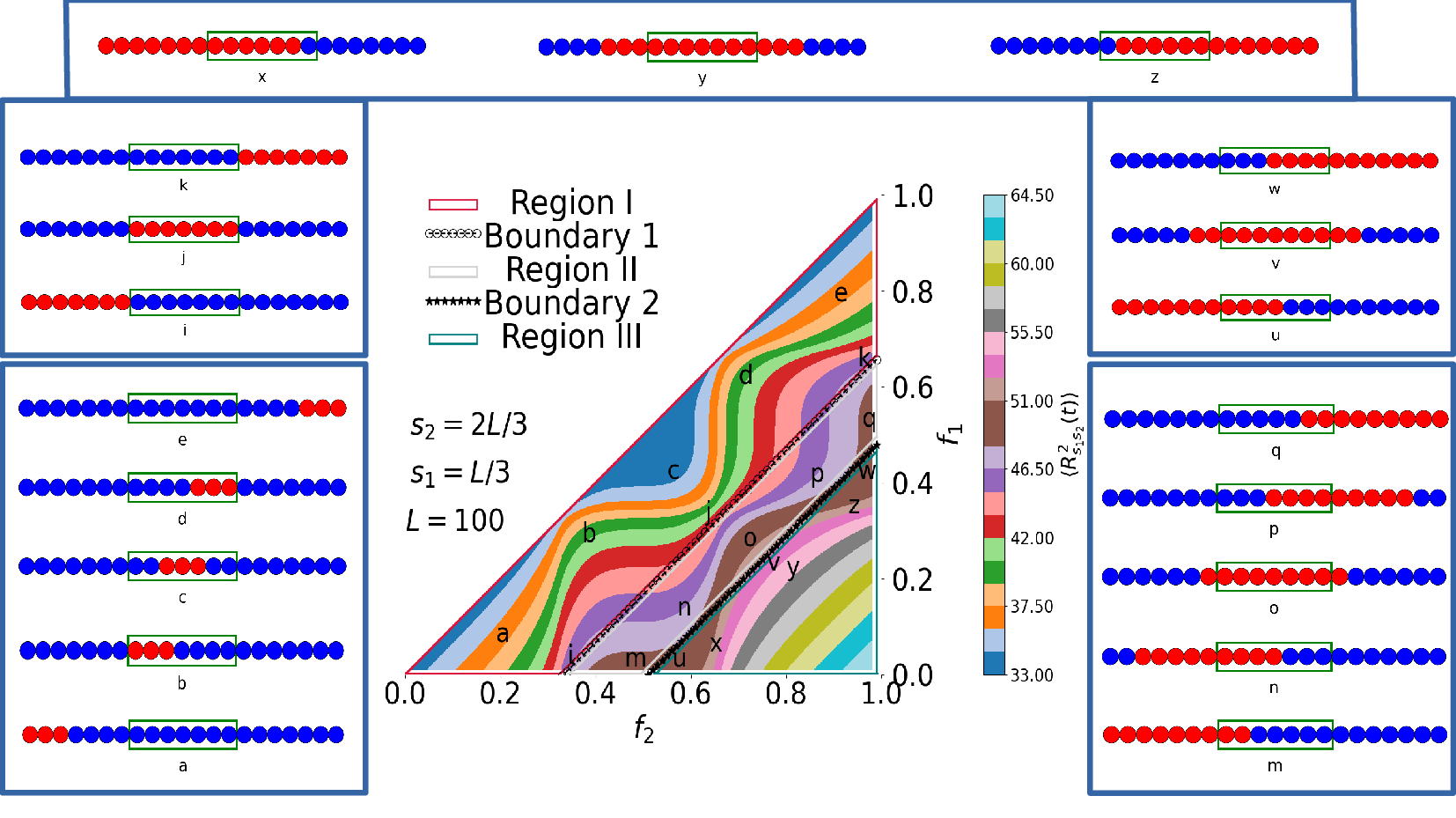}
\caption{$\langle R_{s_1s_2}^2(t) \rangle$ [Eq. (\ref{Rmn_1})] for a middle segment  where $s_1=L/3,\,s_2=2L/3$. Three distinct regions  (I, II, and III) are indicated by three shaded areas, separated by two boundaries. Conformations at various key points ($a$, $b$, ..., $z$) within the regions and boundaries are illustrated with cartoons: red beads represent active segments, blue beads denote passive segments, and beads enclosed within the green box belong to the observation window  $R_o$ where $\langle R_{s_1s_2}^2(t) \rangle$ is observed.  
The values of other parameters are the same as in Fig. \ref{fig:rg}.}
\label{fig:R2mn_c}
\end{figure*}

For further insights, we consider a middle segment of length $L/3$  as $R_o$, extending from $s_1=L/3$ to $s_2=2L/3$ along the contour, as illustrated in Fig.  \ref{fig:R2mn_c}.   In terms of the length of the active segment $|s_A|=(f_2-f_1)L$,  we can distinguish three regions separated by two boundaries located at $|s_A|=L/3$ and $|s_A|=L/2$, denoted by $L_1$ and  $L_2$, respectively. The three regions are classified as region I $(0<|s_A|< L/3)$, region II $(L/3<|s_A|< L/2)$, and region III $(L/2<|s_A|\leq L.)$

In region I, $|s_A|$ is less than one-third of the polymer length.
When $s_A$ is positioned at either boundary of the window $R_o$, such as $b$ and $d$ in Fig. \ref{fig:R2mn_c}, the MSS reaches its highest values, whereas it is minimized when $s_A$ is  moved to the midpoint of this domain,  as $c$ in Fig. \ref{fig:R2mn_c}.
 From $a$ to $e$, the MSS exhibits two maxima at  $b$ and $d$ separated by a minimum $c$. Notably, at the boundary $|s_A|=L_1$, the MSS within $R_o$ shows minimal variation, regardless of the active segment’s position.

Moving to region II, where $|s_A|$ exceeds the size of the observation window $|R_o|$ but is less than half the polymer chain length, part of the active segment always overlaps with $R_o$, resulting in three maxima for the MSS at $m$, $o$, and $q$ in Fig. \ref{fig:R2mn_c}.
Similar to region I, two maxima occur at $m$ and $q$ when  $s_A$ is near the polymer’s two ends. When $s_A$ moves towards the polymer center, the MSS decreases, passing through minima at $n$ and $p$, then rises again afterward. When $s_A$ completely covers $R_o$, the MSS reaches another maximum at $o$, as all points within $R_o$ are actively driven, thereby exhibiting the highest activity in that region. However, the MSS value at $o$ is slightly lower or equal to the values for the other two maxima at $m$ and $q$.

When $s_A$ has a length of $L/2$ and extends from one end of the polymer to the midpoint of $R_o$, as shown at $u$ and $w$ in Fig. \ref{fig:R2mn_c}, its MSS reaches a minimum, with a value slightly lower than the maximum at $v$ when $s_A$ and $R_o$ share the same center. 
Finally, beyond $|s_A|=L/2$, the  maxima induced by tail activity disappear, leaving a single maximum when $s_A$ is centrally positioned, as shown at $y$.   We observed in Fig. \ref{fig:r2e} that the full polymer always extends less when $s_A$ moves toward the center. However, moving from $x$ or $z$ to $y$ in Fig. \ref{fig:R2mn_c}, we observe an increased extension within the window $R_o$ instead. This shows that in this region, reducing the separation between the centers of $s_A$ and $R_o$ greatly facilitates the extension within the window, overcompensating for the reduction of the overall swelling. When $|s_A|$ approaches the total contour length $L$, the MSS values converge to a fixed value, and the placement of $s_A$ has a negligible influence.

 In Fig. \ref{fig:rmn_L}, we consider the MSS as a function of $L$ for different segments.
Like other conformational quantities,  $\langle R_{s_1s_2}^2 \rangle$ exhibits similar scaling with $L$, where $\langle R_{s_1s_2}^2 \rangle \propto L^{\alpha}$ with $\alpha>1$ for short chains and $\alpha=1$ for long chains. This anomalous scaling arises due to the persistence in motion, as noted earlier [cf. Eqs. (\ref{rg0_05})-(\ref{rg0_051})].

\section{Dynamical properties \label{sec-dyn}}
The mean square displacement (MSD) of a point at arc length $s$ is defined as 
\begin{align}
  \Delta^2(s,t) =\langle\left[\bs{r}(s,t)-\bs{r}(s,0)\right]^2\rangle.\label{msd_s}
\end{align}
Plugging Eqs. (\ref{r(s,t)}) and (\ref{normal_corr})  into Eq. (\ref{msd_s}), 
the  MSD can be rewritten as 
\begin{align}
 &\Delta^2(s,t)\nonumber\\
 &=\Delta_{cm}^2(t)+4 \sum_{p=1}^{p=\infty} \text{cos}\Bigg(\frac{p\pi s}{L}\Bigg)\left[2\rho_{p0}(0)-\rho_{p0}(t)-\rho_{0p}(t)\right]\nonumber\\
 &+4  \sum_{p,q=1}^{p,q=\infty} \text{cos}\Bigg(\frac{p\pi s}{L}\Bigg)\text{cos}\Bigg(\frac{q\pi s}{L}\Bigg)\left[2\rho_{pq}(0)-\rho_{pq}(t)-\rho_{qp}(t)\right]\label{msd_sth_t}.
\end{align}

The MSD of the CM, given in Eq. (\ref{msd_cm}) and shown in panel (a) of Fig. \ref{fig:msd_cm}, resembles that of an active Ornstein-Uhlenbeck particle, which corresponds to translational motion in the context of the polymer \cite{osmanovic2017dynamics,gosw19,goswami2022reconfiguration}. For the passive case, the dynamics are always Fickian. For a partially active polymer, however, intermediate times reveal superdiffusive behavior, with a peak  in $\alpha(t)$ occurring at the timescale $t \approx \tau_A$. Notably, for the same persistence time but with different activity strengths, the peak height increases at $t = \tau_A$. Eventually, the dynamics transition to diffusive behavior at long timescales, $t\gg \tau_A$.
It is also noteworthy that the peak height at  the superdiffusive region depends solely on the arc length of the active segment, rather than its location, similar to the  behavior associated with the energy of the polymer.

\begin{figure*}[htp]
    \centering
    \begin{subfigure}[b]{0.495\textwidth}
    \caption{ }
        \centering
        \includegraphics[width=\textwidth]{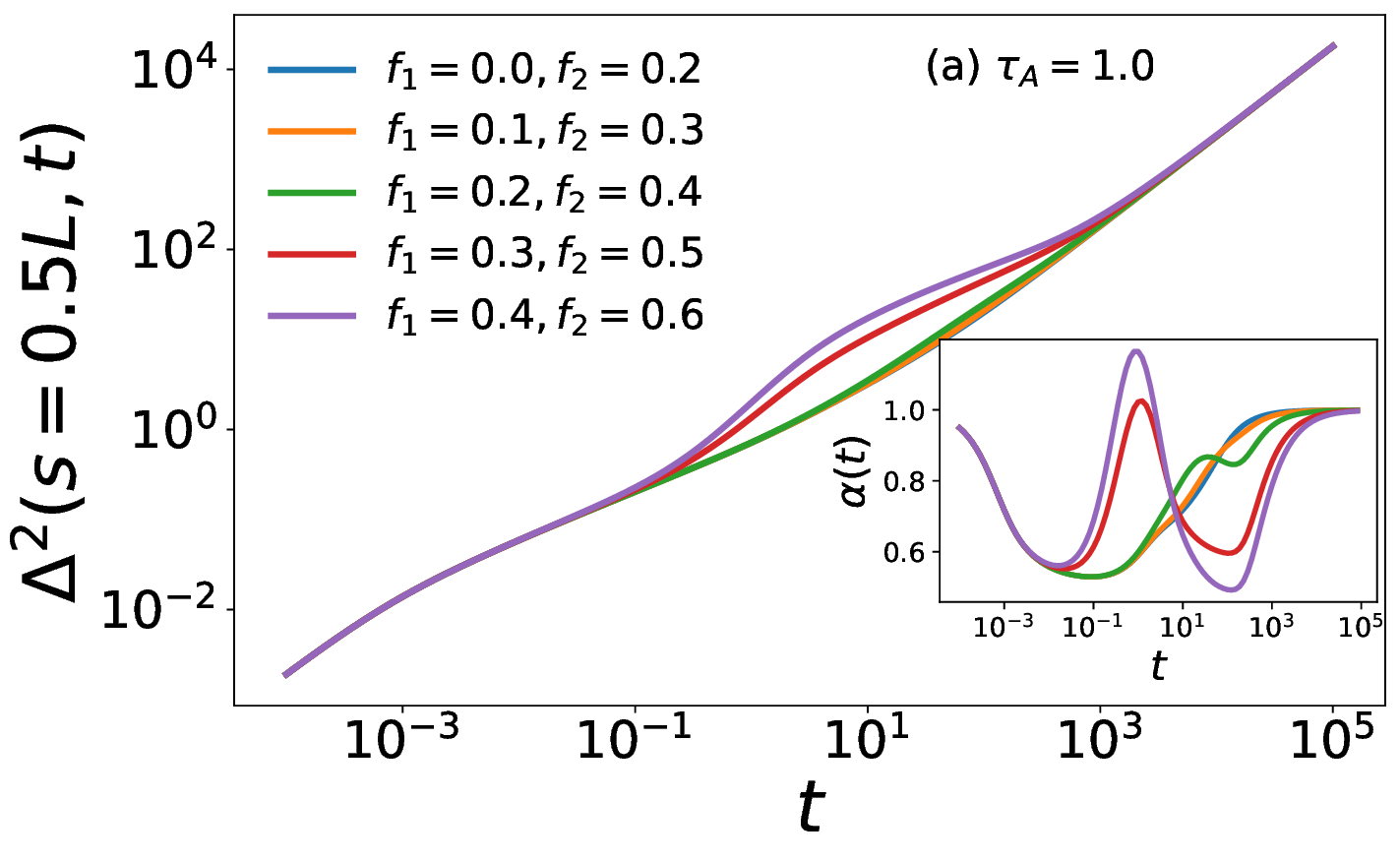}
    \end{subfigure}
    \hfill
    \begin{subfigure}[b]{0.495\textwidth}
    \caption{ }
        \centering
        \includegraphics[width=\textwidth]{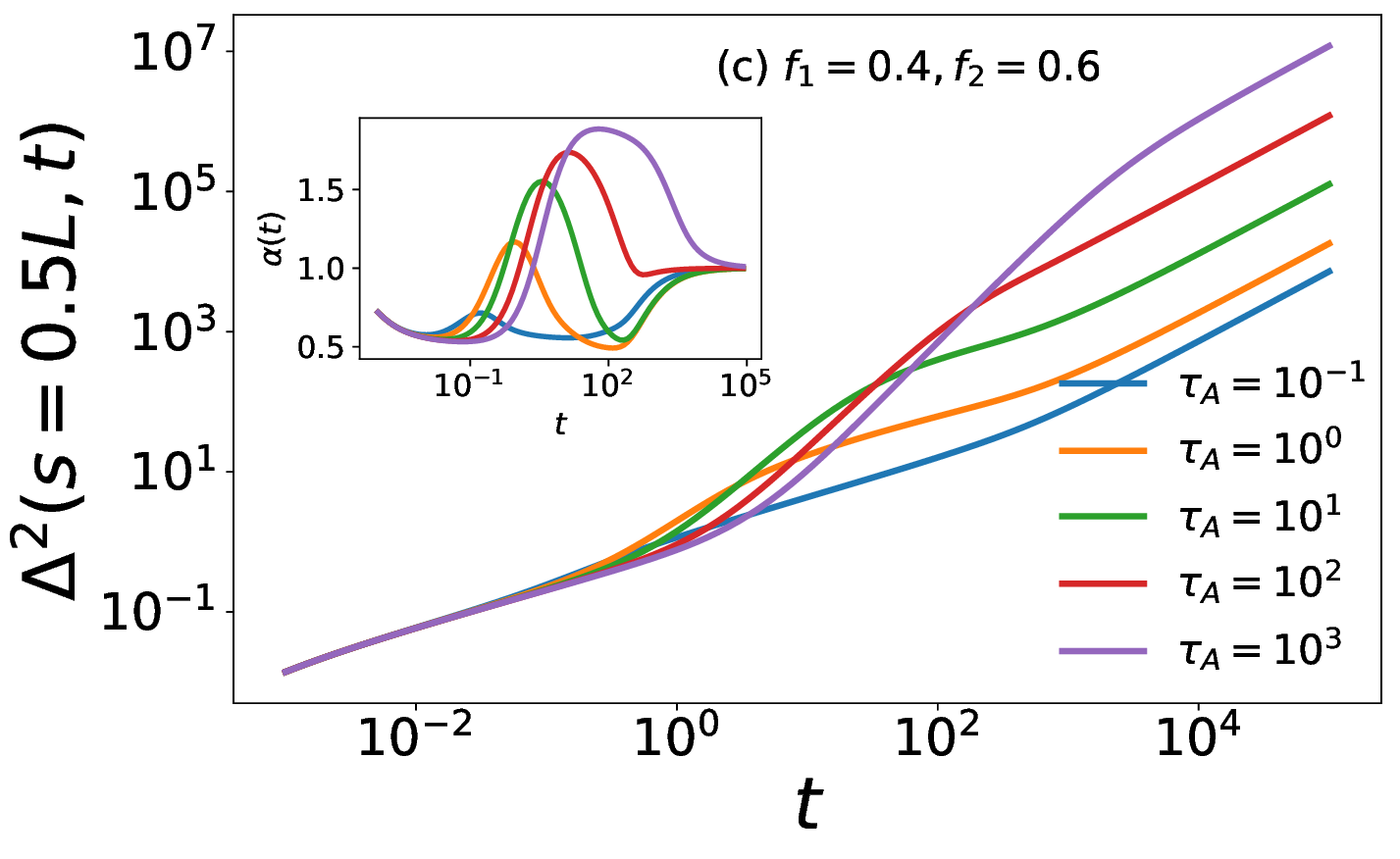}
    \end{subfigure}
        \vspace{0.2cm}
        \centering
        \begin{subfigure}[b]{0.495\textwidth}
    \caption{ }
        \centering
        \includegraphics[width=\textwidth]{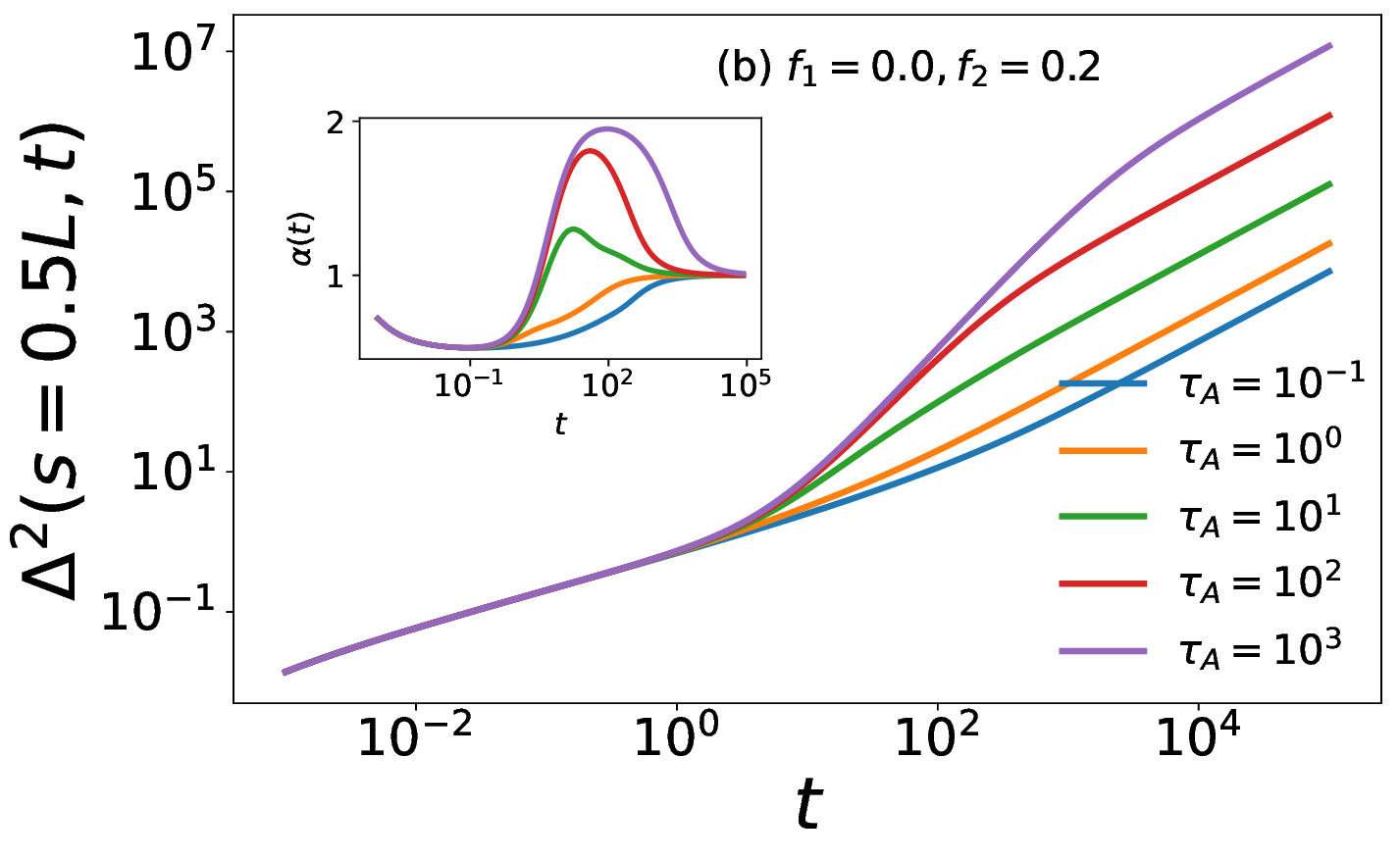}
    \end{subfigure}
    \caption{Log-log plot of $\Delta^2(s=0.5L,t)$ as a function of time $t$,  with the inset showing the local exponent. The values of other parameters are $\kappa=3.0,\,\zeta=1.0,\,k_BT=1.0,\,L=100$,  and $F_0^2=10.0$.}
   \label{fig:delta2_t}
\end{figure*}

The MSD of a tagged point located at the midpoint of the contour,  $s=L/2$, is computed numerically using Eq. (\ref{msd_sth_t}) and is shown  in Fig. \ref{fig:delta2_t}. For small persistence times $\tau_A\ll \tau_1$, the MSD exhibits distinct behavior depending on the location of the active region along the contour. We call an active segment  ``tagged" if it contains the tagged point.  When $s_A$ is tagged, the diffusive behavior of the polymer transitions as follows: diffusive → subdiffusive → superdiffusive → subdiffusive → diffusive.
When $s_A$ is situated farther from the tagged point, the intermediate superdiffusive behavior, characterized by a higher scaling exponent $\alpha$ [as shown in the inset of Fig. \ref{fig:delta2_t}(a)], diminishes, and if $s_A$ is even further away, only subdiffusive behavior is observed. Similar dynamic characteristics have been reported recently for the effect of point activity on a Rouse chain \cite{dutta2024effect}.

\begin{figure*}[htp]
    \centering
    \begin{subfigure}[b]{0.495\textwidth}
     \caption{ }
        \centering
        \includegraphics[width=\textwidth]{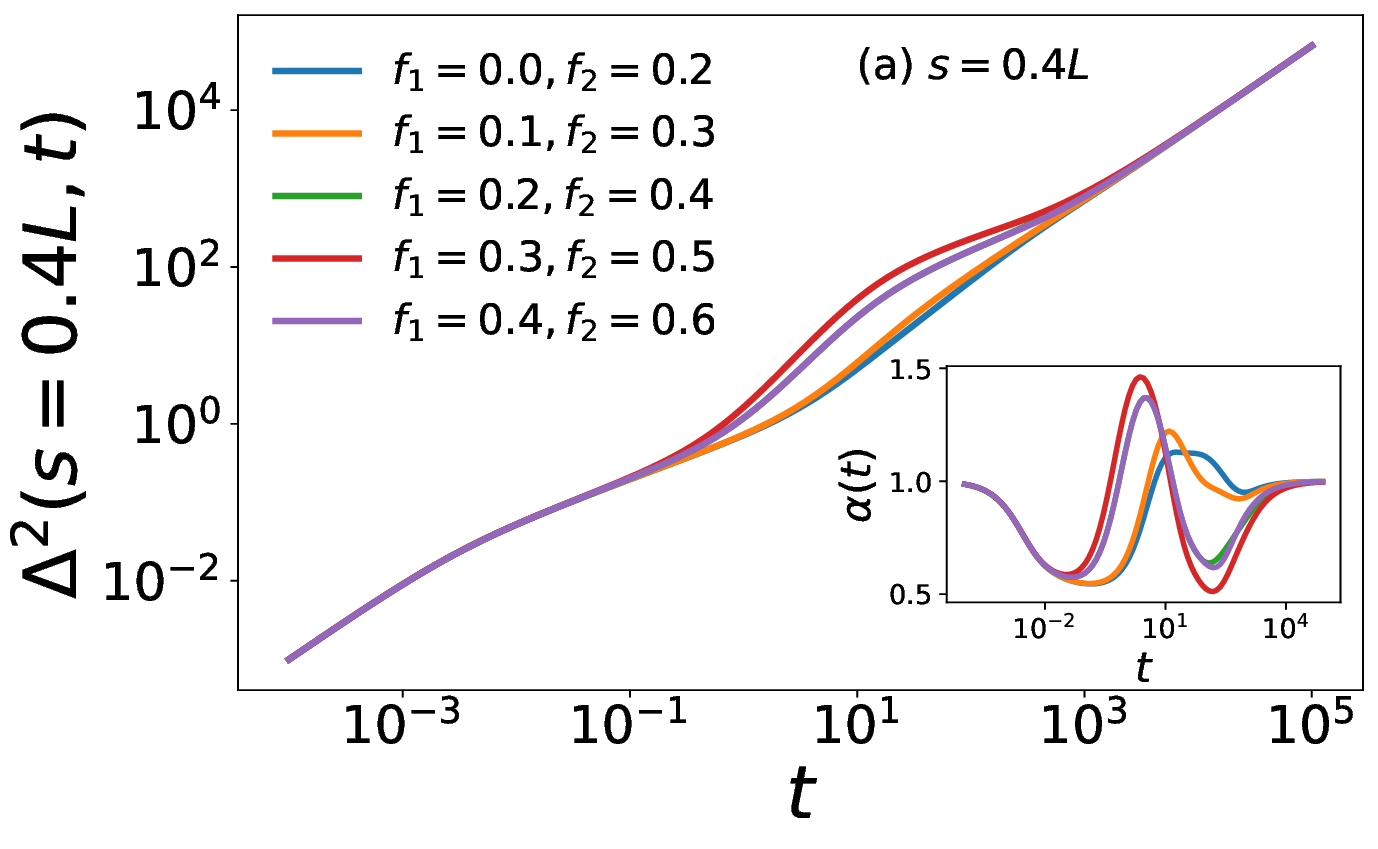}
    \end{subfigure}
    \hfill
    \begin{subfigure}[b]{0.495\textwidth}
     \caption{ }
        \centering
        \includegraphics[width=\textwidth]{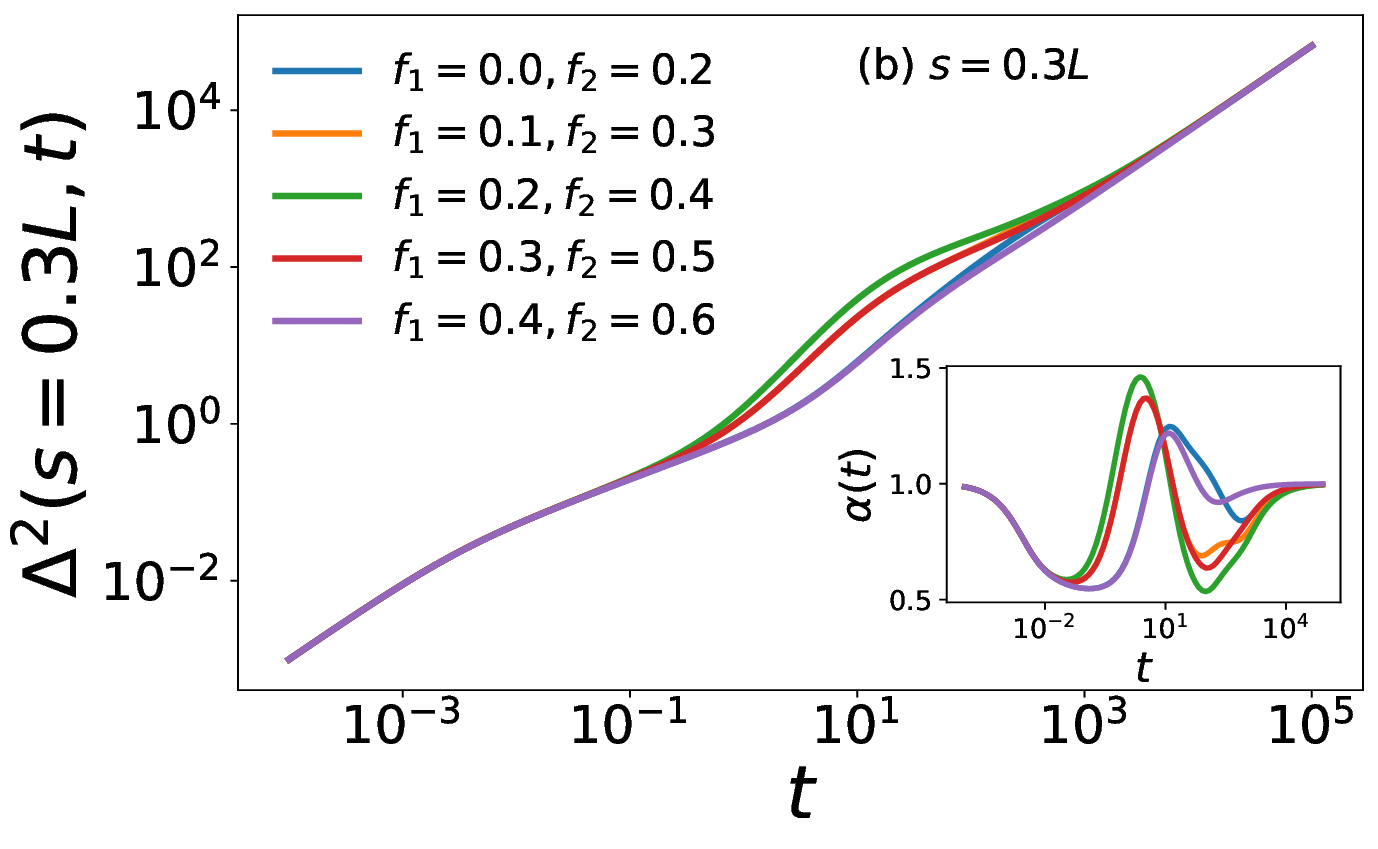}
    \end{subfigure}
    \caption{Plot of $\Delta^2(s,t)$ as a function of time $t$ in log-log scale for $\tau_A=5.0$,  with the inset showing the local exponent. The values of other parameters are the same as in Fig. \ref{fig:delta2_t}. }
   \label{fig:delta2_t_end}
\end{figure*}

As the persistence time increases, notably beyond the Rouse time $\tau_1$, the second subdiffusive regime disappears, even in the case of a tagged active segment, as shown in panel (b) of Fig. \ref{fig:delta2_t}. For distant active regions, as  exemplified in Fig. \ref{fig:delta2_t}(c),
we observe a later superdiffusive behavior instead of subdiffusion. Additionally, as the persistence time increases, the peak of $\alpha$ becomes more pronounced and occurs at progressively longer times. Similar results have been observed for caged active tracers in polymeric gel networks \cite{goswami2024}, though in that case, tracers exhibit complete caging in the initial phase (i.e., $\alpha \approx 0$ in the subdiffusive regime), likely due to the larger size effect of the tracer.

To generalize these observations, we also consider different locations of the tagged point along the chain, such as $s=0.4L$ and $0.3L$ in Figs. \ref{fig:delta2_t_end}(a) and (b), respectively. It is evident that for a tagged active segment, there is an initial phase of enhanced superdiffusion, followed by a more pronounced subdiffusive behavior at later times. Notably, this effect is most prominent when the tagged point is located at the center of the segment, as demonstrated in Fig.  \ref{fig:delta2_t_end} for $(f_1,f_2)=(0.3,0.5)$ in panel (a), and for $(f_1,f_2)=(0.2,0.4)$ in panel (b). In contrast, active regions that are not proximal to the tagged point, as exemplified by the curve for $(f_1,f_2)=(0.0,0.2)$ in panel (a) and $(f_1,f_2)=(0.4,0.6)$ in panel (b) of Fig. \ref{fig:delta2_t_end}, exhibit mild superdiffusive behavior (with $\alpha$ close to 1), followed by slight subdiffusion (with $\alpha \approx 1$) at intermediate to long times in the limit of small persistence time, such as at $\tau_A=5.0$ in this plot. Thus, by analyzing the dynamics of a tagged point, one can effectively map the activity profile along the chain.

If the end-to-end distance $\bs{R}_e(t)$ is a column vector, then its autocorrelation can be expressed in terms  of the covariance between two mode amplitudes as
\begin{align}
\Phi_{corr}(t,t')&= \Phi_{corr}(|t-t'|)=\langle \bs{R}_e(t)^{\rm T}\bs{R}_e(t') \rangle\nonumber\\
&= 48 \sum_{p,q=1}^{\infty} \, \langle \chi_{2p-1}(t) \chi_{2q-1}(t') \rangle , \label{phi_corr_00}
\end{align}
and can be decomposed into passive and active contributions, as given by Eq. (\ref{phi_corr_0}).
We define the normalized correlation function of the end-to-end distance as 
\begin{align}
\Tilde{\Phi}_{corr}(t)= \frac{\Phi_{corr}(t)}{\Phi_{corr}(0)}\label{phi_corr},    
\end{align}
where, by construction, $\Tilde{\Phi}_{corr}(0)=1.$ 
In Fig. \ref{fig:corr_nu_f12}, the decay of the normalized correlation function $\Tilde{\Phi}_{corr}(t)$ over time is presented. In the two opposing  limits $\tau_1 \gg \tau_A$ and $\tau_A \gg \tau_1$, the decay is primarily governed by $\tau_1$ and $\tau_A$, respectively.
Specifically, in the later limit,
the polymer exhibits slower  relaxation due to the persistence introduced by activity.
As a comparison between Figs. \ref{fig:corr_nu_f12}(a) and \ref{fig:corr_nu_f12}(b), when activity is localized at the end segments, the relaxation dynamics of the end-to-end distance is more significantly influenced, whereas activity at the midsegment has minimal impact on overall relaxation. The regime where $\tau_1 \sim \tau_A$ is particularly interesting, as the interplay between these timescales introduces competition between two factors [see Eq. (\ref{phi_corr_0A})]: The longer persistence caused by active forces slows the initial decay of $\Tilde{\Phi}_{corr}$, and more rapid decay over the timescale $\tau_1$  is driven  by the thermal fluctuations.
Ultimately, at longer times, the decay is governed by the Rouse relaxation time $\tau_1$.
When $s_A$ is gradually elongated from one end to the other,  as shown in Figs.  \ref{fig:corr_nu_f12}(a), (c), to (d), the decay of  $\Tilde{\Phi}_{corr}(t)$ over time becomes slower for larger $\tau_A$ due to persistence. In the figures, this behavior is particularly evident for $\tau_A=500.$

\section{Reconfiguration and looping kinetics \label{sec-loop}}
\begin{figure*}[htp]
    \centering
    \begin{subfigure}[b]{0.495\textwidth}
     \caption{$(f_1,\,f_2)=(0.0,\,0.2).$ }
        \centering
        \includegraphics[width=\textwidth]{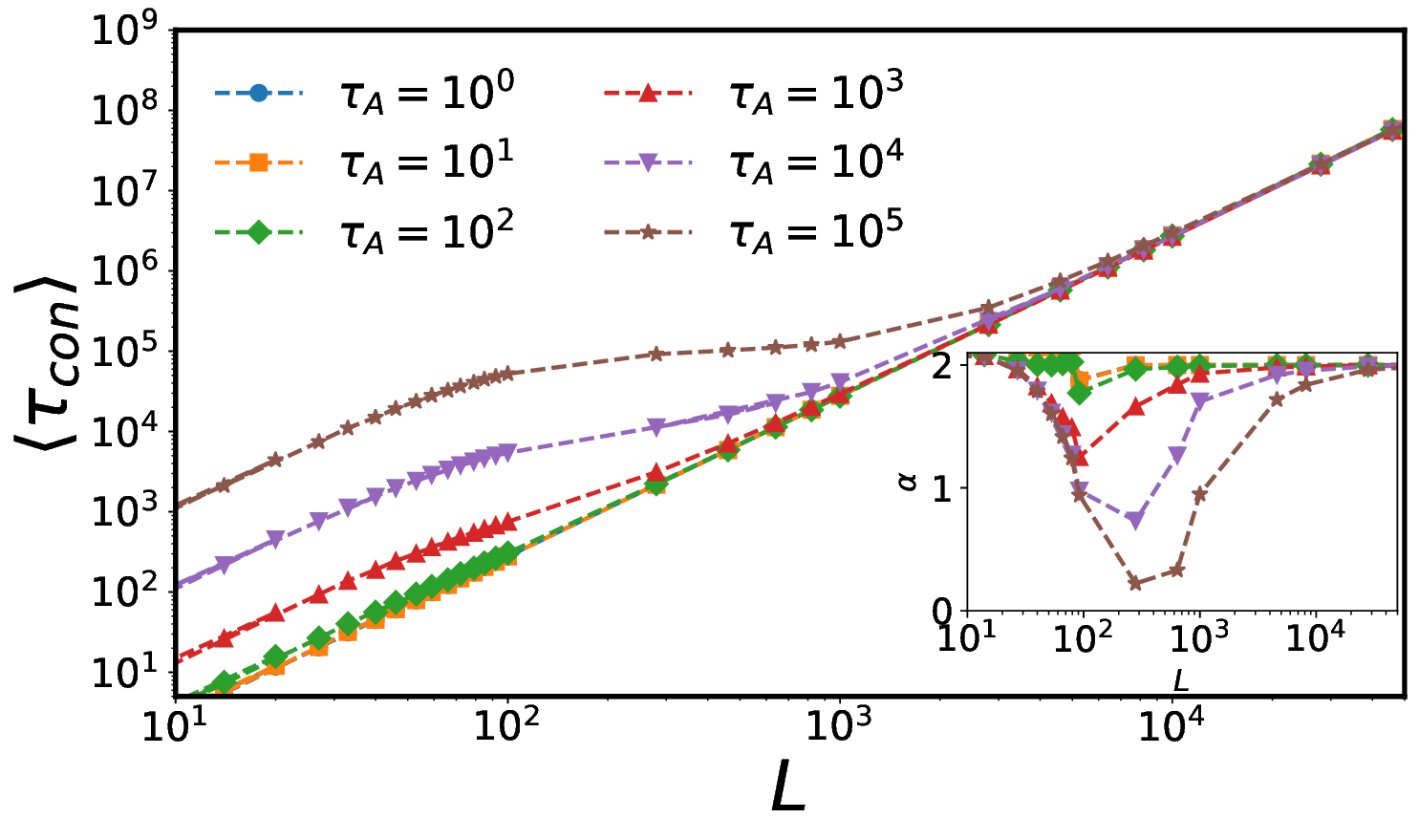}
    \end{subfigure}
    \hfill
    \begin{subfigure}[b]{0.495\textwidth}
     \caption{$(f_1,\,f_2)=(0.4,\,0.6).$}
        \centering
        \includegraphics[width=\textwidth]{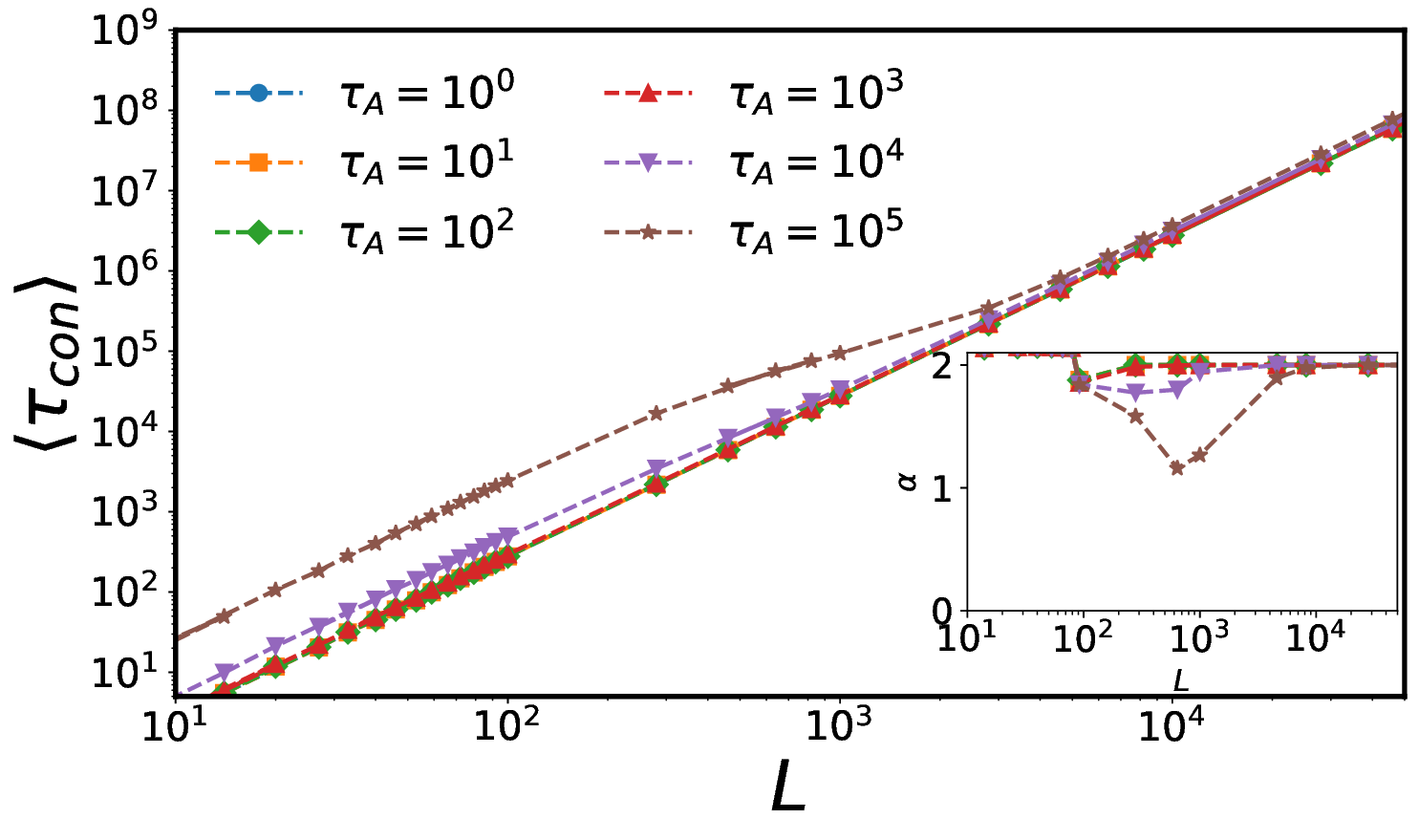}
    \end{subfigure}
        \vspace{0.2cm}
        \centering
    \begin{subfigure}[b]{0.495\textwidth}
     \caption{$(f_1,\,f_2)=(0.0,\,0.5).$ }
        \centering
        \includegraphics[width=\textwidth]{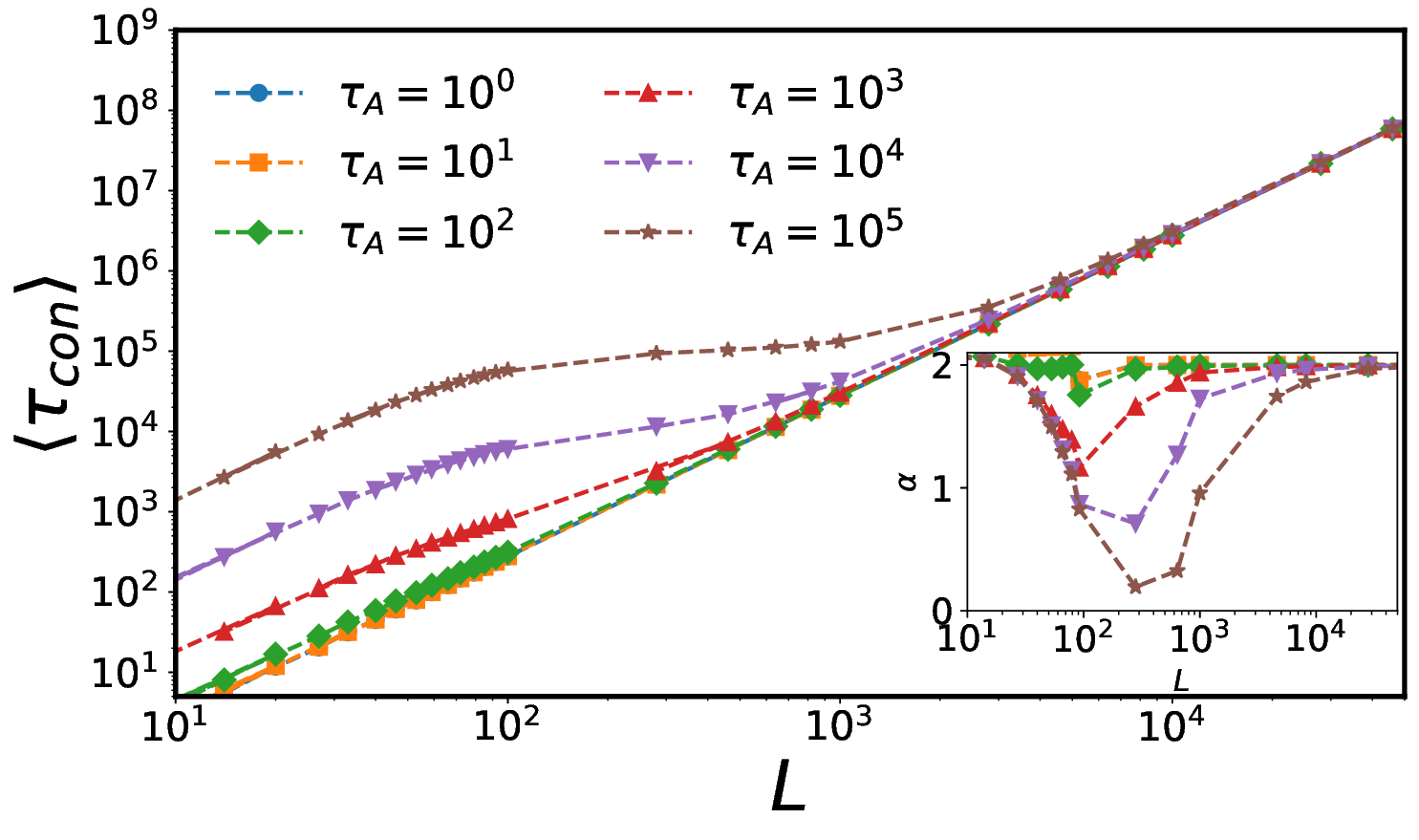}
    \end{subfigure}
    \hfill
    \begin{subfigure}[b]{0.495\textwidth}
    \caption{$(f_1,\,f_2)=(0.0,\,1.0).$}
        \centering
        \includegraphics[width=\textwidth]{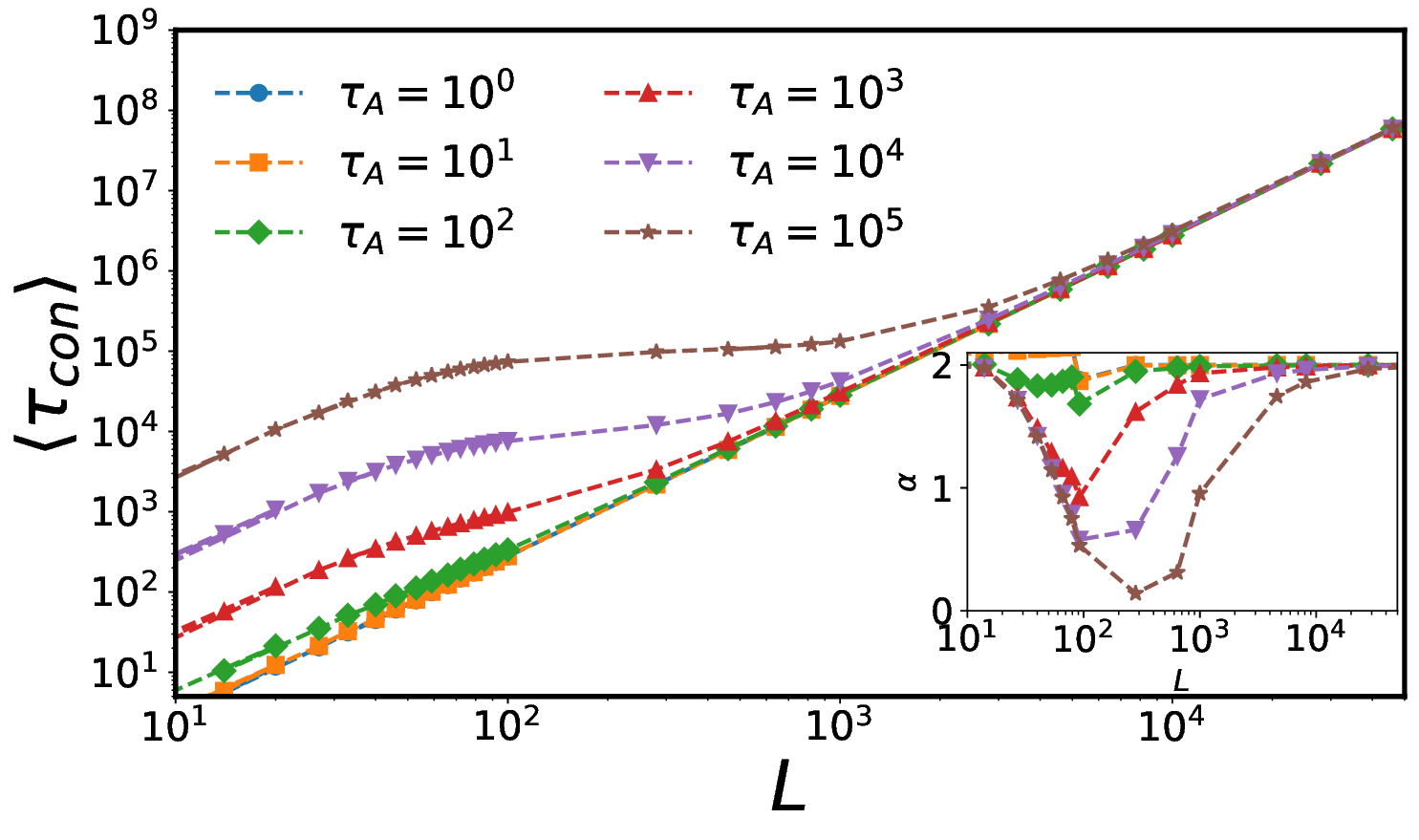}
    \end{subfigure}
\caption{Reconfiguration time versus contour length of the polymer, plotted in log-log scale. Each panel considers specific values of $f_1$ and $f_2$, as indicated in the respective captions, for various values of $\tau_A$. Here, $ \langle \tau_{\text{con}}\rangle \propto L^{\alpha}$, with the exponent $\alpha$ shown as a function of $L$ in the insets. The values of other parameters are $\kappa=3.0$, $\zeta=1.0$, $k_BT=1.0$, and 
$F_0=0.1$.}
    \label{fig:tcon_l}
\end{figure*}

The overall relaxation of a polymer's end-to-end distance can be quantified by the reconfiguration time which  is defined as 
\begin{align}
 \langle \tau_{con}\rangle=\int_{0}^{\infty}dt\,\Tilde{\Phi}_{corr}(t).\label{t_con}   
\end{align}
The time  $\langle \tau_{con}\rangle$  primarily corresponds to the slowest decay timescale in the correlation function, $\Tilde{\Phi}_{corr}(t)$. 
For a passive polymer, the reconfiguration time of the relaxation dynamics, $\langle \tau_{con} \rangle = \langle \tau_{con} \rangle_T$, follows Eq. (\ref{t_con_passive}), which depends solely on $\tau_1$ and, therefore, scales as $L^2$ for any polymer length [see Eq. (\ref{tau1})].
For a partially active polymer, the scaling of $\langle \tau_{con} \rangle$ in Fig. \ref{fig:corr_nu_f12} deviates from $L^2$, suggesting that both timescales, $\tau_1$ and $\tau_A$, contribute to this behavior.
According to Eq. (\ref{t_con_act}), $\langle \tau_{con} \rangle_A \sim \tau_1 \sim L^2$ for short polymers, while $\langle \tau_{con} \rangle_A \sim \tau_1^0$ for long polymers.
Since $\langle \tau_{con} \rangle$ for a partially active polymer is a linear combination of $\langle \tau_{con} \rangle_T$ and $\langle \tau_{con} \rangle_A$, its scaling is $L^2$ in both the short and long polymer limits.
For intermediate polymer lengths  $L \approx \sqrt\frac{\tau_A \kappa \pi^2}{\zeta}$, where $\tau_1 \approx \tau_A$,  $\langle \tau_{con} \rangle_A$, and consequently $\langle \tau_{con} \rangle$, scale as $L^\alpha$, with $0 < \alpha < 2$.
When $s_A$ is localized at the polymer ends, the deviation of $\alpha$ from 2 becomes more significant and the minimum of $\alpha$ shifts toward longer length scale, as shown in panels (a), (c), and (d) of Fig. \ref{fig:tcon_l}.
In contrast, for polymers with $s_A$ localized in the middle segment, $\langle \tau_{con} \rangle$ is less influenced by activity, with $\alpha$ remaining near 2, except at very high $\tau_A$, as seen in Fig. \ref{fig:tcon_l}(b).

\begin{figure*}[htp]
    \centering
    \begin{subfigure}[b]{0.495\textwidth}
    \caption{$(f_1,\,f_2)=(0.0,\,0.2).$ }
        \centering
        \includegraphics[width=\textwidth]{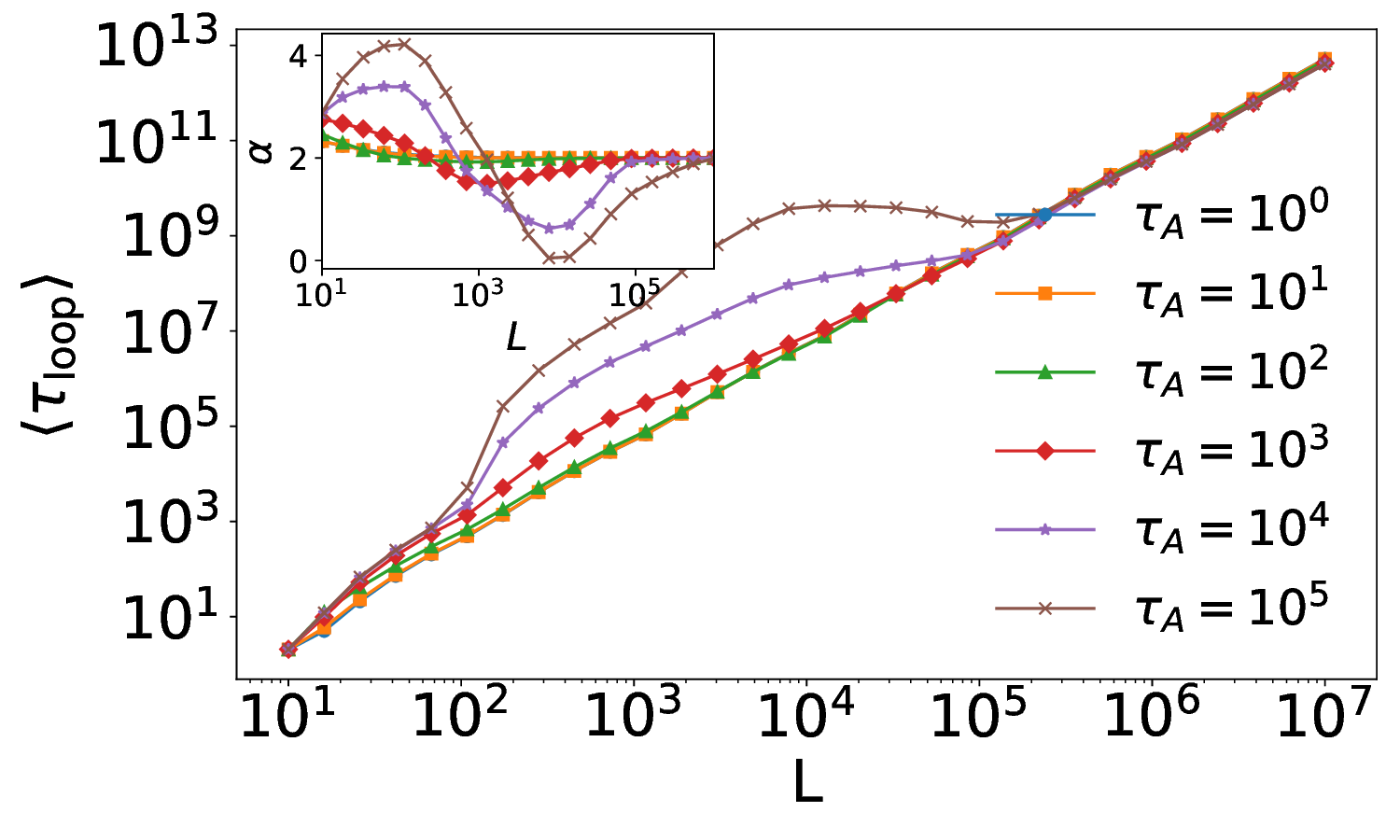}
    \end{subfigure}
    \hfill
        \begin{subfigure}[b]{0.495\textwidth}
    \caption{$(f_1,\,f_2)=(0.01,0.21).$}
        \centering
        \includegraphics[width=\textwidth]{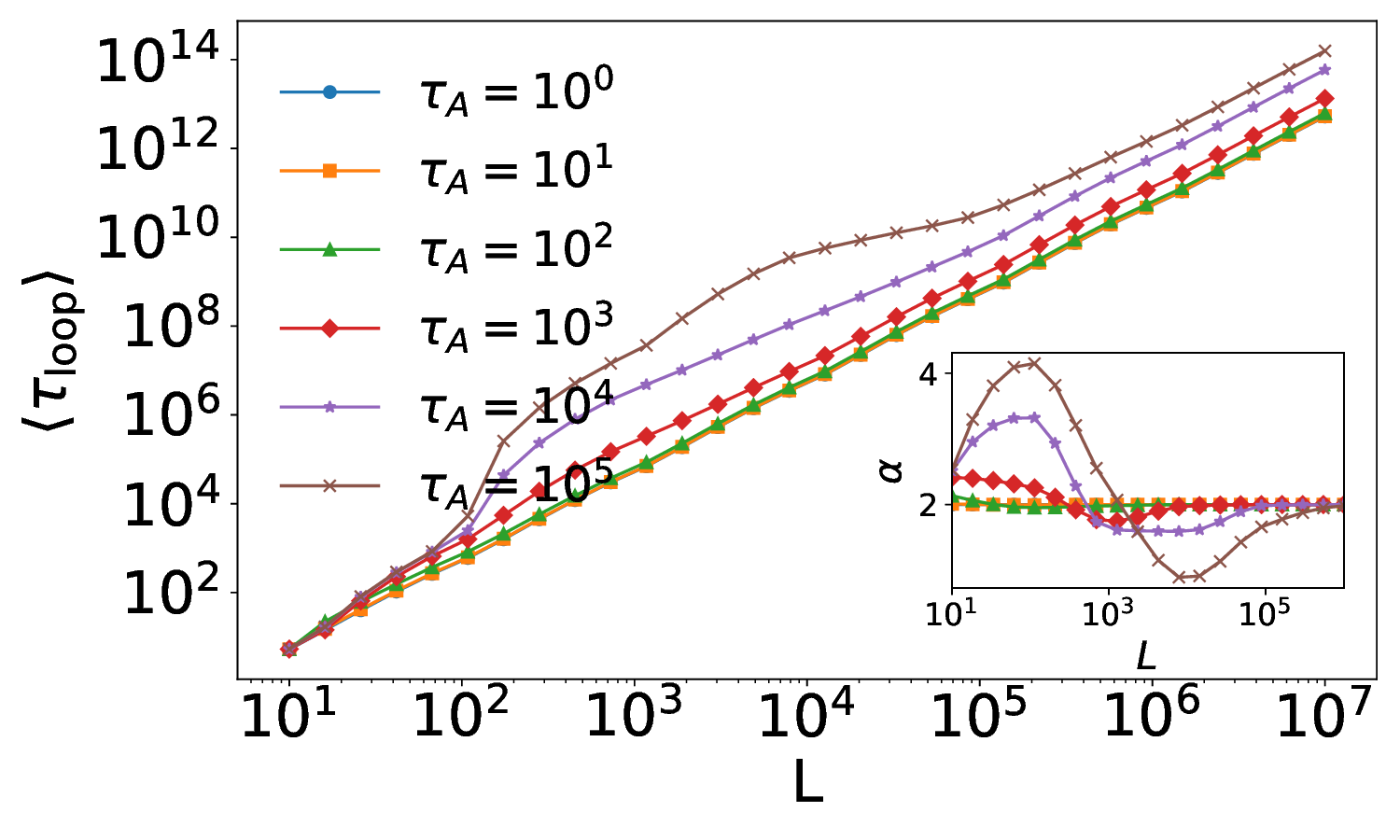}
    \end{subfigure}
    \vspace{0.2cm}
        \centering
   \begin{subfigure}[b]{0.495\textwidth}
     \caption{$(f_1,\,f_2)=(0.2,\,0.4).$}
        \centering
        \includegraphics[width=\textwidth]{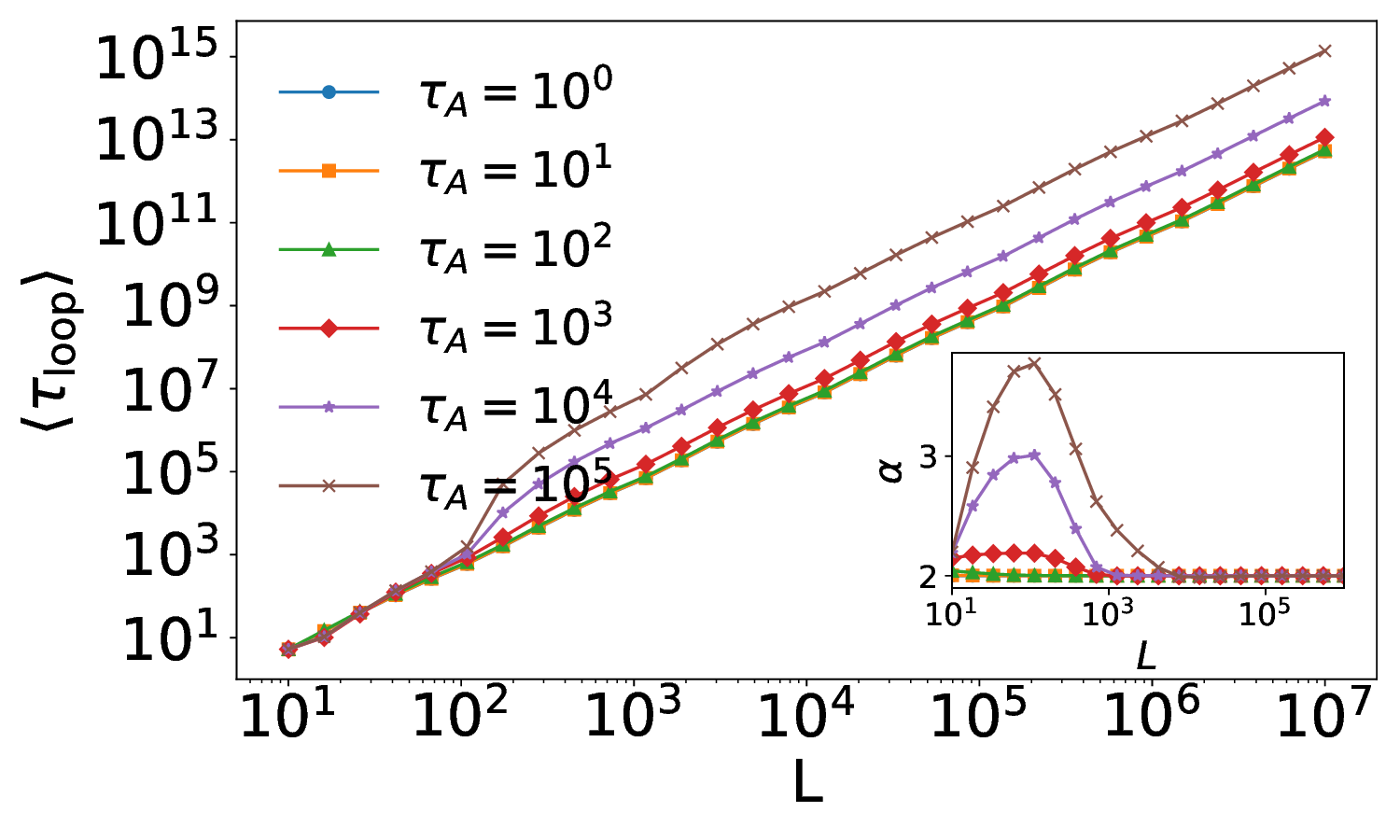}
    \end{subfigure}
        \hfill
    \begin{subfigure}[b]{0.495\textwidth}
    \caption{$(f_1,\,f_2)=(0.4,\,0.6).$}
        \centering
        \includegraphics[width=\textwidth]{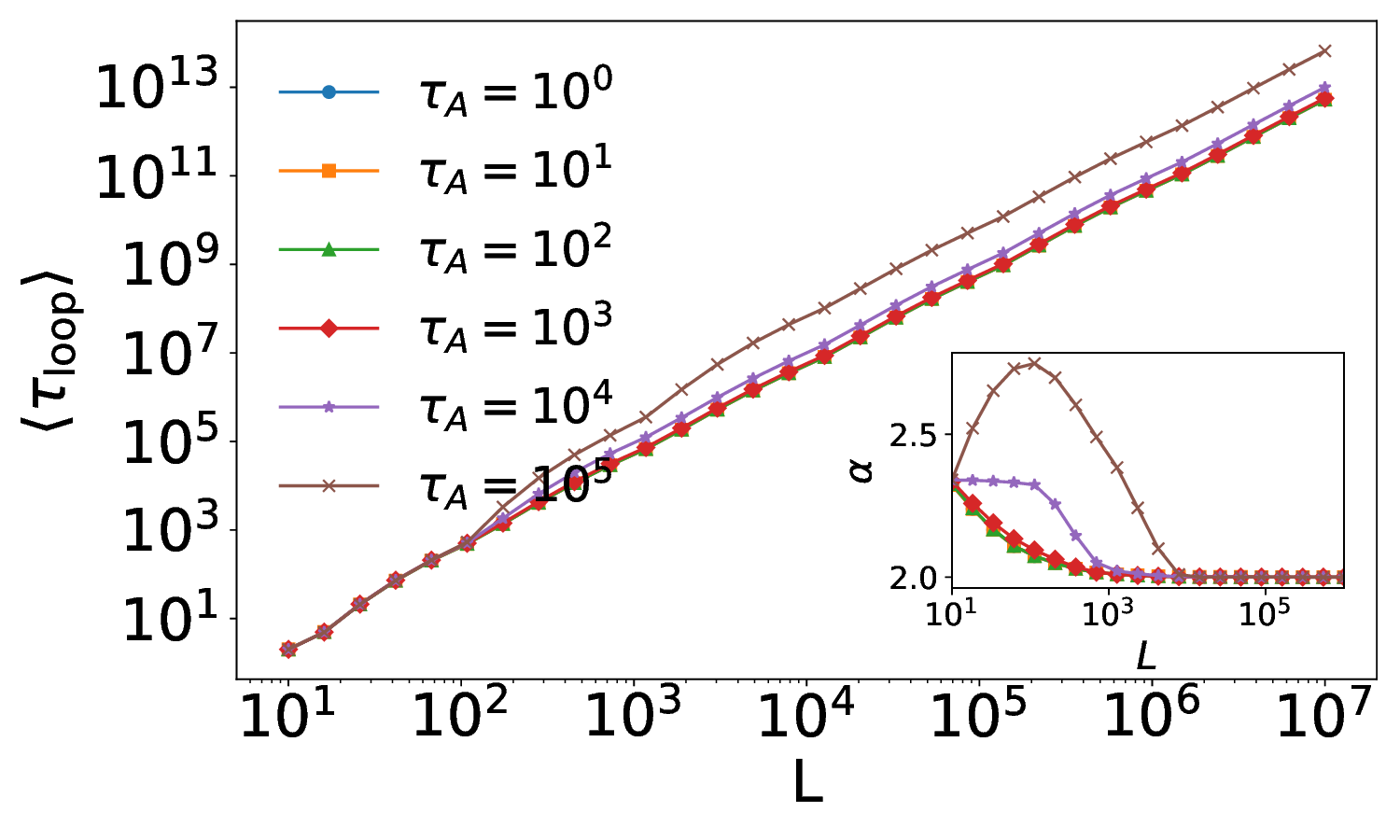}
    \end{subfigure}
    \vspace{0.2cm}
        \begin{subfigure}[b]{0.495\textwidth}
    \caption{$(f_1,\,f_2)=(0.0,\,0.5).$}
        \centering
        \includegraphics[width=\textwidth]{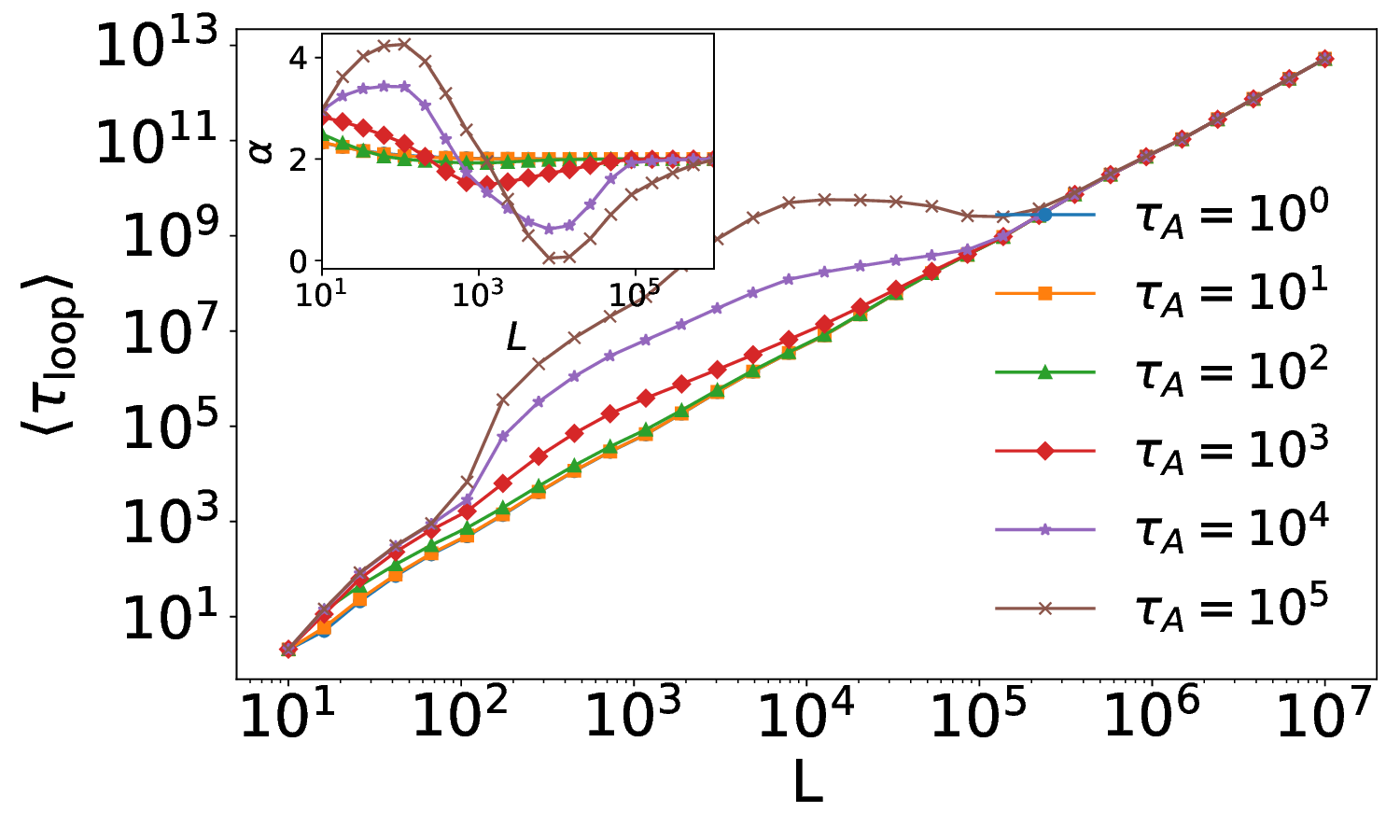}
    \end{subfigure}
    \hfill
        \centering
    \begin{subfigure}[b]{0.495\textwidth}
     \caption{$(f_1,\,f_2)=(0.0,\,1.0).$}
        \centering
        \includegraphics[width=\textwidth]{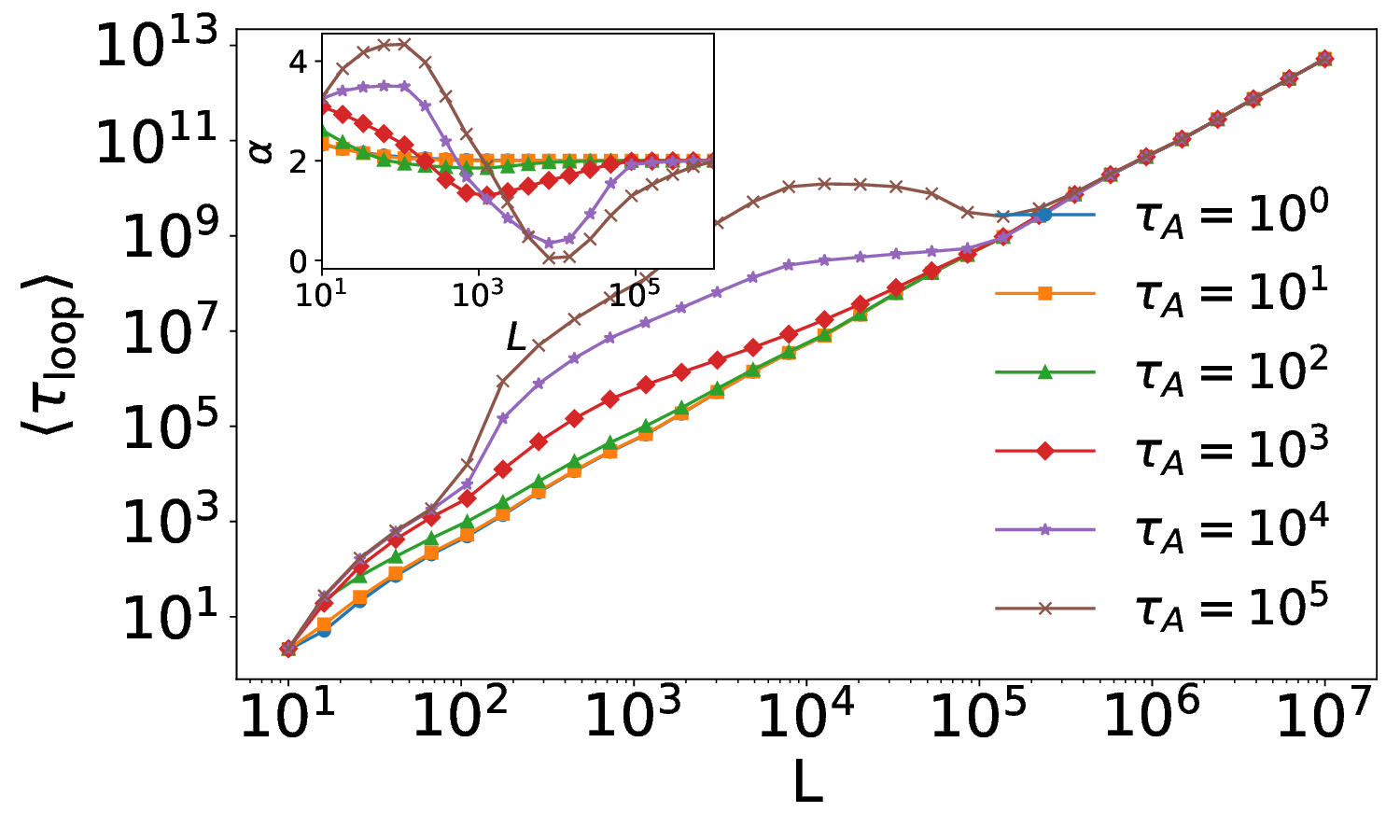}
    \end{subfigure}
    \caption{Log-log plot of looping time as a function of $L$. In insets, $\alpha$ is plotted  as a function of $L$ obtained from the relation  $ \langle \tau_{\text{con}}\rangle \propto L^{\alpha}$. The values of other parameters  are the same as in Fig. \ref{fig:tcon_l}. Here, $a_0$ is set to $a_0=0.01.$
    }
   \label{fig:looping_l}
\end{figure*}

When the two ends of a polymer come within a distance $a_0$—commonly referred to as the reaction radius—looping occurs between the end points. 
Within the Wilemski-Fixman (WF) approximation \cite{wilemski1973general}, the looping time can be expressed as [refer to Eq. (\ref{t_loop1})]
\begin{align}
 \langle \tau_{\text{loop}}\rangle = \int_{0}^{\infty} dt_1 \left(\frac{\mathcal{C}_{ss}(t_1)}{\mathcal{C}_{ss}(\infty)}-1\right),\label{t_loop}
\end{align}
where the sink-sink  correlation $\mathcal{C}_{ss}(t)$ is given by Eq. (\ref{sink-sink-corr}). 
Assuming $a_0$ is very small, the looping time can be approximated by Eq. (\ref{t_loop5}). For the convenience of the reader, it is restated here,
\begin{align}
 \langle \tau_{\text{loop}}\rangle = \tau_1\int_{0}^{1} \frac{dy}{y} \left[\frac{1}{\left(1-\Tilde{\Phi}_{corr}^2(y)\left[1-\frac43 z_0\right] \right)^{3/2}}-1\right],  \label{t_loop50}
 \end{align}
 where $z_0=\frac{3a_0^2}{2\Phi_{corr}(0)}.$ For a Rouse (passive) chain, the looping time scales as $\langle \tau_{\text{loop}}\rangle \propto L^{2}$, similar to reconfiguration time, as both are primarily determined by the relaxation of the first Rouse mode. 

To gain deeper insight, Fig. \ref{fig:looping_l} demonstrates the scaling behavior of $\langle \tau_{\text{loop}} \rangle$ for polymers with various active segments, revealing significant deviations from the Rouse scaling for short and intermediate chain lengths. In the small $L$ limit,  where $\tau_1 \ll \tau_A$, the looping time scales as $L^\alpha$ with $\alpha > 2$, due to activity-induced swelling that delays looping.
In the intermediate $L$ range, particularly in the range $L\approx\sqrt\frac{\tau_A \kappa \pi^2}{\zeta}$, the timescales $\tau_1$ and $\tau_A$ are comparable, so that the persistence due to activity and the relaxation of all Rouse modes—driven by random fluctuations—become competitive: The swelling effect is balanced by the relaxation of all Rouse modes.  The looping time is significantly enhanced, specifically when the end segment is active.  A typical example, as shown in Fig. \ref{fig:looping_l}(a), is the sharp rise of $\langle \tau_{\text{loop}} \rangle$ at $\tau_A = 10^5$, deviating significantly from the straight-line behavior observed for $\tau_A = 1.0$ beyond $L = 10^2$.  
For even longer chains where $\tau_1>\tau_A$, the swelling effect is suppressed by the thermal effect in Rouse mode relaxation, leading to a reduction in $\alpha$ below 2. For very large values of $L$, i.e., when $\tau_1 \gg \tau_A$, $\tau_1$ gradually becomes the sole factor determining the scaling behavior of $\langle \tau_{\text{loop}} \rangle$. In particular, for polymers with an active end segment, the looping times for different $\tau_A$ values eventually converge to a single curve, as illustrated in panels (a), (e), and (f) of Fig. \ref{fig:looping_l}. Correspondingly, the scaling exponent $\alpha$ returns to the Rouse limit, as seen in the insets.
As $s_A$  is positioned farther from the polymer’s end, the enhancement of the looping time in the intermediate $L$ regime diminishes. This is indicated by a decreasing maximum value of $\alpha$, as shown in the insets of panels (b), (c), and (d) in Fig. \ref{fig:looping_l}. 
Unlike the case with $s_A$ at one end, in the limit $\tau_1 \gg \tau_A,$ the looping times for a specific polymer length no longer overlap for different $\tau_A$ values. Instead, the looping time depends directly on $\tau_A$.

To assess the impact of swelling on looping, we examine the looping time for active segments of specific lengths positioned at various locations along the chain. In the regime where $\tau_1 \approx \tau_A$, $\langle \tau_{\text{loop}} \rangle$ in Fig. \ref{fig:loop_f1}(a) follows a nonmonotonic trend, reaching a minimum when $s_A$ is centrally located. This behavior, along with panel (d), aligns with the swelling pattern in Fig. \ref{fig:rg}, providing a clear explanation for the observed variations in looping time. However, when $\tau_1\gg \tau_A$, the looping time no longer follows the swelling trend. Instead, it exhibits a different nonmonotonic behavior, as shown in panels (b) and (d). For longer chains, the influence of persistence on the looping process becomes increasingly evident and variable. While the exact physical mechanism underlying this behavior remains an open question, it can be inferred that activity at an end segment facilitates the looping process between the two end points of a long polymer, whereas an active segment positioned farther from any end may either hinder or facilitate the process, depending on its specific location.

\section{Conclusions \label{sec-con}}
In this paper, we systematically investigate the impact of localized active segments on the conformation and dynamics of polymers. While it is well-established that increased activity along the polymer backbone can induce swelling, the influence of the position of activity along the chain remains unexplored. Here, we demonstrate that positioning an active segment near the polymer’s ends induces greater swelling than placing it centrally. This pattern emerges even with an active force that is delta-correlated in time, suggesting that persistence in the force is not a necessary condition. 

The effect of local activity on a specific segment $R_o$ of the polymer is also notable. Depending on the length of the active segment as compared to $R_o$, maximum or minimum fluctuations within the domain $R_o$ can occur. This is mirrored in the dynamics of a tagged point along the contour: when the point is situated within the active segment, its intermediate-time dynamics exhibit enhanced superdiffusion followed by pronounced subdiffusion. However, for points positioned far from the active segment, this characteristic dynamic diminishes, reflecting the reduced influence of activity at a distance.

Both reconfiguration time and end-to-end looping time are sensitive to the polymer’s length. For short chains, where all normal modes relax rapidly, these timescales are primarily governed by the persistence time of the active force, with higher persistence significantly delaying both reconfiguration and looping. In this regime, the looping time scales as $L^{\alpha}$ with $\alpha>2$, deviating from the typical Rouse scaling. Such scaling aligns with the swelling behavior: polymers with active segments located centrally exhibit less pronounced length dependence. For very long chains, thermal fluctuations dominate, and the Rouse scaling $L^{2}$ is recovered. While persistence effects are negligible for long chains with end-segment activity, the looping time consistently increases with persistence time when the active segment is centrally located. However, in the former case, within the intermediate length regime where the Rouse relaxation time is comparable to the persistence time, we observe a competition between swelling (due to activity) and relaxation (due to thermal fluctuations).  This interplay results in a transition from $\alpha > 2$ to $\alpha < 2$, accompanied by a significant enhancement in the looping time with increasing persistence.
Notably, our noise model, characterized by a correlation strength independent of persistence time, introduces a notable contrast to another active noise model, described by Eq. (\ref{tempo_corr_2}), where the noise strength typically scales inversely with persistence time. This distinction causes the intermediate length regime for looping to progressively shift toward longer lengths as persistence time increases [see Fig. \ref{fig:looping_l_model2}], revealing a fundamentally different behavior from that observed in our primary model.

Our study sheds light on how the position and extent of a local active segment impact both local fluctuations and the overall behavior in partially active polymers. We explore experimentally relevant metrics, including the radius of gyration, mean segment separation, and mean squared displacement of tagged points along the polymer, offering a foundation for potential experimental validation. While our analysis centers on the simplified Rouse polymer model, expanding this framework to include polymers with finite rigidity, excluded volume interactions, and realistic interaction effects in polymer melts would be of significant interest. Such extensions could reveal finer details of chromatin dynamics and provide insights into scaling laws associated with chromatin behavior, setting the stage for future investigations.

\section*{Author contributions}
KG, CHC, and HYS conceived the study. KG performed the main calculations and simulations. NH, CHC, and HYS validated the results. All authors contributed to the writing of the manuscript.

\section*{Data availability}
All data supporting the findings are included either in the main text or in the appendices.

\section*{Conflicts of Interest}

There are no conflicts of interest to declare.

\section*{Acknowledgments} 
KG would like to acknowledge funding from the National Science and Technology Council, the Ministry of Education (Higher Education Sprout Project NTU-113L104022), and the National Center for Theoretical Sciences of Taiwan.
 This work was supported by the National Science and Technology Council, Taiwan, under Grants No.  NSTC 113-2112-M-A49-019 (CHC), No. 109-2112-M-001 -017 -MY3 (HYS) and No. 111-2112-M-001 -027 -MY3 (HYS). 
CHC acknowledges the support of CTCP, NYCU.

\appendix
\setcounter{figure}{0}
\section{Normal mode analysis}
Here, we compute the correlations and moments of the normal modes, which are used in the main text to determine the key properties of the polymer.

The zeroth  and $p^{\text{th}}$ mode ($p \neq 0$) amplitudes can be expressed as 
\begin{align}
& \bs{\chi}_0(t) = \int_{-\infty}^{t}dt_1 \left(\bs{\eta}^{\rm T}_0(t_1)+\bs{\sigma}^{\rm A}_0(t_1)\right),\label{chi0_t}\\
& \bs{\chi}_p(t) = e^{-\frac{t}{\tau_p}} \int_{-\infty}^{t}dt_1 \,e^{\frac{t_1}{\tau_p}}\left(\bs{\eta}^{\rm T}_p(t_1)+\bs{\sigma}^{\rm A}_p(t_1)\right).
\label{chip_t}
\end{align}
The active and thermal components of the mode amplitudes are uncorrelated, and thus the total correlation  is the sum of these two components: 
\begin{align}
\rho_{pq}(t,t')&=\rho_{pq}(t-t')=\langle  \chi_p(t)  \chi_q(t')\rangle\nonumber\\
&=\langle  \chi_p(t)  \chi_q(t')\rangle_T+\langle  \chi_p(t)  \chi_q(t')\rangle_A\nonumber\\
&=\rho^T_{pq}(t,t')+\rho^A_{pq}(t,t').\label{normal_corr}
\end{align}
The thermal part of the correlation is given by  $(p \neq 0,\,q\neq 0)$
\begin{align}
& \rho^T_{pq}(t,t')=\langle  \chi_p(t)  \chi_q(t')\rangle_T\nonumber\\
&=  \int_{-\infty}^{t}dt_1 \,  \int_{-\infty}^{t'}dt_2 \,e^{-\frac{t}{\tau_p}+\frac{t_1}{\tau_p}-\frac{t'}{\tau_q}+\frac{t_2}{\tau_q}}
\langle \eta_p^T(t_1) \eta_q^T(t_2)\rangle\nonumber\\
&=  \Bigg(\frac{k_B T}{L\zeta}\Bigg)(1+\delta_{p0})\int_{-\infty}^{t}dt_1 \,  \int_{-\infty}^{t'}dt_2 \,\nonumber\\
& \qquad\qquad e^{-\frac{t}{\tau_p}+\frac{t_1}{\tau_p}-\frac{t'}{\tau_q}+\frac{t_2}{\tau_q}}\delta_{pq}\delta(t_1-t_2)\nonumber\\
&= \Bigg(\frac{k_B T}{L\zeta}\Bigg)\frac{\tau_p}{2}\delta_{pq}\,e^{-\frac{|t-t'|}{\tau_p}}. \label{normal_corr_t}
\end{align}
Thus, the mean square amplitude turns out to be 
\begin{align}
& \langle  \chi_p(t)  \chi_q(t)\rangle_T=\rho^T_{pq}(0)= \Bigg(\frac{k_B T}{L\zeta}\Bigg)\frac{\tau_p}{2}\delta_{pq}. \label{normal_corr_t1}
\end{align}
On the other hand, using Eq. (\ref{rho_pq}) in Eq. (\ref{active_corr2}), the active part of the correlation  can be computed as follows  
\begin{align}
&\rho^A_{pq}(t,t')=\langle  \chi_p(t)  \chi_q(t')\rangle_A\nonumber\\
&=  \int_{-\infty}^{t}dt_1 \,  \int_{-\infty}^{t'}dt_2 \,e^{-\frac{t}{\tau_p}+\frac{t_1}{\tau_p}-\frac{t'}{\tau_q}+\frac{t_2}{\tau_q}}\langle \sigma_p^A(t_1) \sigma_q^A(t_2)\rangle\nonumber\\
&= \frac{\mathcal{I}_{pq}}{L^2\zeta^2}\int_{-\infty}^{t}dt_1 \,  \int_{-\infty}^{t'}dt_2\,  e^{-\frac{t}{\tau_p}+\frac{t_1}{\tau_p}-\frac{t'}{\tau_q}+\frac{t_2}{\tau_q}} \mathcal{K}(t_1,t_2)\nonumber\\
&=\frac{ \mathcal{I}_{pq} \tau_p\tau_q\tau_A }{L^2\zeta^2 (\tau_A-\tau_p)}\Bigg[\frac{\tau_A\, e^{-\frac{|t-t'|}{\tau_A}}}{(\tau_A+\tau_q)}-\frac{2 \tau_p^2\,e^{-\frac{|t-t'|}{\tau_p}}}{(\tau_A+\tau_p)(\tau_p+\tau_q)}\Bigg].\label{rho_pq_A}
\end{align}
Similarly, one has 
\begin{align}
&\rho^A_{qp}(t,t')=\langle  \chi_q(t)  \chi_p(t')\rangle_A\nonumber\\  
&=\frac{ \mathcal{I}_{qp} \tau_q\tau_p\tau_A }{L^2\zeta^2 (\tau_A-\tau_q)}\Bigg[\frac{\tau_A\, e^{-\frac{|t-t'|}{\tau_A}}}{(\tau_A+\tau_p)}-\frac{2 \tau_q^2\,e^{-\frac{|t-t'|}{\tau_q}}}{(\tau_A+\tau_q)(\tau_q+\tau_p)}\Bigg].\label{rho_qp_A}
\end{align}
Taking $p=q$ in the above equation, the autocorrelation function  reads
\begin{align}
&\rho^A_{pp}(t,t')=\langle  \chi_p(t)  \chi_p(t')\rangle_A\nonumber\\
& =\frac{\mathcal{I}_{pp}}{L^2\zeta^2}\Bigg[\frac{\tau_A^2 \tau_p\tau_p}{(\tau_A-\tau_p)(\tau_A+\tau_p)}e^{-\frac{|t-t'|}{\tau_A}}\nonumber\\
&\qquad\qquad-\frac{2\tau_A \tau_p^3\tau_p}{(\tau_A^2-\tau_p^2)(\tau_p+\tau_p)}e^{-\frac{|t-t'|}{\tau_p}}\Bigg]\nonumber\\
&=\frac{\mathcal{I}_{pp}\tau_A}{L^2\zeta^2} \frac{\tau_p^2\left(\tau_Ae^{-\frac{|t-t'|}{\tau_A}}-\tau_pe^{-\frac{|t-t'|}{\tau_p}}\right)}{\tau_A^2-\tau_p^2} .\label{rho_pp_A}
\end{align}
From the above, it follows
\begin{align}
&\rho^A_{pp}(t,t)=\rho^A_{pp}(0)=\langle  \chi_p(t)^2\rangle_A \nonumber\\
&= \frac{\mathcal{I}_{pp}}{L^2\zeta^2}\Bigg[\frac{\tau_A^2 \tau_p^2}{(\tau_A^2-\tau_p^2)}-\frac{\tau_A \tau_p^3}{(\tau_A^2-\tau_p^2)}\Bigg] =\frac{\tau_A\tau_p^2}{L^2\zeta^2\left(\tau_A+\tau_p\right)}\mathcal{I}_{pp}.\label{chip2}
\end{align}
Similarly, one obtains
\begin{align}
& \rho^A_{pq}(t,t)=\rho^A_{pq}(0) \nonumber\\
&=\frac{ \mathcal{I}_{pq} \tau_p\tau_q\tau_A }{L^2\zeta^2 (\tau_A-\tau_p)}\Bigg[\frac{\tau_A}{(\tau_A+\tau_q)}-\frac{2 \tau_p^2}{(\tau_A+\tau_p)(\tau_p+\tau_q)}\Bigg]\nonumber\\
&=\frac{ \mathcal{I}_{pq}\tau_A \tau_p \tau_q}{L^2\zeta^2(\tau_p+\tau_q)}\Bigg[\frac{\tau_p}{(\tau_A+\tau_p)}+\frac{\tau_q}{(\tau_A+\tau_q)}\Bigg].\label{rho_pq_A0} 
\end{align}
Using Eq. (\ref{rho_p0}),  it yields 
\begin{align}
&\rho^A_{p0}(t,t') =\langle  \chi_p(t)  \chi_0(t')\rangle_{A} \nonumber\\
&=\int_{-\infty}^{t}dt_1 \,  \int_{-\infty}^{t'}dt_2 \,e^{-\frac{t}{\tau_p}+\frac{t_1}{\tau_p}} \langle\sigma_p^A(t_1) \sigma_0^A(t_2)\rangle\nonumber\\
& = \frac{\mathcal{I}_{p0}}{L^2\zeta^2} \int_{-\infty}^{t}dt_1  \int_{-\infty}^{t'}dt_2\, e^{-\frac{t}{\tau_p}+\frac{t_1}{\tau_p}}\mathcal{K}(t_1,t_2)\nonumber\\
& = \frac{ \mathcal{I}_{p0}}{L^2\zeta^2}\Bigg[ \frac{\tau_A^2 \tau_p}{\tau_A-\tau_p}e^{-\frac{|t-t'|}{\tau_A}}+\frac{2\tau_A \tau_p^3}{\tau_p^2-\tau_A^2}e^{-\frac{|t-t'|}{\tau_p}}\Bigg].\label{rho_p0_1}
\end{align}
In a similar fashion, from Eq. (\ref{rho_0p}) one obtains
\begin{align}
 &\rho^A_{0p}(t,t')=\langle  \chi_0(t)  \chi_p(t')\rangle_A \nonumber\\
 &=  \int_{-\infty}^{t}dt_1 \,  \int_{-\infty}^{t'}dt_2 \,e^{-\frac{t'}{\tau_p}+\frac{t_2}{\tau_p}}\langle \sigma_0^A(t_1) \sigma_p^A(t_2)\rangle\nonumber\\
& = \frac{\mathcal{I}_{0p}}{L^2\zeta^2} \int_{-\infty}^{t}dt_1  \int_{-\infty}^{t'}dt_2\, e^{-\frac{t'}{\tau_p}+\frac{t_2}{\tau_p}}\mathcal{K}(t_1,t_2)\nonumber\\
&= \frac{\mathcal{I}_{0p}}{L^2\zeta^2}\Bigg[2 \tau_A \tau_p-\frac{\tau_A^2 \tau_p}{\tau_A+\tau_p}e^{-\frac{|t-t'|}{\tau_A}}\Bigg]. \label{rho_0p_1}
\end{align}

\section{Derivation of $\langle R_g^2 \rangle$ and $\langle R_e^2 \rangle$}
By virtue of Eqs. (\ref{normal_corr_t1}) and (\ref{chip2}),  the MSRG in Eq. (\ref{rg}) becomes 
\begin{align}
 \langle R_g^2 \rangle &= \frac{3k_B T}{L\zeta}\sum_{p=1}^{\infty}\frac{\zeta L^2}{\kappa \pi^2}\frac{1}{p^2}+ \frac{6 \tau_A}{L^2\zeta^2}\sum_{p=1}^{\infty} \frac{\tau_p^2}{\tau_A+\tau_p}\mathcal{I}_{pp} \nonumber\\
 & =  \frac{k_B T L}{2\kappa}+\frac{6  \tau_1^2}{L^2\zeta^2}\sum_{p=1}^{\infty}\frac{\mathcal{I}_{pp}}{p^2(p^2+\tau_1/\tau_A)}\label{rg_1},
\end{align}
 which is sometimes denoted as $\langle R_g^2 \rangle_{(f_1,f_2)}$ to specify its dependence on $f_1$ and $f_2$.
The summation is computed numerically, considering $p$ and $q$ within the range $[1,500]$, since the contributions  from modes with higher values of $p$ or $q$ are negligible. The results obtained for various values of $f_1$ and $f_2$ are plotted in Fig. \ref{fig:rg}. For a fully active polymer where $f_1=0$ and $f_2=1$, the MSRG can be computed exactly, and it is given by 
\begin{align}
\langle R_g^2 \rangle_{(0,1)}  & =  \frac{k_B T L}{2\kappa}+ \frac{\pi^2  F_A^2 \tau_A \tau_1}{2L^2\zeta^2} \nonumber\\
& +\frac{3  F_A^2 \tau_A^2}{2L^2\zeta^2}\left[1-\sqrt{\frac{\pi^2 \tau_1}{\tau_A}}\coth{\sqrt{\frac{\pi^2\tau_1}{\tau_A}}}\right]. \label{rg_full}    
\end{align}

\begin{figure}[htp]
\centering
\includegraphics[width=1\linewidth]{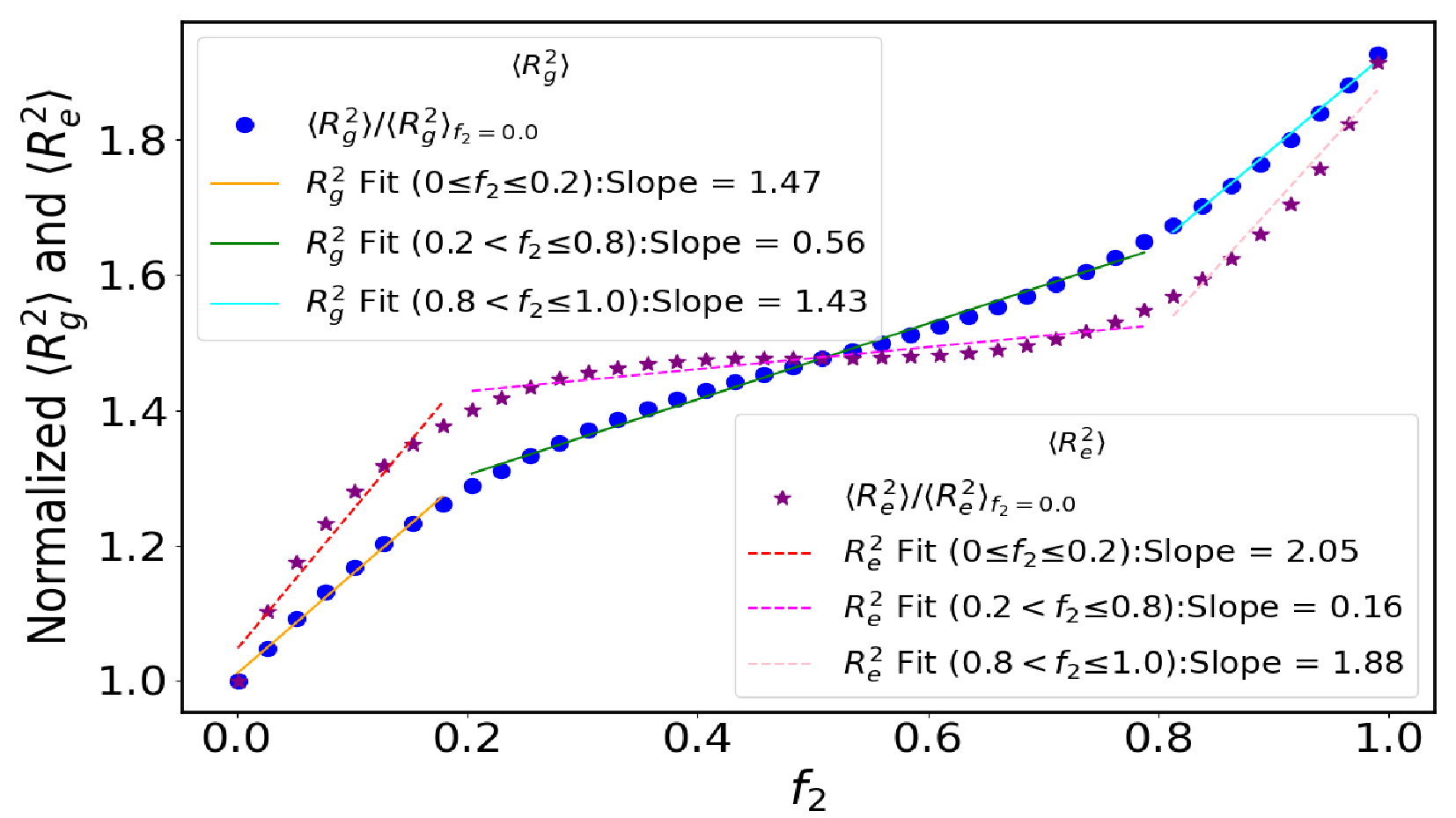}
\caption{Plot of the MSRG and MSED, normalized by their values at $f_2=0$, as a function of $f_2$ with $f_1$  set to $f_1=0,$ and  the values of other parameters are  the same as in  Figs. \ref{fig:rg}-\ref{fig:r2e}. 
}
    \label{fig:rgre_f10f2}
\end{figure}

The mean square end-to-end distance given in Eq. (\ref{phi_corr0}) can be rewritten  as 
\begin{align}
\Phi_{corr}(0)& = \langle \bs{R}_e^2 \rangle = 48 \sum_{p,q=1}^{\infty} \, \langle \chi_{2p-1}(t) \chi_{2q-1}(t) \rangle \nonumber\\
&=\Phi_{corr,T}(0)+\Phi_{corr,A}(0), \label{phi_corr0_1}
\end{align}
where the thermal part is \begin{align}
\Phi_{corr,T}(0)  =\Bigg(\frac{3 k_B T }{\kappa }\Bigg)L. \label{phi_corr0_t}  \end{align}
With the help of Eq. (\ref{rho_pq_A}), one can obtain 
\begin{align}
 \Phi_{corr,A}(0) & =  \frac{48 }{L^2\zeta^2} \sum_{p,q=1}^{\infty} \,\mathcal{I}_{(2p-1)(2q-1)}  \nonumber\\
&\times\Bigg[\frac{\tau_A^2 \tau_{2p-1}\tau_{2q-1}}{(\tau_A-\tau_{2p-1})(\tau_A+\tau_{2q-1})} \nonumber \\
&-\frac{2\tau_A \tau_{2p-1}^3\tau_{2q-1}}{(\tau_A^2-\tau_{2p-1}^2)(\tau_{2p-1}+\tau_{2q-1})}\Bigg] . \label{phi_corr0_a}
\end{align}   

\begin{figure*}[htp]
    \centering
    \begin{subfigure}[b]{0.495\textwidth}
    \caption{ $\tau_A=100.0$}
        \centering
        \includegraphics[width=\textwidth]{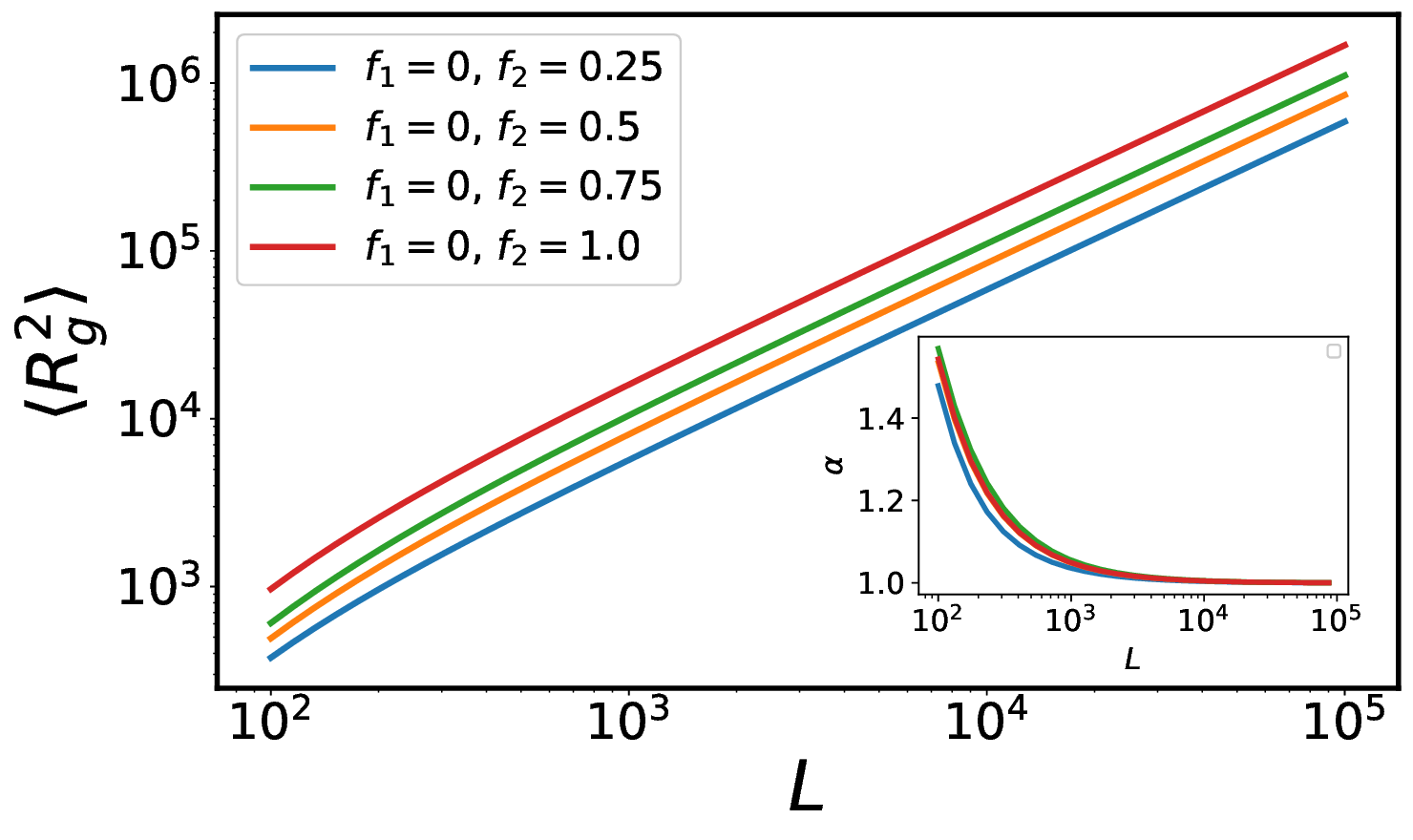}
        \label{fig:rg_tau1a1}
    \end{subfigure}
    \hfill
    \begin{subfigure}[b]{0.495\textwidth}
    \caption{$\tau_A=100.0$}
        \centering
        \includegraphics[width=\textwidth]{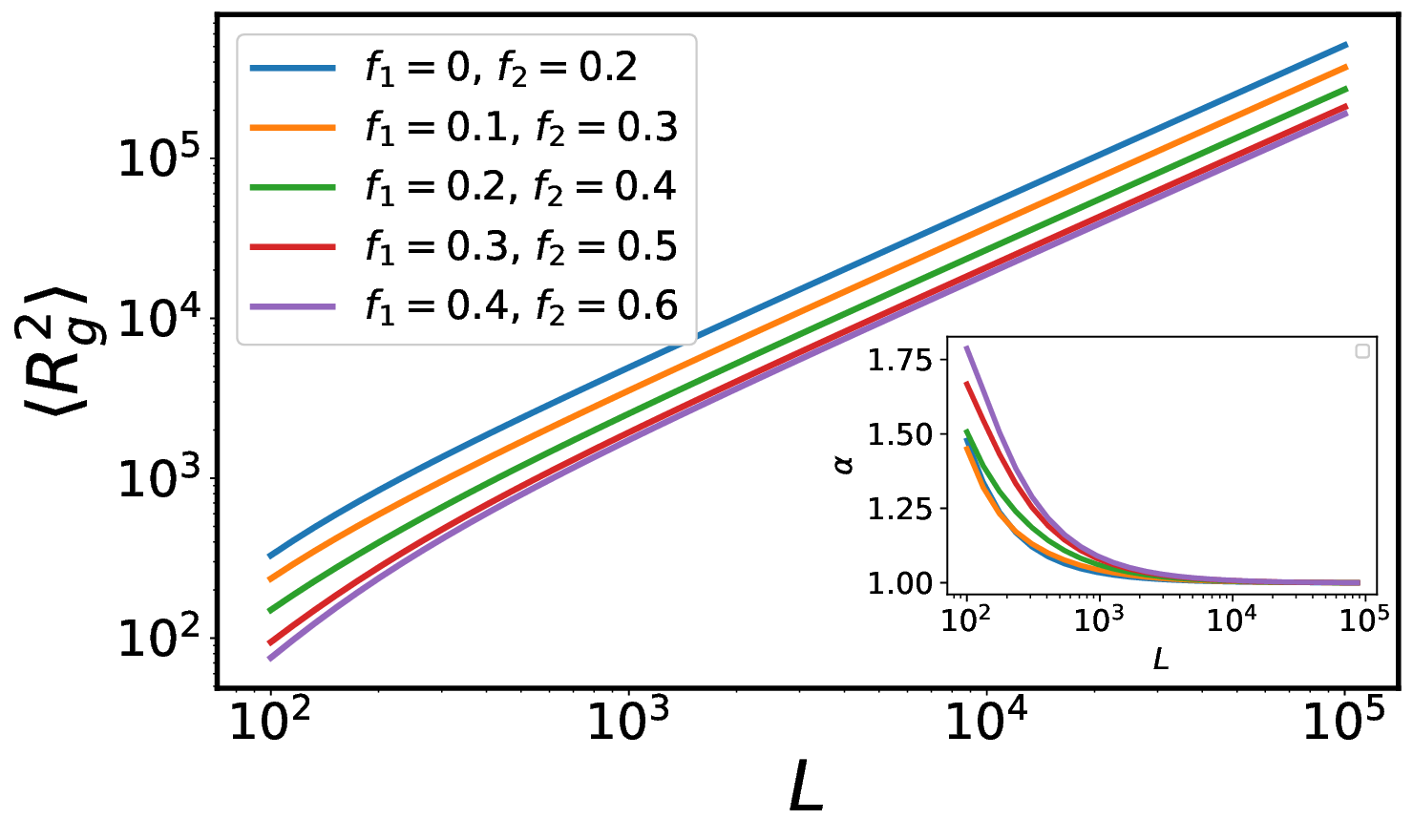}
         \label{fig:rg_tau1a2}
    \end{subfigure}
  \vspace{0.2cm}  
  
    \begin{subfigure}[b]{0.495\textwidth}
    \caption{ $(f_1,\,f_2)=(0.4,\,0.6).$}
        \centering
        \includegraphics[width=\textwidth]{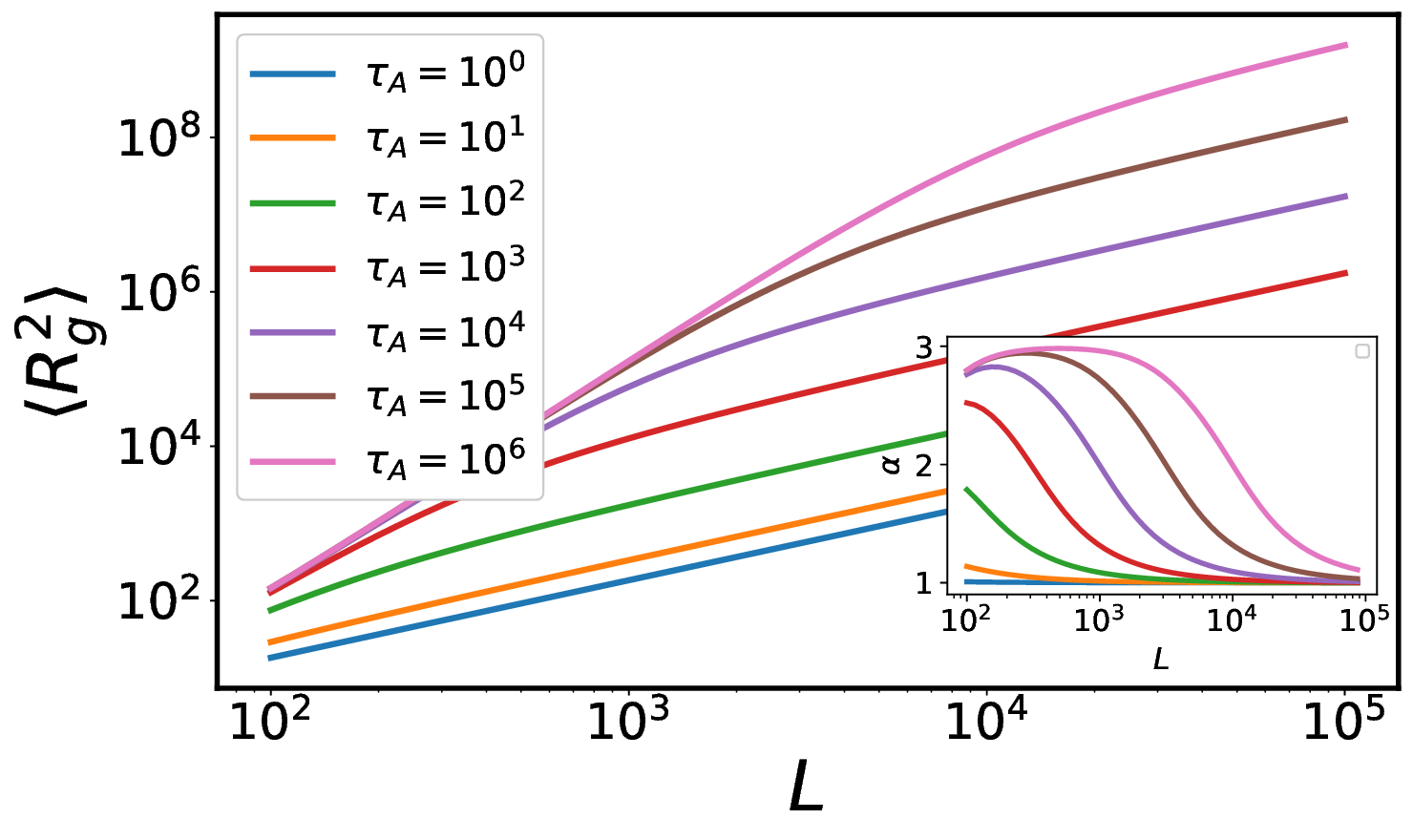}
    \label{fig:rg_tau1a3}
    \end{subfigure}
    \caption{Log-log plot of radius of gyration $\langle R_g^2 \rangle$ [Eq. (\ref{rg_1})]  as functions of $L$ 
     for various values of $f_1$, $f_2$, and $\tau$, shown in panels (a), (b), and (c). In inset: Log-log Plot of exponent $\alpha$ extracted from $\langle R_g^2 \rangle \propto L^{\alpha}$ as a function of $L$. The values of other parameters are $\kappa=3.0,\,\zeta=1.0,\,k_BT=1.0$, and $F_0=1.0.$} 
    \label{fig:rg_alls}
\end{figure*}

\begin{figure*}[htp]
    \centering
    \begin{subfigure}[b]{0.495\textwidth}
    \caption{  $\tau_A=100.0.$}
        \centering
        \includegraphics[width=\textwidth]{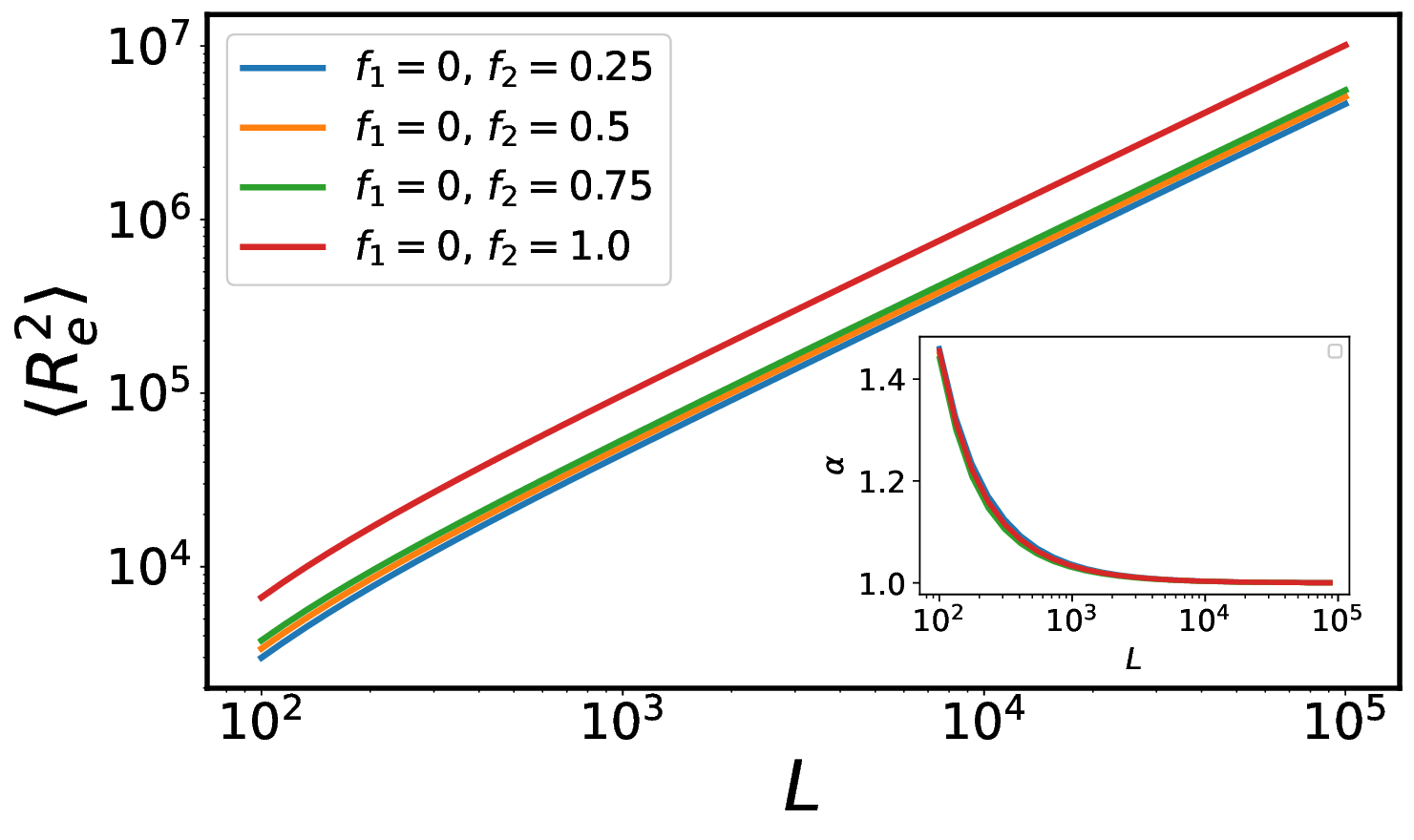}
        \label{fig:re_tau1a1}
    \end{subfigure}
    \hfill
    \begin{subfigure}[b]{0.495\textwidth}
    \caption{$\tau_A=100.0.$}
        \centering
        \includegraphics[width=\textwidth]{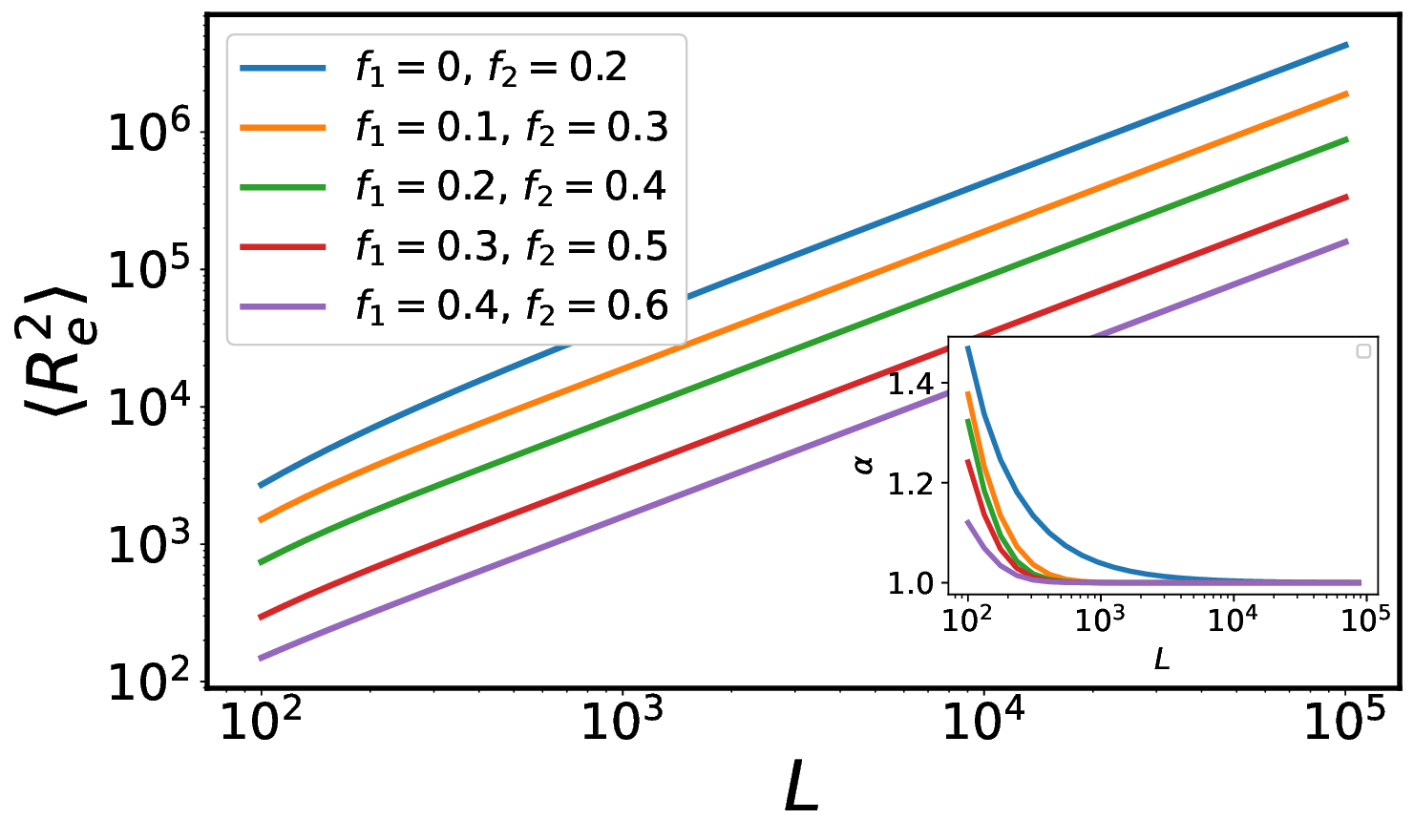}
         \label{fig:rg_tau1a2}
    \end{subfigure}
  \vspace{0.2cm}  
  
    \begin{subfigure}[b]{0.495\textwidth}
        \centering
         \caption{$(f_1,\,f_2)=(0.4,\,0.6).$}
        \includegraphics[width=\textwidth]{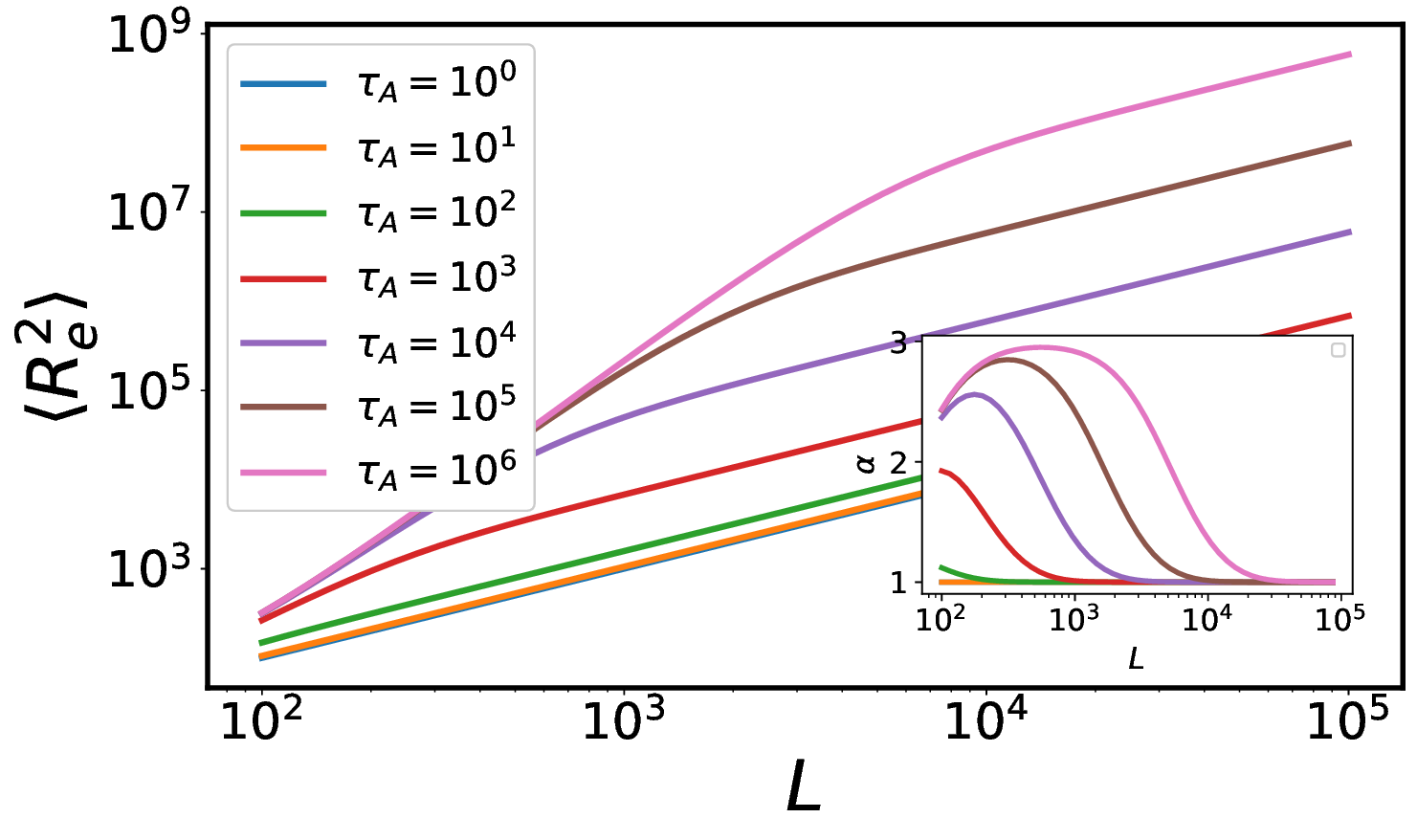}
    \label{fig:rg_tau1a3}
    \end{subfigure}
    \caption{Log-log plot of end-to-end distance $\langle R_e^2 \rangle$ [Eq. (\ref{phi_corr0_1})]  as functions of $L$. In inset: Plot of exponent $\alpha$ extracted from $\langle R_e^2 \rangle \propto L^{\alpha}$ as a function of $L$ in log-log scale. The values of other parameters are $\kappa=3.0$, $\zeta=1.0$, $k_BT=1.0$, and $F_0=1.0$.}
    \label{fig:re_alls}
\end{figure*}

\subsection{Results for temporally delta-correlated active noise}
\begin{figure}[htp]
\centering
\includegraphics[width=1\linewidth]{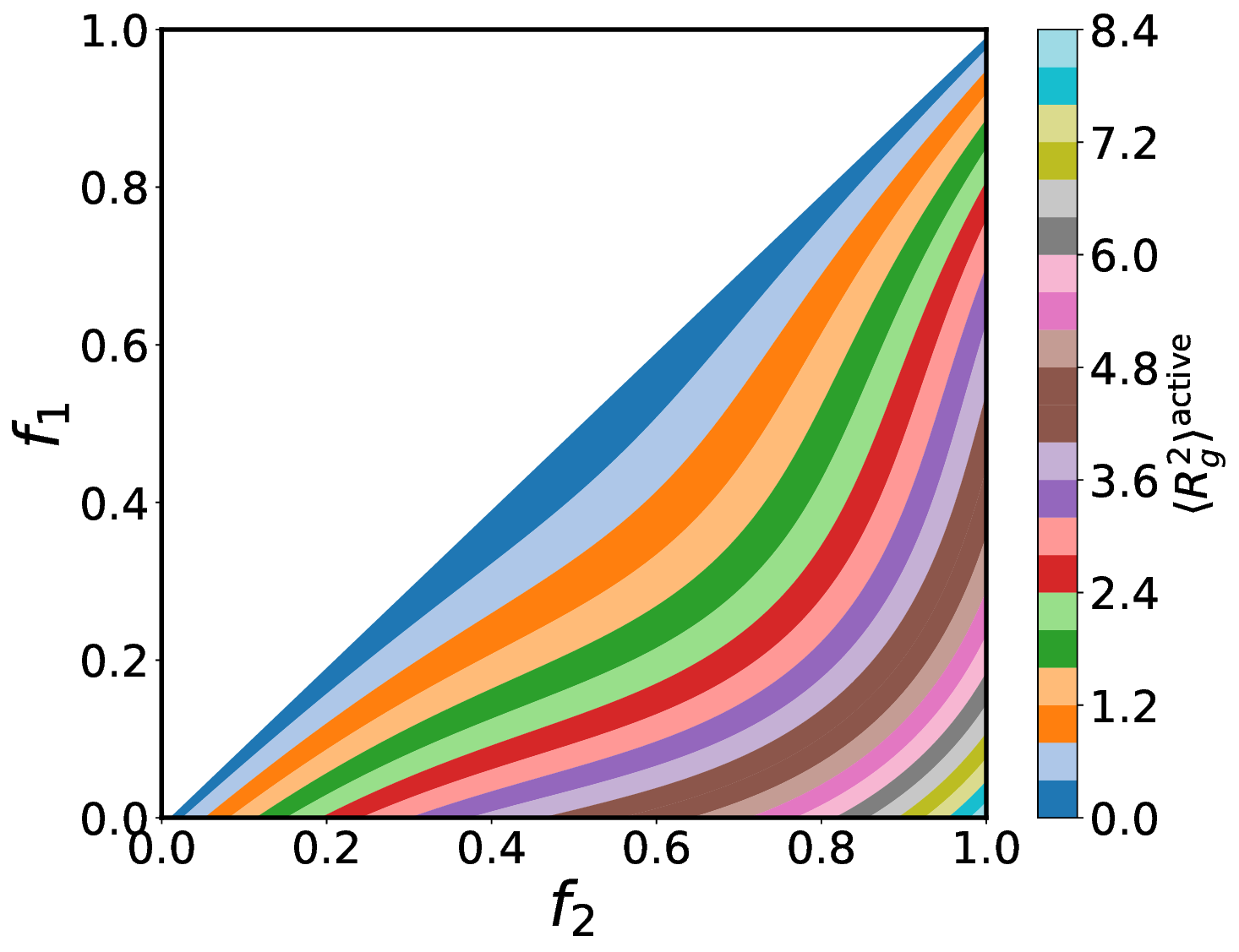}
\caption{Contour plot of the active part of MSRG $\langle R_g^2 \rangle^{\rm active}$  [Eq. (\ref{rg_A1})]  as functions of $f_1$ and $f_2$ for $\mathcal{K}(t_1,t_2)=\,\delta(t_1-t_2)$. The values of other parameters are $\kappa=3.0$, $\zeta=1.0,\,k_BT=1.0,\,\,L=100$, and $F_0=1$. 
}
\label{fig:rga}
\end{figure}

\begin{figure}[htp]
\centering
\includegraphics[width=1\linewidth]{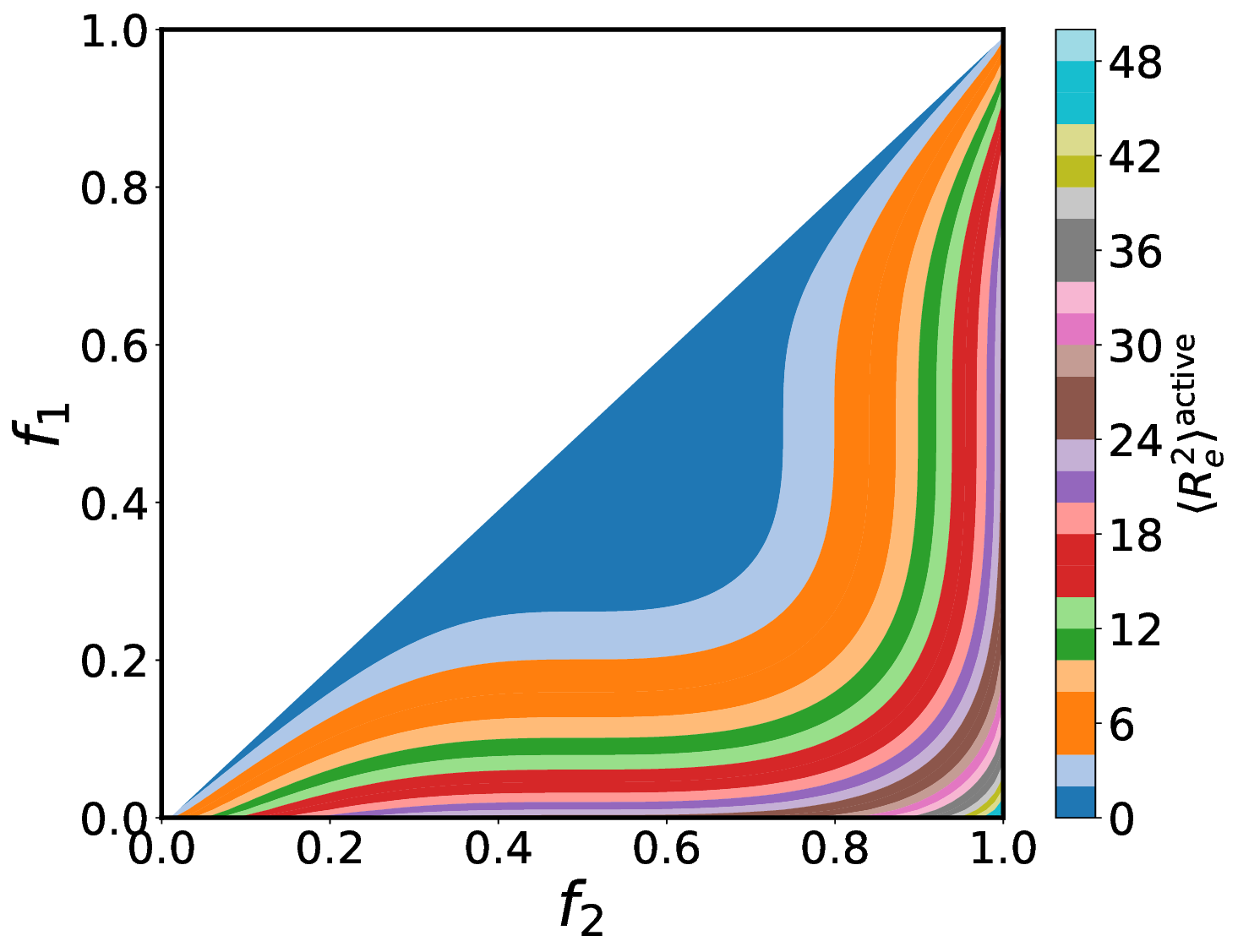}
\caption{Contour plot of  the active part of MSED $\langle R_e^2 \rangle^{\rm active}$  [Eq. (\ref{phi_corr0_a_delta})]  as functions of $f_1$ and  $f_2$ for $\mathcal{K}(t_1,t_2)=\,\delta(t_1-t_2)$. The values of other parameters are $\kappa=3.0$, $\zeta=1.0,\,k_BT=1.0,\,\,L=100$, and $F_0=1.0$. 
}
\label{fig:rea}
\end{figure}
Here, we consider a simpler model for the  active force, where its temporal part is delta-correlated, acting as an additional energy source.
Thus, the correlation can be expressed as $\mathcal{K}(t_1,t_2)=\,\delta(t_1-t_2)$. Using this, one can obtain the covariance between two mode amplitudes as follows: 
\begin{align}
&\rho^A_{pq}(t,t')=\langle  \chi_p(t)  \chi_q(t')\rangle_A\nonumber\\
&=  \int_{-\infty}^{t}dt_1 \,  \int_{-\infty}^{t'}dt_2 \,e^{-\frac{t}{\tau_p}+\frac{t_1}{\tau_p}-\frac{t'}{\tau_q}+\frac{t_2}{\tau_q}}\langle \sigma_p^A(t_1) \sigma_q^A(t_2)\rangle\nonumber\\
&= \frac{\mathcal{I}_{pq}}{L^2\zeta^2}\int_{-\infty}^{t}dt_1 \,  \int_{-\infty}^{t'}dt_2 \, e^{-\frac{t}{\tau_p}+\frac{t_1}{\tau_p}-\frac{t'}{\tau_q}+\frac{t_2}{\tau_q}} \mathcal{K}(t_1,t_2)\nonumber\\
&=  \frac{\mathcal{I}_{pq}}{L^2\zeta^2}\int_{-\infty}^{t}dt_1 \,  \int_{-\infty}^{t}dt_2\,\Theta(t-t')\Theta(t'-t_2) \,\nonumber\\
& \qquad\qquad\qquad\qquad e^{-\frac{t}{\tau_p}+\frac{t_1}{\tau_p}-\frac{t'}{\tau_q}+\frac{t_2}{\tau_q}}\delta(t_1-t_2)\nonumber\\
&= \frac{\mathcal{I}_{pq}}{L^2\zeta^2}\frac{\tau_p\tau_q}{\tau_p+\tau_q}\,e^{-\frac{|t-t'|}{\tau_p}}.\label{rho_pq_AA0}
\end{align}
 As a result,
\begin{align}
&\rho^A_{pp}(t,t)= \langle  \chi_p^2(t)\rangle_A=\frac{\mathcal{I}_{pp}}{L^2\zeta^2}\frac{\tau_p}{2}.\label{rho_pp_AA}   
\end{align}
From Eq. (\ref{rho_pp_AA}),  it follows the active part of the MSRG
\begin{align}
& \langle R_g^2 \rangle^{\rm active} = \frac{6}{L^2\zeta^2}\sum_{p=1}^{\infty} \mathcal{I}_{pp}\frac{\tau_p}{2}\nonumber\\
&= \frac{3F_A^2 \tau_1}{2L^2\zeta^2}\sum_{p=1}^{\infty}\left[\frac{f_2-f_1}{p^2}+\frac{\text{sin}(2f_2 p \pi )-\text{sin}(2f_1 p \pi )}{2 \pi p^3 }\right]\nonumber\\ &=\frac{3F_A^2 \tau_1}{2L^2\zeta^2}\Bigg[\frac{i \left(\text{Li}_3(e^{-i 2f_2 \pi})-\text{Li}_3(e^{i 2f_2 \pi})\right)}{4\pi}\nonumber\\
&\quad - \frac{i \left(\text{Li}_3(e^{-i 2f_1 \pi})-\text{Li}_3(e^{i 2f_1 \pi})\right)}{4\pi}+ \frac{(f_2-f_1)\pi^2}{6}\Bigg]\label{rg_A1},
\end{align}
where $\text{Li}_3(x)$ is the polylogarithm function defined as $\text{Li}_n(x)=\sum_{k=1}^{\infty}\frac{x^k}{k^n}.$ The MSRG given in the above equation is shown in Fig. \ref{fig:rga}.

Using Eq. (\ref{rho_pq_AA0}), the active part of the MSED  turns out to be
\begin{align}
\langle \bs{R}_e^2 \rangle^{\rm active} &=   \Phi_{corr,A}(0) \nonumber\\
& = \frac{48 }{L^2\zeta^2}\sum_{p,q=1}^{\infty} \frac{\tau_1\,\mathcal{I}_{(2p-1)(2q-1)}}{(2p-1)^2+(2q -1)^2} .\label{phi_corr0_a_delta}
\end{align}
The term $\langle \bs{R}_e^2 \rangle^{\rm active}$ given in Eq. (\ref{phi_corr0_a_delta}) is computed numerically and plotted in Fig. \ref{fig:rea}.

\begin{figure*}[htp]
    \centering
    \begin{subfigure}[b]{0.495\textwidth}
    \caption{}
        \centering
        \includegraphics[width=\textwidth]{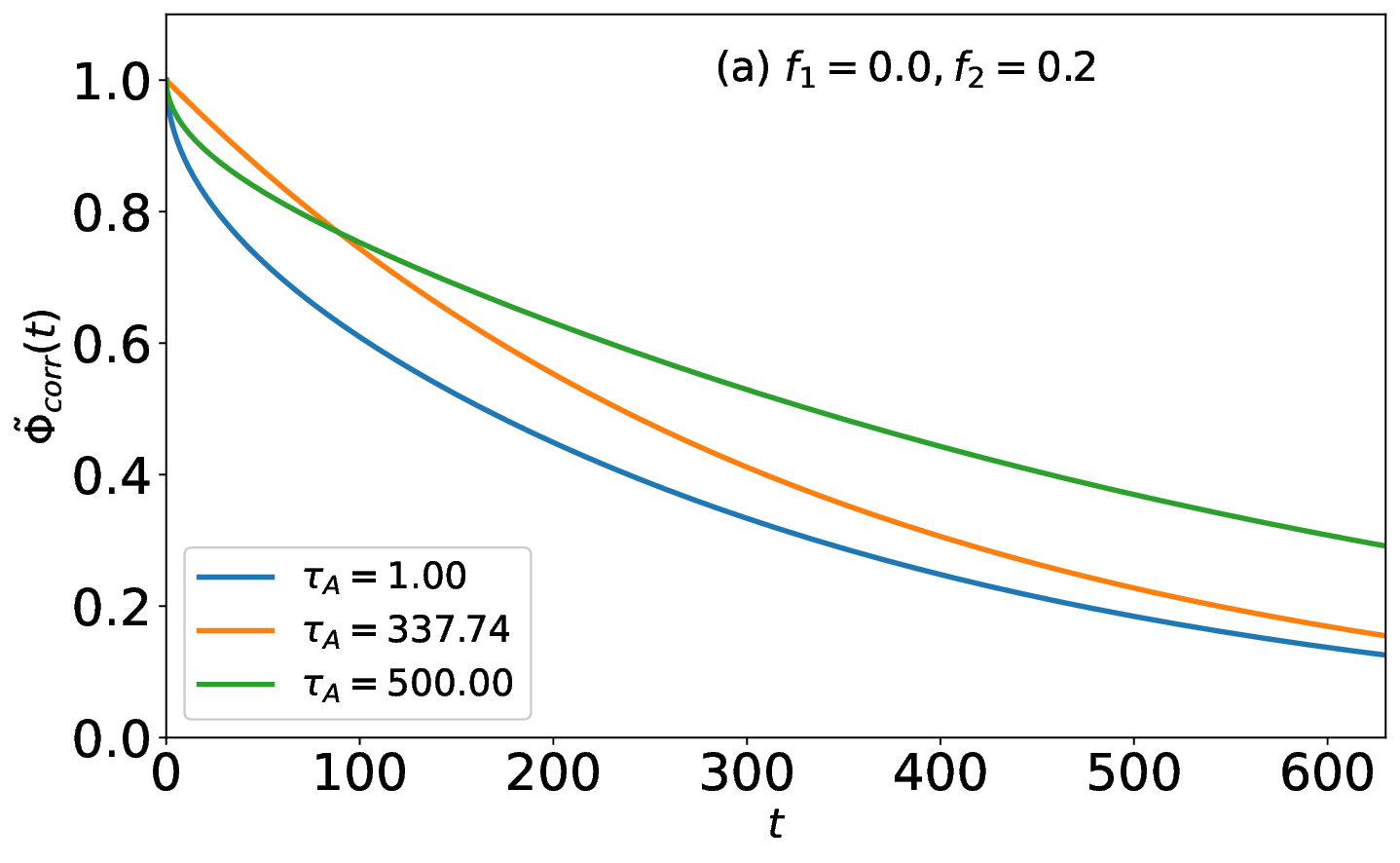}
    \end{subfigure}
    \hfill
    \begin{subfigure}[b]{0.495\textwidth}
    \caption{}
        \centering
        \includegraphics[width=\textwidth]{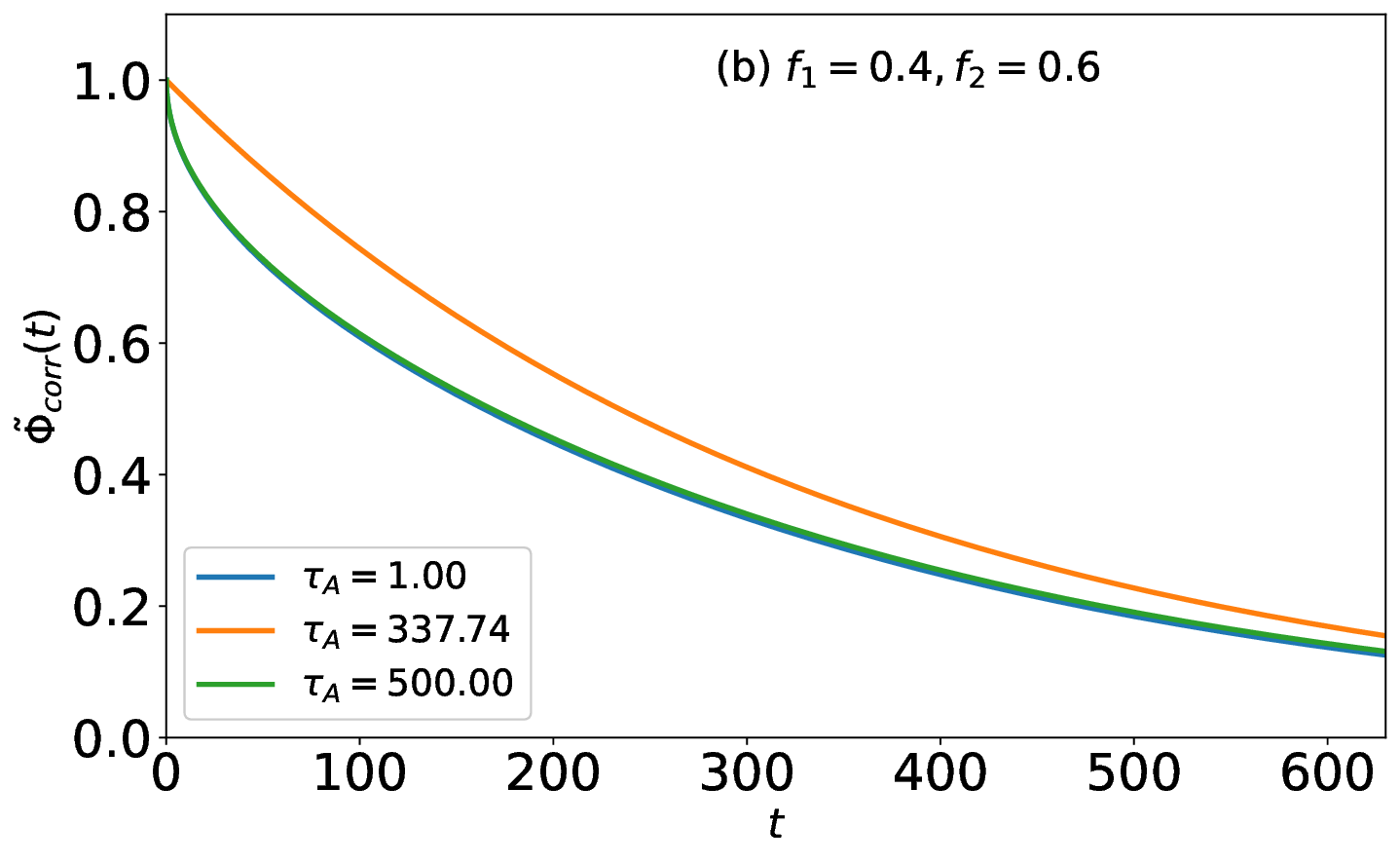}
    \end{subfigure}
  \vspace{0.2cm}  
    \begin{subfigure}[b]{0.495\textwidth}
    \caption{}
        \centering
        \includegraphics[width=\textwidth]{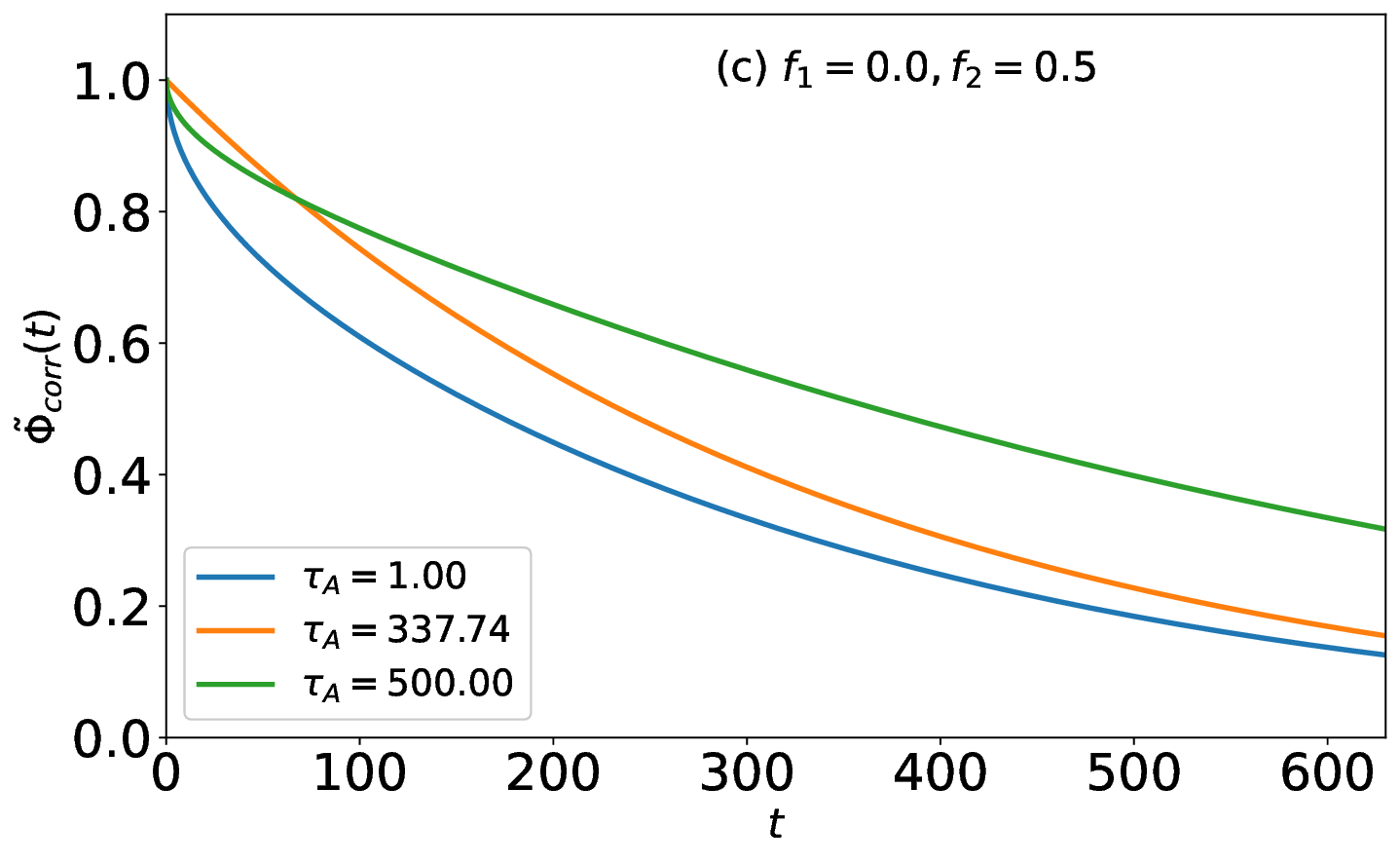}
    \end{subfigure}
        \hfill
    \begin{subfigure}[b]{0.495\textwidth}
      \caption{}
        \centering
        \includegraphics[width=\textwidth]{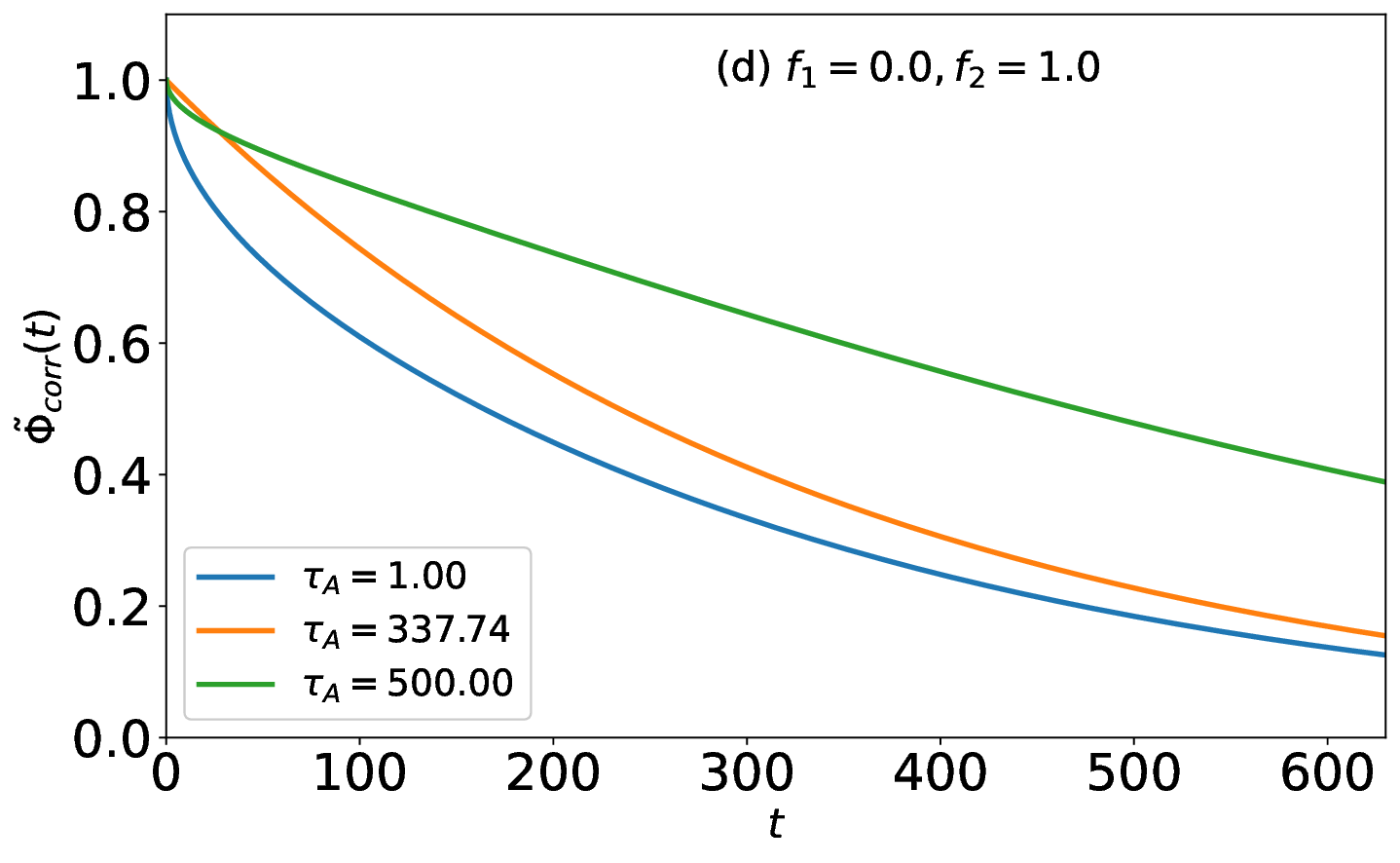}
    \end{subfigure}
\caption{Plot of normalized correlation function of the end-to-end distance $\Tilde{\Phi}_{corr}(t)$ [Eq. (\ref{phi_corr})] as a function of time $t$ for three values of $\tau_A$ $(\tau_1>\tau_A, \,\tau_1 \approx \tau_A,\, \tau_1<\tau_A)$, as indicated in the plots. The values of other parameters are $\kappa=3.0$, $\zeta=1.0$, $k_BT=1.0$, $F_0=0.1$, $L=100$,  and $\tau_1 \approx 337.74.$ 
}
\label{fig:corr_nu_f12}
\end{figure*}

\section{Derivation of $\langle R_{s_1s_2}^2\rangle$}
Let us consider $\langle R_{s_1s_2}^2\rangle$ for a segment bounded between the points of arc lengths $f_3L$ and $f_4 L$.
Using Eq. (\ref{normal_corr_t1}) and Eq. (\ref{rho_pq_A0}), Eq. (\ref{Rmn}) can be explicitly rewritten as ($s_2=f_4L,\,s_1=f_3L$) 
\begin{align}
&\langle R_{s_1s_2}^2(t) \rangle  = \Bigg(\frac{6k_B T}{L\zeta}\Bigg) \sum_{p=1}^{\infty} \tau_p \Bigg[\text{cos}^2\Bigg(p\pi f_4\Bigg) +\text{cos}^2\Bigg(p\pi f_3\Bigg)\nonumber\\
& -2\text{cos}\Bigg(p\pi f_4\Bigg)\text{cos}\Bigg(p\pi f_3\Bigg) \Bigg] +\frac{12 \tau_A }{L^2\zeta^2} \sum_{p,q=1}^{\infty} \frac{\mathcal{I}_{pq} \tau_p\tau_q}{(\tau_p+\tau_q)}\nonumber\\
 & \times \Bigg[\frac{\tau_p}{(\tau_A+\tau_p)}+\frac{\tau_q}{(\tau_A+\tau_q)}\Bigg]\nonumber\\
&\times \Bigg[\text{cos}\Bigg(p\pi f_4\Bigg)\text{cos}\Bigg(q\pi f_4\Bigg) +\text{cos}\Bigg(p\pi f_3\Bigg)\text{cos}\Bigg(q\pi f_3\Bigg) \nonumber\\
& -\text{cos}\Bigg(p\pi f_4\Bigg)\text{cos}\Bigg(q\pi f_3\Bigg)-\text{cos}\Bigg(p\pi f_3\Bigg)\text{cos}\Bigg(q\pi f_4\Bigg) \Bigg]\label{Rmn_1}.
\end{align}

For a bounded domain in the middle, where $s_1=L/4$ and $s_2=3L/4$,
the MSS for the passive case can be simplified to
\begin{align}
&\langle R_{s_1s_2}^2(t) \rangle  = \Bigg(\frac{6k_B T}{L\zeta}\Bigg) \sum_{p=1}^{\infty} \tau_p \Bigg[\text{cos}^2\Bigg(3p\pi/4\Bigg) +\text{cos}^2\Bigg(p\pi/4\Bigg)\nonumber\\
& \qquad\qquad\qquad-2\text{cos}\Bigg(3p\pi/4\Bigg)\text{cos}\Bigg(p\pi/4\Bigg) \Bigg] \nonumber\\
&= \Bigg(\frac{12k_B T}{L\zeta}\Bigg) \sum_{p=1}^{\infty} \tau_{2p+1}=\Bigg(\frac{3 k_B T }{2\kappa }\Bigg)L=\frac12\Phi_{corr,T}(0).
\label{Rmn_passive}
\end{align}

\begin{figure}[htp]
\centering
\includegraphics[width=1\linewidth]{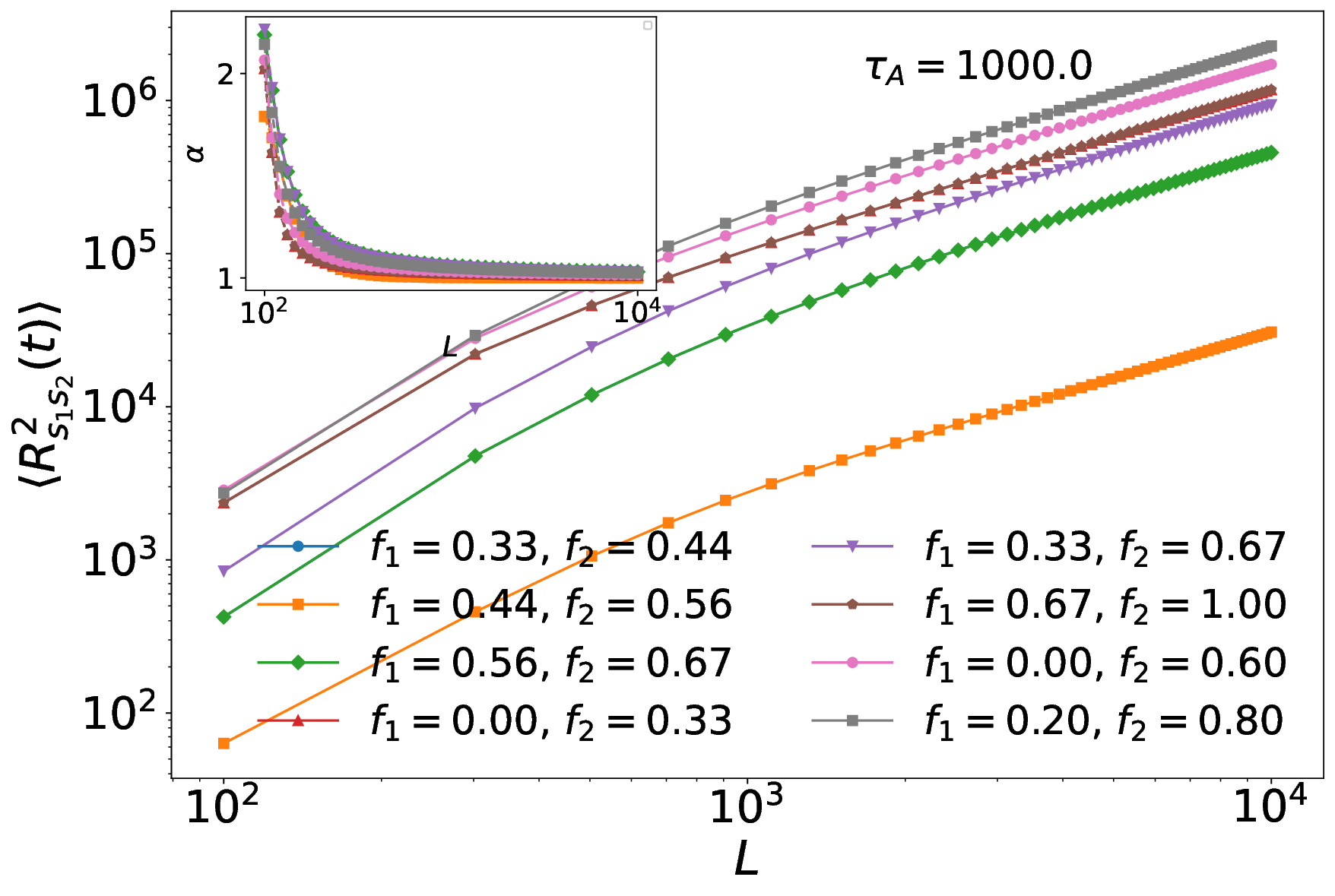}
\caption{Log-log plot of MSS $(s_1=L/3,\,s_2=2L/3)$  as a  function of $L$ for different values of $f_1$ and $f_2$, with $\tau_A=500$. In inset, the exponent $\alpha$ from the relation $\langle R_{s_1s_2}^2(t) \rangle \propto L^{\alpha}$ is shown.  The values of other  parameters are the same as in  Fig. \ref{fig:rg}. }
    \label{fig:rmn_L}
\end{figure}

\section{Derivation of MSD of CM and  $s^{\text{th}}$ point}
\begin{figure*}[htp]
    \centering
    \begin{subfigure}[b]{0.495\textwidth}
    \caption{}
        \centering
        \includegraphics[width=\textwidth]{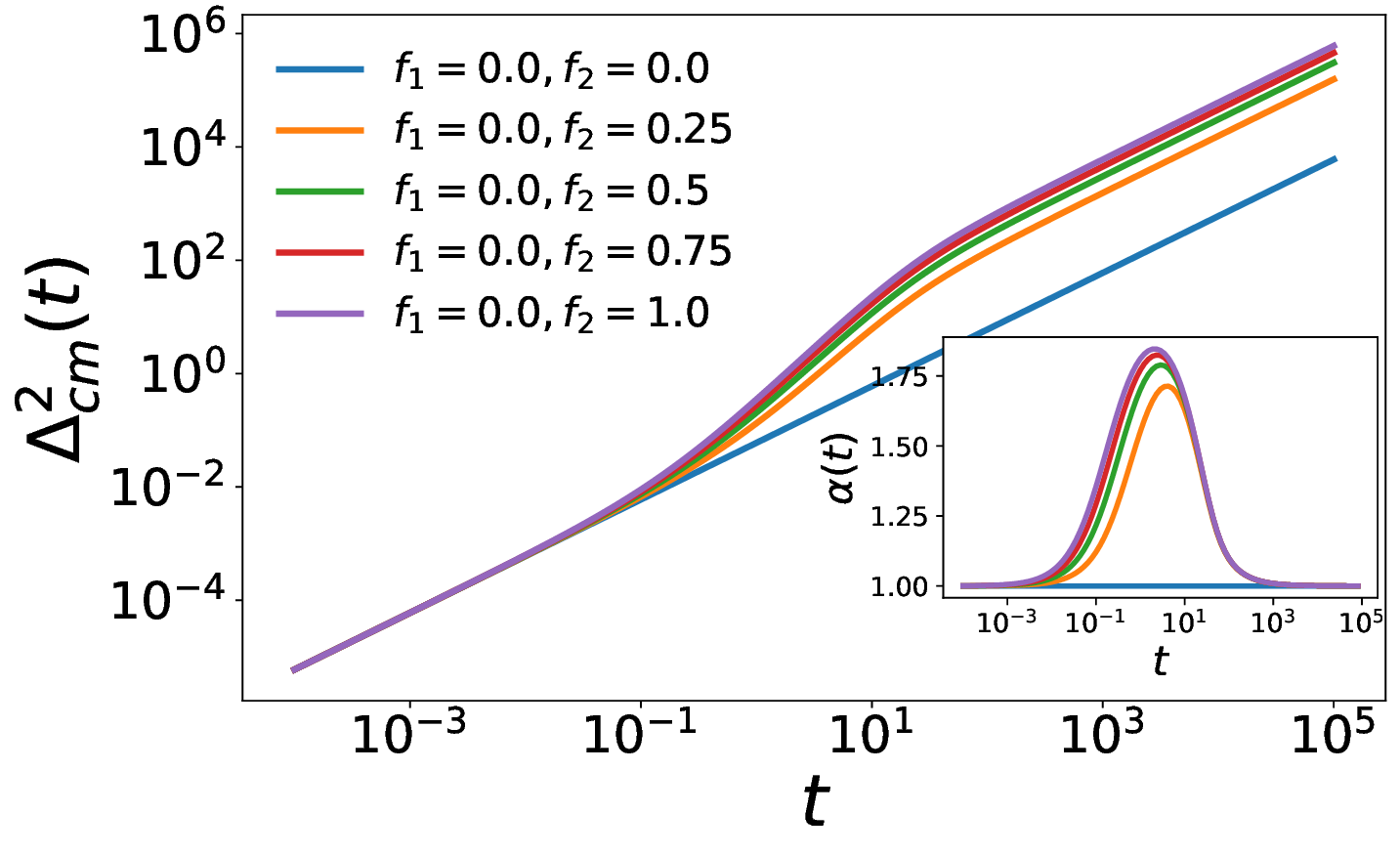}
    \end{subfigure}
    \hfill
    \begin{subfigure}[b]{0.495\textwidth}
    \caption{}
        \centering
        \includegraphics[width=\textwidth]{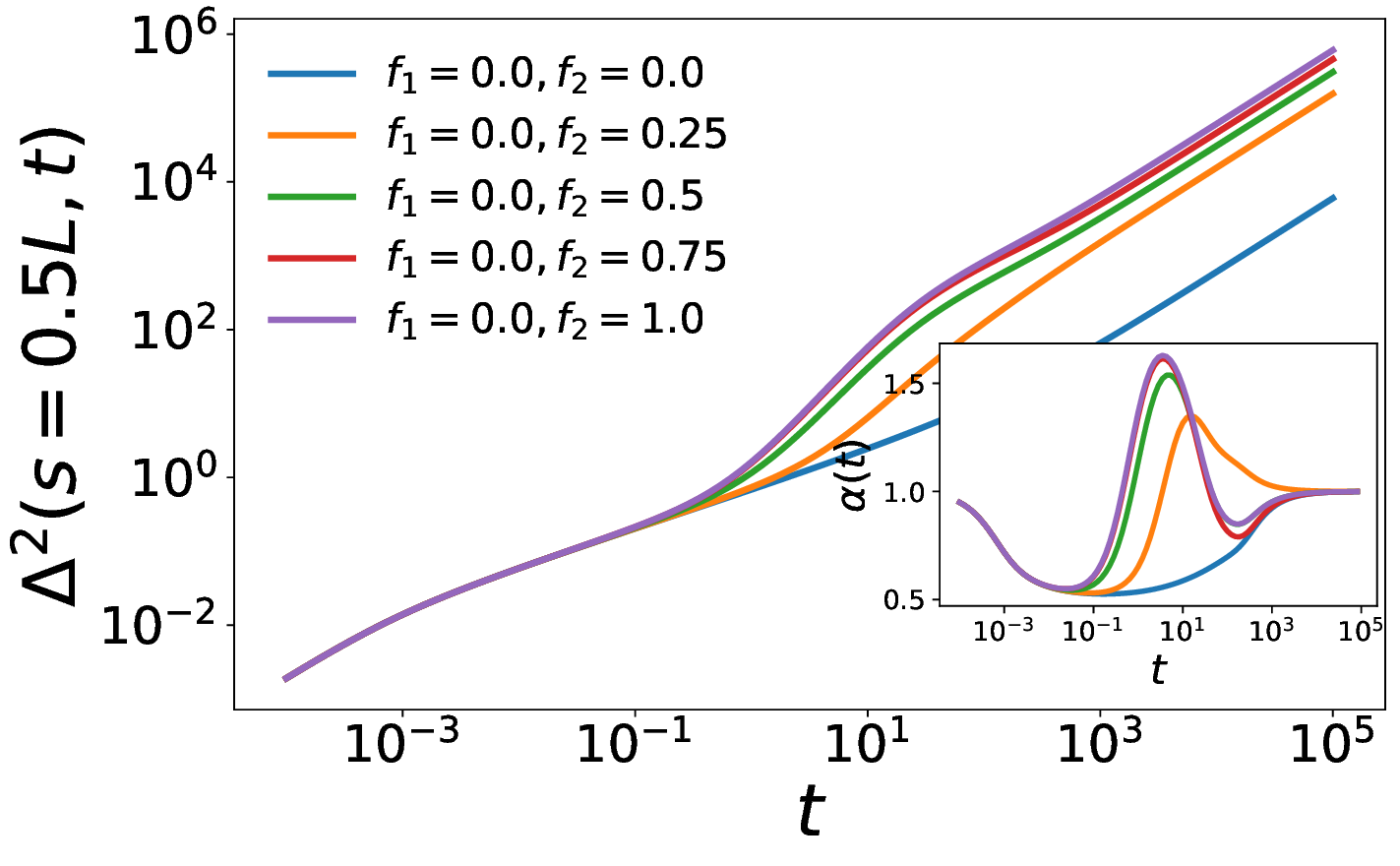}
    \end{subfigure}
\caption{(a) MSD of CM [Eq. (\ref{msd_cm})], and (b) MSD of the middle segment $(s=L/2)$ [Eq. (\ref{msd_sth_t})] versus time $t$ in log-log scale for various values of $f_1$ and $f_2$. In inset: corresponding time exponent obtained from the relation $\Delta_{cm}^2(t) \propto t^{\alpha}$ as a function of time. The values of other parameters are $\kappa=3.0$, $\zeta=1.0$, $k_BT=1.0$, $\tau_A=10.0$, $L=100$ , and $F_0^2=10.0.$}
\label{fig:msd_cm}
\end{figure*}

By virtue of Eq. (\ref{chi0_t}) and Eq. (\ref{Ipq}),  the MSD of CM is found to be 
\begin{align}
 \Delta_{cm}^2(t)&= \langle\left[\bs{r}_0(t)-\bs{r}_0(0)\right]^2\rangle= 3\langle\left[\chi_0(t)-\chi_0(0)\right]^2\rangle\nonumber\\
 & =  \frac{6 k_B T}{L\zeta} t+ \frac{6F_A^2(f_2-f_1)\tau_A}{L^2\zeta^2}\left[t-\tau_A\left(1-e^{-\frac{t}{\tau_A}}\right)\right].\label{msd_cm}
\end{align}
The MSD of CM is plotted in Fig. \ref{fig:msd_cm}. It shows that at intermediate times, comparable to the persistence time, the dynamics exhibits a superdiffusive behavior, which becomes more pronounced with an increase in the length of the active segment.

Using Eq. (\ref{msd_sth_t}), the MSD of the middle segment $s=L/2$ simplifies to 
\begin{align}
 & \Delta^2(s=L/2,t) \nonumber\\
 & =\Delta_{cm}^2(t)+4 \sum_{p=1}^{p=\infty}(-1)^{p}\left[2\rho_{2p0}(0)-\rho_{2p0}(t)-\rho_{02p}(t)\right]\nonumber\\
 &+4  \sum_{p,q=1}^{p=\infty} (-1)^{p+q}\left[2\rho_{2p2q}(0)-\rho_{2p2q}(t)-\rho_{2q2p}(t)\right].\label{msd_s0}
\end{align}

For the thermal part, $2\rho_{p0}(0)-\rho_{p0}(t)-\rho_{0p}(t)=0,$ whereas for the active part, due to spatial correlation, a non-zero quantity is obtained as given below:
\begin{align}
& 2\rho_{p0}(0)-\rho_{p0}(t)-\rho_{0p}(t)\nonumber\\
&=\frac{2\mathcal{I}_{p0}}{L^2\zeta^2}\Bigg[ \frac{\tau_A^2 \tau_p}{\tau_A-\tau_p}+\frac{2\tau_A \tau_p^3}{\tau_p^2-\tau_A^2}\Bigg]- \frac{2 }{L^2\zeta^2}\mathcal{I}_{p0}\nonumber\\
&\times\Bigg[ \tau_A \tau_p+\frac{\tau_A \tau_p^2}{\tau_p^2-\tau_A^2}\left(\tau_p\,e^{-\frac{|t-t'|}{\tau_p}}-\tau_A\,e^{-\frac{|t-t'|}{\tau_A}}\right)\Bigg].
\end{align}

The third term on the  right-hand side of Eq. (\ref{msd_s0}) can be obtained by evaluating the following expression:
\begin{align}
& 2\rho_{pq}(0)-\rho_{pq}(t)-\rho_{qp}(t) \nonumber\\
&=\Bigg(\frac{k_B T}{L\zeta}\Bigg)\tau_p\delta_{pq}\left[1-e^{-\frac{|t-t'|}{\tau_p}}\right] +\frac{2}{L^2\zeta^2}\frac{\tau_A^2 \tau_p\tau_q \mathcal{I}_{pq}}{(\tau_A-\tau_p)(\tau_A+\tau_q)}\nonumber\\
&\times \Bigg[1-\frac{\left(\tau_A^2-\tau_p\tau_q\right)e^{-\frac{|t-t'|}{\tau_A}}}{(\tau_A-\tau_q)(\tau_A+\tau_p)}\Bigg]-\frac{4}{L^2\zeta^2}\frac{\tau_A \tau_p^3\tau_q\,\mathcal{I}_{pq}}{(\tau_A^2-\tau_p^2)(\tau_p+\tau_q)}\nonumber\\
&\times \Bigg[1-\frac{e^{-\frac{|t-t'|}{\tau_p}}}{2}-\frac12\frac{\tau_q^2(\tau_A^2-\tau_p^2)}{\tau_p^2(\tau_A^2-\tau_q^2)}e^{-\frac{|t-t'|}{\tau_q}}\Bigg].
\end{align}

\section{Key integrals used in computing the correlation function}
The integration used in Eq. (\ref{rho_p0_1}) is evaluated as follows:
\begin{align}
& \, \mathcal{I}_1(t,t')=\int_{-\infty}^{t}dt_1  \int_{-\infty}^{t'}dt_2\, e^{-\frac{t}{\tau_p}+\frac{t_1}{\tau_p}}\mathcal{K}(t_1,t_2)\nonumber\\
&=  \int_{-\infty}^{t}dt_1  \int_{-\infty}^{t'}dt_2\, e^{-\frac{t}{\tau_p}+\frac{t_1}{\tau_p}}\exp\left(-\frac{|t_1-t_2|}{\tau_A}\right)\nonumber\\
&= \int_{-\infty}^{t}dt_1  \int_{-\infty}^{t}dt_2\, e^{-\frac{t}{\tau_p}+\frac{t_1}{\tau_p}-\frac{t_1}{\tau_A}+\frac{t_2}{\tau_A}}\Theta(t-t') \Theta(t'-t_2)\nonumber\\
& \times \Theta(t_1-t_2)+ \int_{-\infty}^{t}dt_1  \int_{-\infty}^{t}dt_2\, e^{-\frac{t}{\tau_p}+\frac{t_1}{\tau_p}-\frac{t_2}{\tau_A}+\frac{t_1}{\tau_A}}\nonumber\\
&\times \Theta(t-t')\Theta(t'-t_2)\Theta(t_2-t_1)\nonumber\\
&= \int_{-\infty}^{t}dt_2 \Theta(t-t')\Theta(t'-t_2)  \int_{t_2}^{t}dt_1\, e^{-\frac{t}{\tau_p}+\frac{t_1}{\tau_p}-\frac{t_1}{\tau_A}+\frac{t_2}{\tau_A}}\nonumber\\
& +  \int_{-\infty}^{t}dt_2 \Theta(t-t')\Theta(t'-t_2) \int_{-\infty}^{t_2}dt_1\, e^{-\frac{t}{\tau_p}+\frac{t_1}{\tau_p}-\frac{t_2}{\tau_A}+\frac{t_1}{\tau_A}}.
\end{align}
Performing the integration over $t_1$ simplifies the above expression to
\begin{align}
& \mathcal{I}_1(t,t')=  \int_{-\infty}^{t'}dt_2\frac{\tau_A \tau_p}{\tau_A-\tau_p}\Bigg( e^{-\frac{t-t_2}{\tau_A}}-e^{-\frac{t-t_2}{\tau_p}}\Bigg)\nonumber\\
& \qquad\qquad +  \int_{-\infty}^{t'}dt_2\frac{\tau_A \tau_p}{\tau_A+\tau_p} e^{-\frac{t-t_2}{\tau_p}}\nonumber\\
& =\frac{\tau_A^2 \tau_p}{\tau_A-\tau_p}e^{-\frac{|t-t'|}{\tau_A}}+\frac{2\tau_A \tau_p^3}{\tau_p^2-\tau_A^2}e^{-\frac{|t-t'|}{\tau_p}}\label{rho_p0}.
\end{align}

To compute Eq. (\ref{rho_0p_1}), we evaluate the following integral, as shown below:
\begin{align}
& \mathcal{I}_2(t,t')= \int_{-\infty}^{t}dt_1  \int_{-\infty}^{t'}dt_2\, e^{-\frac{t'}{\tau_p}+\frac{t_2}{\tau_p}}\mathcal{K}(t_1,t_2)\nonumber\\
&=  \int_{-\infty}^{t}dt_1  \int_{-\infty}^{t'}dt_2\, e^{-\frac{t'}{\tau_p}+\frac{t_2}{\tau_p}}\exp\left(-\frac{|t_1-t_2|}{\tau_A}\right)\nonumber\\
&= \int_{-\infty}^{t}dt_1  \int_{-\infty}^{t}dt_2\, e^{-\frac{t'}{\tau_p}+\frac{t_2}{\tau_p}-\frac{t_1}{\tau_A}+\frac{t_2}{\tau_A}}\Theta(t-t')\Theta(t'-t_2)\nonumber\\
&\times \Theta(t_1-t_2)+ \int_{-\infty}^{t}dt_1  \int_{-\infty}^{t}dt_2\, e^{-\frac{t'}{\tau_p}+\frac{t_2}{\tau_p}-\frac{t_2}{\tau_A}+\frac{t_1}{\tau_A}}\nonumber\\
&\times \Theta(t-t')\Theta(t'-t_2)\Theta(t_2-t_1)\nonumber\\
&= \int_{-\infty}^{t}dt_2 \Theta(t-t')\Theta(t'-t_2)  \int_{t_2}^{t}dt_1\, e^{-\frac{t'}{\tau_p}+\frac{t_2}{\tau_p}-\frac{t_1}{\tau_A}+\frac{t_2}{\tau_A}}\nonumber\\
& +  \int_{-\infty}^{t}dt_2 \Theta(t-t')\Theta(t'-t_2) \int_{-\infty}^{t_2}dt_1\, e^{-\frac{t'}{\tau_p}+\frac{t_2}{\tau_p}-\frac{t_2}{\tau_A}+\frac{t_1}{\tau_A}}
\end{align}
This equation can be further simplified to
\begin{align}
& \mathcal{I}_2(t,t')= \int_{-\infty}^{t'}dt_2\,\tau_A e^{-\frac{t'}{\tau_p}+\frac{t_2}{\tau_p}} \Bigg(1- e^{-\frac{t-t_2}{\tau_A}}\Bigg)\nonumber\\
& +  \int_{-\infty}^{t'}dt_2\,\tau_A e^{-\frac{t'-t_2}{\tau_p}}\nonumber\\
&=2 \tau_A \tau_p-\frac{\tau_A^2 \tau_p}{\tau_A+\tau_p}e^{-\frac{|t-t'|}{\tau_A}}.\label{rho_0p}
\end{align}
To evaluate Eq. (\ref{rho_pq_A}), we need to compute the following integral, with the detailed steps as follows:
\begin{align}
&  \mathcal{I}_3(t,t')= \int_{-\infty}^{t}dt_1  \int_{-\infty}^{t'}dt_2\, e^{-\frac{t}{\tau_p}+\frac{t_1}{\tau_p}-\frac{t'}{\tau_q}+\frac{t_2}{\tau_q}}\,\mathcal{K}(t_1,t_2)\nonumber\\
&=  \int_{-\infty}^{t}dt_1  \int_{-\infty}^{t'}dt_2\, e^{-\frac{t}{\tau_p}+\frac{t_1}{\tau_p}-\frac{t'}{\tau_q}+\frac{t_2}{\tau_q}}\,\exp\left(-\frac{|t_1-t_2|}{\tau_A}\right)\nonumber\\
&= \int_{-\infty}^{t}dt_1  \int_{-\infty}^{t}dt_2\, e^{-\frac{t}{\tau_p}+\frac{t_1}{\tau_p}-\frac{t'}{\tau_q}+\frac{t_2}{\tau_q}-\frac{t_1}{\tau_A}+\frac{t_2}{\tau_A}}\nonumber\\
& \times \Theta(t-t')\Theta(t'-t_2)\Theta(t_1-t_2)+ \int_{-\infty}^{t}dt_1  \int_{-\infty}^{t}dt_2 \nonumber\\
&\, e^{-\frac{t}{\tau_p}+\frac{t_1}{\tau_p}-\frac{t'}{\tau_q}+\frac{t_2}{\tau_q}-\frac{t_2}{\tau_A}+\frac{t_1}{\tau_A}}\Theta(t-t')\Theta(t'-t_2)\Theta(t_2-t_1)\nonumber\\
&= \int_{-\infty}^{t}dt_2\, \Theta(t-t')\Theta(t'-t_2) \int_{t_2}^{t}dt_1\,\nonumber\\
&  e^{-\frac{t}{\tau_p}+\frac{t_1}{\tau_p}-\frac{t'}{\tau_q}+\frac{t_2}{\tau_q}-\frac{t_1}{\tau_A}+\frac{t_2}{\tau_A}} +  \int_{-\infty}^{t}dt_2 \Theta(t-t')\Theta(t'-t_2)\nonumber\\
& \times \int_{-\infty}^{t_2}dt_1\, e^{-\frac{t}{\tau_p}+\frac{t_1}{\tau_p}-\frac{t'}{\tau_q}+\frac{t_2}{\tau_q}-\frac{t_2}{\tau_A}+\frac{t_1}{\tau_A}}.
\end{align}
Further integration over $t_2$ simplifies the above expression to yield
\begin{align}
& \mathcal{I}_3(t,t')= \int_{-\infty}^{t'}dt_2\frac{\tau_A \tau_p}{\tau_A-\tau_p}\Bigg( e^{-\frac{t-t_2}{\tau_A}-\frac{t'-t_2}{\tau_q}}-e^{-\frac{t-t_2}{\tau_p}-\frac{t'-t_2}{\tau_q}}\Bigg)\nonumber\\
&\qquad\qquad +  \int_{-\infty}^{t'}dt_2\frac{\tau_A \tau_p}{\tau_A+\tau_p} e^{-\frac{t-t_2}{\tau_p}-\frac{t'-t_2}{\tau_q}}\nonumber\\
&=\frac{\tau_A^2 \tau_p\tau_q}{(\tau_A-\tau_p)(\tau_A+\tau_q)}e^{-\frac{|t-t'|}{\tau_A}}-\frac{2\tau_A \tau_p^3\tau_q}{(\tau_A^2-\tau_p^2)(\tau_p+\tau_q)}e^{-\frac{|t-t'|}{\tau_p}}.\label{rho_pq}
\end{align}

\section{Autocorrelation of end-to-end distance }
The autocorrelation function of the end-to-end distance $R_e(t)$ can be decomposed into thermal and active components as follows
\begin{align}
\Phi_{corr}(t,t')&= \langle \bs{R}_e(t)^{\rm T}\bs{R}_e(t') \rangle = 48 \sum_{p,q=1}^{\infty} \, \langle \chi_{2p-1}(t) \chi_{2q-1}(t') \rangle \nonumber\\
&= \Phi_{corr,T}(t,t') +\Phi_{corr,A}(t,t'), \label{phi_corr_0} 
\end{align}
where the thermal and active parts of the correlation are given, respectively, as
\begin{subequations}
 \begin{align}
& \Phi_{corr,T}(t,t') = 48\Bigg(\frac{k_B T}{L\zeta}\Bigg) \sum_{p=1}^{\infty} \,\frac{\tau_{2p-1}}{2}\,e^{-\frac{|t-t'|}{\tau_{2p-1}}} \label{phi_corr_0T} \\
 &\Phi_{corr,A}(t,t')= \frac{48}{L^2\zeta^2} \sum_{p,q=1}^{\infty} \, \Bigg[\frac{\tau_A^2 \tau_{2p-1}\tau_{2q-1}\,e^{-\frac{|t-t'|}{\tau_A}}}{(\tau_A-\tau_{2p-1})(\tau_A+\tau_{2q-1})}\nonumber\\
&\qquad-\frac{2\tau_A \tau_{2p-1}^3\tau_{2q-1}}{(\tau_A^2-\tau_{2p-1}^2)(\tau_{2p-1}+\tau_{2q-1})}e^{-\frac{|t-t'|}{\tau_{2p-1}}}\Bigg]\,\mathcal{I}_{(2p-1)(2q-1)}.\label{phi_corr_0A} 
 \end{align}   
\end{subequations}
Taking $t=t'$,  one obtains the initial value of the correlation function, which is equivalent to the end-to-end distance at the steady state, as given in Eq. (\ref{phi_corr0_1}).

\section{Analytical Results of Reconfiguration Time} 
To compute the reconfiguration time $\langle \tau_{con}\rangle$, as given in Eq. (\ref{t_con}),  it is essential to derive the following terms: 
\begin{subequations}
 \begin{align}
 &\int_{0}^{\infty}dt\, \Phi_{corr,T}(t)  =  \frac{24k_B T}{L\zeta} \sum_{p=1}^{\infty} \,\tau_{2p-1}^2 = \frac{k_B T \pi^4}{4 L\zeta}\tau_1^2,
 \end{align} 
  \begin{align}
 & \Phi_{corr,T}(0)  =  \frac{24k_B T}{L\zeta} \sum_{p=1}^{\infty} \,\tau_{2p-1} = \frac{3k_B T \pi^2}{L\zeta}\tau_1.
 \end{align} 
\end{subequations}
Thus, for a passive polymer: 
\begin{align}
 \langle \tau_{con}\rangle_T =\frac{\int_{0}^{\infty}dt\, \Phi_{corr,T}(t)}{\Phi_{corr,T}(0)} \sim \tau_1 \sim L^2.\label{t_con_passive}   
\end{align}
For general values of $f_1$ and $f_2$, the active part does not simplify easily. However, for a fully active polymer in the absence of thermal noise, the active part simplifies to  
\begin{subequations}
  \begin{align}
 &\frac{\int_{0}^{\infty}dt\, \Phi_{corr,A}(t)}{\Phi_{corr,A}(0)}  =  \frac{48F_A^2}{L^2\zeta^2} \sum_{p,q=1}^{\infty} \, \Bigg[\frac{\tau_A^3 \tau_{2p-1}\tau_{2q-1}}{(\tau_A-\tau_{2p-1})(\tau_A+\tau_{2q-1})}\nonumber\\
&\qquad-\frac{2\tau_A \tau_{2p-1}^4\tau_{2q-1}}{(\tau_A^2-\tau_{2p-1}^2)(\tau_{2p-1}+\tau_{2q-1})}\Bigg]\frac{1}{2\Phi_{corr,A}(0)}.\\
&\langle \tau_{con}\rangle_A= \frac{\tau_1\pi^2}{12} \frac{1}{1-\frac{2}{\pi}\sqrt{\frac{\tau_A}{\tau_1}}\tanh{\left(\frac{\pi}{2}\sqrt{\frac{\tau_1}{\tau_A}}\right)}},\label{t_con_act}
 \end{align} 
\end{subequations}
where the correlation function due to activity at $t=0$ is expressed as
\begin{align}
&\Phi_{corr,A}(0)=  \frac{48F_A^2}{2L^2\zeta^2} \sum_{p,q=1}^{\infty} \, \Bigg[\frac{\tau_A^2 \tau_{2p-1}\tau_{2q-1}}{(\tau_A-\tau_{2p-1})(\tau_A+\tau_{2q-1})}\nonumber\\
&\qquad\qquad-\frac{2\tau_A \tau_{2p-1}^3\tau_{2q-1}}{(\tau_A^2-\tau_{2p-1}^2)(\tau_{2p-1}+\tau_{2q-1})}\Bigg]. 
\end{align}
From Eq. (\ref{t_con_act}), it can be seen that in the limit $\tau_1/\tau_A\gg 1$, i.e., for long chains,   $\langle \tau_{con}\rangle_A \sim \tau_1 \sim L^2.$ In the opposite limit, or for small $L$ limit, $\langle \tau_{con}\rangle_A \sim L^{0}.$

\section{Wilemski–Fixman Approximation for Looping Kinetics}
Here, we outline the formalism developed by Wilemski and Fixman (WF) to study the cyclization process of polymers, specifically the loop formation between two end segments \cite{wilemski1973general}. Suppose the conditional probability (or propagator) of the end-to-end distance $\bs{R}$ at time $t$, given that it had a value $\bs{R}_0$ at time $t=0$, is denoted by $\mathcal{G}(\bs{R},t|\bs{R}_0,0)$. This propagator can be obtained from the Smoluchowski equation associated with the end-to-end distance dynamics, which is given by $\Bigg[\frac{\partial}{\partial t}-\mathcal{L}\Bigg]\mathcal{G}(\bs{R},t|\bs{R}_0,0)=\delta(t)\delta(\bs{R}-\bs{R}_0)$, where $\mathcal{L}$ is the evolution operator. 
The probability $P_0(\bs{R},t)$ of having an end-to-end distance $\bs{R}$ at any time $t$ obeys the equation $\Bigg[\frac{\partial}{\partial t}-\mathcal{L}\Bigg]P_0(\bs{R},t)=0$. To describe the irreversible cyclization between the two ends, we can introduce a sink term $\mathcal{S}(\bs{R})$ into the evolution equation, which accounts for the loss of probability whenever the two end segments approach a reaction radius $a_0$, resulting in loop formation. The evolution equation in the presence of the sink term then reads \begin{align}
    \frac{\partial}{\partial t} P(\bs{R},t)=\mathcal{L} P(\bs{R},t)-\mathcal{S}(\bs{R})P(\bs{R},t).
\end{align}
The sink-sink correlation function, denoted by $\mathcal{C}_{ss}(t)$, can be defined within this framework, and the looping time $\langle \tau_{\text{loop}}\rangle$ can be computed according to the WF approximation as
\begin{align}
 \langle \tau_{\text{loop}}\rangle = \int_{0}^{\infty} dt_1 \left(\frac{\mathcal{C}_{ss}(t_1)}{\mathcal{C}_{ss}(\infty)}-1\right) , \label{t_loop1}
\end{align}
where 
\begin{align}
&\mathcal{C}_{ss}(t)=\nonumber\\
&\int d\bs{R}_0 \int d\bs{R}\, \mathcal{S}(\bs{R}_0)\,\mathcal{G}\left(\bs{R},t|\bs{R}_0,0\right) \,\mathcal{S}(\bs{R})P_0^{st}(\bs{R}) .   \label{sink-sink-corr}
\end{align}
and $P_0^{st}(\bs{R})$ is the steady-state distribution of end-to-end distance. Since all the stochastic forces acting on the polymer are Gaussian, the propagator for the end-to-end distance can still be written as  
\begin{align}
 \mathcal{G}\left(\bs{R},t|\bs{R}(0)=\bs{R}_0,0\right)&=\left(\frac{3}{2\pi \Phi_{corr}(0)\left[1-\Tilde{\Phi}_{corr}^2(t)\right] }\right)^{3/2}\nonumber\\
 &\exp\left(-\frac{3\left(\bs{R}-\Tilde{\Phi}_{corr}(t)\,\bs{R}(0)\right)^2}{2 \Phi_{corr}(0)\left[1-\Tilde{\Phi}_{corr}^2(t)\right] }\right),
\end{align}
where the terms $ \Phi_{corr}(0)$ and $\Tilde{\Phi}_{corr}$ are given in  Eq. (\ref{phi_corr0_1})  and Eq. (\ref{phi_corr}), respectively. Assuming that looping occurs regardless of the direction in which the two end segments bind, the sink is only a function of the distance $R$. For irreversible looping, the sink can be modeled as a delta function, $ \mathcal{S}(R)=\delta(R-a_0),$ which simplifies the expression for $\mathcal{C}_{ss}(t)$ to 
\begin{align}
\mathcal{C}_{ss}(t) &= \frac{54}{\pi\Phi_{corr}^2(0)\sqrt{1-\Tilde{\Phi}_{corr}^2(t)}}\int_{0}^{\infty}dR_0\,R_0^2 \delta(R_0-a_0)& \nonumber\\
& \times\int_{0}^{\infty}dR\,R^2 \delta(R-a_0)\,\frac{\sinh{\left(\frac{3RR_0\Tilde{\Phi}_{corr}(t)}{\Phi_{corr}(0)\left(1-\Tilde{\Phi}_{corr}^2(t)\right)}\right)}}{3RR_0\Tilde{\Phi}_{corr}(t)}&\nonumber\\
&\times\exp\left(-\frac{3\left(R^2+R_0^2\right)}{2\Phi_{corr}(0)\left(1-\Tilde{\Phi}_{corr}^2(t)\right)}\right)&
\end{align}
and $z_0=\frac{3a_0^2}{2\Phi_{corr}(0)}.$
After integration over $R_0$ and $R$,  one arrives at
\begin{align}
\mathcal{C}_{ss}(t) =\frac{8 z_0^2}{\pi a_0^2}\,\frac{\sinh{\left(\frac{2z_0 \Tilde{\Phi}_{corr}(t)}{1-\Tilde{\Phi}_{corr}^2(t)}\right)}}{\Tilde{\Phi}_{corr}(t)\sqrt{1-\Tilde{\Phi}_{corr}^2(t)}}\,\exp\left(-\frac{2z_0}{1-\Tilde{\Phi}_{corr}^2(t)}\right).\label{c_ss}
\end{align}
Thus the normalized correlation function can be expressed as 
\begin{align}
  \frac{\mathcal{C}_{ss}(t)}{\mathcal{C}_{ss}(\infty)}=  \frac{\sinh{\left(\frac{2z_0 \Tilde{\Phi}_{corr}(t)}{1-\Tilde{\Phi}_{corr}^2(t)}\right)}}{2z_0\Tilde{\Phi}_{corr}(t)\sqrt{1-\Tilde{\Phi}_{corr}^2(t)}}\,\exp\left(-\frac{2z_0\Tilde{\Phi}_{corr}^2(t)}{1-\Tilde{\Phi}_{corr}^2(t)}\right).\label{c_ss_norm}
\end{align}

By virtue of Eq. (\ref{c_ss_norm}) and Eq. (\ref{t_loop1}), the looping time can be computed as
\begin{align}
&\langle \tau_{\text{loop}}\rangle =\nonumber\\
&\int_{0}^{\infty} dt'\left[\frac{\sinh{\left(\frac{2z_0 \Tilde{\Phi}_{corr}(t')}{1-\Tilde{\Phi}_{corr}^2(t')}\right)}\,\exp\left(-\frac{2z_0\Tilde{\Phi}_{corr}^2(t')}{1-\Tilde{\Phi}_{corr}^2(t')}\right)}{2z_0\Tilde{\Phi}_{corr}(t')\sqrt{1-\Tilde{\Phi}_{corr}^2(t')}}-1\right].\label{t_loop2}
\end{align}
Note that in the above expression, the integrand becomes divergent as $t\rightarrow 0$ since $\Tilde{\Phi}_{corr}(0)\rightarrow 1$, making it challenging to compute directly. However, following the approach in Ref. \cite{pastor1996diffusion}, one can expand the integrand around $z_0$ and retain terms up to first order. This yields the following expression, which is similar to the one obtained when considering a Heaviside sink function, viz.
\begin{align}
 &\frac{\sinh{\left(\frac{2z_0 \Tilde{\Phi}_{corr}(t')}{1-\Tilde{\Phi}_{corr}^2(t')}\right)}\,\exp\left(-\frac{2z_0\Tilde{\Phi}_{corr}^2(t')}{1-\Tilde{\Phi}_{corr}^2(t')}\right)}{2z_0\Tilde{\Phi}_{corr}(t')\sqrt{1-\Tilde{\Phi}_{corr}^2(t')}}\nonumber\\
 & \approx  \left(1-\Tilde{\Phi}_{corr}^2(t')+\frac43 z_0 \Tilde{\Phi}_{corr}^2(t') \right)^{-3/2}.\label{t_loop3}
\end{align}
Thus, with the above approximation, the looping time can be re-expressed as
 \begin{align}
 \langle \tau_{\text{loop}}\rangle = \int_{0}^{\infty} dt' \left[\frac{1}{\left(1-\Tilde{\Phi}_{corr}^2(t')\left[1-\frac43 z_0\right] \right)^{3/2}}-1\right].   \label{t_loop4}
 \end{align}

By taking $y=e^{-t/\tau_1}$, the above can be recast as 
 \begin{align}
 \langle \tau_{\text{loop}}\rangle = \tau_1\int_{0}^{1} \frac{dy}{y} \left[\frac{1}{\left(1-\Tilde{\Phi}_{corr}^2(y)\left[1-\frac43 z_0\right] \right)^{3/2}}-1\right],  \label{t_loop5}
 \end{align}
where [see Eq. (\ref{phi_corr_0} )]
\begin{subequations}
 \begin{align}
& \Tilde{\Phi}_{corr}(y) = \Phi_{corr}(y)/\Phi_{corr}(1),\nonumber\\
&\Phi_{corr}(y)= \Phi_{corr,T}(y) +\Phi_{corr,A}(y),\nonumber\\ 
&  \Phi_{corr,T}(y) = 24\Bigg(\frac{k_B T}{L\zeta}\Bigg) \sum_{p=1}^{\infty} \,\tau_{2p-1}\,y^{(2p-1)^2},\\
&\Phi_{corr,A}(y)= \frac{48}{L^2\zeta^2} \sum_{p,q=1}^{\infty} \, \Bigg[\frac{\tau_A^2 \tau_{2p-1}\tau_{2q-1}\,y^{\frac{\tau_1}{\tau_A}}}{(\tau_A-\tau_{2p-1})(\tau_A+\tau_{2q-1})}\nonumber\\
&\qquad-\frac{2\tau_A \tau_{2p-1}^3\tau_{2q-1}\,y^{(2p-1)^2}}{(\tau_A^2-\tau_{2p-1}^2)(\tau_{2p-1}+\tau_{2q-1})}\Bigg]\,\mathcal{I}_{(2p-1)(2q-1)}.\label{phi_corr_y} 
 \end{align}  
\end{subequations}

\begin{figure*}[htp]
    \centering
    \begin{subfigure}[b]{0.495\textwidth}
    \caption{}
        \centering
        \includegraphics[width=\textwidth]{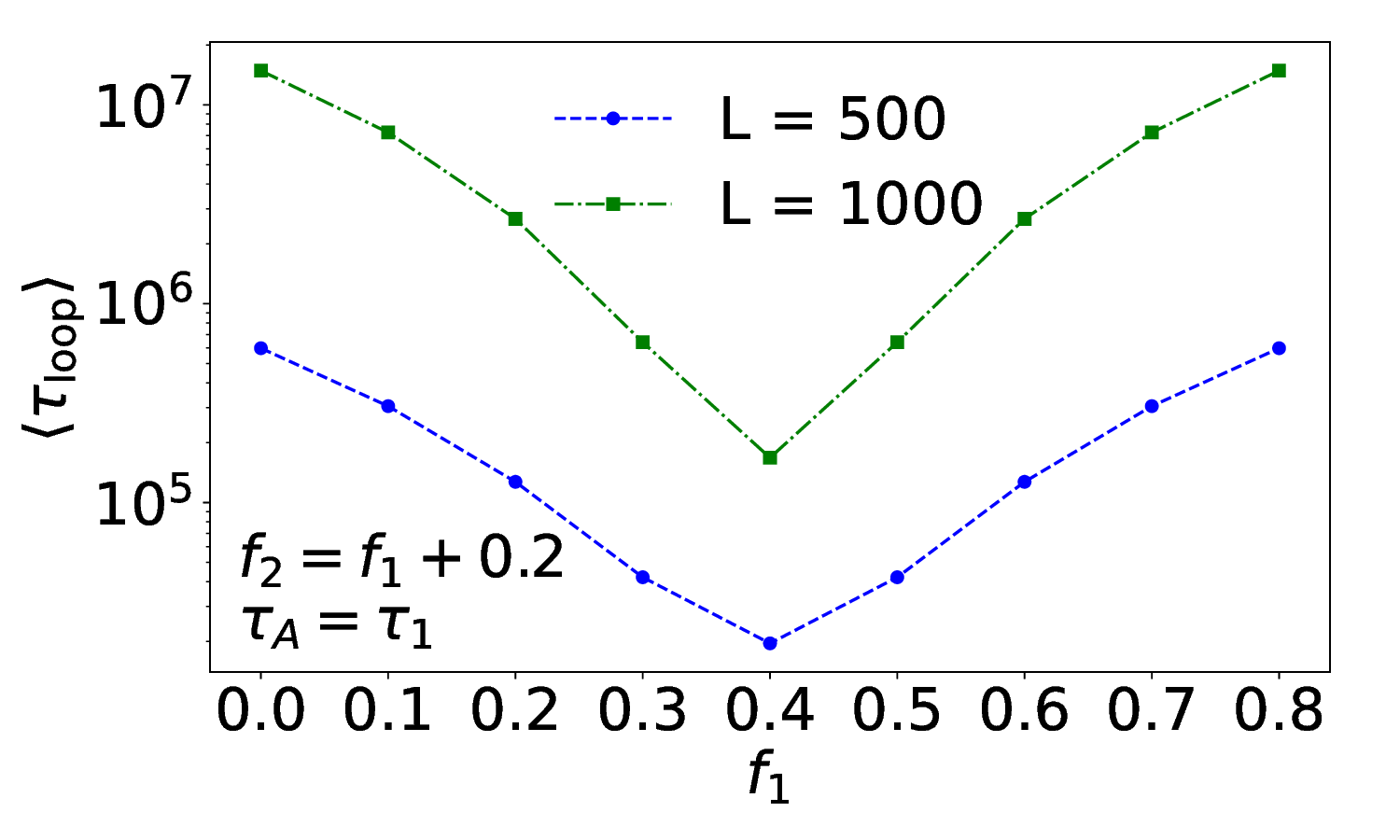}
    \end{subfigure}
    \hfill
    \begin{subfigure}[b]{0.495\textwidth}
    \caption{}
        \centering
        \includegraphics[width=\textwidth]{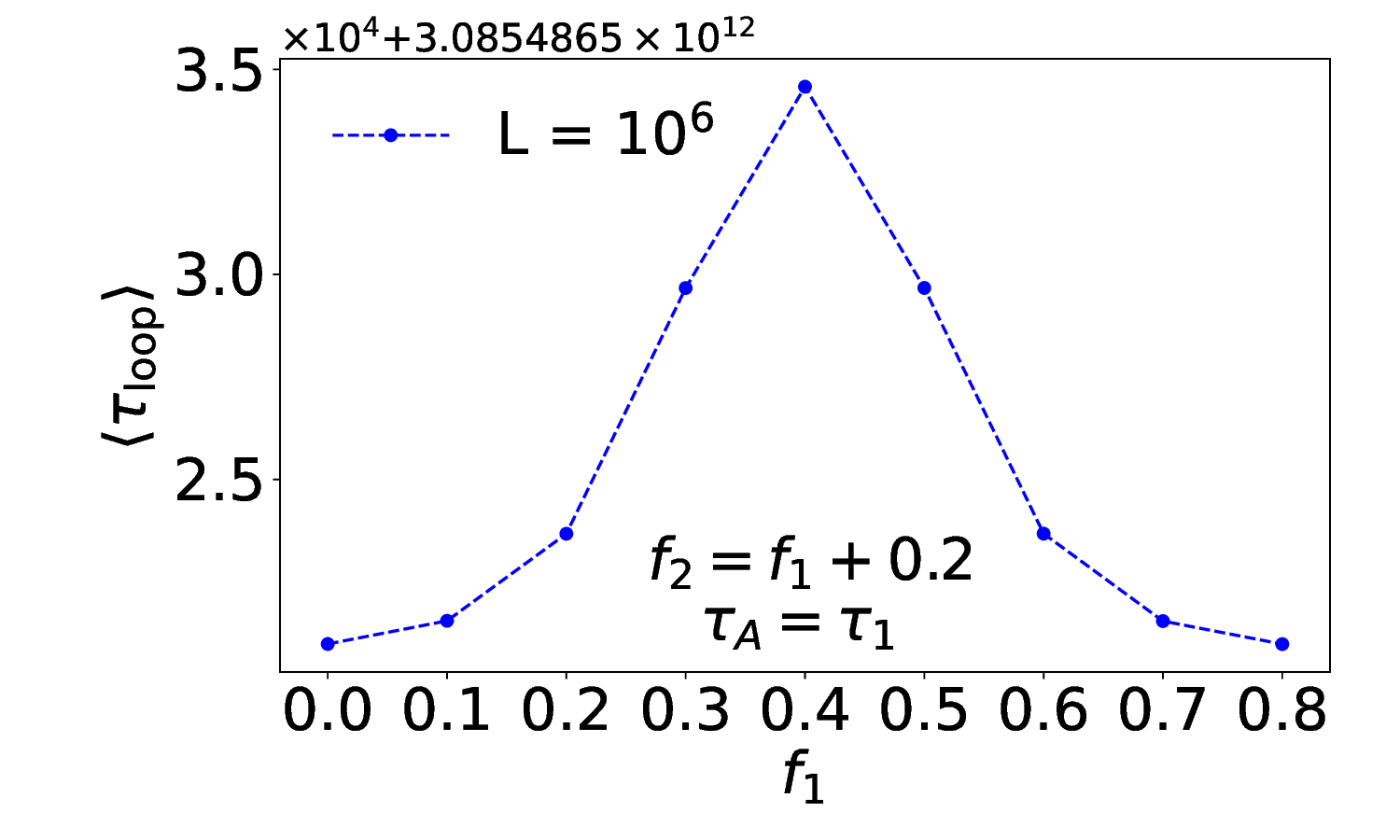}
    \end{subfigure}
        \centering
    \begin{subfigure}[b]{0.495\textwidth}
    \caption{}
        \centering
        \includegraphics[width=\textwidth]{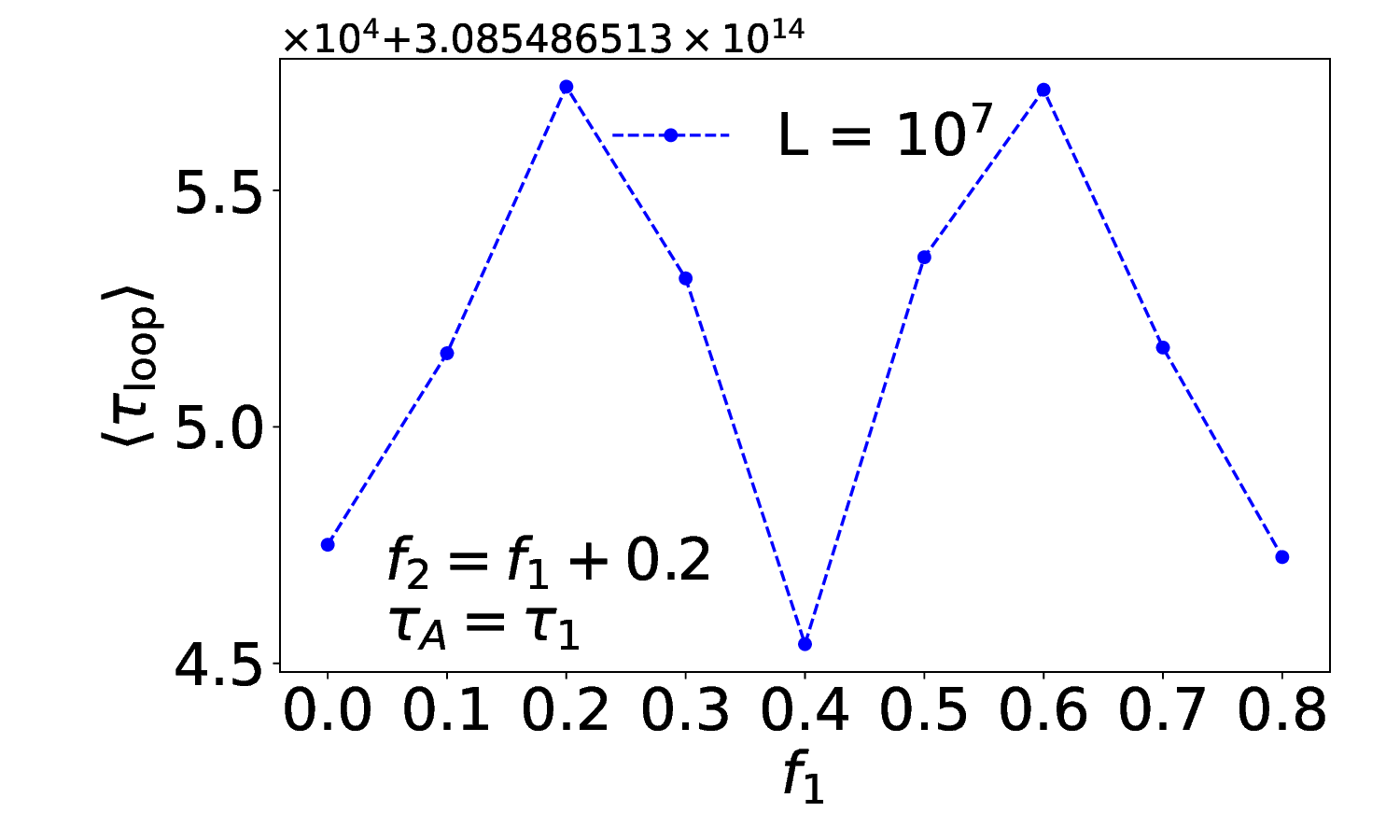}
    \end{subfigure}
    \hfill
    \begin{subfigure}[b]{0.495\textwidth}
    \caption{}
        \centering
        \includegraphics[width=\textwidth]{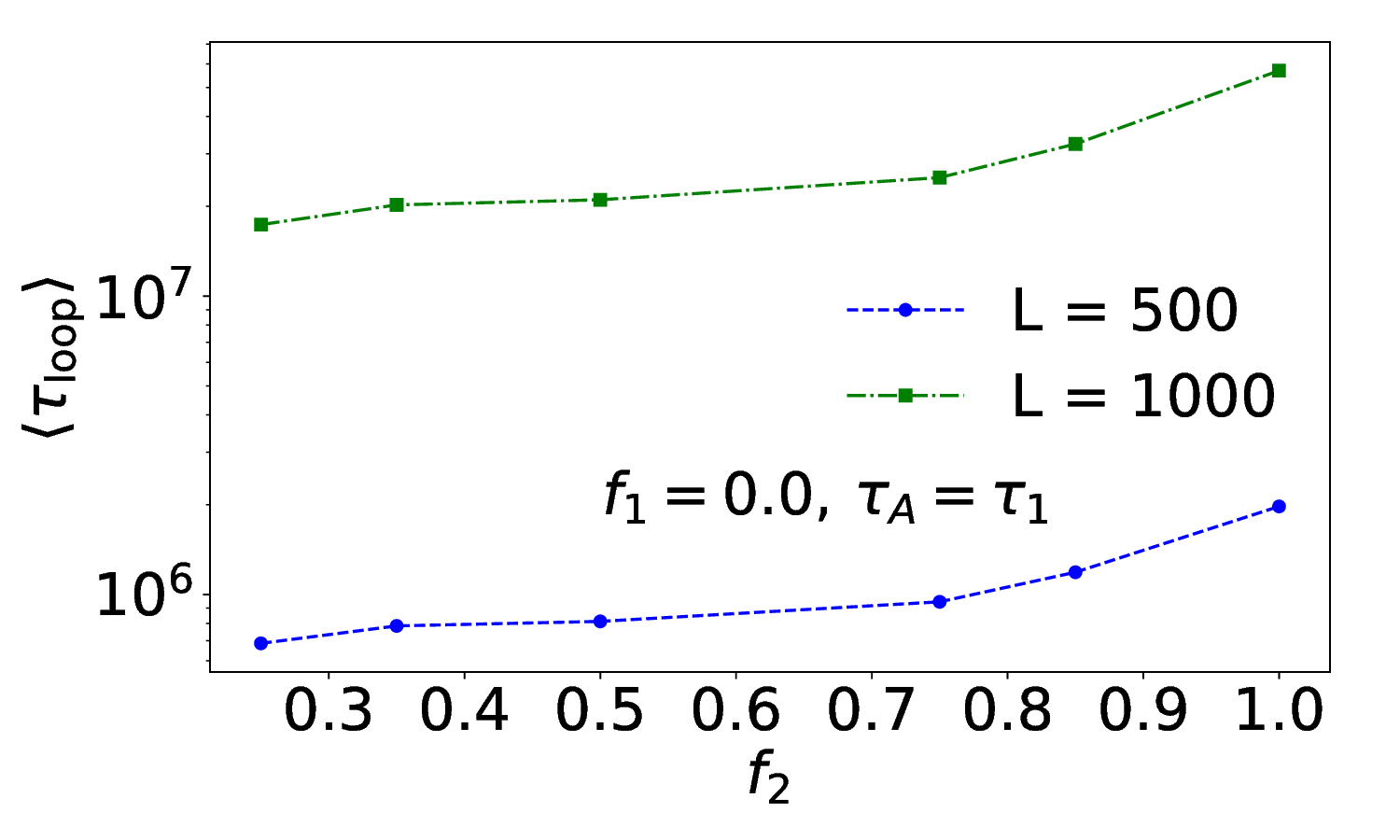}
    \end{subfigure}
\caption{Logarithmic plot of the looping time (a)-(c) as a function of $f_1$, with a fixed separation $f_2 - f_1 = 0.2$, and (b) as a  function of $f_2$, keeping $f_1=0.$  Here,  $\tau_A = \tau_1$, and the values of other parameters are the same as in  Fig. \ref{fig:looping_l}. }
    \label{fig:loop_f1}
\end{figure*}

\subsection{Results of Looping Time  for an Alternative Model of Correlated Active Force}
\begin{figure*}[htp]
    \centering
    \begin{subfigure}[b]{0.495\textwidth}
        \centering
        \caption{$(f_1,\,f_2)=(0.0,\,0.2).$ }
        \includegraphics[width=\textwidth]{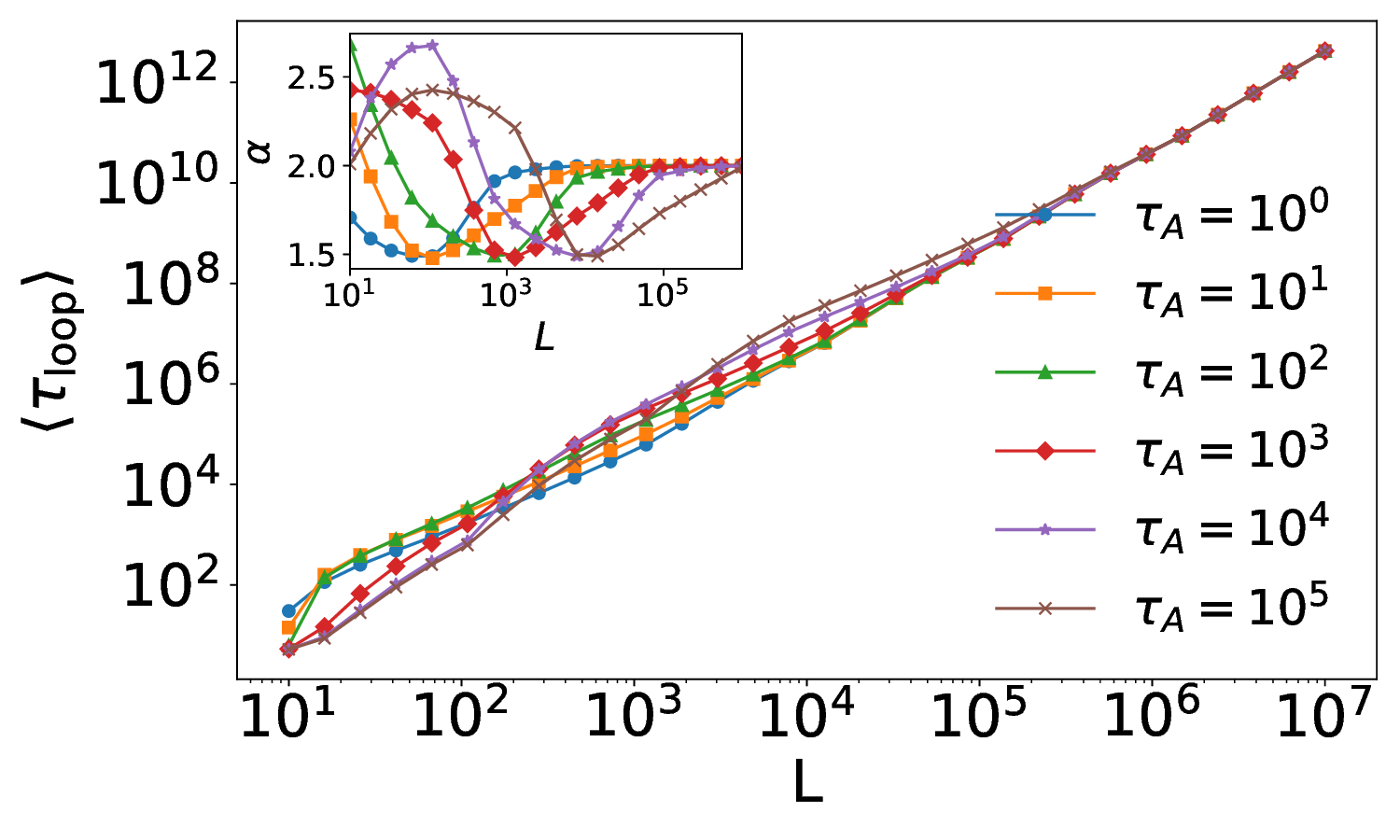}
    \end{subfigure}
    \hfill
    \begin{subfigure}[b]{0.495\textwidth}
        \centering
        \caption{$(f_1,\,f_2)=(0.4,\,0.6).$}
        \includegraphics[width=\textwidth]{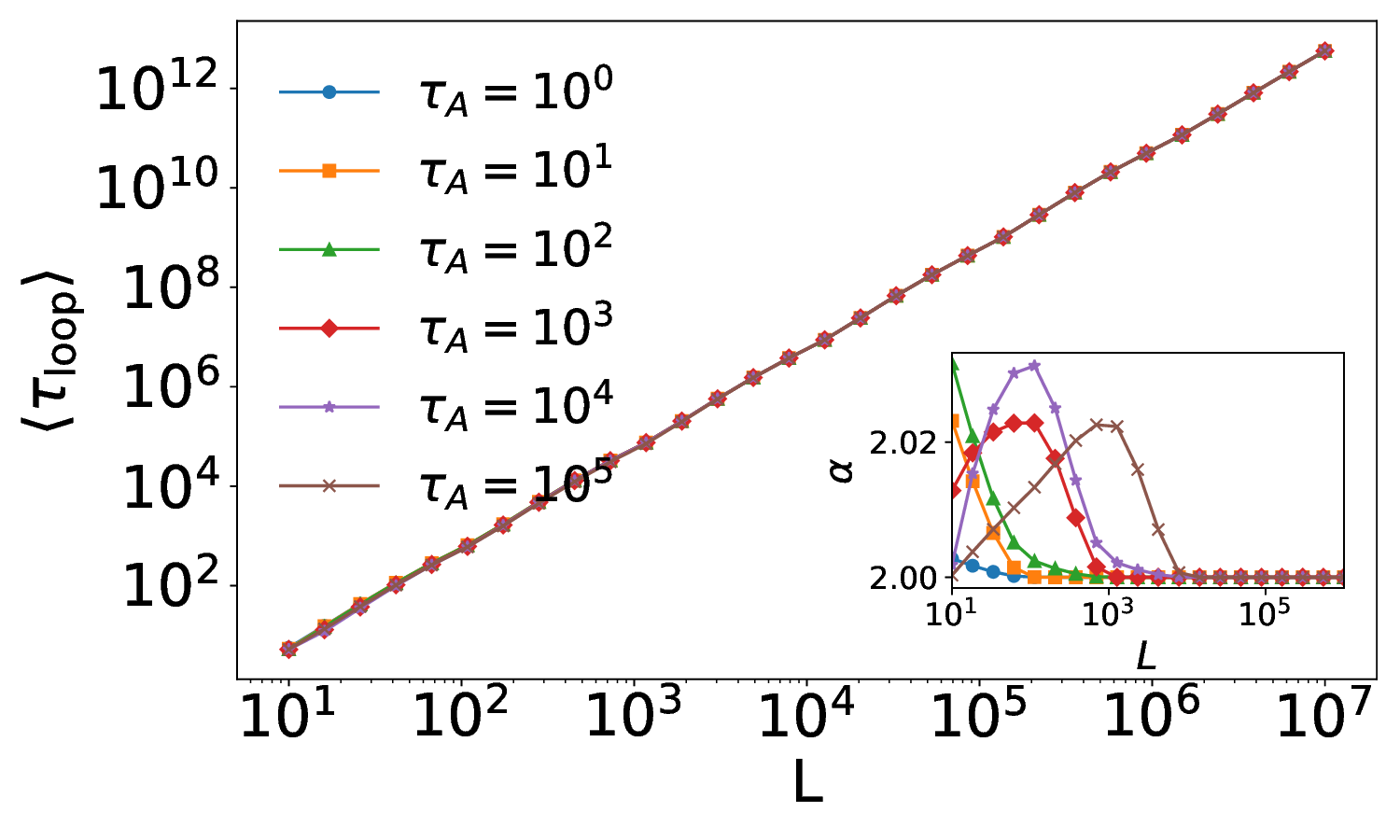}
    \end{subfigure}
    \vspace{0.2cm}
        \centering
    \begin{subfigure}[b]{0.495\textwidth}
        \centering
        \caption{$(f_1,\,f_2)=(0.0,\,0.5).$}
        \includegraphics[width=\textwidth]{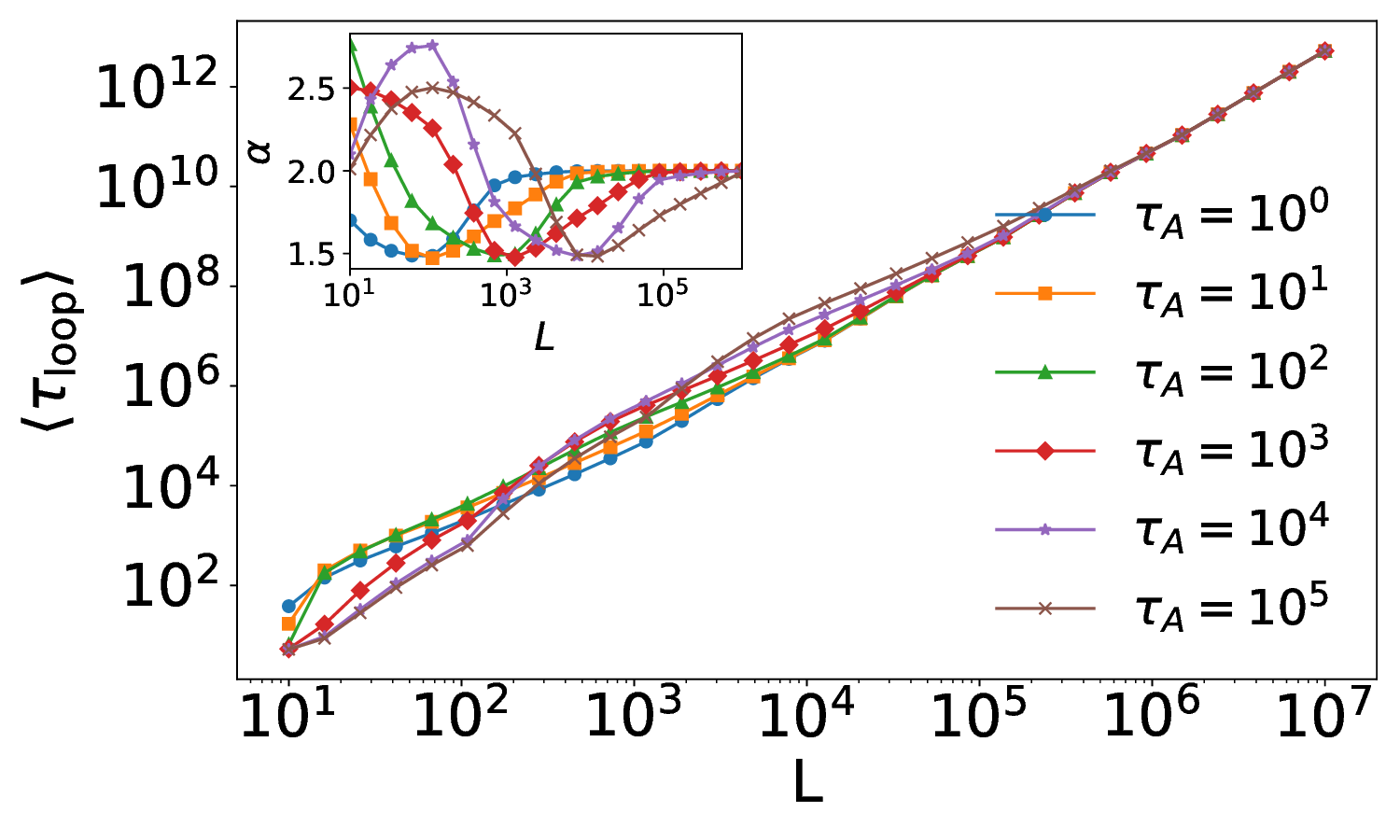}
    \end{subfigure}
    \hfill
    \begin{subfigure}[b]{0.495\textwidth}
        \centering
        \caption{$(f_1,\,f_2)=(0.0,\,1.0).$}
        \includegraphics[width=\textwidth]{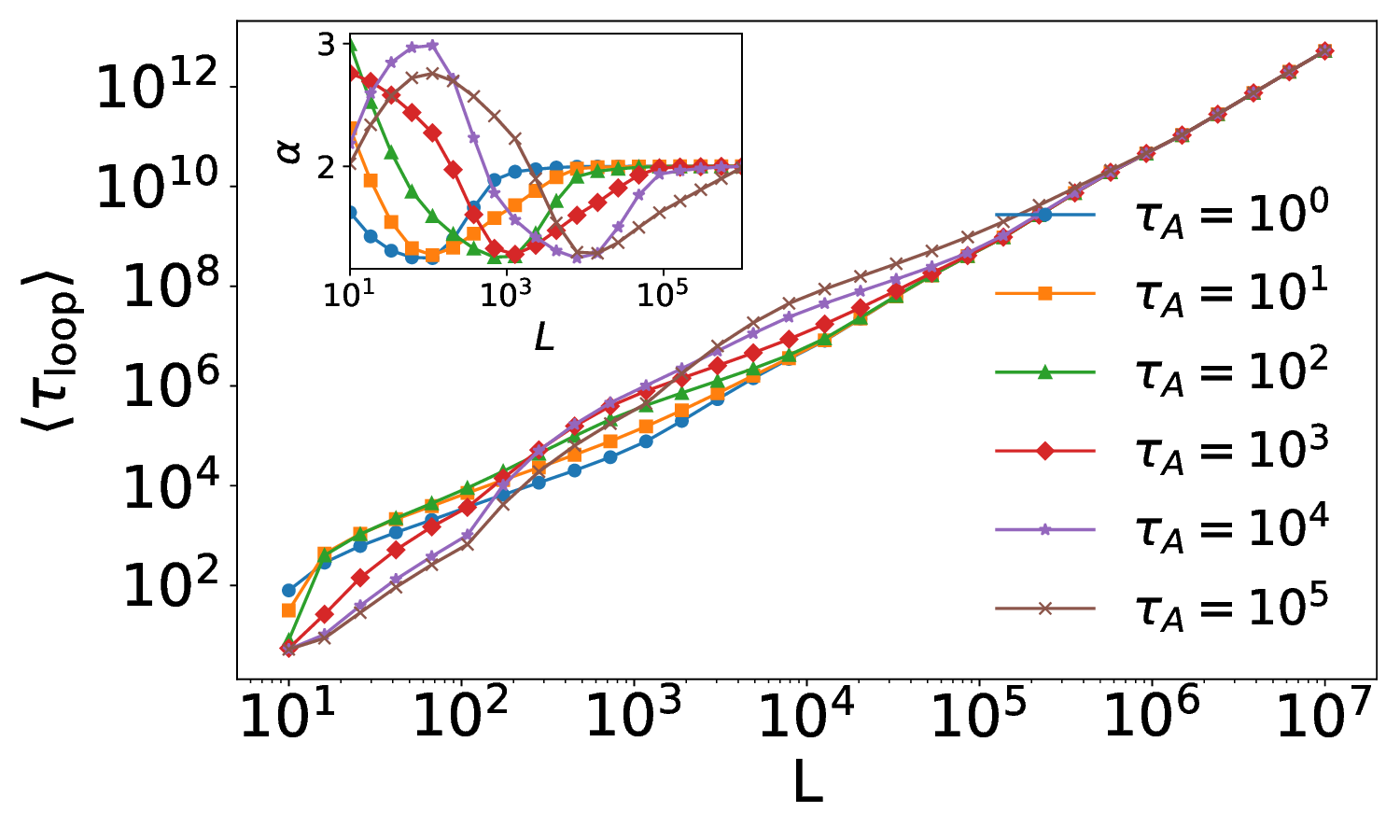}
    \end{subfigure}
\caption{Log-log plot of the looping time as a function of $L$ for an alternative model of correlated active noise [see Eq. (\ref{tempo_corr_2})]. In the insets, the exponent $\alpha$ is plotted as a function of $L$, obtained from the scaling relation $\langle \tau_{\text{con}} \rangle \propto L^{\alpha}$. The values of other parameters are $\kappa=3.0$, $\zeta=1.0$, $k_B T=1.0$, and $F_0^2=10.0$. Here, the reaction radius is set to $a_0=0.01$.}

   \label{fig:looping_l_model2}
\end{figure*}

\begin{figure*}[htp]
    \centering
    \begin{subfigure}[b]{0.495\textwidth}
    \caption{}
        \centering
        \includegraphics[width=\textwidth]{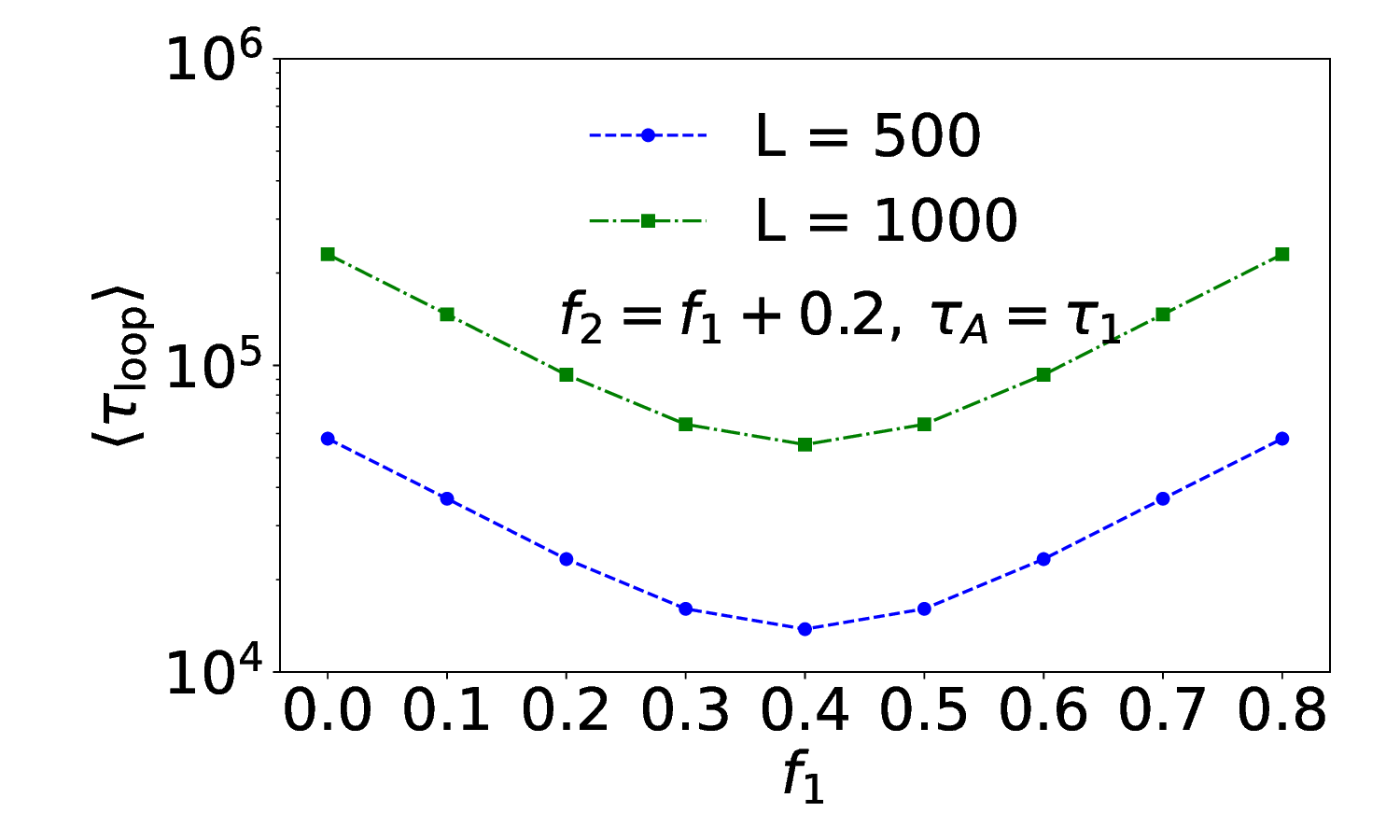}
    \end{subfigure}
    \hfill
    \begin{subfigure}[b]{0.495\textwidth}
    \caption{}
        \centering
        \includegraphics[width=\textwidth]{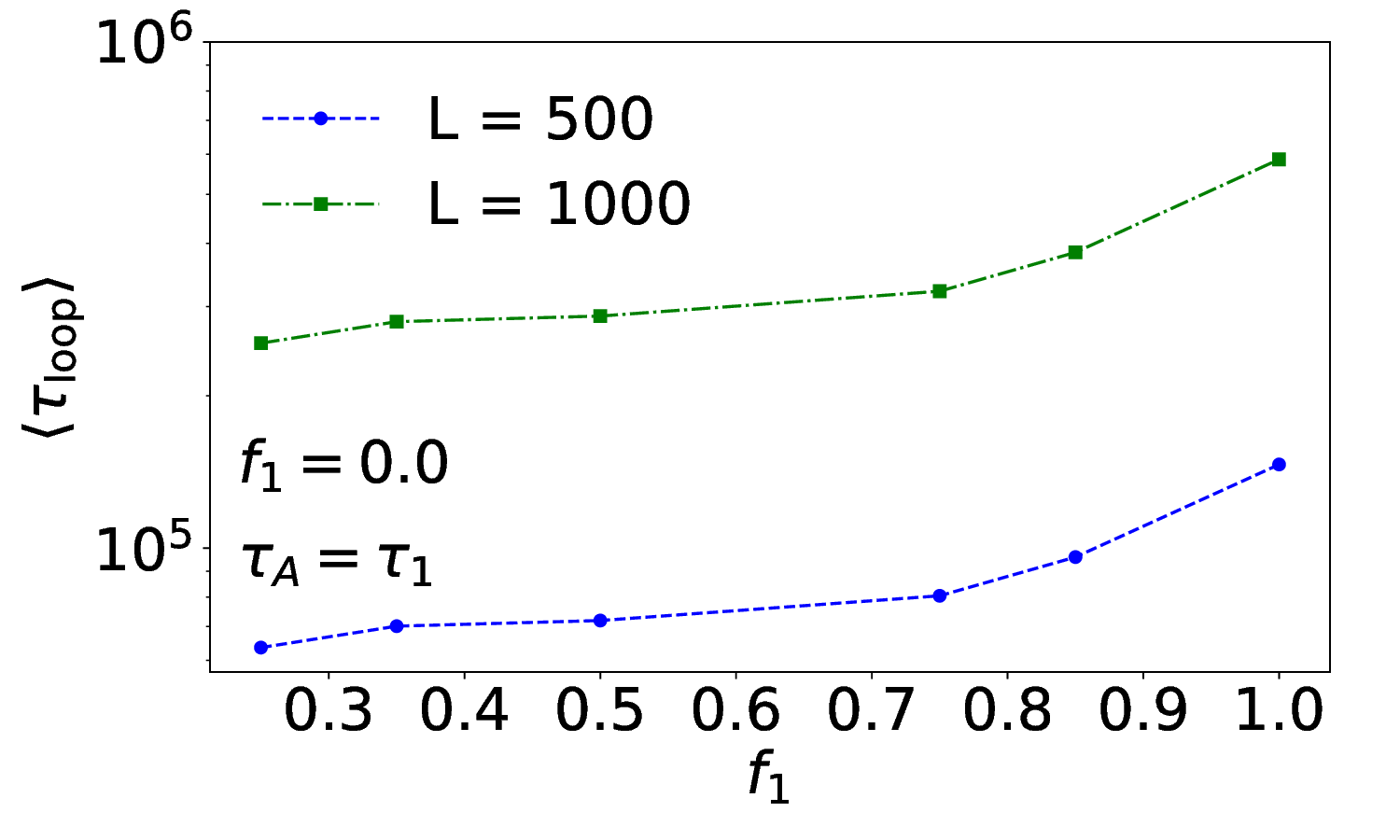}   
    \end{subfigure}
\caption{Logarithmic plot of the looping time for an alternative model of correlated active noise [see Eq. (\ref{tempo_corr_2})], (a)  as a function of $f_1$ with a fixed separation $f_2 - f_1 = 0.2$, and (b)  as a function of $f_2$ keeping $f_1=0.$ Here, we set $\tau_A = \tau_1$, and the values of other parameters are the same as in Fig. \ref{fig:looping_l_model2}.}
    \label{fig:loop_f1_model2}
\end{figure*}

Here, we model the active force as an exponentially correlated noise, where the strength of the correlation depends on the persistence time. Specifically, we define the correlation function as \cite{gos23eff}
\begin{align} \mathcal{K}(t_1, t_2) = \frac{1}{\tau_A} \exp\left(-\frac{|t_1 - t_2|}{\tau_A}\right), \label{tempo_corr_2} \end{align} and incorporate this expression to redefine $\mathcal{I}_{pq}$ in Eq. (\ref{Ipq}) with  
$F_A^2=F_0^2 \frac{L}{\tau_A}.$ Using this model, we compute the looping time, and the results are presented in Figs.  \ref{fig:looping_l_model2}- \ref{fig:loop_f1_model2}. We observe that as the persistence time of the noise increases, the peak corresponding to the largest $\alpha$ value, derived from fitting $\langle \tau_{\text{loop}}\rangle \propto L^{\alpha}$, shifts to longer length scales, as recently reported for fully active polymers in Ref. \cite{ghosh2022active}. However, if the activity is localized in the middle segment, the looping time remains largely unaffected.

Figure \ref{fig:loop_f1_model2} illustrates that, similar to the noise model [given by  Eq. (\ref{tempo_corr})] presented in the main text, $\langle \tau_{\text{loop}}\rangle$ varies nonmonotonically as a function of $f_1$ for an active segment of fixed length. However, the extent of this variation is less pronounced compared to the primary model in Eq. (\ref{tempo_corr}). For the polymer, as activity is introduced progressively along the chain from one end to the other, the looping time also increases, albeit at a slower rate.

\section{Simulation methods and results \label{appen:simulation}}
To corroborate our analytical results, we perform Brownian dynamics simulations of the discretized version of Eq. \ref{polymer_dynamics}, which describes a partially active Rouse polymer. We solve the stochastic differential equation using an improved Runge–Kutta scheme for a polymer consisting of $N=100$ beads with Kuhn length $b_0=1$, so that $L=Nb_0=100$. The simulation is run with a time step of $dt=10^{-3}$ for a total time $\mathcal{T}=2000$, over 2000 independent trajectories. The initial configuration of the polymer is a random Rouse chain with no activity. To generate the active force, Eq. (\ref{ouprocess}) is integrated numerically to obtain the temporal part of the force, while Gaussian random values, corresponding to the strength of the force, are assigned to all beads within the active segment for the spatial component. Excluded volume interactions are neglected, and other parameters are kept the same as in analytical calculations. The parameters are set to $\kappa=3.0$, $\zeta=1.0$, $k_BT=1.0$, $\tau_A=1.0$, and  $F_0=1.0.$  All conformational properties reach plateau values and fluctuate around these levels in the long-time limit, indicating that the system has reached a steady state.  The results of these properties are presented in the following plots.

The conformational properties defined in Sec. \ref{sec-conform} for the continuum Rouse model can be written in their discretized forms, which we use to compute these properties from the simulation data. The center of mass for the polymer at time $t$ is given by $$
\mathbf{r}_{\text{cm}}(t) = \frac{1}{N} \sum_{i=1}^{N} \mathbf{x}_i(t).
$$
The mean square radius of gyration for a single trajectory is calculated as $$
R_g^2(t) = \frac{1}{N} \sum_{i=1}^{N} \left( \sum_{j=1}^{3} \left( x_{i,j}(t) - r_{\text{cm},j}(t) \right)^2 \right),
$$ where $x_{ij}(t)$ is the position of the $i^{\text{th}}$ bead in the $j^{\text{th}}$ spatial dimension. 
Therefore, the ensemble-averaged quantity can be computed by summing over all trajectories as follows $$
\langle R_g^2(t) \rangle =  \frac{1}{N_t} \sum_{k=1}^{N_t}\frac{1}{N} \sum_{i=1}^{N} \left( \sum_{j=1}^{3} \left( x_{i,j}(t) - r_{\text{cm},j}(t) \right)^2 \right),
$$ where $N_t$ is the number of all trajectories. The ensemble averaged mean square end-to-end distance can be expressed as 
$$
\langle R_e^2(t) \rangle =   \frac{1}{N_t}\sum_{k=1}^{N_t}\sum_{j=1}^{3} \left( x_{N,j}(t) - x_{1,j}(t) \right)^2.
$$
For arbitrary points  $n_1$ and $n_2$ along the polymer, the square distance between them is
$$
\langle R_{n_1n_2}^2(t) \rangle =  \frac{1}{N_t}\sum_{k=1}^{N_t} \sum_{j=1}^{3} \left( x_{n_2,j}(t) - x_{n_1,j}(t) \right)^2.
$$
The potential energy is stored in the elastic interaction between consecutive beads and is computed based on the squared difference between their positions over all trajectories as follows
$$
\langle E(t) \rangle = \frac{\kappa}{2}\sum_{k=1}^{N_t} \frac{1}{N_t}\sum_{i=2}^{N} \sum_{j=1}^{3} \left( x_{i,j} - x_{i-1,j} \right)^2.
$$

For a harmonically bound single active particle, the mean squared displacement at the steady state
is given by \cite{goswami2019heat} $\langle x^2\rangle=\left(b_0^2+F_0^2\frac{\tau_A}{1+\kappa \tau_A}\right).$
 Analogous to the equipartition theorem, the energy of the polymer can be approximated as 
\begin{align}
 \langle E(t) \rangle \approx \frac{\kappa}{2}N\left(b_0^2++F_0^2\frac{ \tau_A(f_2-f_1)}{1+\kappa \tau_A}\right).\label{energy_ana} 
\end{align}
 While the analytical results generally follow the trend observed in simulation data, they tend to  slightly underestimate the simulation results. This discrepancy may arise because the equipartition result for a single particle cannot be directly applied to a chain of interacting particles in a straightforward manner.

\begin{figure*}
\centering
\includegraphics[width=0.495\linewidth]{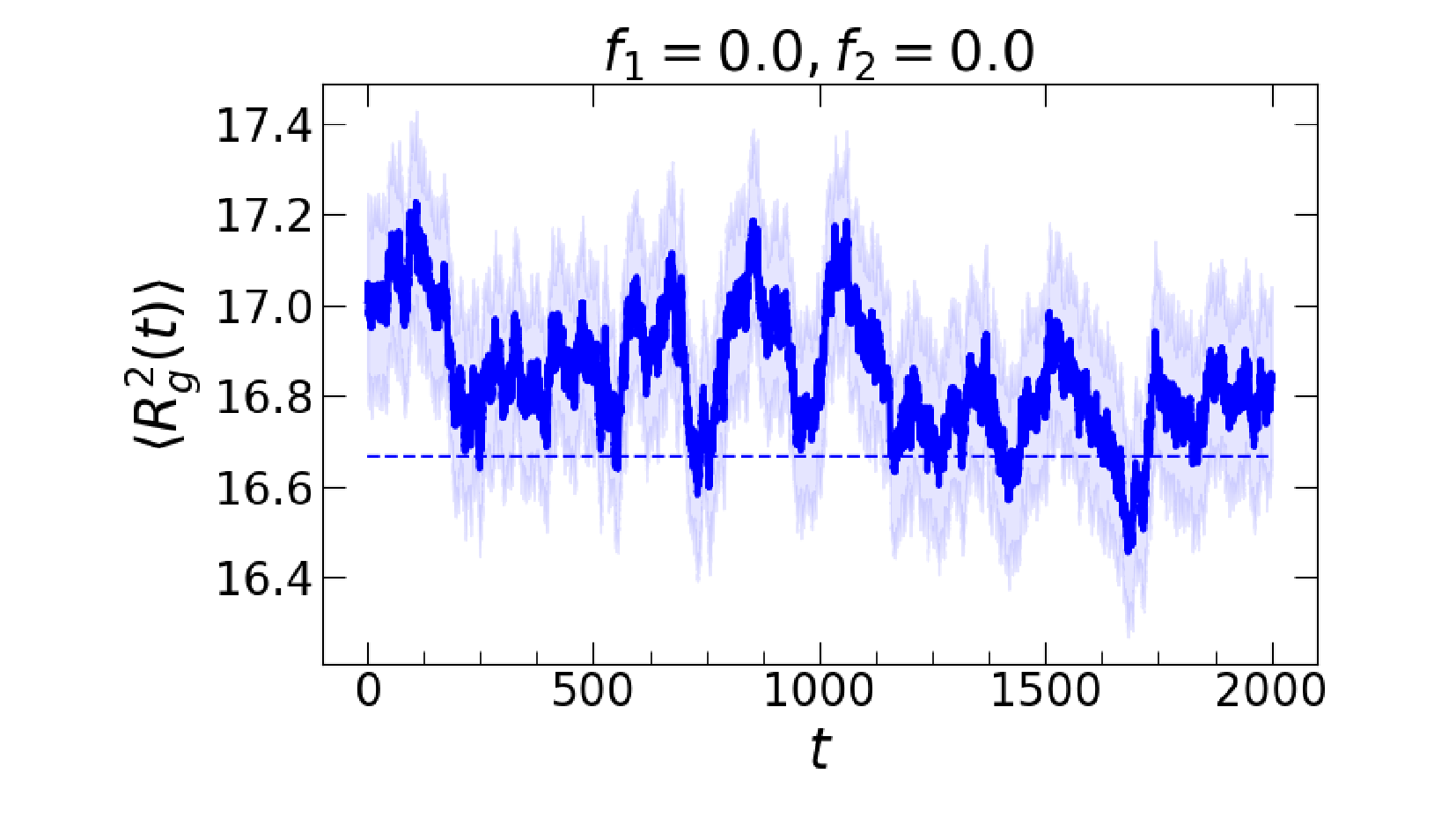}
\includegraphics[width=0.495\linewidth]{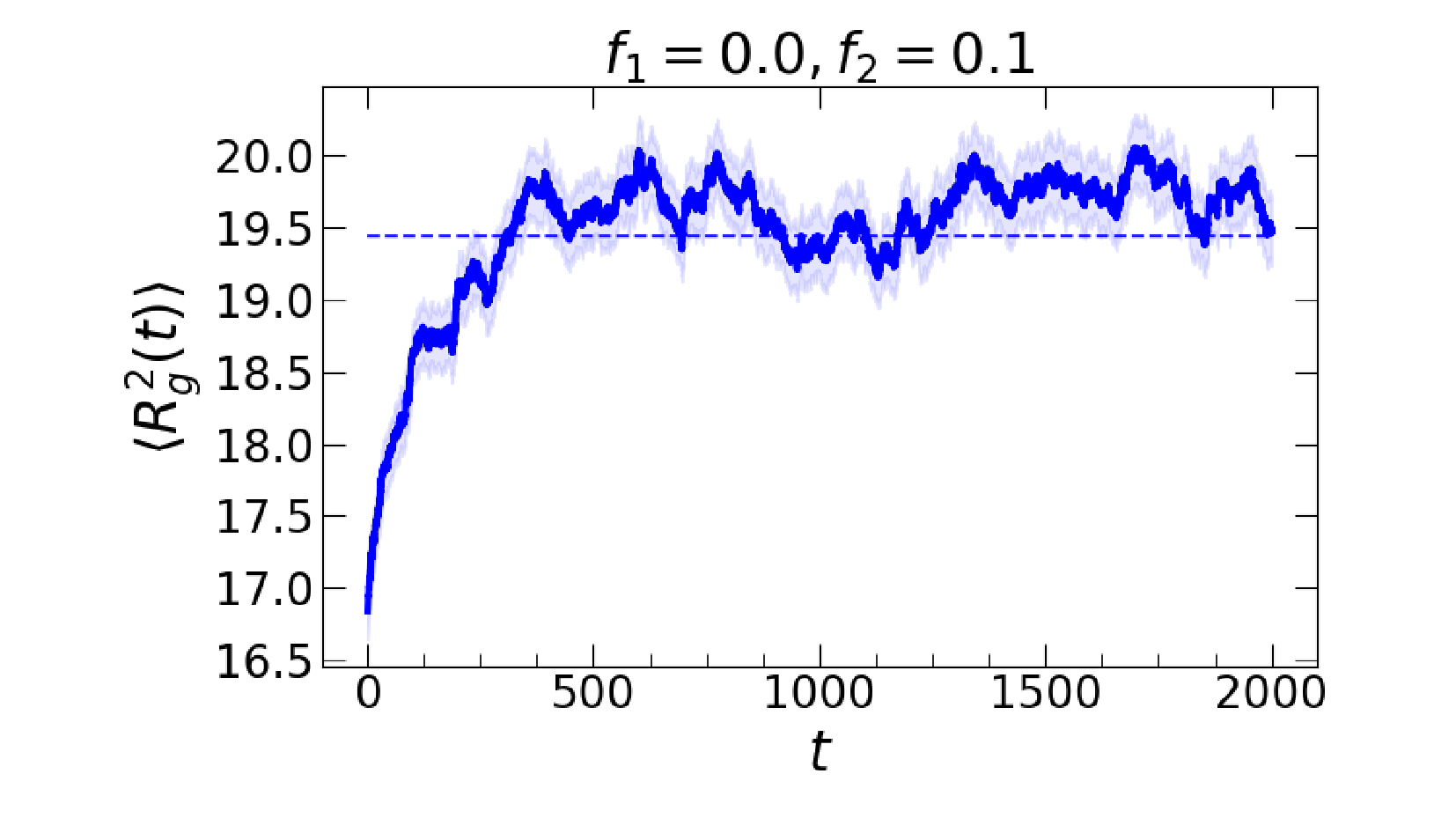}
\includegraphics[width=0.495\linewidth]{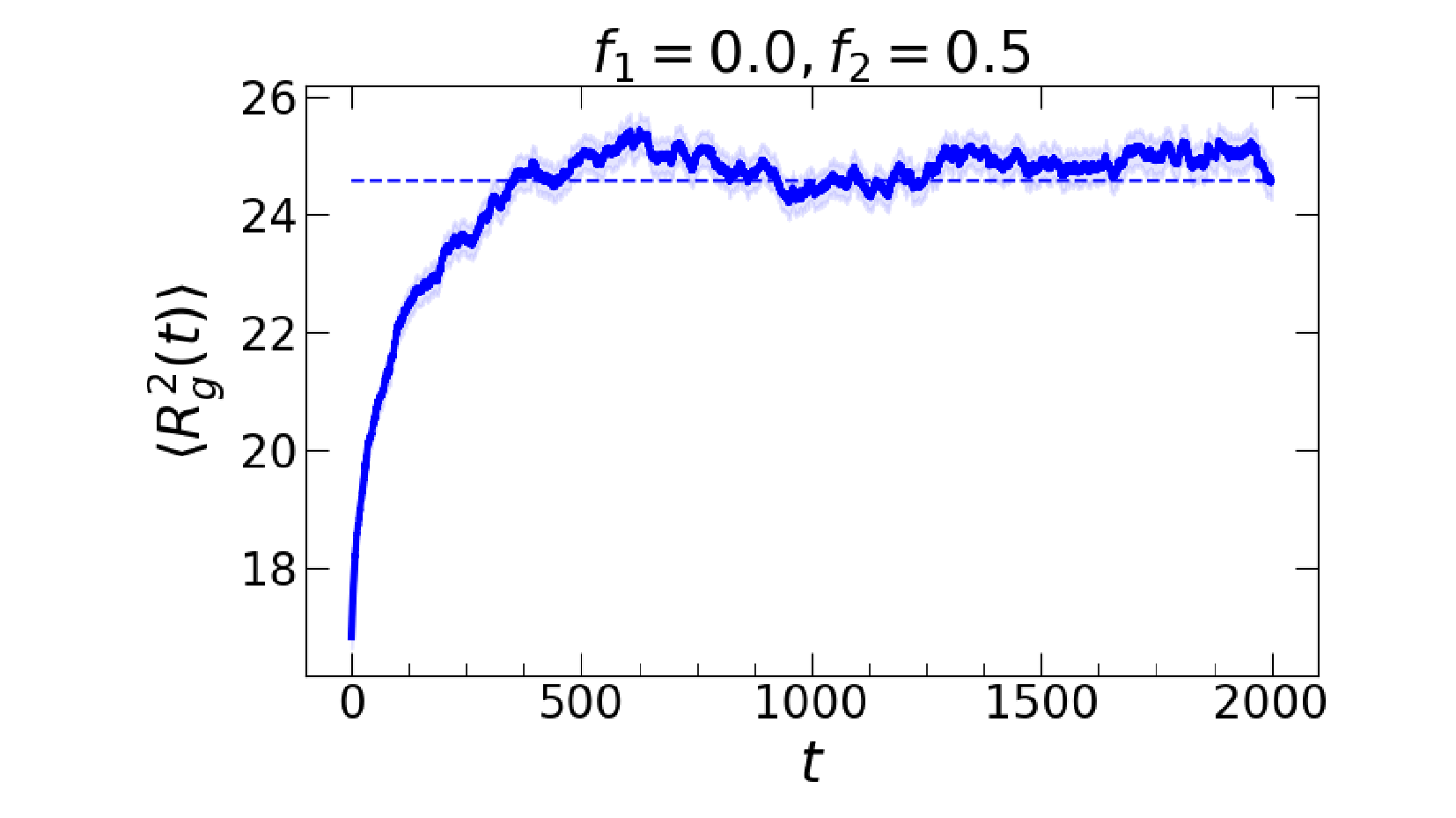}
\includegraphics[width=0.495\linewidth]{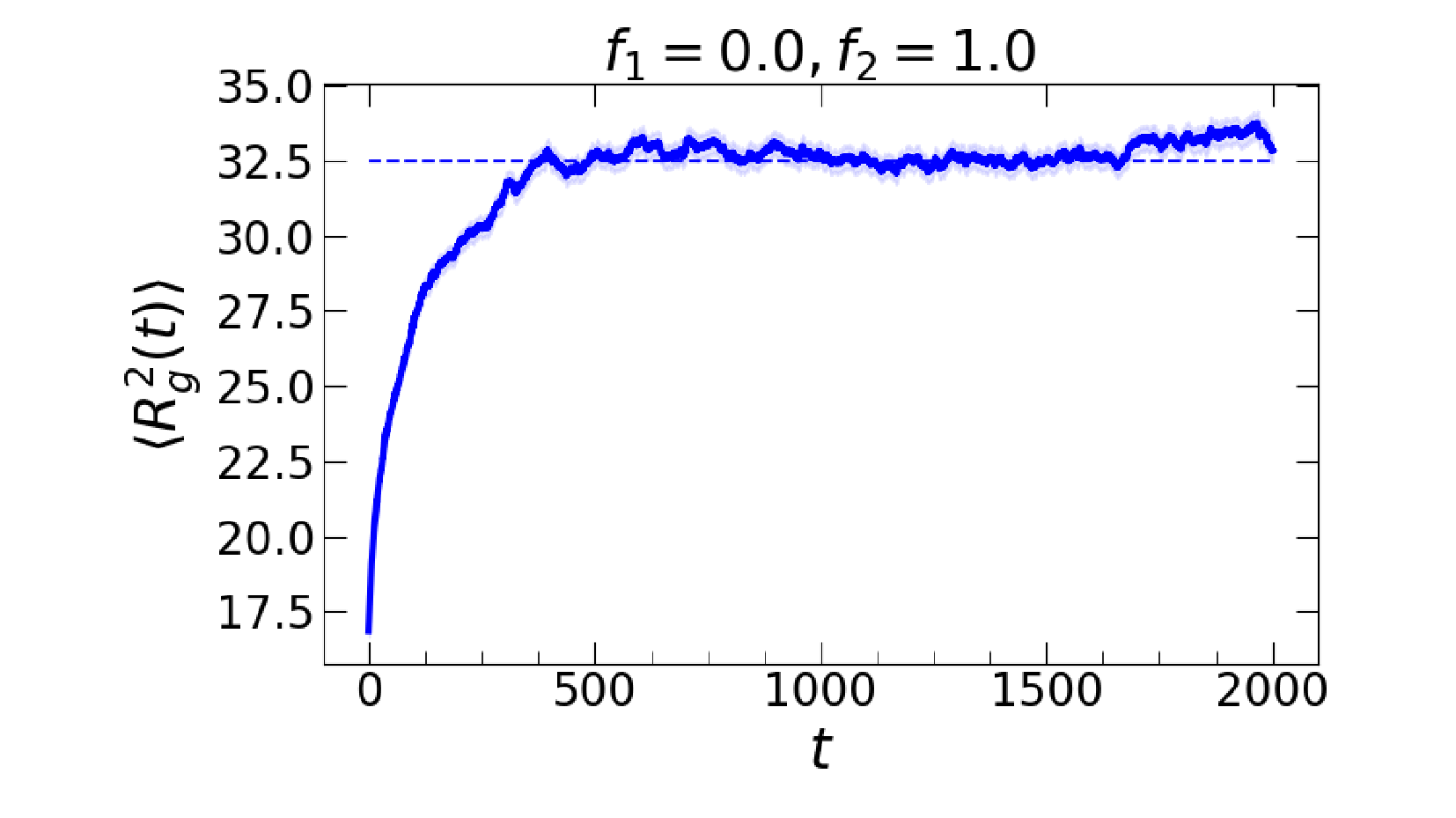}
\includegraphics[width=0.495\linewidth]{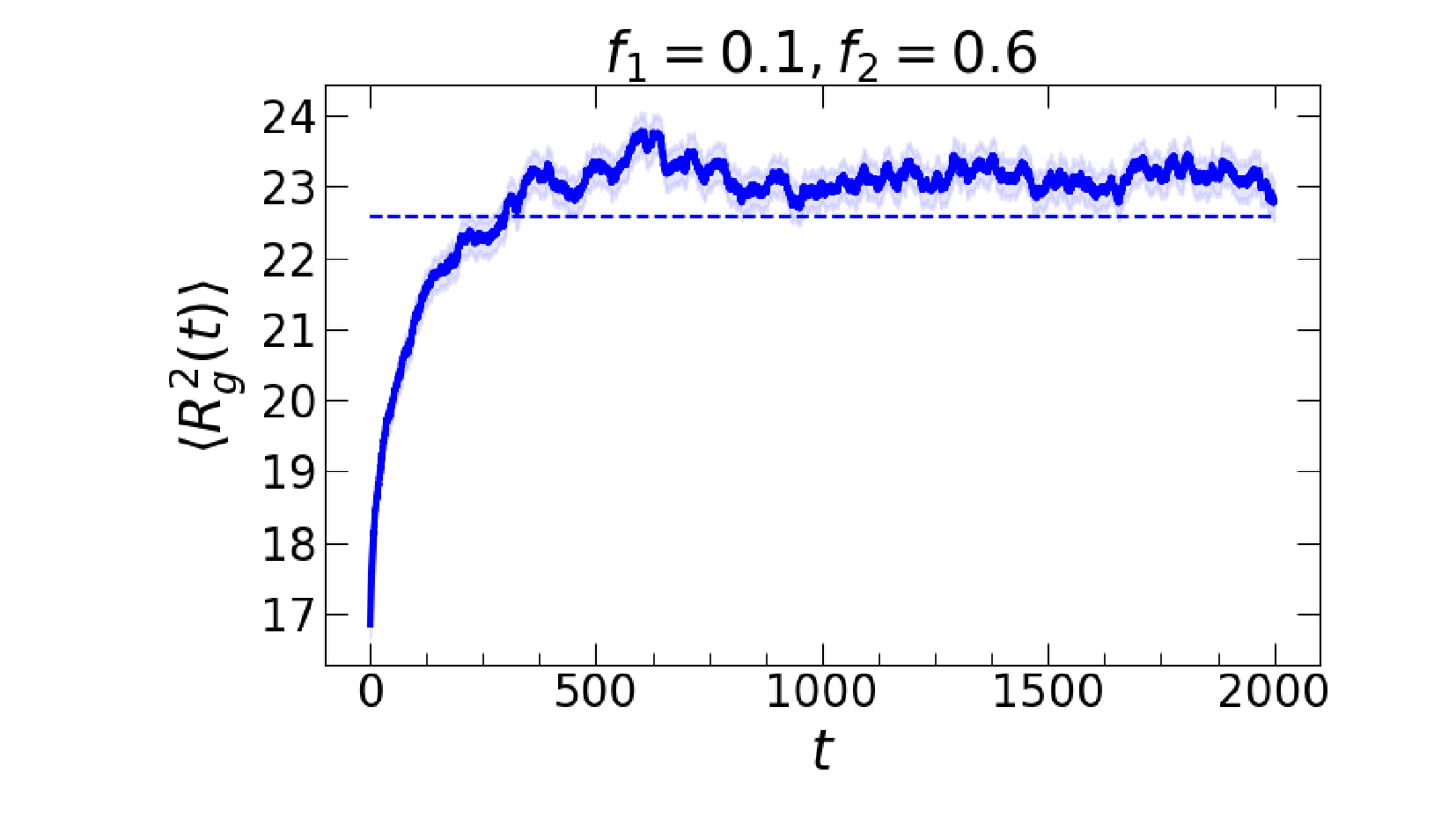}
\includegraphics[width=0.495\linewidth]{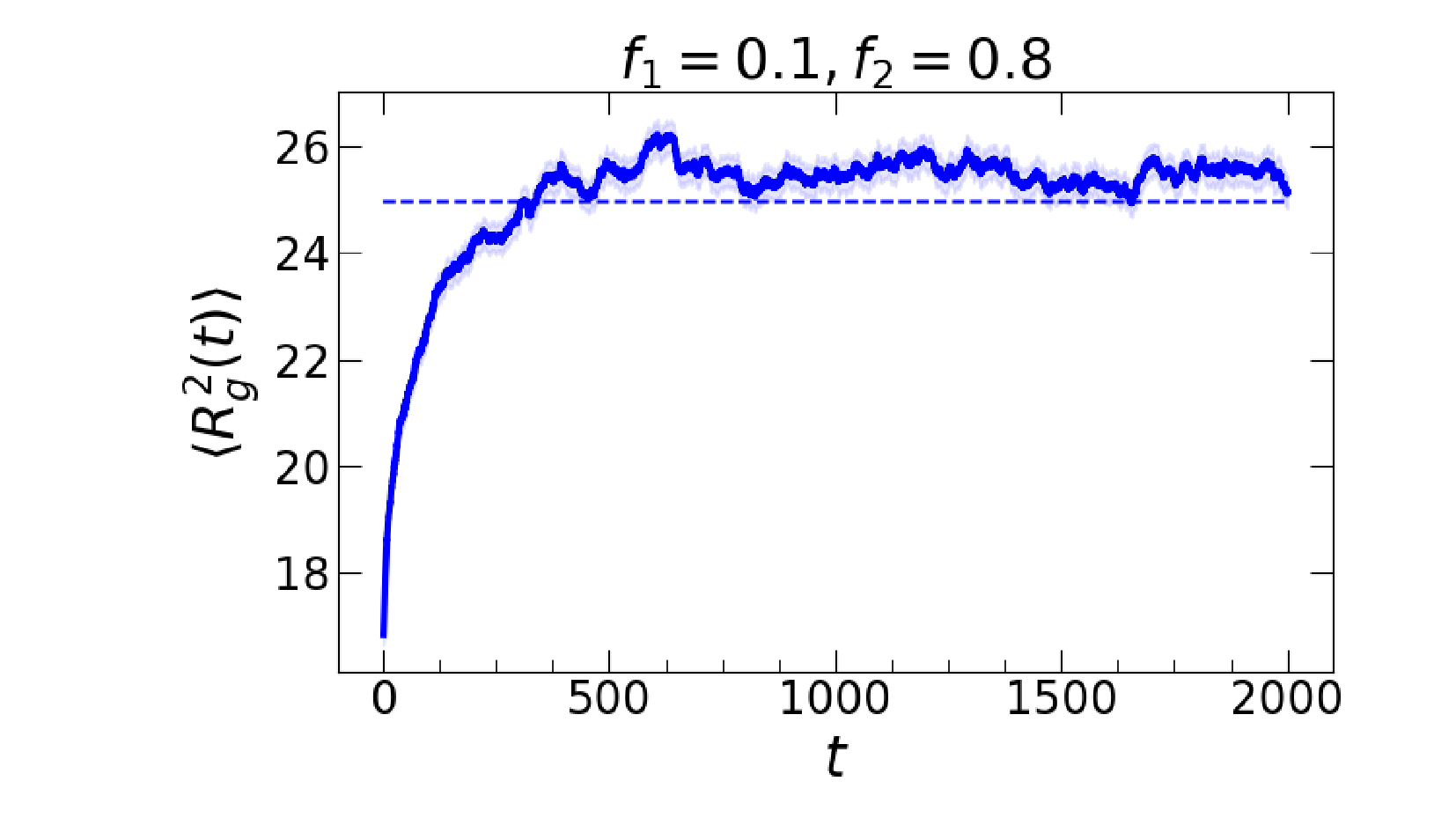}
\includegraphics[width=0.495\linewidth]{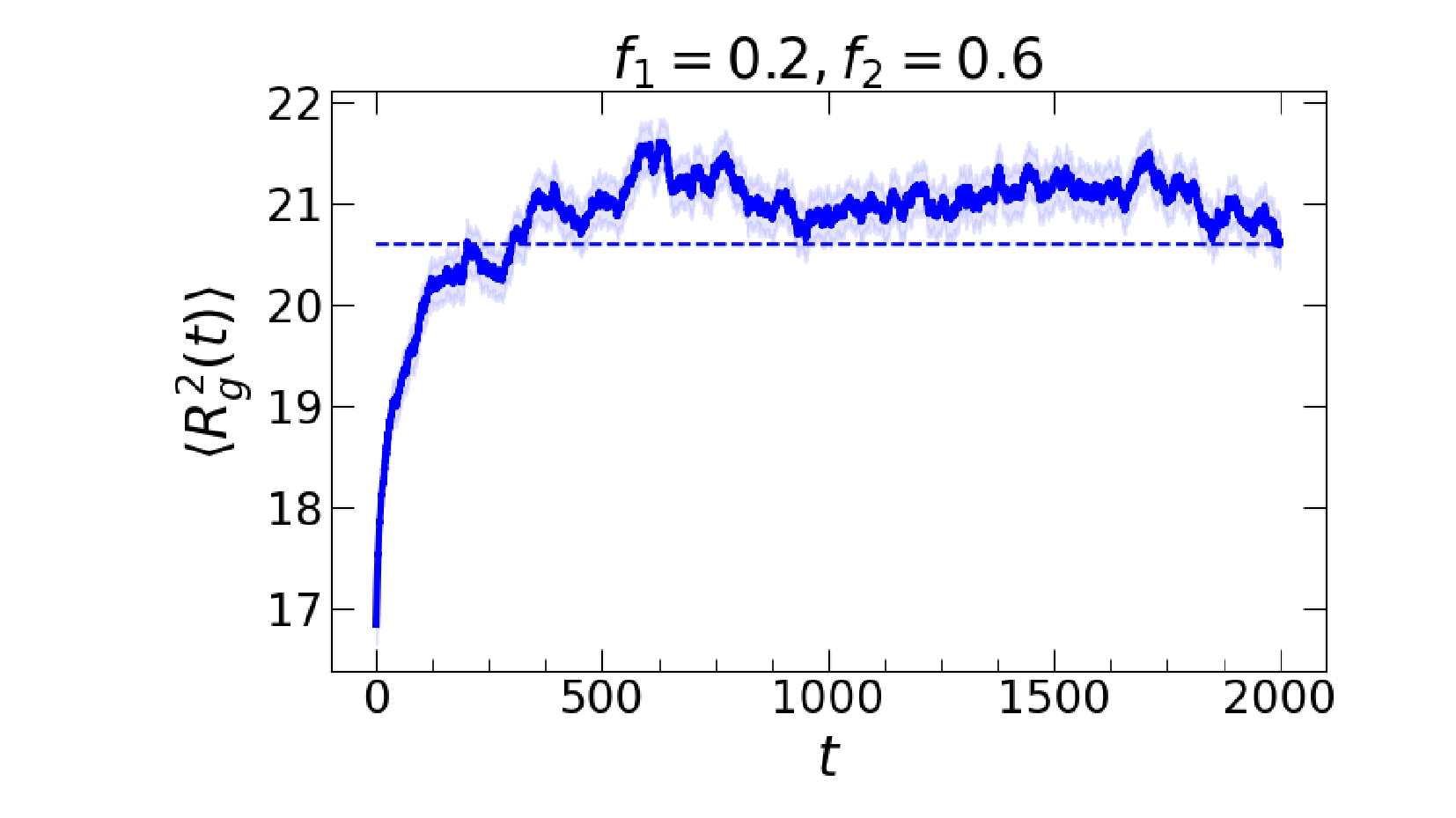}
\includegraphics[width=0.495\linewidth]{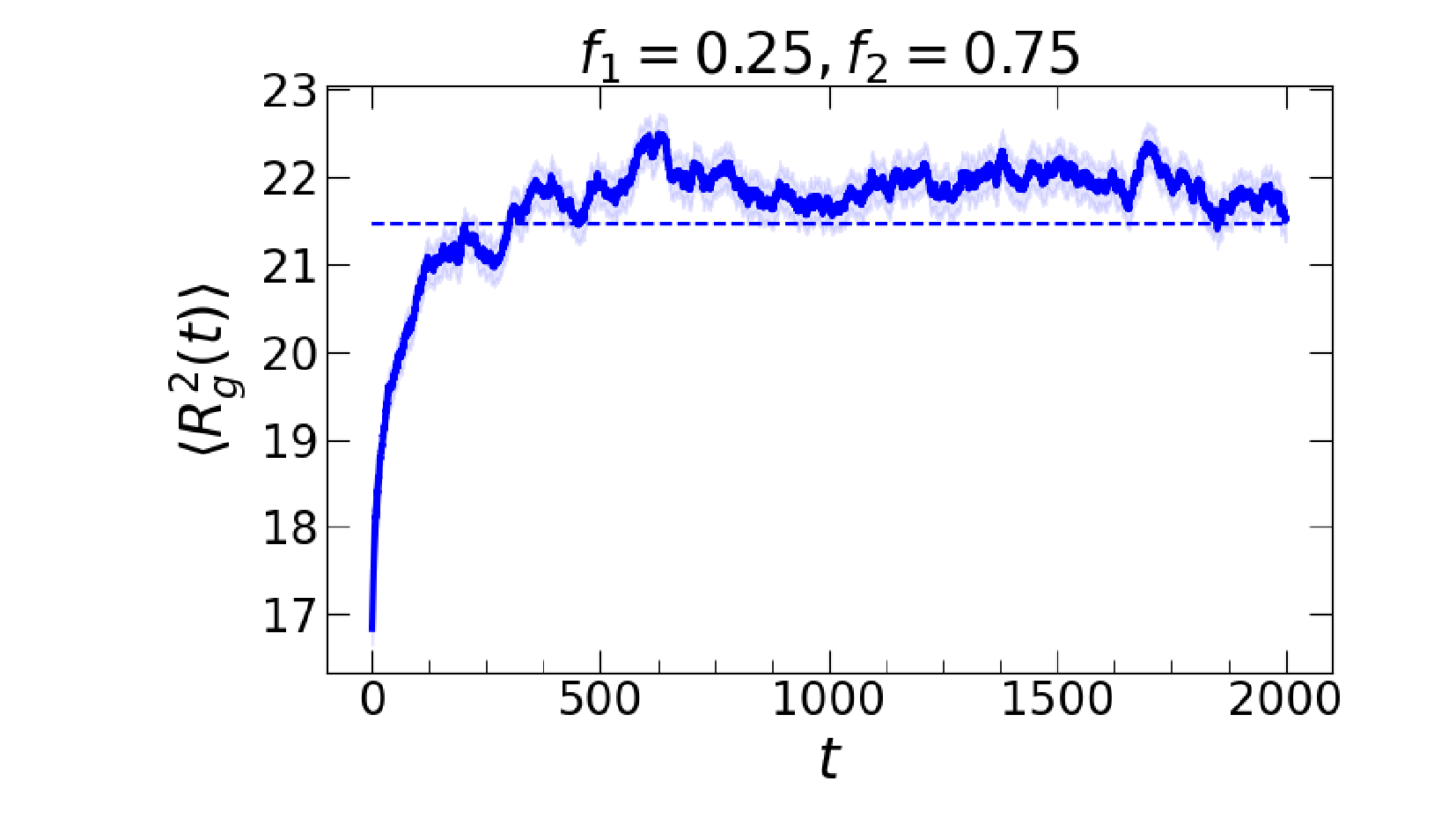}
\caption{The mean square radius of gyration as a function of time is presented for different fractions of active segments, as indicated in the title. The dashed line corresponds to the steady-state value of $\langle R_g^2(t) \rangle$  derived from the analytical solution in Eq. (\ref{rg_1}). At long times, $\langle R_g^2(t) \rangle$ from the simulation data reaches a plateau, which agrees closely with the analytical result. 
The light blue-colored shaded area represents the standard error, highlighting the extent of fluctuations around the mean in the simulation data. Details of the simulation and the parameters used are provided at the beginning of Appendix \ref{appen:simulation}.
}\label{fig:Rg_sim}
\end{figure*}

\begin{figure*}
\centering
\includegraphics[width=0.495\linewidth]{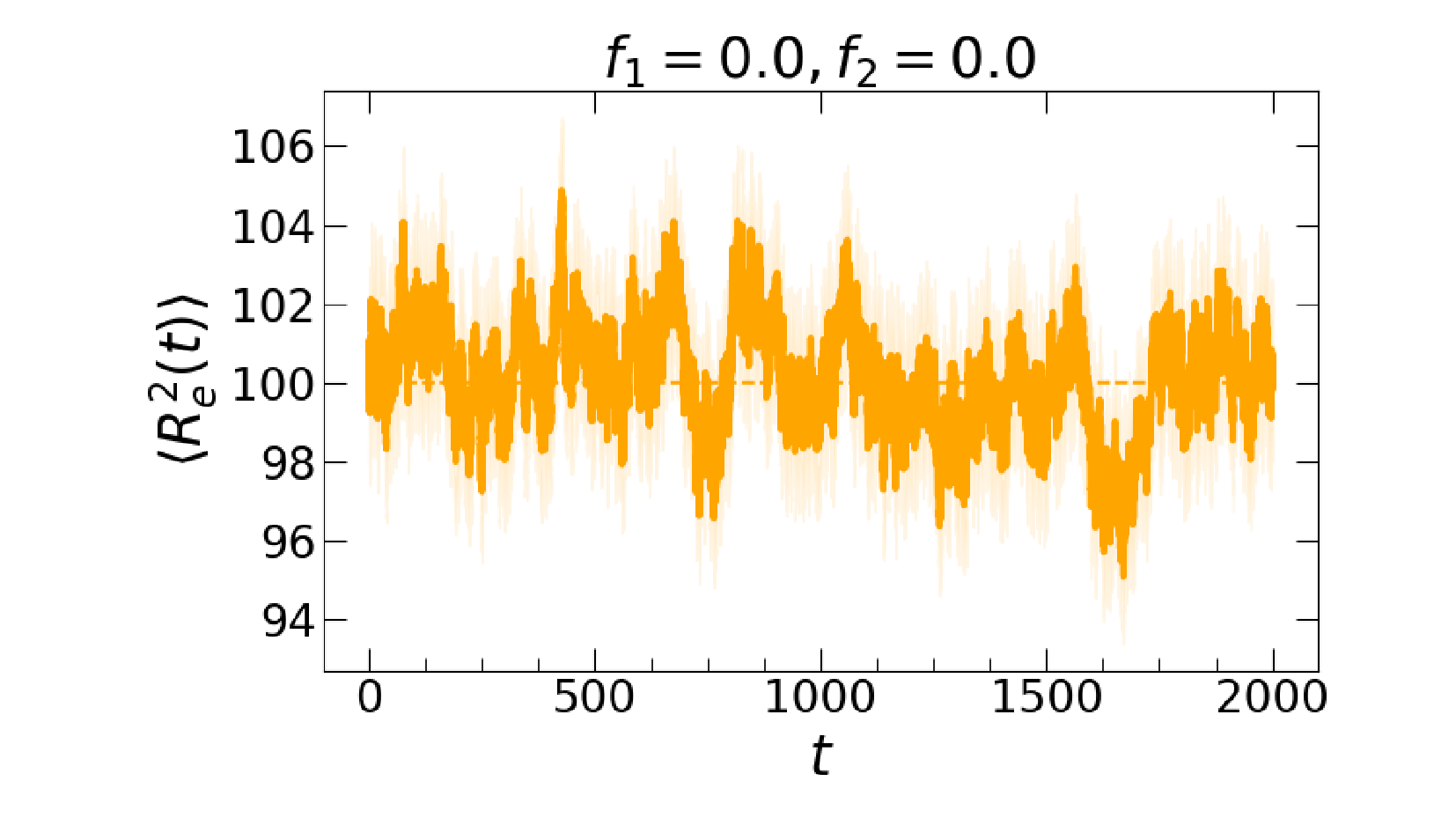}
\includegraphics[width=0.495\linewidth]{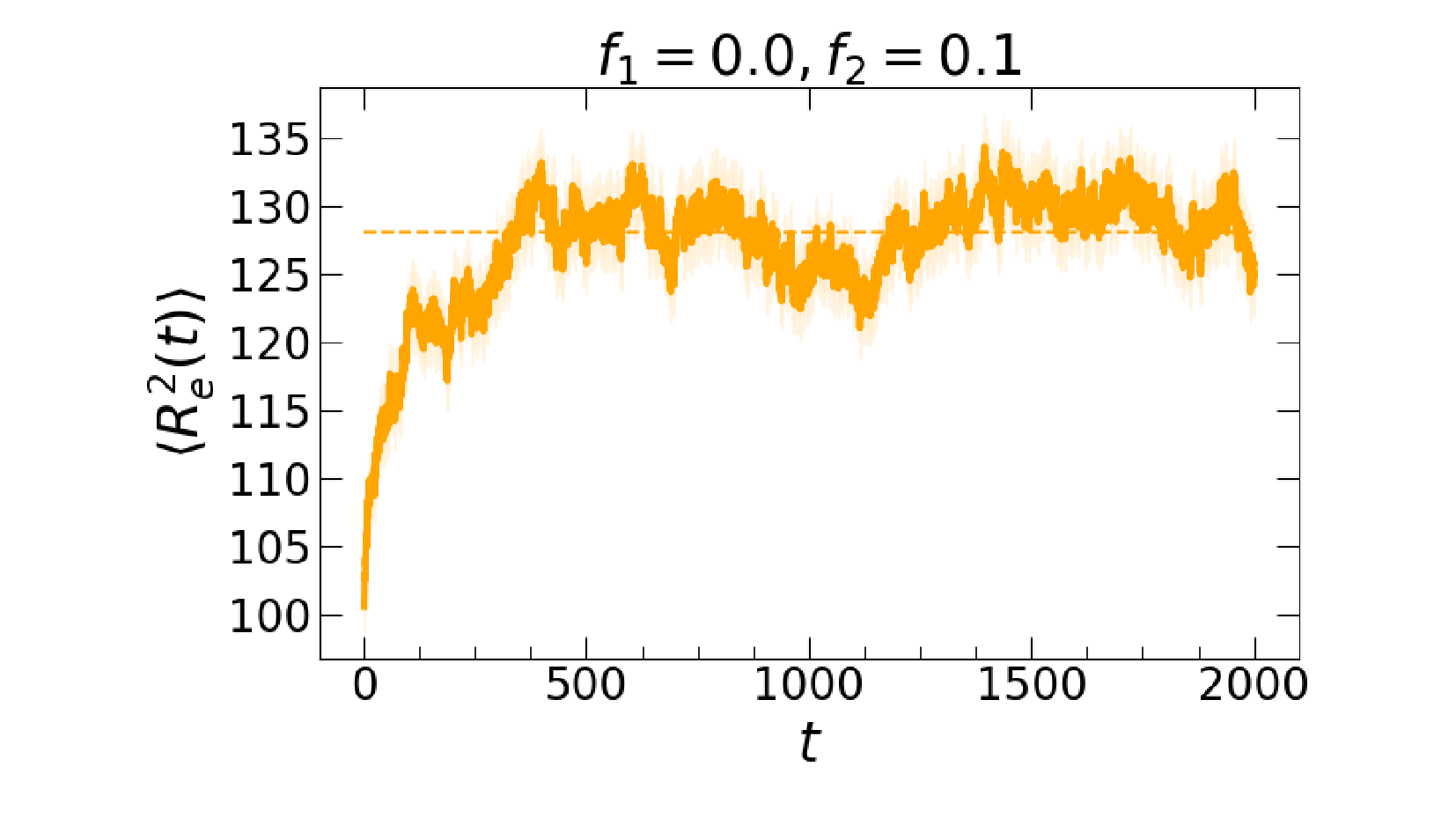}
\includegraphics[width=0.495\linewidth]{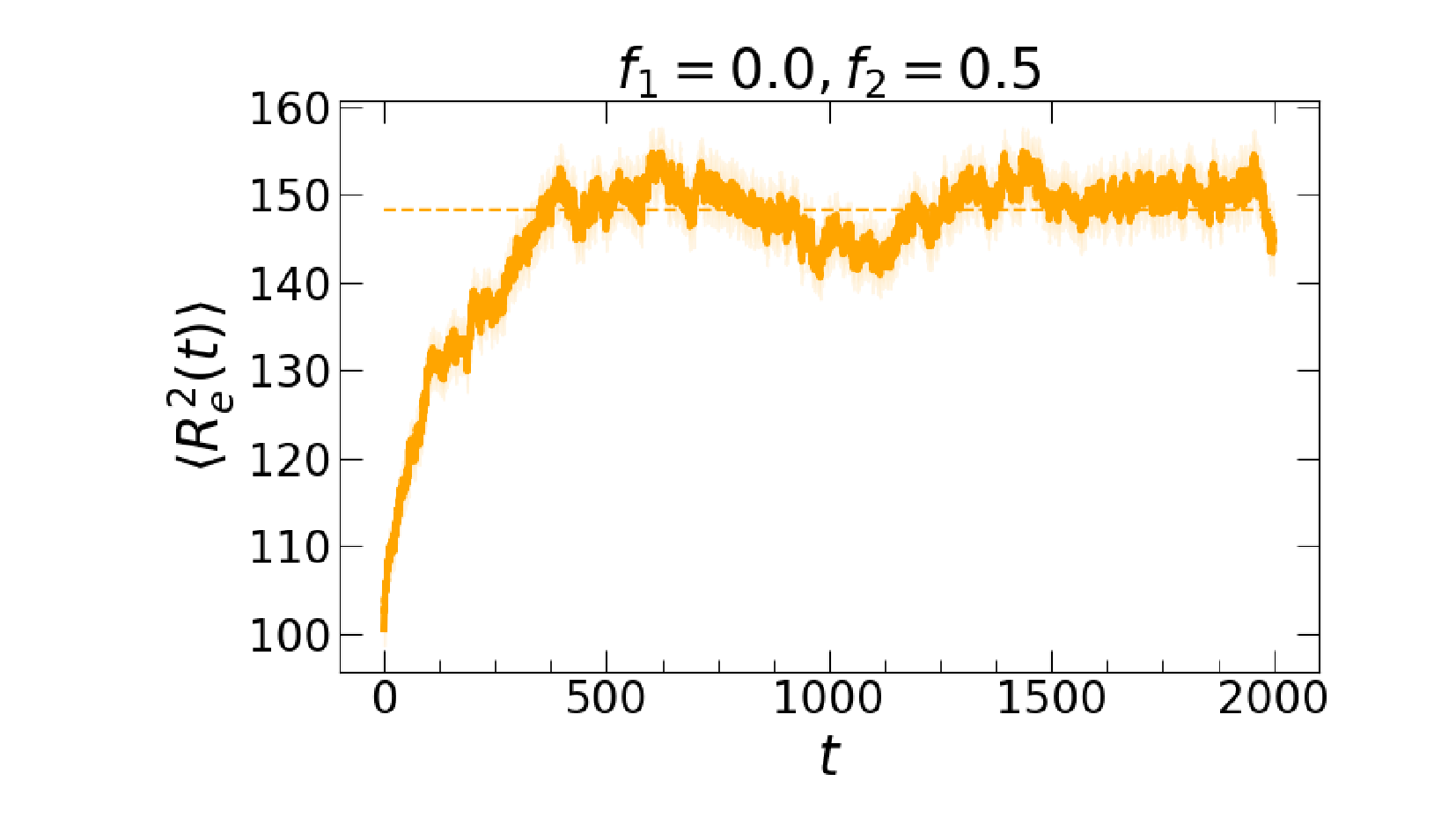}
\includegraphics[width=0.495\linewidth]{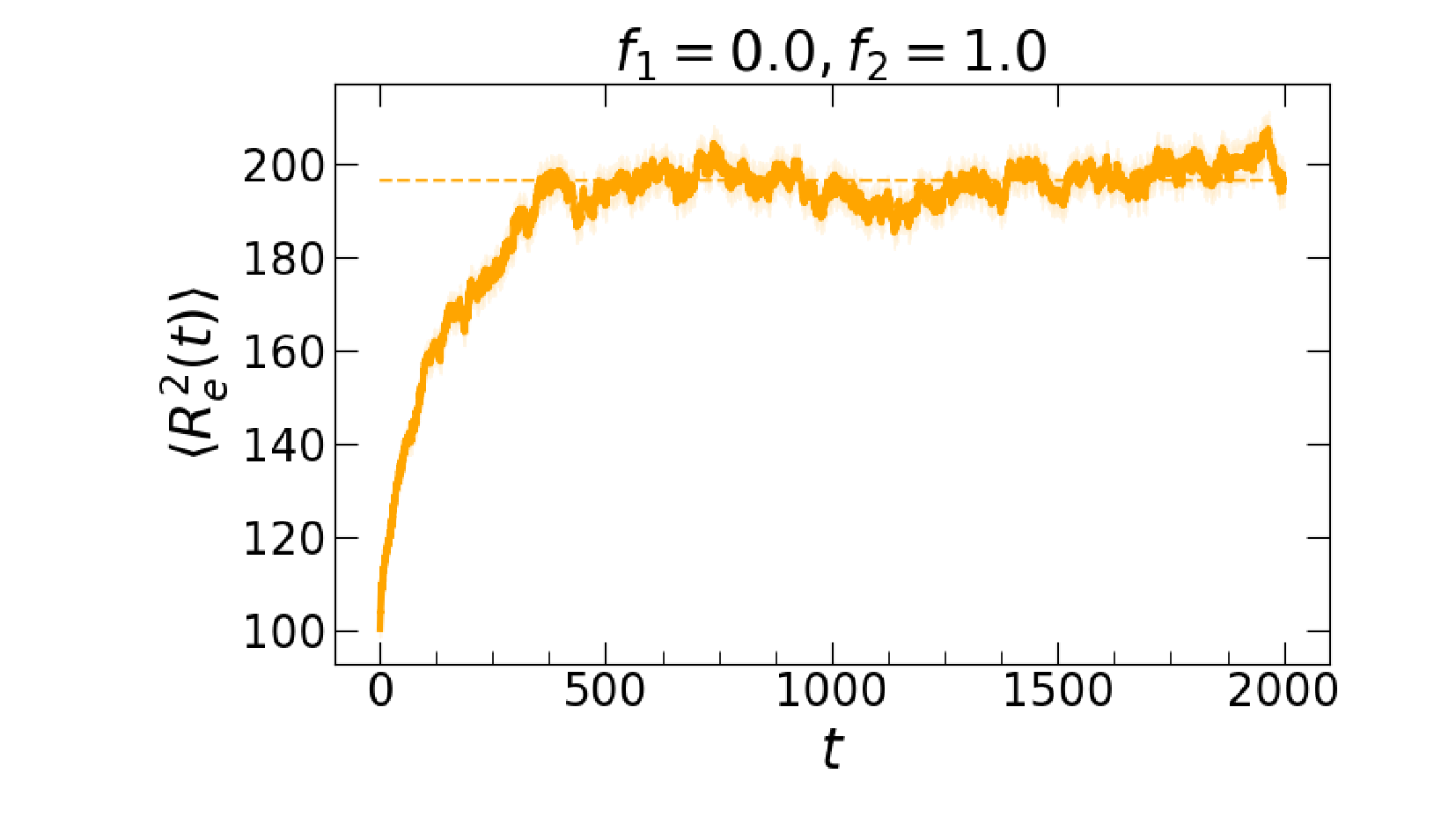}
\includegraphics[width=0.495\linewidth]{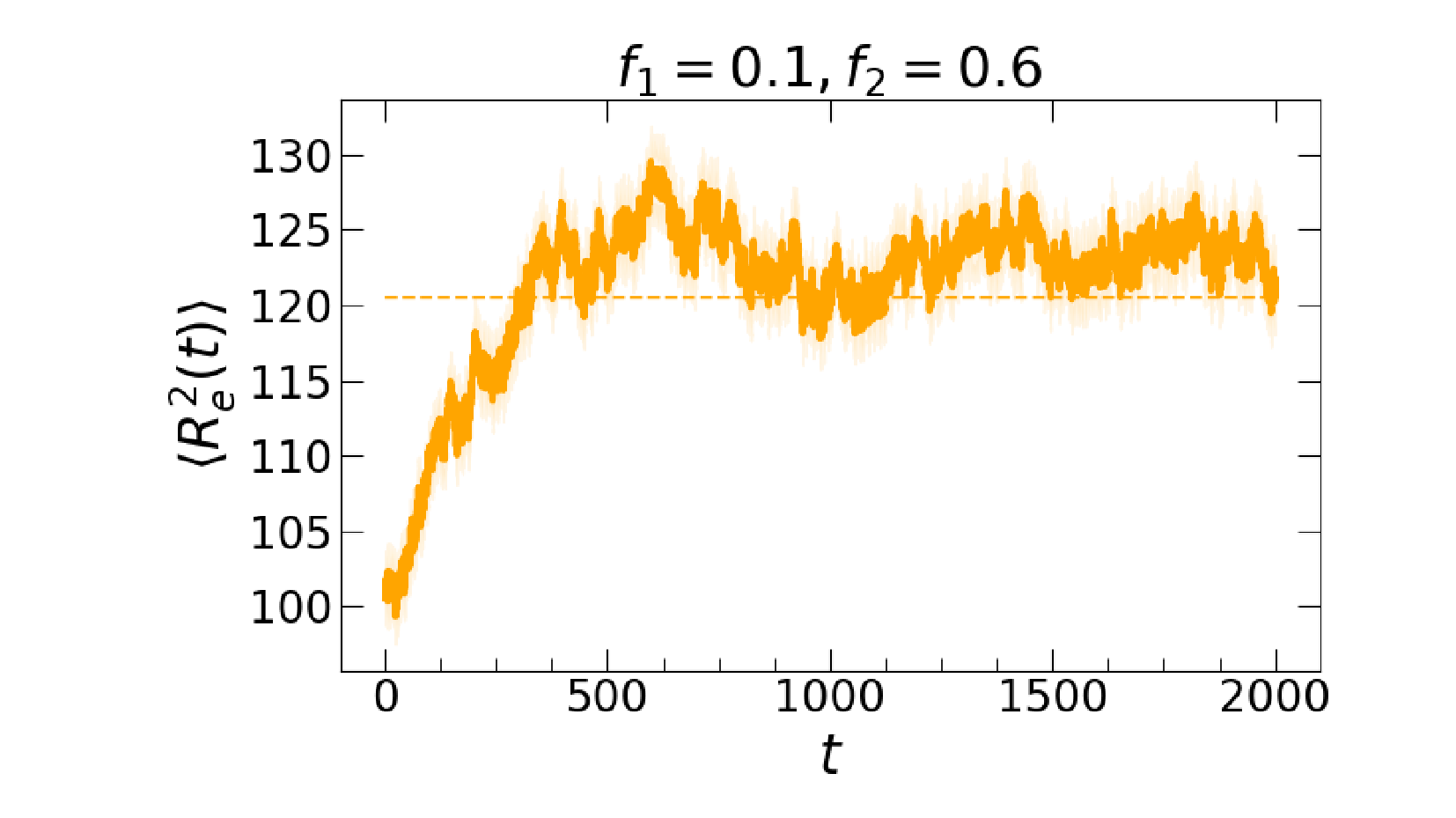}
\includegraphics[width=0.495\linewidth]{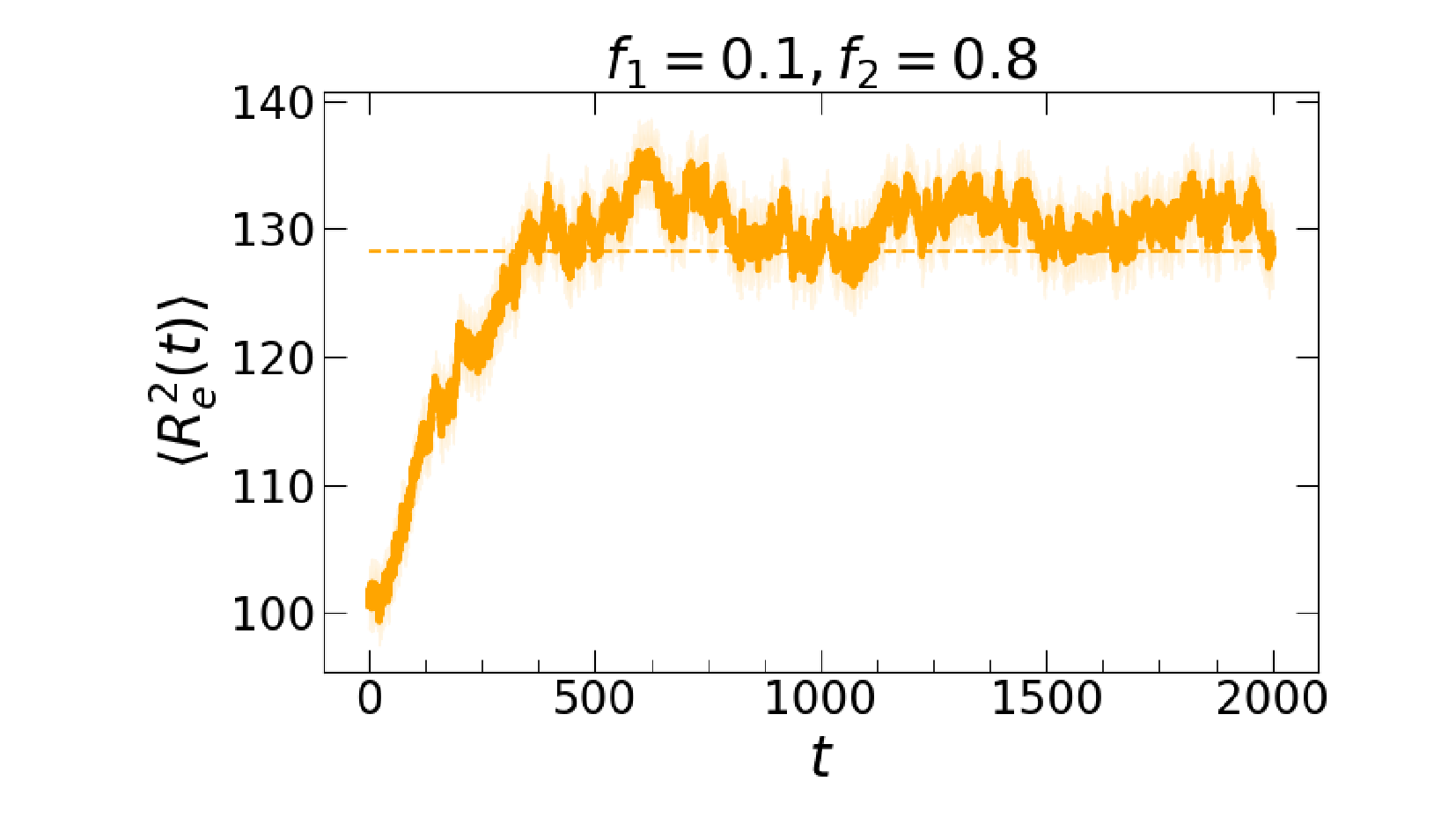}
\includegraphics[width=0.495\linewidth]{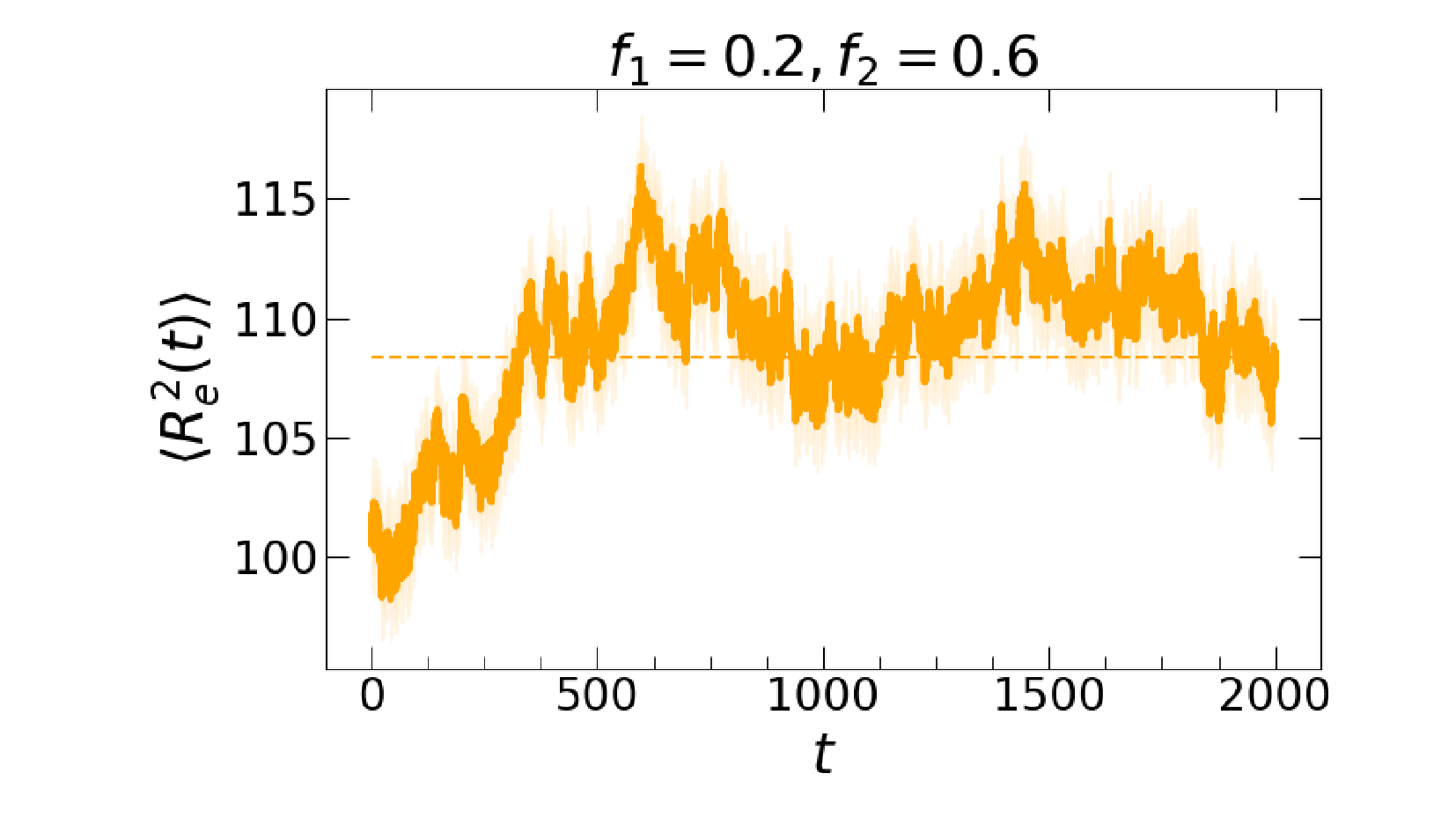}
\includegraphics[width=0.495\linewidth]{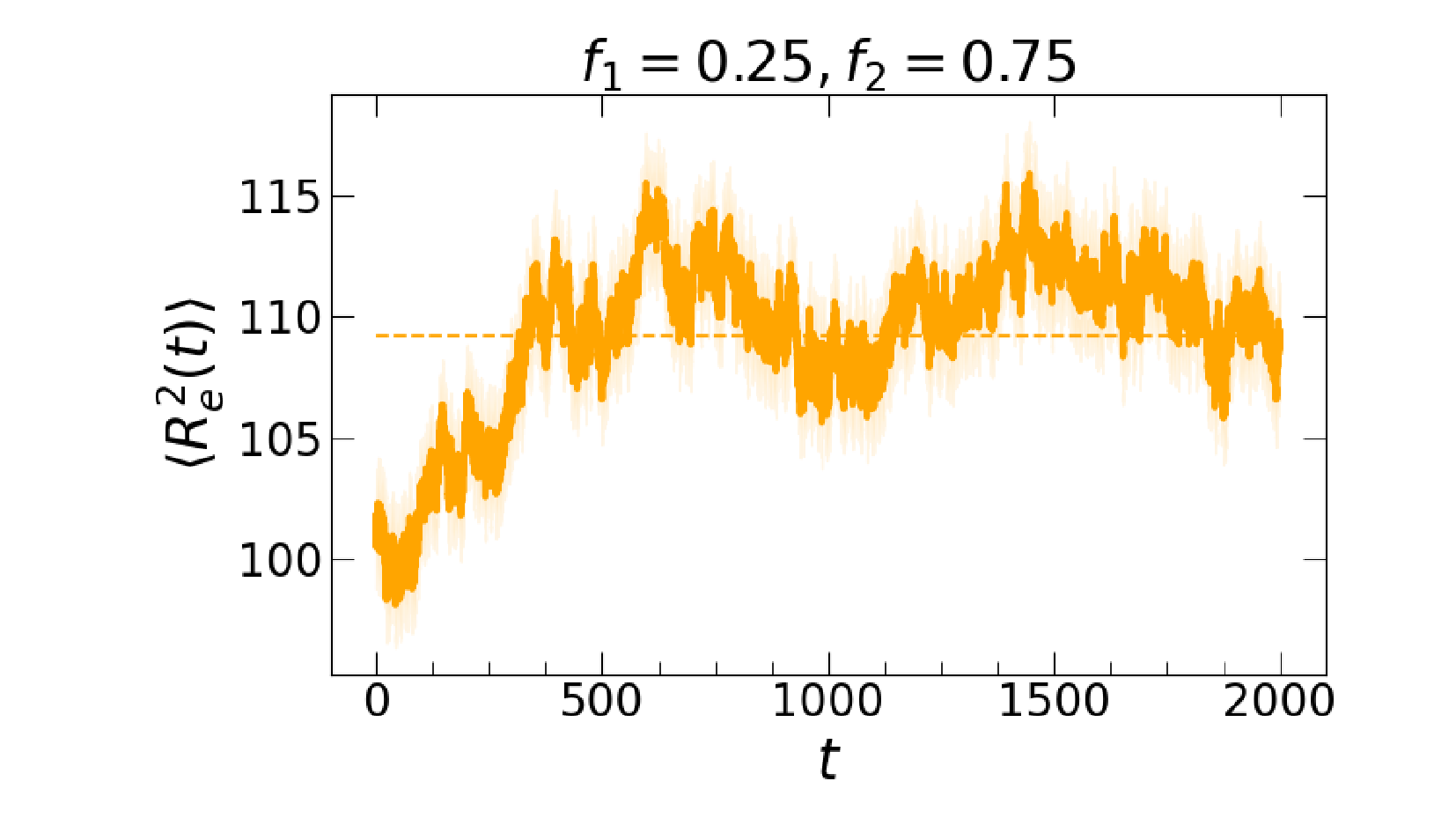}
\caption{Mean square end-to-end distance $\langle R_e^2(t) \rangle$ as a function of time. The dashed line corresponds to Eq. (\ref{phi_corr0_1}), and other details are the same as in Fig. \ref{fig:Rg_sim}.}\label{fig:Re_sim}
\end{figure*}

\begin{figure*}
\centering
\includegraphics[width=0.495\linewidth]{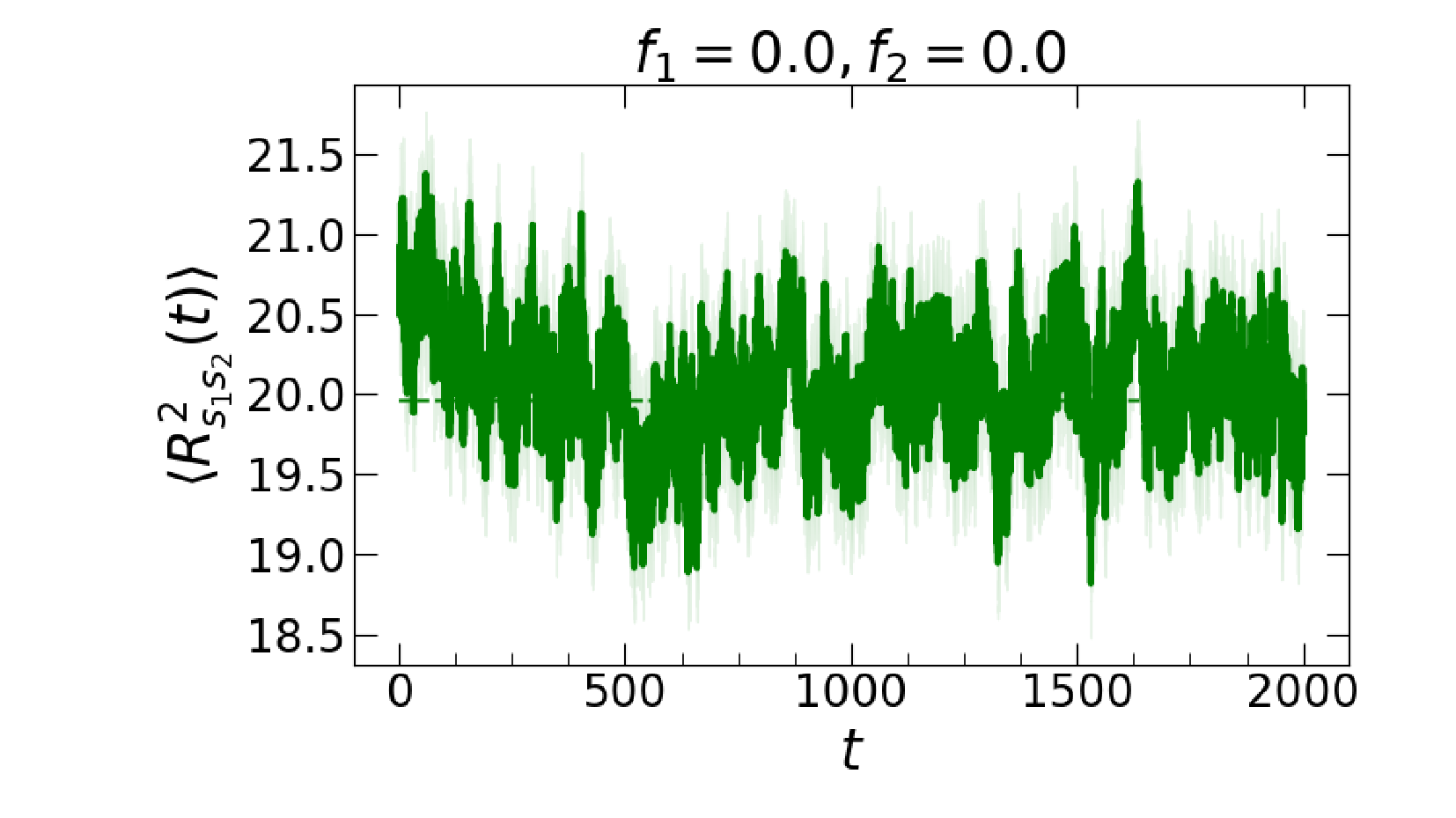}
\includegraphics[width=0.495\linewidth]{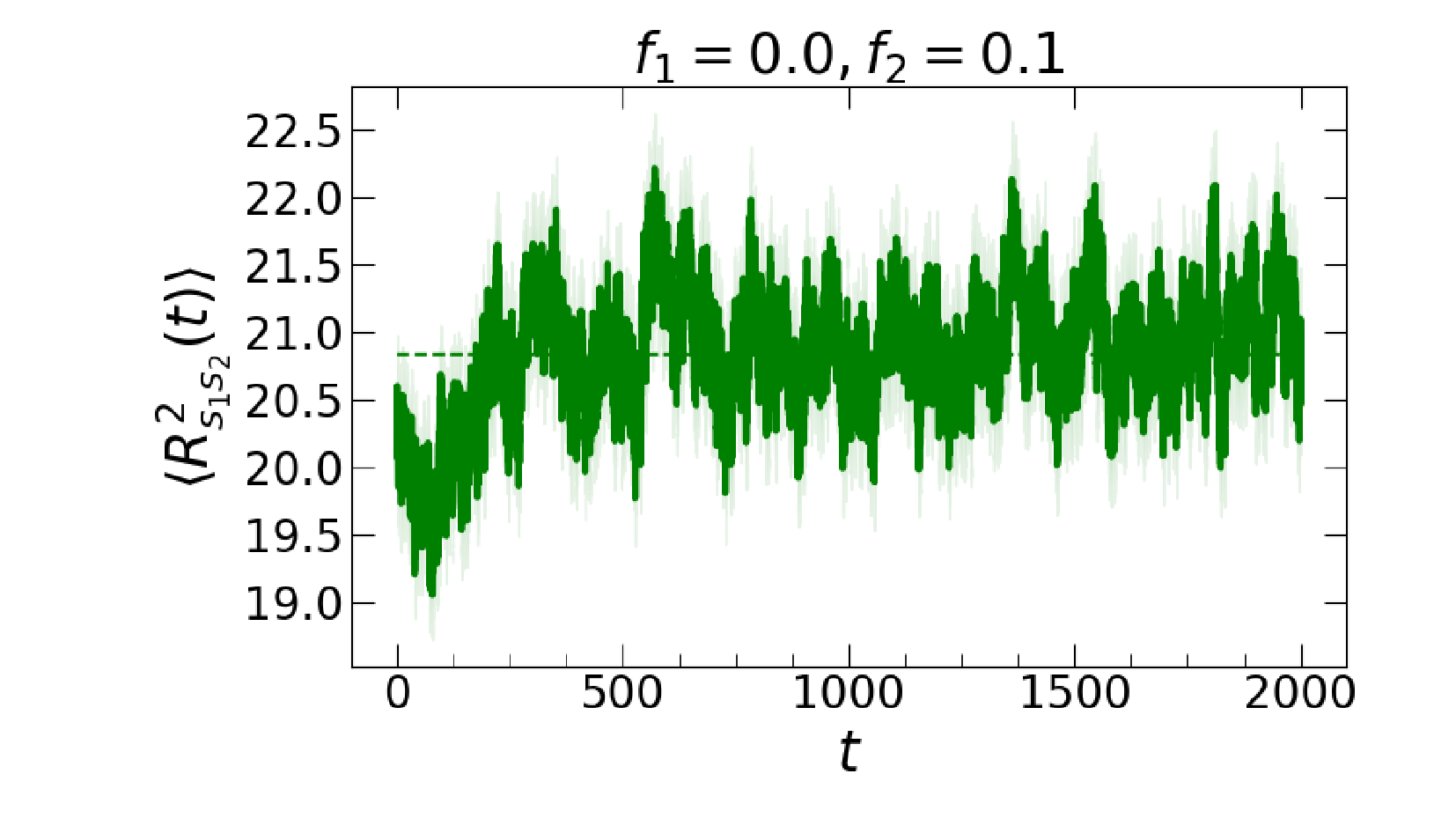}
\includegraphics[width=0.495\linewidth]{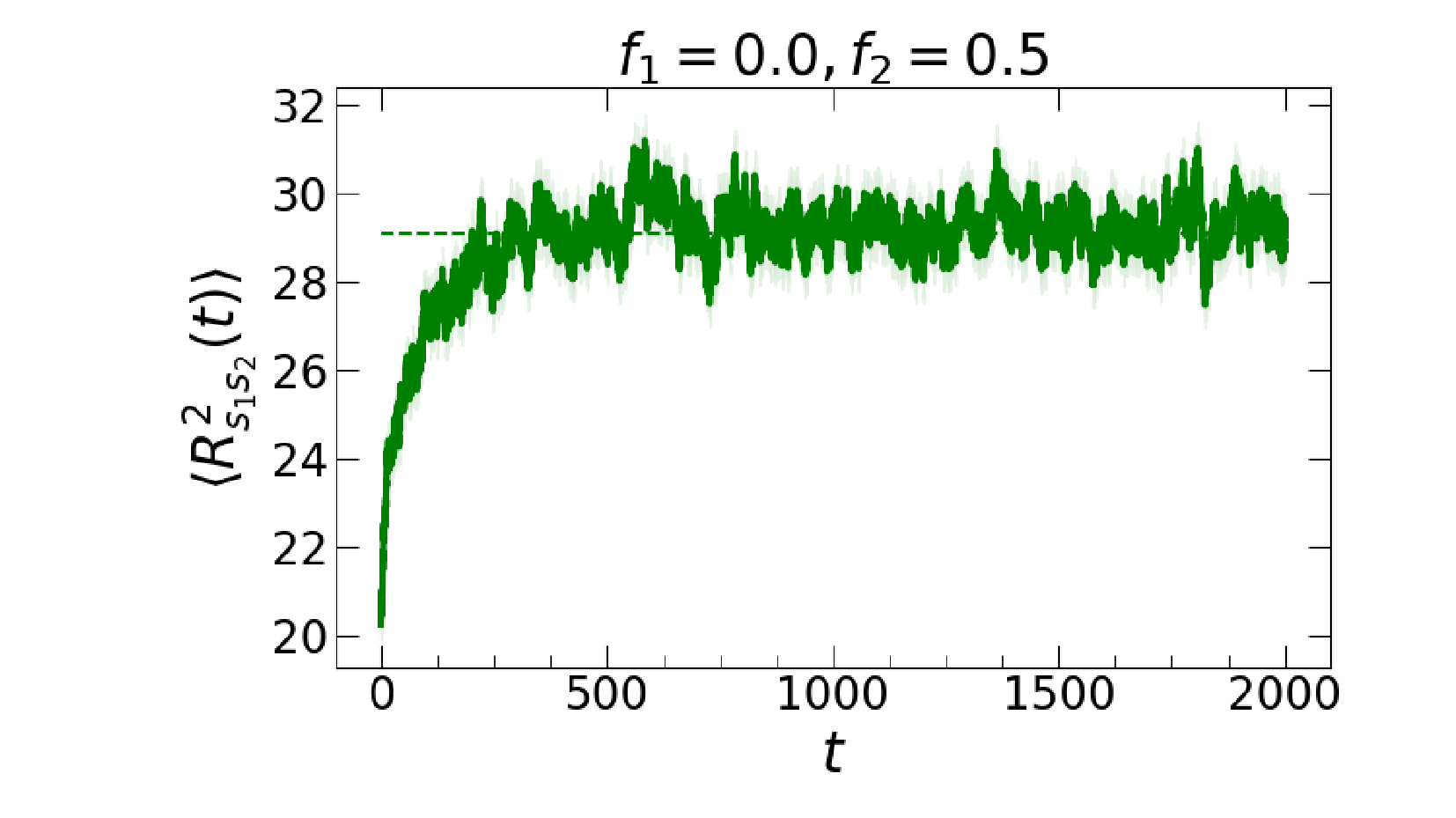}
\includegraphics[width=0.495\linewidth]{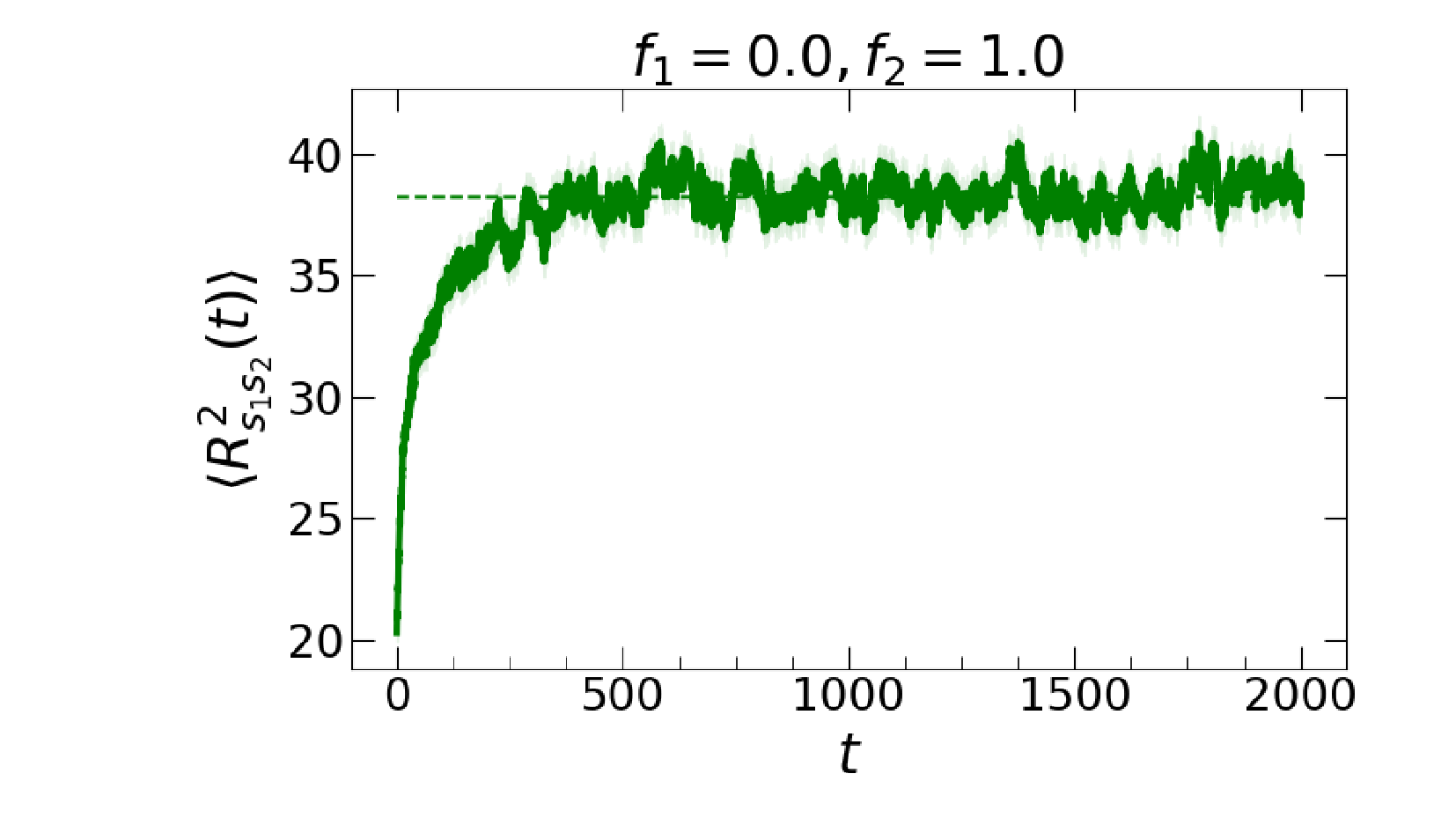}
\includegraphics[width=0.495\linewidth]{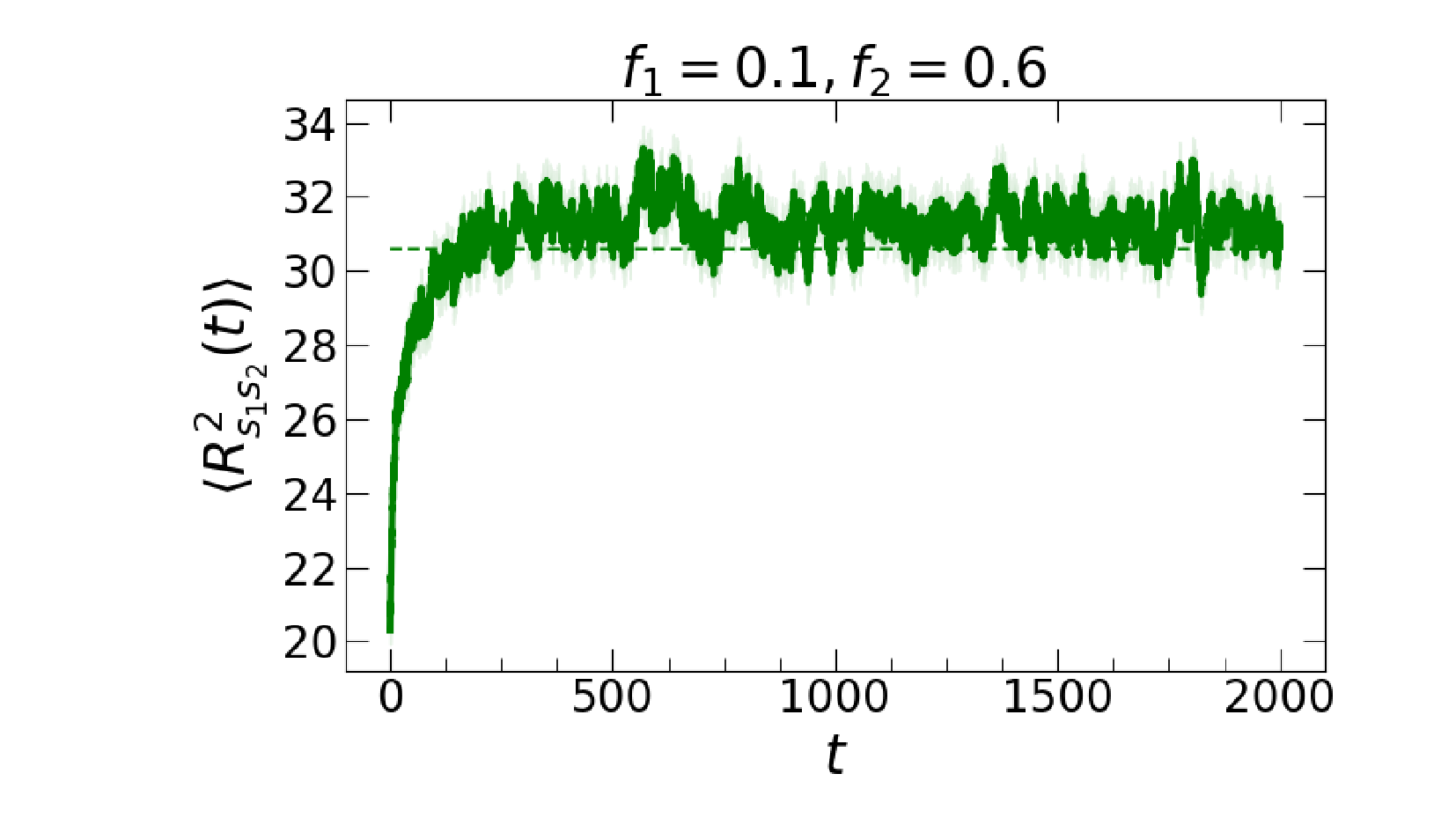}
\includegraphics[width=0.495\linewidth]{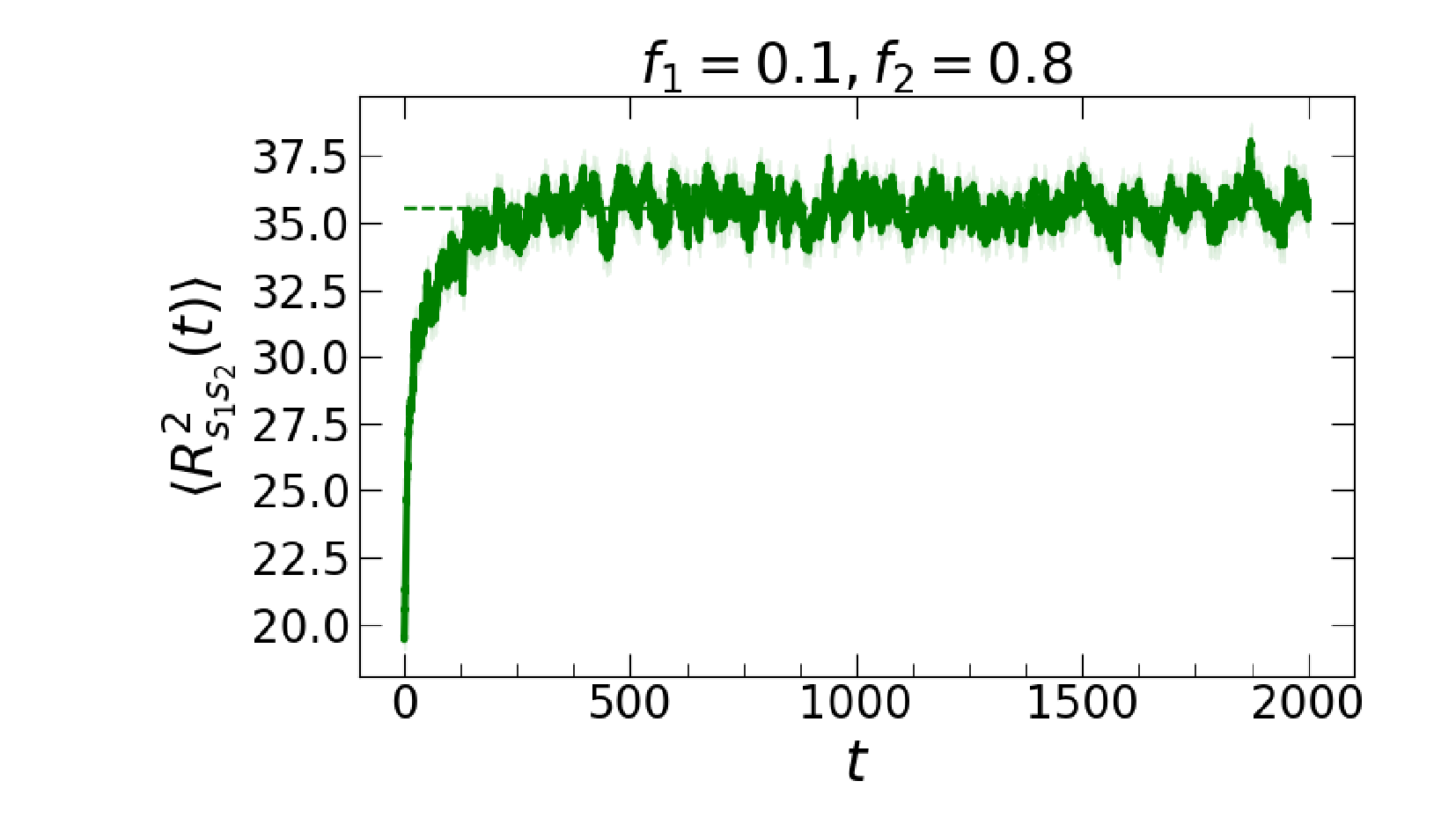}
\includegraphics[width=0.495\linewidth]{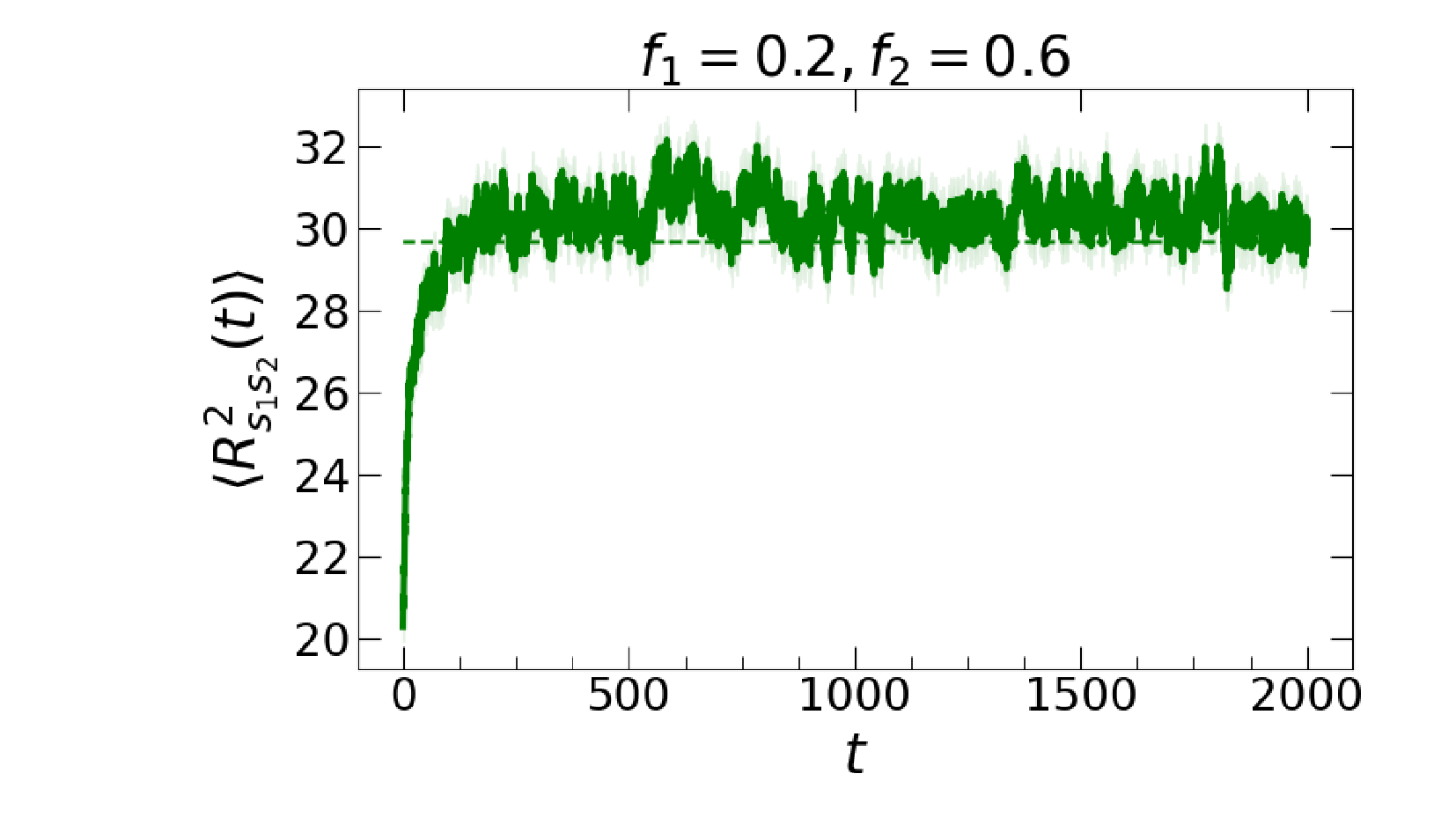}
\includegraphics[width=0.495\linewidth]{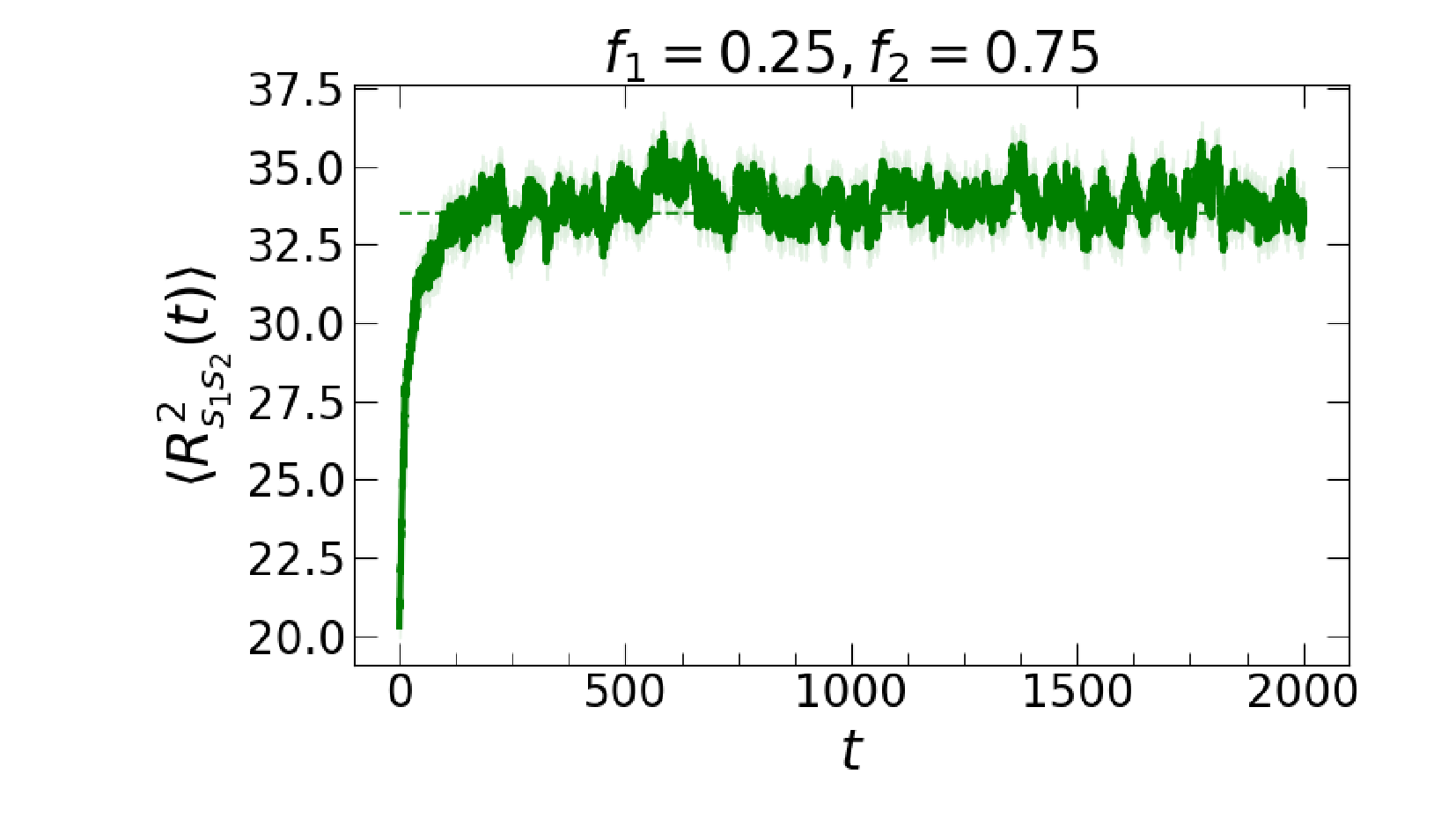}
\caption{Mean square separation between two beads located at arc lengths $0.4L$ and $0.6L$.
The steady-state value of the separation obtained from the simulation data is compared with the analytical result provided in Eq. (\ref{Rmn_1}). Other details are the same as in Fig. \ref{fig:Rg_sim}.}\label{fig:Rmn_sim}
\end{figure*}

\begin{figure*}
\centering
\includegraphics[width=0.495\linewidth]{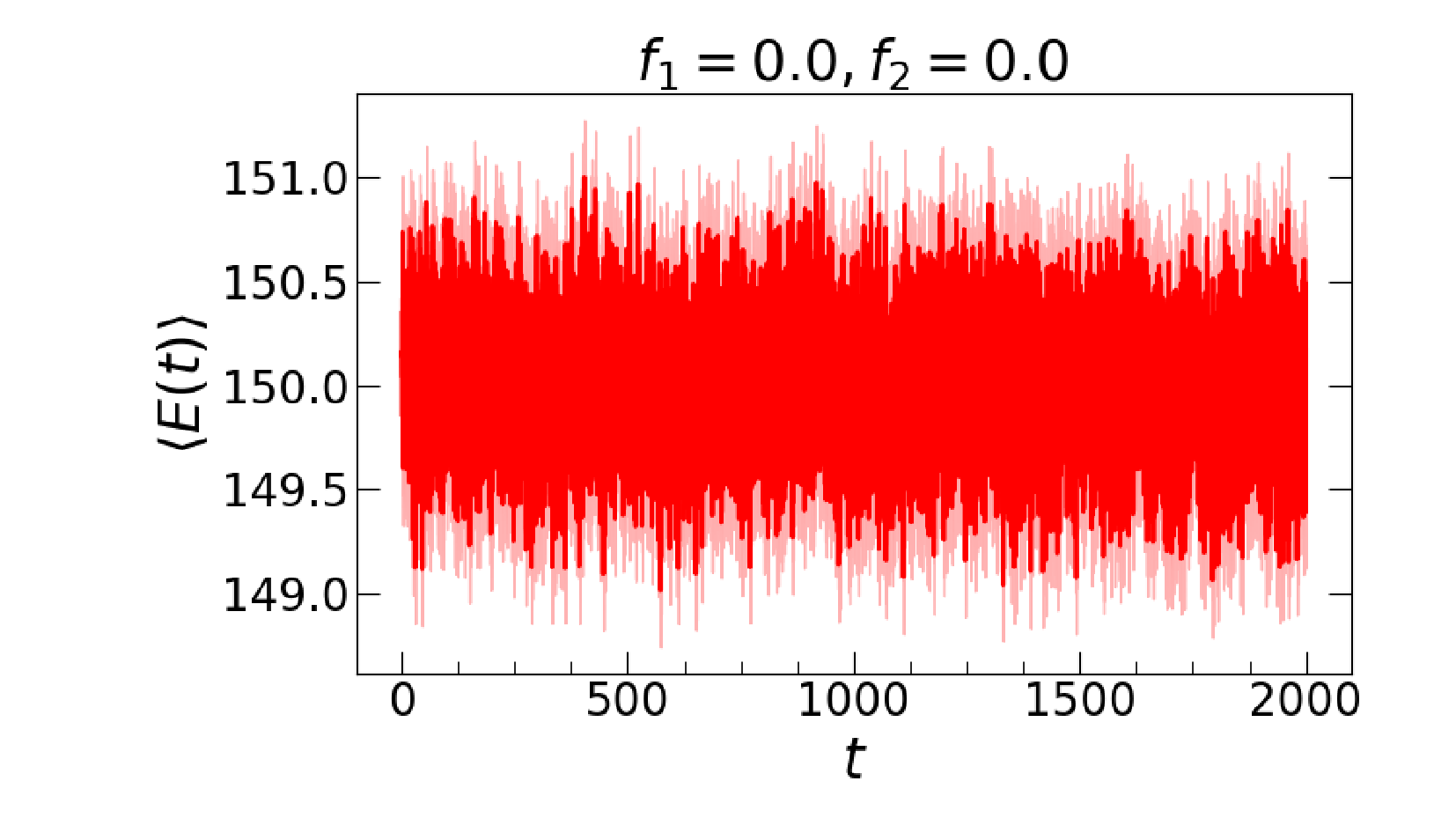}
\includegraphics[width=0.495\linewidth]{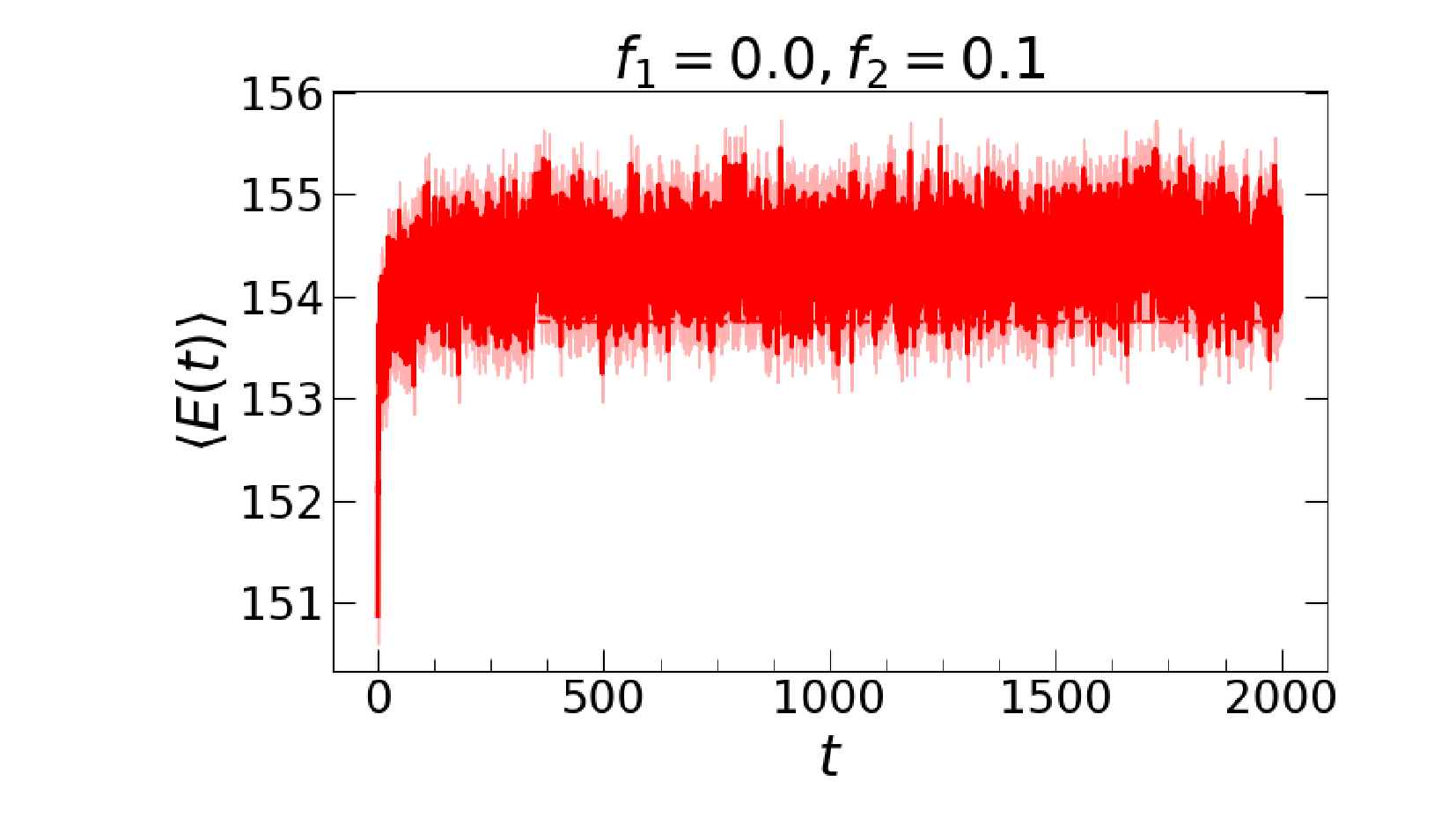}
\includegraphics[width=0.495\linewidth]{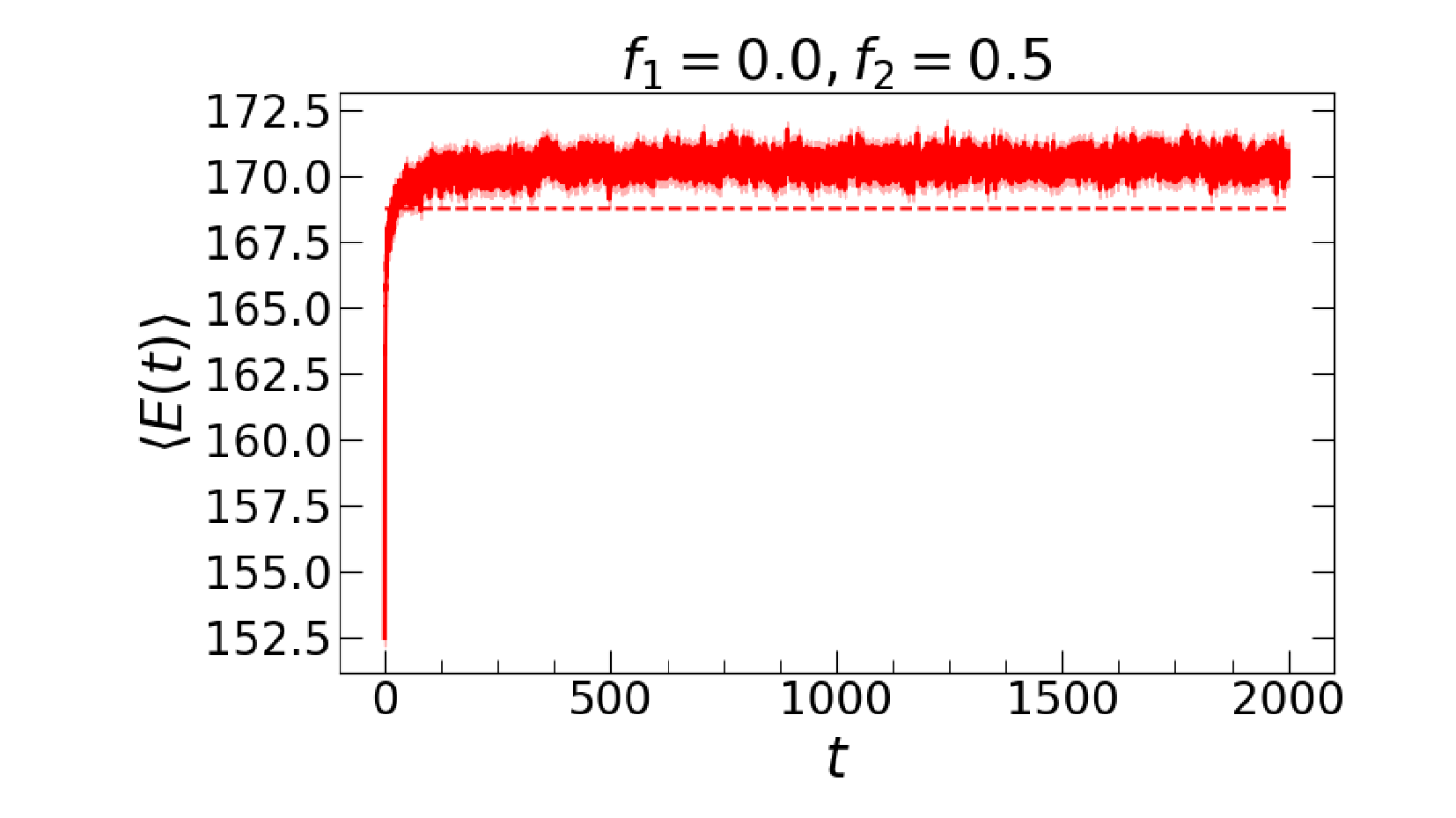}
\includegraphics[width=0.495\linewidth]{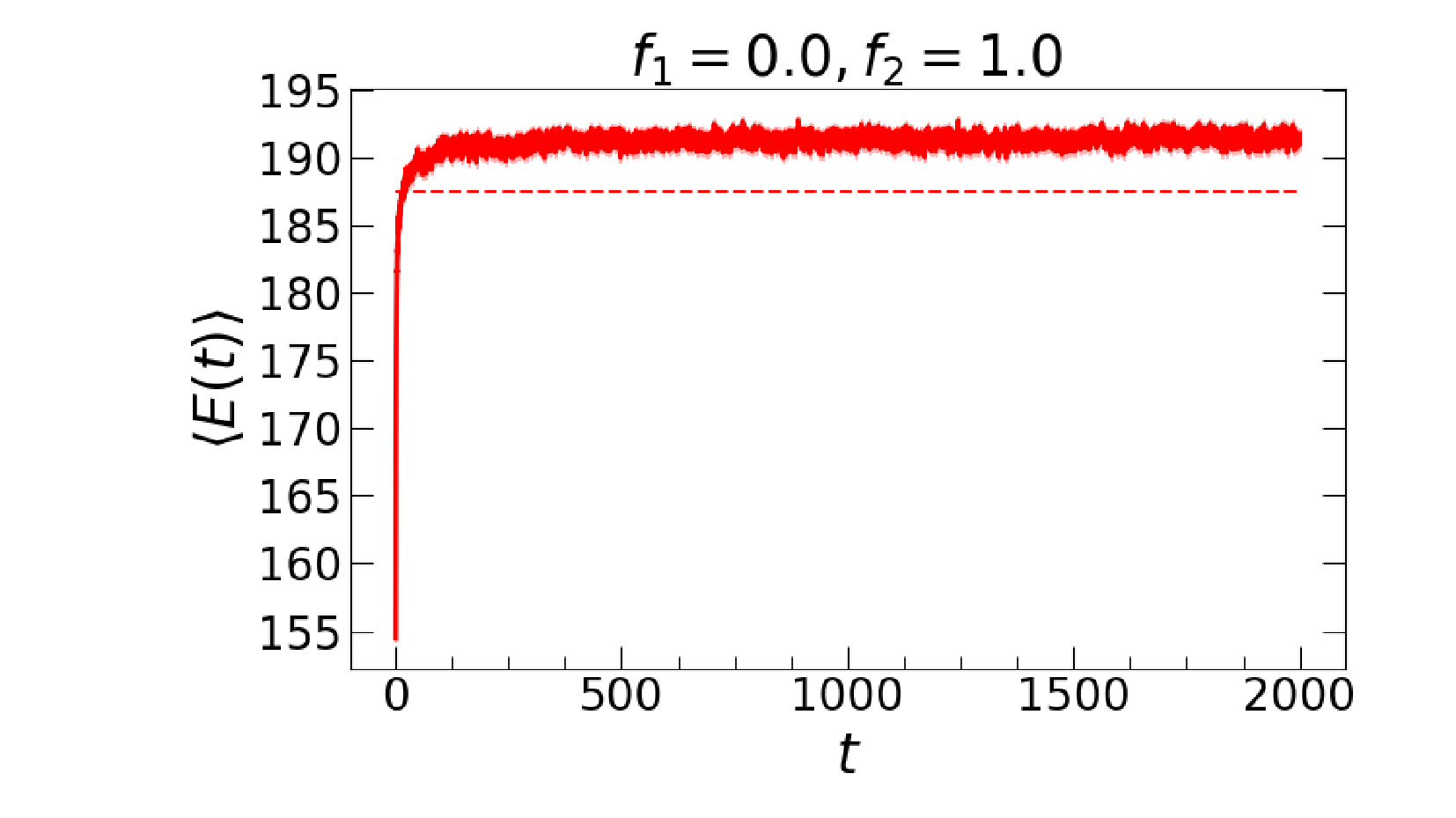}
\includegraphics[width=0.495\linewidth]{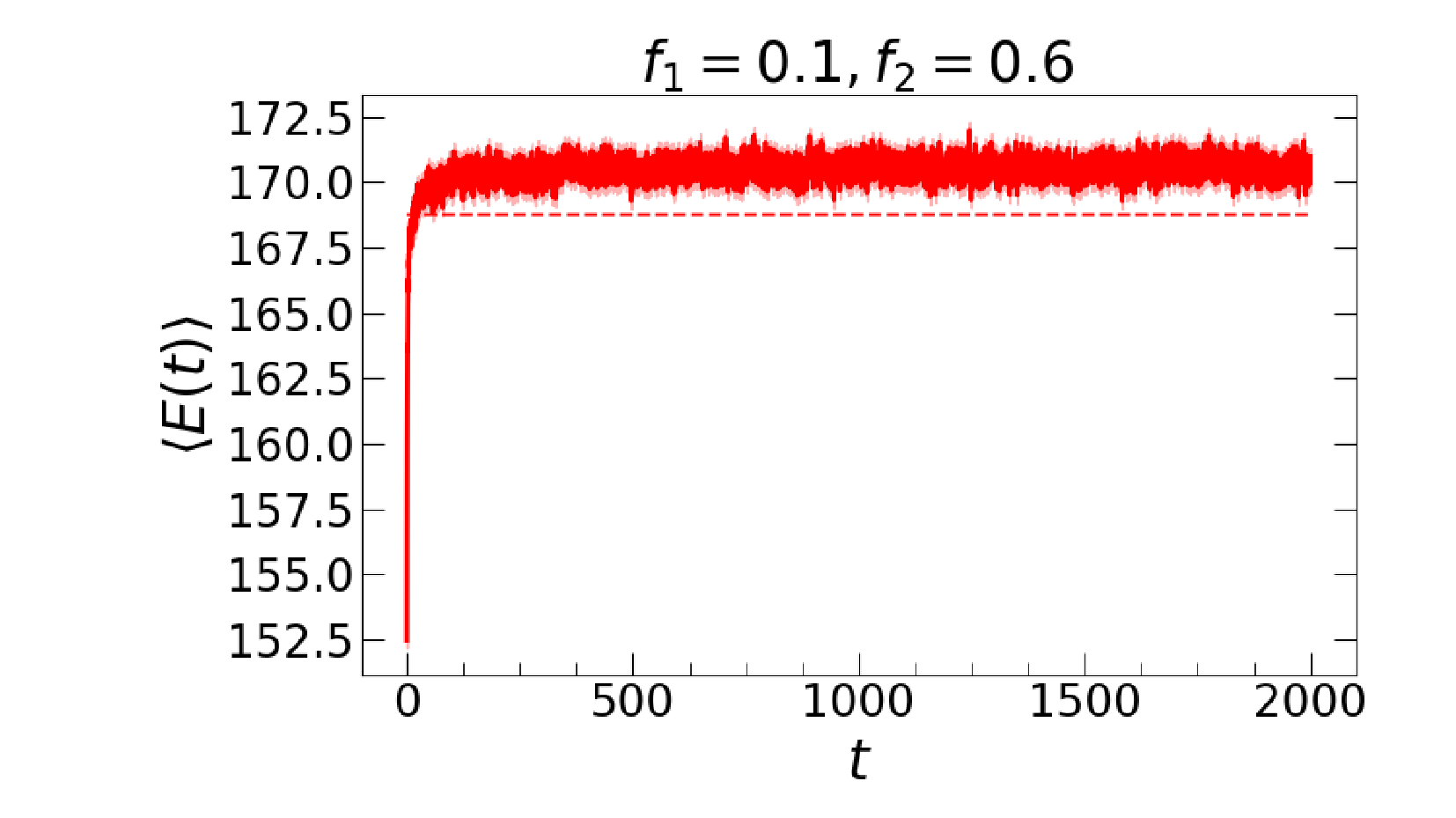}
\includegraphics[width=0.495\linewidth]{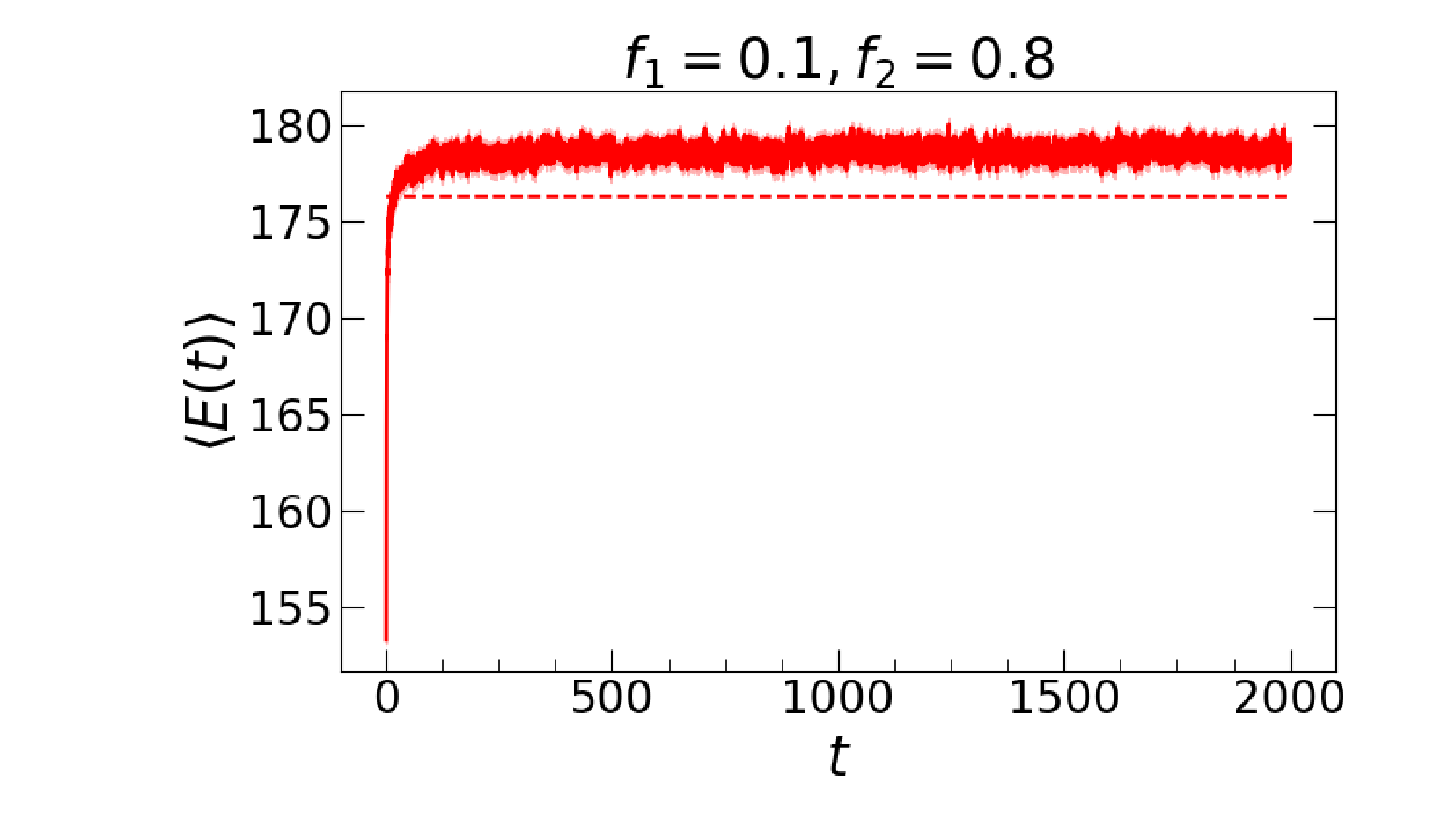}
\includegraphics[width=0.495\linewidth]{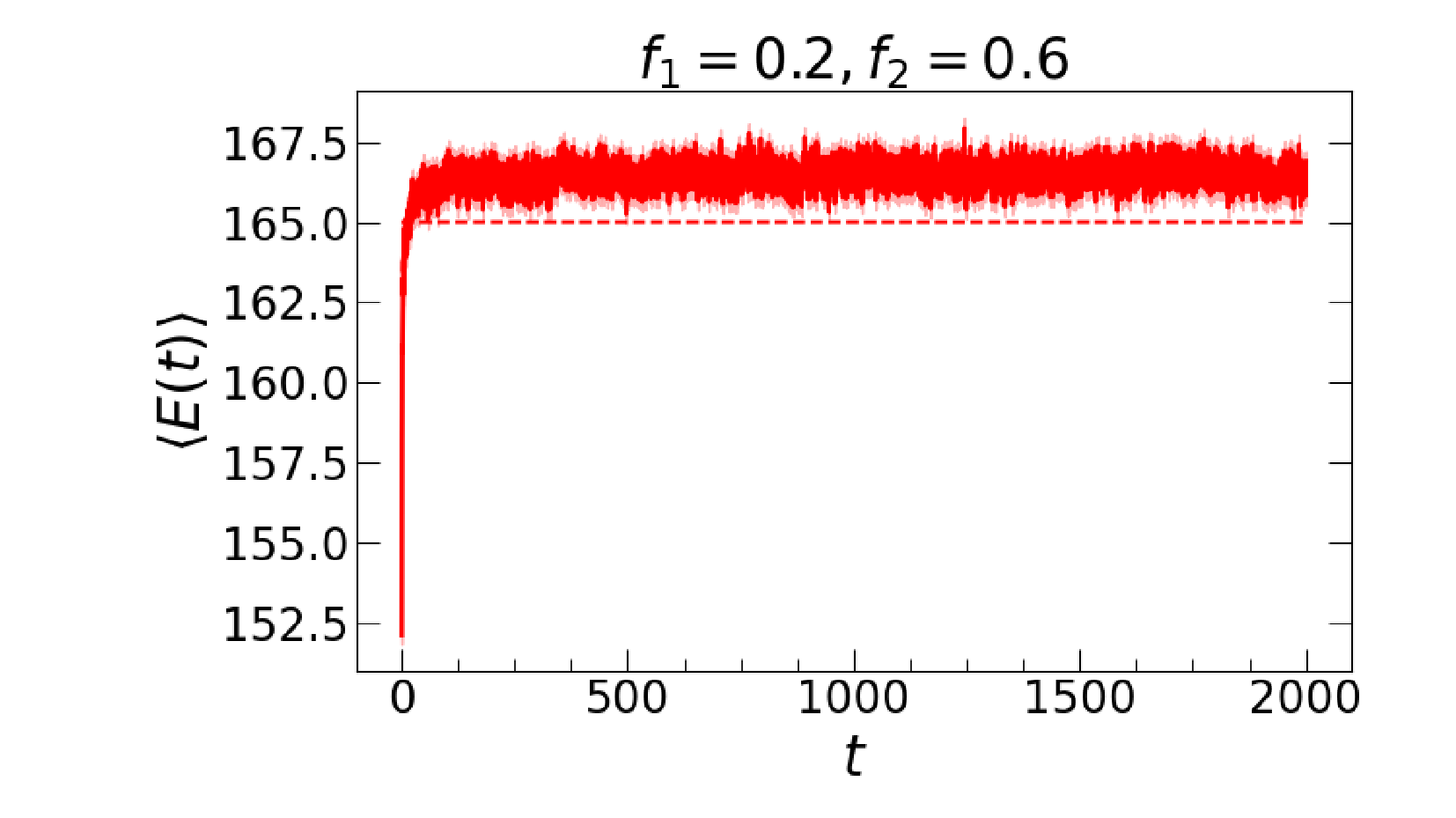}
\includegraphics[width=0.495\linewidth]{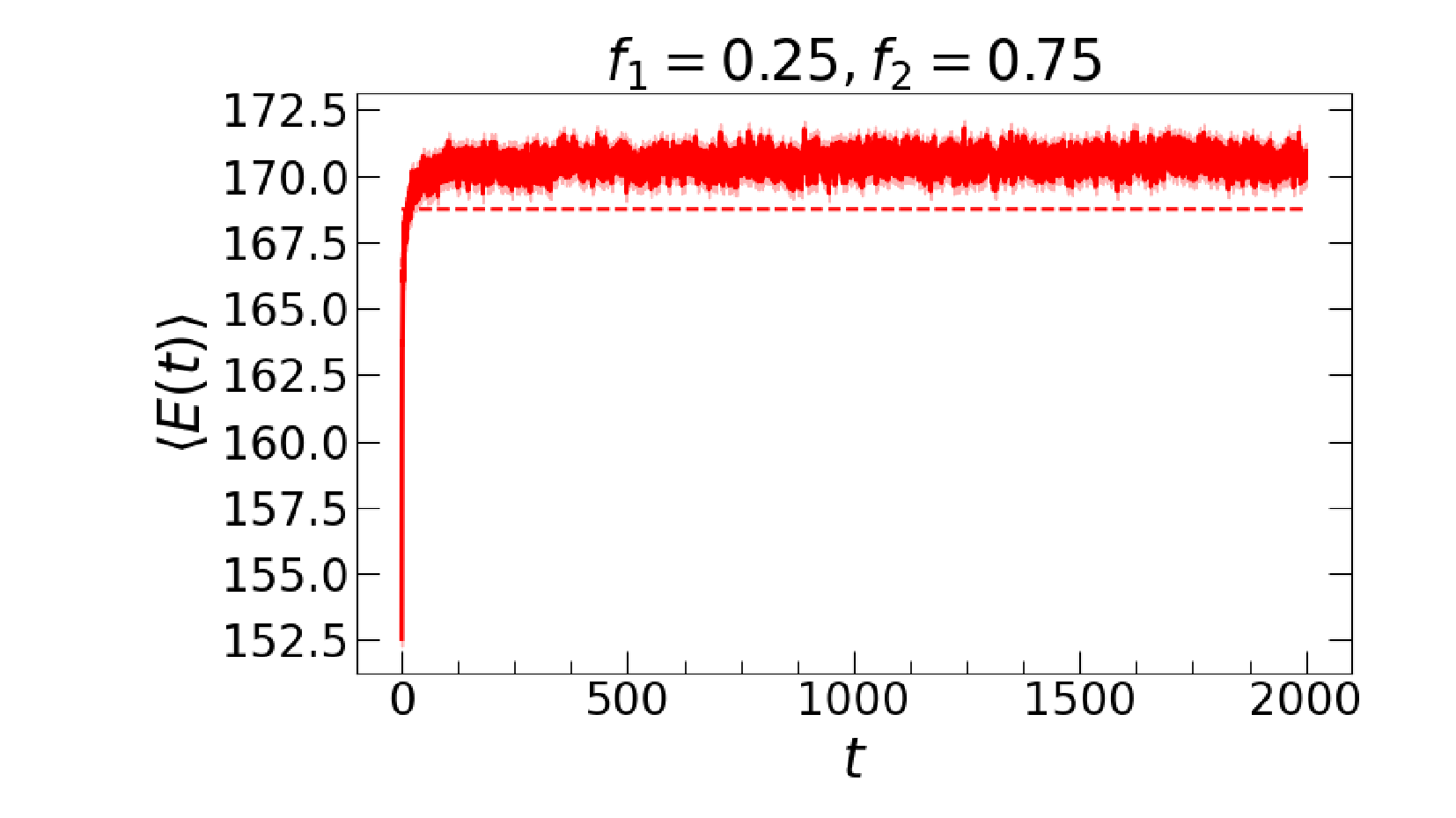}
\caption{Energy
$\langle E(t) \rangle$  is plotted as a function of time $t$ for different partially active polymers. The dashed line represents the analytical result given in Eq. (\ref{energy_ana}), compared with the fluctuating simulation data. 
Other details are consistent with those in Fig. \ref{fig:Rg_sim}.}\label{fig:E_sim}
\end{figure*}

\clearpage 


%

\end{document}